\documentclass{amsbook}

\usepackage{graphics}

\newcommand{\HH}{{\mathcal H}}
\newcommand{\LL}{{\mathcal L}}

\newcommand{\vth}{\vartheta}
\newcommand{\CC}{{\mathbb C}}
\newcommand{\RR}{{\mathbb R}}
\newcommand{\ZZ}{{\mathbb Z}}

\newcommand{\tDl}{\tilde{\Delta}}
\newcommand{\tla}{\tilde{\lambda}}
\newcommand{\vph}{\varphi}

\newcommand{\vq}{\vec{q}}

\newcommand{\e}{\epsilon}
\newcommand{\U}{{\mathcal U}}
\newcommand{\tu}{\tilde{u}}
\newcommand{\tq}{\tilde{q}}
\newcommand{\tQ}{\tilde{Q}}
\newcommand{\vQ}{\vec{Q}}

\renewcommand{\k}{\kappa}
\newcommand{\ga}{\gamma}
\newcommand{\Ga}{\Gamma}

\newcommand{\hk}{\hat{k}}

\newcommand{\hS}{\hat{S}}

\newcommand{\dl}{\delta}
\newcommand{\Dl}{\Delta}
\renewcommand{\th}{\theta}
\newcommand{\ra}{\rightarrow}
\newcommand{\al}{\alpha}
\newcommand{\be}{\beta}
\newcommand{\sg}{\sigma}
\newcommand{\Sg}{\Sigma}

\newcommand{\pa}{\partial}
\newcommand{\z}{\zeta}
\newcommand{\Z}{\ZZ^2/\{0\}}

\newcommand{\hy}{\hat{y}}
\newcommand{\hz}{\hat{z}}

\newcommand{\La}{\Lambda}

\newcommand{\la}{\lambda}
\newcommand{\bq}{\bar{q}}
\newcommand{\bp}{\bar{p}}

\newcommand{\nid}{\noindent}
\newcommand{\F}{{\mathcal F}}
\newcommand{\N}{{\mathcal N}}

\newcommand{\tF}{\tilde{F}}
\newcommand{\om}{\omega}
\newcommand{\Om}{\Omega}
\newcommand{\na}{\nabla}
\newcommand{\lag}{\langle}
\newcommand{\rag}{\rangle}
\newcommand{\tx}{\tilde{x}}

\newcommand{\tz}{\tilde{z}}
\newcommand{\vtQ}{\vec{\tilde{Q}}}

\newcommand{\htau}{\hat{\tau}}
\newcommand{\hrho}{\hat{\rho}}
\newcommand{\hvth}{\hat{\vartheta}}
\newcommand{\tb}{\tilde{b}}
\newcommand{\W}{{\mathcal W}}
\renewcommand{\O}{{\mathcal O}}

\newcommand{\R}{\widehat{R}}

\newcommand{\im}{\mathop{\rm Im}\nolimits}

\newcommand{\BD}{{B\"acklund-Darboux transformations}}

\def\maprightu#1{\smash{
    \mathop{\longrightarrow}\limits^{#1}}}
\def\maprightd#1{\smash{
    \mathop{\longrightarrow}\limits_{#1}}}
\def\mapdownl#1{
    \llap{$\vcenter{\hbox{$\scriptstyle#1$}}$}\Big\downarrow}
\def\mapdownr#1{\Big\downarrow
    \rlap{$\vcenter{\hbox{$\scriptstyle#1$}}$}}

\newtheorem{theorem}{Theorem}[chapter]
\newtheorem{lemma}[theorem]{Lemma}
\newtheorem{corollary}[theorem]{Corollary}
\newtheorem{proposition}[theorem]{Proposition}
\theoremstyle{definition}
\newtheorem{definition}[theorem]{Definition}

\theoremstyle{remark}
\newtheorem{remark}[theorem]{Remark}

\numberwithin{section}{chapter}
\numberwithin{equation}{chapter}

\begin{document}
\frontmatter
\title{Chaos in Partial Differential Equations}

\author{Y. Charles Li}
\address{Department of Mathematics, University of Missouri, 
Columbia, MO 65211}






\maketitle

\setcounter{page}{6}
\tableofcontents

\chapter*{Preface}

The area: Chaos in Partial Differential Equations, is at its fast developing 
stage. Notable results have been obtained in recent years. The present book 
aims at an overall survey on the existing results. On the other hand, we 
shall try to make the presentations introductory, so that beginners can 
benefit more from the book. 

It is well-known that the theory of chaos in finite-dimensional dynamical 
systems has been well-developed. That includes both discrete maps and systems 
of ordinary differential equations. Such a theory has produced important 
mathematical theorems and led to important applications in physics, 
chemistry, biology, and engineering etc.. For a long period of time, 
there was no theory on chaos in partial differential equations. On the other 
hand, the demand for such a theory is 
much stronger than for finite-dimensional systems. Mathematically, 
studies on infinite-dimensional systems pose much more challenging problems. 
For example, as phase spaces, Banach spaces possess much more structures 
than Euclidean spaces. In terms of applications, most of important natural
phenomena are described by partial differential equations -- nonlinear wave 
equations, Maxwell equations, Yang-Mills equations, and Navier-Stokes 
equations, to name a few. Recently, the author and collaborators have 
established a systematic theory on chaos in nonlinear wave equations.

Nonlinear wave equations are the most important class of equations in 
natural sciences. They describe a wide spectrum of phenomena -- motion of 
plasma, nonlinear optics (laser), water waves, vortex motion, to name a 
few. Among these nonlinear wave equations, there is a class of equations 
called soliton equations. This class of equations describes a variety of 
phenomena. In particular, the same soliton equation describes several 
different phenomena. Mathematical theories on soliton equations have been well 
developed. Their Cauchy problems are completely solved through inverse 
scattering transforms. Soliton equations are integrable Hamiltonian 
partial differential equations which are the natural counterparts of 
finite-dimensional integrable Hamiltonian systems. We have established a 
standard program for proving the existence of chaos in perturbed soliton 
equations, with the machineries: 1. Darboux transformations for soliton 
equations, 2. isospectral theory for soliton equations under periodic boundary 
condition, 3. persistence of invariant manifolds and Fenichel fibers,
4. Melnikov analysis, 5. Smale horseshoes and symbolic dynamics, 6. shadowing
lemma and symbolic dynamics.

The most important implication of the theory on chaos in partial 
differential equations in theoretical physics will be on the study of 
turbulence. For that goal, we chose the 2D Navier-Stokes 
equations under periodic boundary conditions to begin a dynamical system 
study on 2D turbulence. Since they possess Lax pair
and Darboux transformation, the 2D Euler equations are the 
starting point for an analytical study. The high Reynolds number 2D 
Navier-Stokes equations are viewed as a singular perturbation of the 2D Euler 
equations through the perturbation parameter $\e = 1/Re$ which is the 
inverse of the Reynolds number.

Our focus will be on nonlinear wave equations. New results on shadowing lemma
and novel results related to Euler equations of inviscid fluids will also be 
presented. The chapters on figure-eight structures and Melnikov vectors 
are written in great details. The readers can learn these machineries 
without resorting to other references. In other chapters, details of proofs 
are often omitted. Chapters 3 to 7 illustrate how to prove the existence of 
chaos in perturbed soliton equations. Chapter 9 contains the most recent 
results on Lax pair structures of Euler equations of inviscid fluids. In 
chapter 12, we give brief comments on other related topics. 

The monograph will be of interest to researchers in mathematics, physics, 
engineering, chemistry, biology, and science in general. Researchers who 
are interested in chaos in high dimensions, will find the book of particularly 
valuable. The book is also accessible to graduate students, and can be 
taken as a textbook for advanced graduate courses. 

I started writing this book in 1997 when I was at MIT. This project continued 
at Institute for Advanced Study during the year 1998-1999, and at University 
of Missouri - Columbia since 1999. In the Fall of 2001, I started to rewrite 
from the old manuscript. Most of the work was done in the summer of 2002.
The work was partially supported by an AMS centennial fellowship in 1998, 
and a Guggenheim fellowship in 1999.

Finally, I would like to thank my wife Sherry and my son Brandon for their 
strong support and appreciation.

\mainmatter
\chapter{General Setup and Concepts}

We are mainly concerned with the Cauchy problems of 
partial differential equations, and view them as defining flows
in certain Banach spaces. Unlike the Euclidean space
$\RR^n$, such Banach spaces admit a variety of norms
which make the structures in infinite dimensional dynamical
systems more abundant. The main difficulty in studying infinite
dimensional dynamical systems often comes from the fact that the
evolution operators for the partial differential equations are
usually at best $C^0$ in time, in contrast to finite
dimensional dynamical systems where the evolution operators are
$C^1$ smooth in time. The well-known concepts for finite
dimensional dynamical systems can be generalized to infinite
dimensional dynamical systems, and this is the main task of this
chapter.

\section{Cauchy Problems of Partial Differential
  Equations}

The types of evolution equations studied in this book can be
casted into the general form,
\begin{equation}
  \partial_t Q = G(Q, \partial_xQ, \dots, \partial^{\ell}_x Q)\ ,
\label{geq}
\end{equation}
where $t \in \RR^1$ (time), $x = (x_1, \dots, x_n) \in
\RR^n$, $Q = (Q_1, \dots, Q_m)$ and $G= (G_1, \dots, G_m)$
are either real or complex valued functions, and
$\ell$, $m$ and $n$ are integers. The equation (\ref{geq}) is
studied under certain boundary conditions, for example, 

\begin{itemize}
\item periodic boundary conditions, e.g. $Q$ is periodic in each
  component of $x$ with period $2\pi $,
\item decay boundary conditions, e.g. $Q \ra 0$ as $x \ra \infty$.
\end{itemize}
Thus we have Cauchy problems for the equation (\ref{geq}), and we
would like to pose the Cauchy problems in some Banach spaces 
$\mathcal{H}$, for example,

\begin{itemize}
\item $\mathcal{H}$ can be a Sobolev space $H^k$,
\item $\mathcal{H}$ can be a Solobev space $H^k_{e,p}$ of even
  periodic functions.
\end{itemize}
We require that the problem is well-posed in $\mathcal{H}$, for
example,

\begin{itemize}
\item for any $Q_0 \in \mathcal{H}$, there exists a unique
solution $Q = Q(t,Q_0) \in C^0[(-\infty, \infty); \mathcal{H}]$
or $C^0[[0; \infty), \mathcal{H}]$ to the equation (\ref{geq}) 
such that $Q(0,Q_0)=Q_0$,
\item for any fixed $t_0 \in (-\infty,\infty)$ or $[0,\infty)$, 
$Q(t_0,Q_0)$ is a $C^r$ function of $Q_0$, for $Q_0 \in \mathcal{H}$ 
and some integer $r \geq 0$.
\end{itemize}

{\bf Example:}
Consider the integrable cubic nonlinear Schr\"{o}dinger (NLS) equation,
\begin{equation}
  iq_t = q_{xx} + 2 \left[\mid q \mid^2 - \omega^2 \right] q\,,
\label{inls}
\end{equation}
where $i=\sqrt{-1}$, $t\in \RR^1$, $x \in \RR^1$, $q$ is a 
complex-valued
function of $(t,x)$, and  $\omega$ is a real constant.
We pose the periodic boundary condition,
\begin{displaymath}
  q(t, x +1) = q(t,x)\,.
\end{displaymath}
The Cauchy problem for equation (\ref{inls}) is posed in the
Sobolev space $H^1$ of periodic functions,
\begin{eqnarray*}
   \mathcal{H}& \equiv& \left\{ Q = (q, \bar{q}) \ \bigg| \ q(x
       + 1) = q(x), \ q \in H^1_{[0,1]}: \hbox{the} \right.\\
&&\qquad  \hbox{Sobolev space $H^1$ over the period interval}\ [0,1]
\bigg\}\, ,
\end{eqnarray*}
and is well-posed \cite{Caz89} \cite{Bou93} \cite{Bou94}.

Fact 1: For any $Q_0 \in \mathcal{H}$, there exists a unique solution $Q
= Q(t,Q_0) \in C^0 [ ( -\infty, \infty), \mathcal{H} ]$ to
the equation (\ref{inls}) such that $Q (0,Q_0) = Q_0$.

Fact 2: For any fixed $t_0 \in ( -\infty, \infty)$, $Q (t_0, Q_0)$ is a
$C^2$ function of $Q_0$, for $Q_0 \in \mathcal{H}$.

\section{Phase Spaces and Flows}

For finite dimensional dynamical systems, the phase spaces are
often $\RR^n$ or $\CC^n$. For infinite dimensional
dynamical systems, we take the Banach space $\mathcal{H}$
discussed in the previous section as the counterpart.

\begin{definition}
We call the Banach space $\mathcal{H}$ in which the Cauchy
problem for (\ref{geq}) is well-posed, a \emph{phase
space}. Define an operator $F^t$ labeled by $t$ as
\begin{displaymath}
  Q(t,Q_0) = F^t (Q_0)\,;
\end{displaymath}
then $F^t:\mathcal{H} \to \mathcal{H}$ is called the
\emph{evolution operator} (or flow) for the system (\ref{geq}).
\end{definition}

A point $p \in \mathcal{H}$ is called a \emph{fixed point} if $F^t (p) =
p$ for any $t$. Notice that here the fixed point $p$ is in fact a
function of $x$, which is the so-called stationary solution of
(\ref{geq}). Let $q \in \mathcal{H}$ be a point; then $\ell_q \equiv \{ F^t
(q), \hbox{for all}\ t \}$ is called the orbit with initial point
$q$. An orbit $\ell_q$ is called a \emph{periodic orbit} if there
exists a $T \in(-\infty, \infty)$ such that $F^T(q) = q$. An
orbit $\ell_q$ is called a \emph{homoclinic orbit} if there
exists a point $q_{\ast} \in \mathcal{H}$ such that $F^t(q) \to q_{\ast}$,
as $\mid t\mid \to \infty$, and $q_{\ast}$ is called the asymptotic
point of the homoclinic orbit. An orbit $\ell_q$ is called a
\emph{heteroclinic orbit} if there exist two different points
$q_\pm \in \mathcal{H}$ such that $F^t(q) \to q_{\pm}$, as $t \to \pm
\infty$, and $q_{\pm}$ are called the asymptotic points of the
heteroclinic orbit. An orbit $\ell_q$ is said to be homoclinic to
a submanifold $W$ of $\mathcal{H}$ if $\inf_{Q\in W} \parallel 
F^t(q) -Q \parallel \to 0$, as $\mid t \mid \to \infty$.

{\bf Example 1:}
Consider the same Cauchy problem for the system (\ref{inls}).
The fixed points of (\ref{inls}) satisfy the second order
ordinary differential equation
\begin{equation}
q_{xx} + 2 \left[\mid q \mid^2 - \omega^2 \right] q = 0\,.
\label{ftq}
\end{equation}
In particular, there exists a circle of fixed points $q = \omega
e^{i\gamma}$, where $\gamma \in [0,2 \pi]$. For simple periodic
solutions, we have
\begin{equation}
  q = ae^{i\theta (t)}, \quad \theta(t) = - \left[2 (a^2 -
    \omega^2) t - \gamma \right]\,;
\label{spsl}
\end{equation}
where $a> 0 $, and $\gamma \in [0, 2\pi]$. For
orbits homoclinic to the circles (\ref{spsl}), we have
\begin{eqnarray}
  q &=& \frac{1}{\Lambda} \bigg [ \cos 2 p - \sin p \ \mbox{sech}\ \tau \cos
    2 \pi x - i \sin 2 p \tanh \tau \bigg ] a e^{i \theta
    (t)}\,,\label{horb} \\ \nonumber \\
&&\quad \Lambda = 1 + \sin p \ \mbox{sech}\ \tau \cos 2 \pi x\,,\nonumber
\end{eqnarray}
where $\tau = 4 \pi \sqrt{a^2 - \pi^2} \ t + \rho$, $p = \arctan
\bigg [ \frac{\sqrt{a^2 - \pi^2}}{\pi} \bigg ]$, $\rho \in (-\infty, \infty)$
is the B\"{a}cklund parameter. Setting $a= \omega$ in
(\ref{horb}), we have heteroclinic orbits asymptotic to points on
the circle of fixed points. The expression
(\ref{horb}) is generated from (\ref{spsl}) through a B\"acklund-Darboux 
transformation \cite{LM94}. 

{\bf Example 2:}
Consider the sine-Gordon equation,
\begin{displaymath}
  u_{tt} - u_{xx} + \sin u = 0\ ,
\end{displaymath}
under the decay boundary condition that $u$ belongs to the
Schwartz class in $x$. The well-known ``breather'' solution,
\begin{equation}
  u(t,x) = 4 \arctan \left[ \frac{\tan \nu \cos [(\cos
      \nu)t]}{\cosh [(\sin \nu) x ]} \right]\,,
\label{br}
\end{equation}
where $\nu$ is a parameter, is a periodic orbit. The expression
(\ref{br}) is generated from trivial solutions through a
B\"{a}cklund-Darboux transformation \cite{EFM90}. 

\section{Invariant Submanifolds}

Invariant submanifolds are the main objects in studying phase
spaces. In phase spaces for partial differential equations,
invariant submanifolds are often submanifolds with
boundaries. Therefore, the following concepts on invariance are
important.

\begin{definition}[Overflowing and Inflowing Invariance]
A submanifold $W$ with boundary $\partial W$ is
\begin{itemize}
\item overflowing invariant if for any $t>0$, $\bar{W} \subset
  F^t \circ W$, where $\bar{W} = W \cup \partial W$,
\item inflowing invariant if any $t>0$, $F^t \circ \bar{W}
  \subset W$,
\item invariant if for any $t>0$, $F^t \circ \bar{W} = \bar{W}$.
\end{itemize}
\end{definition}

\begin{definition}[Local Invariance]
A submanifold $W$ with boundary $\partial W$ is locally invariant
if for any point $q \in W$, if $\bigcup\limits_{t \in [0,\infty)} F^t
(q) \not \subset W$, then there exists $T \in (0,\infty)$ such
that $\bigcup\limits_{t \in [0,T)} F^t (q) \subset W$, and $F^T(q) \in
\partial W$; and if $\bigcup\limits_{t \in (-\infty,0]} F^t(q)
\not \subset W$, then there exists $T \in (-\infty,0)$ such that
$\bigcup\limits_{t\in (T,0]} F^t(q) \subset W$, and $F^T(q) \in \partial
W$.
\end{definition}

Intuitively speaking, a submanifold with boundary is locally
invariant if any orbit starting from a point inside the
submanifold can only leave the submanifold through its boundary
in both forward and backward time.

{\bf Example:}
Consider the linear equation,
\begin{equation}
  iq_t= (1 + i) q_{xx} + iq\,,
\label{liq}
\end{equation}
where $i = \sqrt{-1}$, $t\in \RR^1$, $x \in \RR^1$, and $q$ is a
complex-valued function of $(t,x)$, under periodic boundary
condition,
\begin{displaymath}
  q (x + 1) = q(x)\,.
\end{displaymath}
Let $q = e^{\Omega_j t + ik_jx}$; then
\begin{displaymath}
  \Omega_j = (1-k^2_j) + i\, k^2_j\,,
\end{displaymath}
where $k_j = 2 j \pi$, $(j \in \ZZ )$. $\Omega_0 = 1$, and when
$\mid j\mid >0$, $R_e \{ \Omega_j \} <0$. We take the $H^1$ space
of periodic functions of period 1 to be the phase space. Then the
submanifold
\begin{displaymath}
  W_0 = \left\{ q \in H^1 \ \bigg| \ q = c_0, \,\, c_0 \, \hbox{is
      complex and}\, \parallel q \parallel <1 \right\}
\end{displaymath}
is an outflowing invariant submanifold, the submanifold
\begin{displaymath}
  W_1 = \left\{ q \in H^1 \ \bigg| \ q = c_1 e^{ik_1x}, \ \ 
      c_1 \, \hbox{is complex}, \, \hbox{and}\,
        \parallel q \parallel <1 \right\}
\end{displaymath}
is an inflowing invariant submanifold, and the submanifold
\begin{displaymath}
  W = \left\{ q \in H^1 \ \bigg| \ q = c_0 + c_1 e^{ik_1x}, \ \ c_0 \,
      \hbox{and} \, c_1 \, \hbox{are complex}, \, \hbox{and}\,
        \parallel q \parallel <1 \right\}
\end{displaymath}
is a locally invariant submanifold. The \emph{unstable subspace} is
given by 
\begin{displaymath}
W^{(u)} = \left\{ q \in H^1 \ \bigg| \ q = c_0, \ \ c_0 \, \hbox{is
    complex} \right\}\,,
\end{displaymath}
and the \emph{stable subspace} is given by 
\begin{displaymath}
  W^{(s)} = \left\{ q \in H^1 \ \bigg| \ q= \sum_{j \in Z/\{0\}} c_j
    \, e^{ik_jx}, \ \ c_j\hbox{'s are complex} \right\}\,.
\end{displaymath}
Actually, a good way to view the partial differential equation
(\ref{liq}) as defining an infinite dimensional dynamical system
is through Fourier transform, let
\begin{displaymath}
  q(t,x) = \sum_{j \in Z} c_j (t) e^{ik_jx}\,;
\end{displaymath}
then $c_j (t)$ satisfy
\begin{displaymath}
  \dot{c}_j = \left[ (1-k^2_j) + ik^2_j \right] c_j\,, \quad j
  \in \ZZ\,;
\end{displaymath}
which is a system of infinitely many ordinary differential
equations. 

\section{Poincar\'{e} Sections and Poincar\'{e} Maps}

In the infinite dimensional phase space $\mathcal{H}$, Poincar\'{e} 
sections can
be defined in a similar fashion as in a finite dimensional phase
space. Let $l_q$ be a periodic or homoclinic orbit in $\mathcal{H}$ under
a flow $F^t$, and $q_{\ast}$ be a point on $l_q$, then the
Poincar\'{e} section $\Sigma$ can be defined to be any
codimension 1 subspace which has a transversal intersection with
$l_q$ at $q_{\ast}$. Then the flow $F^t$ will induce a Poincar\'e map 
$P$ in the neighborhood of $q_{\ast}$ in $\Sigma_0$. Phase
blocks, e.g. Smale horseshoes, can be defined using the
norm.

\chapter{Soliton Equations as Integrable Hamiltonian PDEs}

\section{A Brief Summary}

Soliton equations are integrable Hamiltonian partial 
differential equations. For example, the Korteweg-de Vries (KdV) equation
\[
u_t = -6uu_x -u_{xxx}\ ,
\]
where $u$ is a real-valued function of two variables $t$ and $x$, can 
be rewritten in the Hamiltonian form
\[
u_t = \pa_x {\dl H \over \dl u} \ ,
\]
where
\[
H= \int \left [ {1 \over 2} u_x^2 - u^3 \right ] dx \ , 
\]
under either periodic or decay boundary conditions. It is integrable in 
the classical Liouville sense, i.e., there exist enough functionally 
independent constants of motion. These constants of motion can be generated 
through isospectral theory or B\"acklund transformations \cite{AI79}. The 
level sets of these constants of motion are elliptic tori \cite{PT87} 
\cite{MM75} \cite{MT76} \cite{FM76}.

There exist soliton equations which possess level sets which are normally 
hyperbolic, for example, the focusing cubic nonlinear Schr\"odinger equation 
\cite{LM94},
\[
iq_t = q_{xx} + 2 |q|^2q\ ,
\]
where $i =\sqrt{-1}$ and $q$ is a complex-valued function of two variables 
$t$ and $x$; the sine-Gordon equation \cite{MO95}, 
\[
u_{tt} = u_{xx} +\sin u \ ,
\]
where $u$ is a real-valued function of two variables $t$ and $x$, etc.

Hyperbolic foliations are very important since they are the sources of chaos 
when the integrable systems are under perturbations. We will investigate 
the hyperbolic foliations of three typical types of soliton equations: 
(i). (1+1)-dimensional soliton equations represented by the focusing 
cubic nonlinear Schr{\"{o}}dinger equation, (ii). soliton lattices 
represented by the focusing cubic nonlinear Schr{\"{o}}dinger lattice, 
(iii). (1+2)-dimensional soliton equations represented by the 
Davey-Stewartson II equation.
\begin{remark}
For those soliton equations which have only elliptic level sets, the 
corresponding representatives can be chosen to be the KdV equation for 
(1+1)-dimensional soliton equations, the Toda lattice for soliton 
lattices, and the KP equation for (1+2)-dimensional soliton equations.
\end{remark}

Soliton equations are canonical equations which model a variety of physical 
phenomena, for example, nonlinear wave motions, nonlinear optics, plasmas, 
vortex dynamics, etc. \cite{AS81} \cite{AC91}. Other typical examples of such 
integrable Hamiltonian partial differential equations are, e.g., the 
defocusing cubic nonlinear Schr\"odinger equation,
\[
iq_t = q_{xx} - 2 |q|^2q\ ,
\]
where $i =\sqrt{-1}$ and $q$ is a complex-valued function of two variables 
$t$ and $x$; the modified KdV equation,
\[
u_t = \pm 6u^2u_x -u_{xxx}\ ,
\]
where $u$ is a real-valued function of two variables $t$ and $x$; the 
sinh-Gordon equation,
\[
u_{tt} = u_{xx} +\sinh u \ ,
\]
where $u$ is a real-valued function of two variables $t$ and $x$; the 
three-wave interaction equations,
\[
{\pa u_i \over \pa t} + a_i {\pa u_i \over \pa x} = b_i \bar{u}_j 
\bar{u}_k \ ,
\]
where $i,j,k=1,2,3$ are cyclically permuted, $a_i$ and $b_i$ are real 
constants, $u_i$ are complex-valued functions of $t$ and $x$; the 
Boussinesq equation,
\[
u_{tt}-u_{xx}+(u^2)_{xx} \pm u_{xxxx} = 0 \ ,
\]
where $u$ is a real-valued function of two variables $t$ and $x$; 
the Toda lattice,
\[
\pa^2 u_n / \pa t^2 = \exp \left \{ -(u_n-u_{n-1})\right \}- 
\exp \left \{ -(u_{n+1}-u_n)\right \} \ ,
\]
where $u_n$'s are real variables; the focusing cubic nonlinear Schr{\"{o}}dinger lattice,
\[
i{\pa q_n \over \pa t} = (q_{n+1}-2q_n+q_{n-1})+|q_n|^2(q_{n+1}+q_{n-1})\ ,
\]
where $q_n$'s are complex variables; the 
Kadomtsev-Petviashvili (KP) equation,
\[
(u_t +6uu_x +u_{xxx})_x = \pm 3 u_{yy}\ ,
\]
where $u$ is a real-valued function of three variables $t$, $x$ and $y$; 
the Davey-Stewartson II equation,
\[
\left \{ \begin{array}{l} 
i \pa_t q = [\pa_x^2 - \pa_y^2]q + [2|q|^2 + u_y]q\ , \cr \cr 
[\pa_x^2 + \pa_y^2]u = -4 \pa_y |q|^2 \ , \cr
\end{array} \right.
\]
where $i =\sqrt{-1}$, $q$ is a complex-valued function of three 
variables $t$, $x$ and $y$; and $u$ is a real-valued function of three 
variables $t$, $x$ and $y$. For more complete list of soliton equations, 
see e.g. \cite{AS81} \cite{AC91}.

The cubic nonlinear Schr\"odinger equation is one of our main focuses 
in this book, which can be written in the Hamiltonian form,
\[
i q_t = {\dl H \over \dl \bar{q}} \ ,
\]
where
\[
H= \int [-|q_x|^2 \pm |q|^4] dx \ ,
\]
under periodic boundary conditions. Its phase space is defined as
\begin{eqnarray*}
\HH^k &\equiv& \bigg \{ \vq = \left ( \begin{array}{c} q \cr r \cr \end{array} 
\right )\ \bigg | \ r=-\bq, \ q(x+1)=q(x), \\
& &  q \in H^k_{[0,1]}:\ 
\mbox{the Sobolev space}\ H^k \ \mbox{over the period interval} \ 
[0,1] \bigg \}\ .
\end{eqnarray*}
\begin{remark}
It is interesting to notice that the cubic nonlinear Schr\"odinger 
equation can also be written in Hamiltonian form in spatial variable, i.e., 
\[
q_{xx}=iq_t \pm 2 |q|^2 q \ ,
\]
can be written in Hamiltonian form. Let $p=q_x$; then
\[
{\pa \over \pa x}\left ( \begin{array}{c} q \\ \bar{q} \\ \bar{p} 
\\ p \\ \end{array} \right ) = J \left ( \begin{array}{c} 
{\dl H \over \dl q} \\ \\ 
{\dl H \over \dl \bq} \\ \\ {\dl H \over \dl \bp} \\ \\ {\dl H \over \dl p}
\\ \end{array} \right ) \ ,
\]
where
\[
J= \left ( \begin{array}{cccc} 0&0&1&0\\ 0&0&0&1 \\ -1&0&0&0 \\ 0&-1&0&0 
\\ \end{array} \right ) \ ,
\]
\[
H=\int [|p|^2 \mp |q|^4 - {i \over 2} (q_t \bq -\bq_t q)] dt \ ,
\]
under decay or periodic boundary conditions. We do not know whether or not 
other soliton equations have this property.
\end{remark}

\section{A Physical Application of the Nonlinear Schr\"odinger Equation}

The cubic nonlinear Schr\"{o}dinger (NLS) equation has many 
different applications, i.e. it describes many different physical phenomena, 
and that is why it is called a canonical equation. Here, as an example, 
we show how the NLS equation describes the motion of a vortex 
filament -- the beautiful Hasimoto derivation \cite{Has72}. Vortex 
filaments in an inviscid fluid are known to preserve their identities. 
The motion of a very thin isolated vortex filament $\vec{X}=\vec{X}(s,t)$ 
of radius $\e$ in an incompressible inviscid unbounded fluid by its 
own induction is described asymptotically by
\begin{equation}
\pa \vec{X}/\pa t =G \k \vec{b}\ , \label{indu}
\end{equation}
where $s$ is the length measured along the filament, $t$ is the time, 
$\k$ is the curvature, $\vec{b}$ is the unit vector in the direction 
of the binormal and $G$ is the coefficient of local induction,
\[
G={\Ga \over 4\pi}[\ln (1/\e) +O(1)]\ ,
\]
which is proportional to the circulation $\Ga$ of the filament and 
may be regarded as a constant if we neglect the second order term. 
Then a suitable choice of the units of time and length reduces 
(\ref{indu}) to the nondimensional form,
\begin{equation}
\pa \vec{X}/\pa t =\k \vec{b}\ . \label{nindu}
\end{equation}
Equation (\ref{nindu}) should be supplemented by the equations of 
differential geometry (the Frenet-Seret formulae)
\begin{equation}
\pa \vec{X}/\pa s =\vec{t}\ , \ \ \pa \vec{t}/\pa s =\k \vec{n}\ , \ \ 
\pa \vec{n}/\pa s =\tau \vec{b}-\k \vec{t}\ , \ \ 
\pa \vec{b}/\pa s =-\tau \vec{n}\ ,
\label{FSf}
\end{equation}
where $\tau$ is the torsion and $\vec{t}$, $\vec{n}$ and $\vec{b}$ 
are the tangent, the principal normal and the binormal unit vectors. 
The last two equations imply that 
\begin{equation}
\pa (\vec{n}+i \vec{b})/\pa s =-i\tau (\vec{n}+i\vec{b})-\k \vec{t}\ ,
\label{apa1}
\end{equation}
which suggests the introduction of new variables
\begin{equation}
\vec{N} = (\vec{n}+i \vec{b})\exp \bigg \{ i \int_0^s \tau ds \bigg \}\ ,
\label{apa2}
\end{equation}
and 
\begin{equation}
q=\k \exp \bigg \{ i \int_0^s \tau ds \bigg \}\ .
\label{apa3}
\end{equation}
Then from (\ref{FSf}) and (\ref{apa1}), we have 
\begin{equation}
\pa \vec{N}/\pa s =-q \vec{t}\ , \ \ 
\pa \vec{t}/\pa s =\mbox{Re} \{ q \overline{\vec{N}} \} = {1 \over 2} 
(\bq \vec{N} + q \overline{\vec{N}})\ .
\label{apa4}
\end{equation}
We are going to use the relation ${\pa^2 \vec{N} \over \pa s \pa t} = 
{\pa^2 \vec{N} \over \pa t \pa s}$ to derive an equation for $q$. 
For this we need to know $\pa \vec{t}/\pa t$ and $\pa \vec{N}/\pa t$ 
besides equations (\ref{apa4}). From (\ref{nindu}) and (\ref{FSf}), 
we have
\begin{eqnarray*}
& & \pa \vec{t}/\pa t = {\pa^2 \vec{X} \over \pa s \pa t}= \pa 
(\k \vec{b})/\pa s =(\pa \k /\pa s) \vec{b} - \k \tau \vec{n} \\
& & = \k \ \mbox{Re} \{ ({1 \over \k} \pa \k / \pa s +i \tau )
(\vec{b}+i \vec{n}) \}\ ,
\end{eqnarray*}
i.e.
\begin{equation}
\pa \vec{t}/\pa t =\ \mbox{Re} \{ i(\pa q / \pa s)\overline{\vec{N}} \} 
={1 \over 2} i [ (\pa q / \pa s)\overline{\vec{N}}-
(\pa q / \pa s)^-\vec{N}]\ .
\label{apa5}
\end{equation}
We can write the equation for $\pa \vec{N}/\pa t$ in the following form:
\begin{equation}
\pa \vec{N}/\pa t =\al \vec{N}+\be \overline{\vec{N}}+\ga \vec{t}\ , 
\label{apa6}
\end{equation}
where $\al$, $\be$ and $\ga$ are complex coefficients to be determined. 
\begin{eqnarray*}
\al +\bar{\al} &=& {1 \over 2} [\pa \vec{N}/\pa t \cdot \overline{\vec{N}} 
+\pa \overline{\vec{N}}/\pa t \cdot \vec{N}] \\ 
&=& {1 \over 2} \pa (\vec{N}\cdot \overline{\vec{N}})/\pa t = 0\ ,
\end{eqnarray*}
i.e. $\al =i R$ where $R$ is an unknown real function.
\begin{eqnarray*}
& & \be ={1 \over 2}\pa \vec{N}/\pa t \cdot \vec{N} = {1 \over 4}
\pa (\vec{N}\cdot \vec{N})/\pa t = 0\ , \\
& & \ga = -\vec{N} \cdot \pa \vec{t}/\pa t = -i \pa q/\pa s \ .
\end{eqnarray*}
Thus
\begin{equation}
\pa \vec{N}/\pa t =i[R\vec{N}-(\pa q/\pa s)\vec{t}\ ]\ . 
\label{apa7}
\end{equation}
From (\ref{apa4}), (\ref{apa7}) and (\ref{apa5}), we have 
\begin{eqnarray*}
{\pa^2 \vec{N} \over \pa s \pa t} &=& -(\pa q/\pa t) \vec{t} - q 
\pa \vec{t}/\pa t \\
&=& -(\pa q/\pa t) \vec{t} - {1 \over 2} iq [ (\pa q / \pa s)
\overline{\vec{N}}-(\pa q / \pa s)^-\vec{N}] \ , \\
{\pa^2 \vec{N} \over \pa t \pa s} &=& i[(\pa R/ \pa s)\vec{N}-Rq
\vec{t}-(\pa^2 q / \pa s^2)\vec{t} \\
& & -{1 \over 2}(\pa q/ \pa s)(\bq \vec{N}+q \overline{\vec{N}})]\ .
\end{eqnarray*}
Thus, we have 
\begin{equation}
\pa q/\pa t =i[\pa^2 q/\pa s^2 + R q]\ ,
\label{apa8}
\end{equation}
and 
\begin{equation}
{1 \over 2}q \pa \bq/ \pa s = \pa R/ \pa s -{1 \over 2}(\pa q/ \pa s)\bq \ .
\label{apa9}
\end{equation}
The comparison of expressions for ${\pa^2 \vec{t} \over \pa s \pa t}$ 
from (\ref{apa4}) and (\ref{apa5}) leads only to (\ref{apa8}). Solving 
(\ref{apa9}), we have 
\begin{equation}
R={1 \over 2}(|q|^2 +A)\ ,
\label{apa10}
\end{equation}
where $A$ is a real-valued function of $t$ only. Thus we have the 
cubic nonlinear Schr\"{o}dinger equation for $q$:
\[
-i\pa q/\pa t =  \pa^2 q/\pa s^2 +{1 \over 2}(|q|^2 +A) q \ .
\]
The term $Aq$ can be transformed away by defining the new variable 
\[
\tilde{q} = q \exp [ -{1 \over 2} i \int_0^t A(t) dt ] \ .
\]

\chapter{Figure-Eight Structures}

For finite-dimensional Hamiltonian systems, figure-eight structures are often
given by singular level sets. These singular level sets are also called 
separatrices. Expressions for such figure-eight structures can be obtained by 
setting the Hamiltonian and/or other constants of motion to special values. 
For partial differential equations, such an approach is not feasible. 
For soliton equations, expressions for figure-eight structures can be obtained 
via B\"acklund-Darboux transformations \cite{LM94} \cite{Li92} \cite{Li00a}.

\section{1D Cubic Nonlinear Schr\"odinger (NLS) Equation \label{1DCNSE}}

We take the focusing nonlinear Schr\"odinger equation (NLS)
as our first example to show how to construct figure-eight structures.
If one starts from the conservational laws of the NLS, it turns out that 
it is very elusive to get the separatrices. On the contrary, starting 
from the B\"acklund-Darboux transformation to be presented, one can find 
the separatrices rather easily. We consider the NLS
\begin{equation}
iq_t = q_{xx} + 2 |q|^{2} q \ , 
\label{NLS}
\end{equation}
under periodic boundary condition $q(x + 2 \pi ) = q(x)$.
The NLS is an integrable system by virtue of the Lax pair \cite{ZS72},
\begin{eqnarray}
\varphi_x &=& U \varphi \ , \label{Lax1}\\
\varphi_t &=& V \varphi \ , \label{Lax2}
\end{eqnarray}
where
\[
U = i \lambda \sigma_3
\, + \, i \left( \begin{array}{cc}
                       0 & q\\
                      -r & 0
                  \end{array} \right)\ ,
\]
\[
V = \, 2\, i\, \lambda^2 \sigma_3
\,+iqr\sigma_3 + \, 
  \left( \begin{array}{cc}
             0 & 2i\lambda q + q_x\\
             \\
             -2i\lambda r+r_x  & 0 
	\end{array} \right)\ ,
\]
where $\sigma_3$ denotes the third Pauli matrix 
$\sigma_3 = \mbox{diag}(1,-1)$, $r=-\bq$, and $\la$ is the spectral parameter.
If $q$ satisfies the NLS, then the compatibility of the over determined 
system (\ref{Lax1}, \ref{Lax2}) is guaranteed.
Let $M=M(x)$ be the fundamental matrix solution to the ODE (\ref{Lax1}),
$M(0)$ is the $2\times 2$ identity matrix. We introduce the so-called 
transfer matrix $T = T(\lambda, \vec{q})$ where $\vec{q} = (q, -\bq)$,
$T= M(2 \pi)$.
\begin{lemma} Let $Y(x)$ be any solution to the ODE (\ref{Lax1}), then
\[
Y(2n\pi )=T^n \ Y(0)\ .
\]
\end{lemma}
Proof: Since $M(x)$ is the fundamental matrix,
\[
Y(x)=M(x)\ Y(0)\ .
\]
Thus,
\[
Y(2 \pi)=T\ Y(0)\ .
\]
Assume that
\[
Y(2 l\pi)=T^l \ Y(0)\ .
\]
Notice that $Y(x+2 l\pi)$ also solves the ODE (\ref{Lax1}); then
\[
Y(x+2 l \pi)=M(x)\ Y(2 l\pi)\ ;
\]
thus,
\[
Y(2 (l+1) \pi)=T\ Y(2 l\pi )=T^{l+1}\ Y(0)\ .
\]
The lemma is proved. Q.E.D.

\begin{definition}
We define the Floquet discriminant $\Dl$ as,
\[
\Delta(\lambda, \vec{q}) =\ \mbox{trace}\ \{ T(\lambda, \vec{q}) \}\ .
\]
We define the periodic and anti-periodic points
$\la^{(p)}$ by the condition
\[
|\Delta(\lambda^{(p)}, \vq)| = 2\ .
\]
We define the critical points $\la^{(c)}$ by the condition
\[
{\pa \Delta(\lambda, \vq) \over \pa \lambda} \bigg|_{\la =\la^{(c)}} = 0\ .
\]
A multiple point, denoted $\lambda^{(m)}$, is a critical 
point for which 
\[
|\Delta(\lambda^{(m)}, \vq)| = 2.
\]
The algebraic multiplicity of $\lambda^{(m)}$ is defined as the order
of the zero of $\Delta(\lambda) \pm 2$.  Ususally it is 2, but it can exceed
2;  when it does equal 2, we call the multiple point a double point, and
denote it by $\lambda^{(d)}$.
The geometric multiplicity of $\lambda^{(m)}$ is defined as the
maximum number of linearly independent solutions to the ODE (\ref{Lax1}),
and is either 1 or 2.
\end{definition}
Let $q(x,t)$ be a solution to the NLS (\ref{NLS}) for which the linear 
system (\ref{Lax1}) has a complex double point $\nu$ of geometric 
multiplicity 2. We denote  two linearly independent solutions of the 
Lax pair (\ref{Lax1},\ref{Lax2}) at $\lambda = \nu$ by 
$(\phi^+, \phi^-)$. Thus, a general solution of the linear systems 
at $(q,\nu)$ is given by 
\begin{equation} 
\phi (x,t) = c_+ \phi^+\ +\ c_- \phi^- \ .
\label{4.2}
\end{equation}
We use $\phi$ to define a Gauge matrix \cite{SZ87} $G$ by
\begin{equation} 
 G  = G (\lambda ; \nu ; \phi ) = N
\left( \begin{array}{cl}
 \lambda-\nu  & \quad	0\\
0 &  \lambda - \bar{\nu}
 \end{array}\right)
N^{-1}\ ,
\label{4.3}
\end{equation}
where
\begin{equation}
N = \left ( \begin{array}{lr}
 \phi_1  & -{\bar{\phi}}_2 \\
\phi_2 &\ \  {\bar{\phi}}_1
 \end{array} \right )\ .
\label{4.4}
\end{equation}
Then we define $Q$ and $\Psi$ by
\begin{equation}
 Q(x,t) = q(x,t) \ + \ 
2 (\nu-{\bar{\nu}})\ 
\frac{\phi_1{\bar{\phi}}_2}{\phi_1 {\bar{\phi}}_1+ 
\phi_2{\bar{\phi}}_2}
\label{4.5}
\end{equation}
and
\begin{equation}
\Psi (x,t; \lambda) \ = \   G(\lambda; \nu ; \phi ) \ \psi (x,t; \lambda)
\label{4.6}
\end{equation}
where $\psi$ solves the Lax pair (\ref{Lax1},\ref{Lax2}) at 
$(q,\nu)$.  
Formulas (\ref{4.5}) and (\ref{4.6}) are the B\"acklund-Darboux 
transformations for the 
potential and eigenfunctions, respectively.  We have the 
following \cite{SZ87} \cite{LM94},
\begin{theorem}
Let $q(x,t)$ be a solution to the NLS equation (\ref{NLS}), 
for which the linear system (\ref{Lax1}) has 
a complex double point $\nu$ of geometric multiplicity 2, 
with eigenbasis $(\phi^+, \phi^-)$ for the Lax pair
(\ref{Lax1},\ref{Lax2}), and define 
$Q(x,t)$ and $\Psi (x,t;\lambda)$ by 
(\ref{4.5}) and (\ref{4.6}).  Then
\begin{enumerate}
\item $Q(x,t)$ is an  solution of NLS, with spatial period $2\pi$,
\item $Q$ and $q$ have the same Floquet spectrum,
\item $Q(x,t)$ is homoclinic to $q(x,t)$ in the
sense that $Q(x,t) \longrightarrow q_{\theta_\pm}(x,t)$,
expontentially as $\exp (-\sigma_\nu|t|)$, as $t
\longrightarrow \pm \infty$, where $q_{\theta_\pm}$ is a
``torus translate'' of $q, \sigma_\nu$ is the nonvanishing growth 
rate associated to the complex double point $\nu$, and explicit 
formulas exist for this growth rate and for the translation
parameters $\theta_\pm$,
\item $\Psi (x,t;\lambda)$ solves the
Lax pair (\ref{Lax1},\ref{Lax2}) at $(Q, \lambda)$.
\end{enumerate}
\label{theorem 4}
\end{theorem}
This theorem is quite general, constructing homoclinic solutions 
from a wide class of starting solutions $q(x,t)$. It's proof is 
one of direct verification \cite{Li92}. 

We emphasize several qualitative features of these homoclinic orbits:  
(i) $Q(x,t)$ is homoclinic to a torus which itself
possesses rather complicated spatial and temporal structure, and is 
not just a fixed point.  (ii)  Nevertheless, the homoclinic orbit 
typically has still more complicated spatial structure than its 
``target torus''.  (iii) When there are several complex double points, each 
with nonvanishing growth rate, one can iterate the B\"acklund-Darboux 
transformations to generate more complicated homoclinic orbits.  
(iv) The number of complex double points with nonvanishing growth 
rates counts the dimension of the unstable manifold of the critical 
torus in that  two unstable directions are coordinatized by the 
complex ratio $c_+/c_-$. Under even symmetry only one real 
dimension satisfies the constraint of evenness, as will be clearly 
illustrated in the following example.  (v)  These B\"acklund-Darboux formulas 
provide global expressions for the stable and unstable manifolds of the 
critical tori, which represent figure-eight structures.

{\bf Example:} As a concrete example, we take $q(x,t)$ to be the special 
solution
\begin{equation} 
q_c = c \exp \left \{ -i [2c^2 t + \gamma ] \right \}\ .
\label{4.7}
\end{equation}
Solutions of the Lax pair (\ref{Lax1},\ref{Lax2}) can 
be computed explicitly:
\begin{equation} 
\phi^{(\pm)}(x,t;\lambda)= e^{\pm i\kappa(x + 2 \lambda t)}
\left( \begin{array}{cl}
 c e^{-i(2 c^2 t +\gamma)/2} \\
(\pm\kappa - \lambda)e^{i(2 c^2 t+\gamma)/2}
 \end{array}\right)  \ ,
\label{4.8}
\end{equation}
where
\[
\kappa = \kappa (\lambda) = \sqrt{c^2 +
\lambda^2}\ . 
\]
With these solutions one can construct the fundamental matrix
\begin{equation} 
M(x;\lambda ; q_c) = 
\Bigg [ \begin{array}{cl}
\cos \kappa x+ i \frac \lambda \kappa \sin \kappa x &\ 
i \frac {q_c}{\kappa} \sin \kappa x \\
i \frac {\overline{q_c}}{\kappa }\sin \kappa x &
\cos \kappa x - i \frac \lambda \kappa \sin \kappa x
\end{array}\Bigg ]\ ,
\label{4.9}
\end{equation}
from which the Floquet discriminant can be computed:
\begin{equation} 
\Delta (\lambda ; q_c)  = 2 \cos (2 \kappa \pi)\ . 
\label{4.10}
\end{equation}
From $\Delta$,  spectral quantities can be 
computed:
\begin{enumerate}
	\item simple periodic points: $\lambda^\pm
	= \pm i \ c\ , $ 
	\item double points: 
	$\kappa(\lambda^{(d)}_j)
	= j/2\ , \ \  j \in \ZZ \ ,\ 
	j \neq 0 \ ,$
	\item critical points:  
	$\lambda^{(c)}_j = \lambda^{(d)}_j\ ,\quad j \in 
	\ZZ\ ,\  \  j \neq 0 \ ,$
	\item simple periodic points: $\lambda^{(c)}_0 = 0\ $.
\end{enumerate}
For this spectral data, there are 2N  purely 
imaginary 
double points,
\begin{equation} 
  (\lambda^{(d)}_j)^2  = j^2/4 - c^2,  \ \
j =1, 2, \cdots , N;
\label{4.11}
\end{equation}
where
\[
\bigg [N^2/4 - c^2 \bigg ] < 0 <
\bigg [(N+1)^2/4 - c^2 \bigg ] \ .
\]
From this spectral data, the
homoclinic orbits can be explicitly computed through B\"acklund-Darboux 
transformation.  
Notice that to have temporal growth (and decay) in the eigenfunctions 
(\ref{4.8}), one needs $\la$ to be complex. Notice also that the 
B\"acklund-Darboux transformation is built with quadratic products in 
$\phi$, thus choosing $\nu = \la_j^{(d)}$ will guarantee periodicity of 
$Q$ in $x$.
When $N=1$, the B\"acklund-Darboux 
transformation at one purely imaginary double point $\la_1^{(d)}$ yields $Q = 
Q(x, t; c, \gamma; c_+/c_-)$ \cite{LM94}:
\begin{eqnarray} 
Q  &=& \bigg [ \cos 2p - \sin  p \ \  \mbox{sech} \tau \ \cos (x +\vth)- 
i \sin 2p \tanh \tau \bigg ] \nonumber \\
& & \bigg [1 + \sin p \ \  \mbox{sech} \tau 
\ \cos (x +\vth)\bigg ]^{-1} ce^{-i(2 c^2 t+\gamma)} \label{4.13} \\ 
& & \ra e^{\mp 2ip}ce^{-i(2 c^2 t + \gamma)}\quad
\mbox{ as } \ \rho \ra \mp \infty ,\nonumber
\end{eqnarray}
where $c_+/c_- \equiv \exp(\rho + i\beta)$ and $p$ is defined by $1/2 + 
i\sqrt{c^2 - 1/4} = c \exp(ip)$, $\tau \equiv \sigma t -\rho$, and 
$\vth = p - (\beta +\pi/2)$.

\begin{figure}
\includegraphics{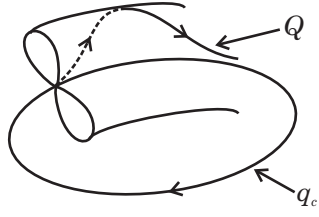}
\caption{An illustration of the figure-eight structure.}
\label{1fig8}
\end{figure}

Several points about this homoclinic orbit need to be made:
\begin{enumerate}
\item The orbit depends 
only upon the ratio $c_+/c_-$, and not upon $c_+$ and $c_-$ 
individually. 

\item $Q$ is homoclinic to the plane wave 
orbit; however,  a phase 
shift of $-4p$ occurs when one compares the asymptotic behavior of 
the orbit as $t \rightarrow \ - \ \infty$ with its behavior as 
$t\rightarrow \ +\ \infty$.  

\item For small p, the formula 
for $Q$ becomes more transparent:
\[ Q \simeq\bigg [ (\cos 2p - i\, 
\sin 2p\, \tanh \tau) - 2 \sin\,
p \ \mbox{sech}\ \tau \cos
(x +\vth)\bigg ] ce^{-i(2 c^2 t + \gamma)}. \]

\item An evenness constraint on $Q$ in $x$ can be  enforced by 
restricting the phase $\phi$ to be one of two values 
\[
\phi  =  0, \pi . \hskip 1truein \mbox{(evenness)} 
\]
In this manner, the even symmetry disconnects the level set. 
Each component constitutes one loop of the figure eight.  While the target 
q is independent of 
$x$, each of these loops has $x$ dependence through the $\cos(x)$.  
One loop has exactly this dependence and can be interpreted as a 
spatial 
excitation located near $x=0$, while the second loop has the 
dependence $\cos (x - \pi )$, which we interpret as spatial 
structure located near $ x = \pi $.  In this example, the disconnected 
nature of the level set is clearly related to distinct spatial 
structures on the individual loops. See Figure \ref{1fig8} for
an illustration.
\item Direct calculation shows that the transformation matrix 
$M(1;\la_1^{(d)};Q)$ is similar to a Jordan form when $t \in (-\infty,\infty)$,
\[
M(1;\la_1^{(d)};Q) \sim \left ( \begin{array}{lr} -1 & 1 \cr 0 & -1 \cr 
\end{array} \right )\ ,
\]
and when $t \ra \pm \infty$, $M(1;\la_1^{(d)};Q) \longrightarrow -I$ (the 
negative of the 2x2 identity matrix). Thus, when $t$ is finite, the 
algebraic multiplicity ($=2$) of $\la =\la_1^{(d)}$ with the potential $Q$ 
is greater than the geometric multiplicity ($=1$).
\end{enumerate} 

In this example the dimension of the loops need not 
be one, but is determined by the number of purely imaginary double 
points which in turn is controlled by the amplitude $c$ of the plane 
wave target  and by the spatial period.  (The dimension of the 
loops increases linearly with the spatial period.) When there are 
several complex double points,  B\"acklund-Darboux transformations must 
be iterated to produce complete
representations. Thus, B\"acklund-Darboux transformations give 
global representations of the figure-eight structures.

\subsection{Linear Instability}

The above figure-eight structure corresponds to the following linear 
instability of Benjamin-Feir type. Consider the uniform solution to the 
NLS (\ref{NLS}),
\[
q_c = c e^{i \th(t)}\ , \ \ \ \ \th(t)=-[2c^2t +\ga ] \ .
\]
Let 
\[
q= [c + \tq ] e^{i \th(t)}\ ,
\]
and linearize equation (\ref{NLS}) at $q_c$, we have 
\[
i \tq_t = \tq_{xx} + 2 c^2 [ \tq + \bar{\tq}]\ .
\]
Assume that $\tq$ takes the form,
\[
\tq = \bigg [ A_j e^{\Om_j t} + B_j e^{\bar{\Om}_j t} \bigg ] \cos k_j x \ ,
\]
where $k_j = 2 j \pi$, ($j=0,1,2, \cdot \cdot \cdot$), $A_j$ and $B_j$ 
are complex constants. Then,
\[
\Om_j^{(\pm)} = \pm k_j \sqrt{4c^2 -k_j^2}\ .
\]
Thus, we have instabilities when $c > 1/2$.

\subsection{Quadratic Products of Eigenfunctions}

Quadratic products of eigenfunctions play a crucial role in 
characterizing the hyperbolic structures of soliton equations. 
Its importance lies in the following aspects: (i). Certain 
quadratic products of eigenfunctions solve the linearized soliton 
equation. (ii). Thus, they are the perfect candidates for building 
a basis to the invariant linear subbundles. (iii). Also, they 
signify the instability of the soliton equation. (iv). Most 
importantly, quadratic products of eigenfunctions can serve as 
Melnikov vectors, e.g., for Davey-Stewartson equation \cite{Li00a}.

Consider the linearized NLS equation at any solution $q(t,x)$ 
written in the vector form:
\begin{eqnarray}
i\pa_t (\dl q) &=& (\dl q)_{xx} +2 [ q^2\overline{\dl q} + 
2 |q|^2 \dl q ]\ ,
\nonumber \\
\label{lNLS} \\
i\pa_t (\overline{\dl q}) &=& -(\overline{\dl q})_{xx} -2 
[ \bq^2\dl q + 2 |q|^2 \overline{\dl q} ]\ ,\nonumber 
\end{eqnarray}
we have the following lemma \cite{LM94}.
\begin{lemma}
Let $\vph^{(j)} = \vph^{(j)}(t,x;\la,q)$ ($j=1,2$) be any two 
eigenfunctions solving the Lax pair
(\ref{Lax1},\ref{Lax2}) at an arbitrary $\la$. Then 
\[
\left ( \begin{array}{c} \dl q \cr \cr \overline{\dl q} \cr 
\end{array} \right )\ , \ \ 
\left ( \begin{array}{c} \vph_1^{(1)} \vph_1^{(2)} \cr \cr 
\vph_2^{(1)} \vph_2^{(2)}\cr \end{array} \right )\ , \ \ 
\mbox{and} \ 
S\left ( \begin{array}{c} \vph_1^{(1)} \vph_1^{(2)} \cr \cr  
\vph_2^{(1)} \vph_2^{(2)}\cr \end{array} \right )^{-}\ ,\ 
\ \mbox{where}\ 
S=\left ( \begin{array}{lr} 0&1 \cr 1&0 \cr \end{array} \right )
\]
solve the same equation (\ref{lNLS}); thus
\[
\Phi = \left ( \begin{array}{c} \vph_1^{(1)} \vph_1^{(2)} \cr \cr
\vph_2^{(1)} \vph_2^{(2)}\cr \end{array} \right ) + 
S\left ( \begin{array}{c} \vph_1^{(1)} \vph_1^{(2)} \cr \cr 
\vph_2^{(1)} \vph_2^{(2)}\cr \end{array} \right )^{-}
\]
solves the equation (\ref{lNLS}) and satisfies the reality 
condition $\Phi_2 = \bar{\Phi}_1$.
\end{lemma}

Proof: Direct calculation leads to the conclusion. Q.E.D.

The periodicity condition $\Phi(x+2\pi ) = \Phi(x)$ can be easily 
accomplished. For example, we can take $\vph^{(j)}$ ($j=1,2$) 
to be two linearly independent Bloch functions $\vph^{(j)} = 
e^{\sg_j x} \psi^{(j)}$ ($j=1,2$), where $\sg_2=-\sg_1$ and 
$\psi^{(j)}$ are periodic functions $\psi^{(j)}(x+2\pi )= 
\psi^{(j)}(x)$. Often we choose $\la$ to be a double point 
of geometric multiplicity $2$, so that $\vph^{(j)}$ are already 
periodic or antiperiodic functions.

\section{Discrete Cubic Nonlinear Schr\"odinger Equation \label{DCNSE}}

Consider the discrete focusing cubic nonlinear Schr\"odinger equation
(DNLS)
\begin{equation}
i \dot{q}_n = {1 \over h^2}[q_{n+1}-2 q_n +q_{n-1}] + |q_n|^2(q_{n+1}+
q_{n-1})
-2 \om^2 q_n \ , \label{DNLS}
\end{equation}
under periodic and even boundary conditions,
\[
q_{n+N}=q_n\ , \ \ \ \ q_{-n}=q_n\ ,
\]
where $i = \sqrt{-1}$, $q_n$'s are complex variables, $n \in \ZZ$, $\om$ 
is a positive parameter, $h=1/N$, and $N$ is a positive integer 
$N \geq 3$. The DNLS is integrable by virtue of the Lax pair
\cite{AL76}:
\begin{eqnarray}
\varphi_{n+1}&=&L^{(z)}_n\varphi_n\ , \label{dLax1} \\
\dot{\varphi}_n&=&B^{(z)}_n\varphi_n\ , \label{dLax2}
\end{eqnarray}
\noindent
where
\begin{eqnarray*}
L^{(z)}_n&=&\left( \begin{array}{cc} 
                                            z& ihq_n \cr
                                            ih\bq_n & 1/z \cr
                                       \end{array} \right),
\\
\\
B^{(z)}_n&=&{i\over h^2}\left( \begin{array}{cc}
  b_n^{(1)} & -izhq_n+(1/z)ihq_{n-1}  \\
  -izh\bq_{n-1}+(1/z)ih\bq_n& b_n^{(4)}
                                 \end{array} \right), \\
& & b_n^{(1)} = 1-z^2+2i\la h-h^2q_n\bq_{n-1}+\om^2 h^2 ,\\
& & b_n^{(4)} = 1/z^2-1+2i\la h+h^2\bq_nq_{n-1}-\om^2 h^2,
\end{eqnarray*}
\noindent
and where $z = \exp(i\la h)$. Compatibility of the over-determined 
system (\ref{dLax1},\ref{dLax2}) gives the ``Lax representation'' 
\[
\dot{L}_n=B_{n+1}L_n-L_nB_n
\]
of the DNLS (\ref{DNLS}). Let $M(n)$ be the fundamental matrix solution 
to (\ref{dLax1}), the {\em Floquet discriminant} is defined as 
\[
\Dl = \ \mbox{trace}\ \{ M(N) \} \ .
\]
Let $\psi^+$ and $\psi^-$ be any two solutions to (\ref{dLax1}), and 
let $W_n(\psi^+,\psi^-)$ be the Wronskian
\[
W_n(\psi^+,\psi^-) = \psi_n^{(+,1)}\psi_n^{(-,2)}-\psi_n^{(+,2)}
\psi_n^{(-,1)} \ .
\]
One has 
\[
W_{n+1}(\psi^+,\psi^-)=\rho_n W_n(\psi^+,\psi^-)\ ,
\]
where $\rho_n = 1 +h^2|q_n|^2$, and 
\[
W_N(\psi^+,\psi^-)=D^2 \ W_0(\psi^+,\psi^-)\ , 
\]
where $D^2 = \prod^{N-1}_{n=0}\rho_n$.
Periodic and antiperiodic points $z^{(p)}$ are defined by
\[
\Delta(z^{(p)})=\pm 2D\ .
\]
A critical point $z^{(c)}$ is defined by the condition
$$
            {d\Delta \over dz}\bigg |_{z=z^{(c)}} = 0.
$$
\noindent
A multiple point $z^{(m)}$ is a critical point which is also a periodic 
or antiperiodic point. The {\em algebraic multiplicity} of 
$z^{(m)}$ is defined as the order of the zero of $\Delta(z)\pm 2D$. Usually
it is $2$, but it can exceed $2$; when it does equal $2$, we call the 
multiple point a {\em double point}, and denote it by $z^{(d)}$. The 
{\em geometric multiplicity} of $z^{(m)}$ is defined as the dimension of 
the periodic (or antiperiodic) eigenspace of (\ref{dLax1}) at $z^{(m)}$, 
and is either $1$ or $2$. 

Fix a solution $q_n(t)$ of the DNLS (\ref{DNLS}), for which (\ref{dLax1})
has a double point $z^{(d)}$ of geometric multiplicity 2, which is not 
on the unit circle.
We denote two linearly independent solutions (Bloch functions) of the 
discrete Lax pair (\ref{dLax1},\ref{dLax2}) at $z=z^{(d)}$ by $(\phi_n^+,\phi_n^-)$. Thus, a general solution of the discrete Lax pair 
(\ref{dLax1};\ref{dLax2}) at $(q_n(t),z^{(d)})$ is given by
\[
\phi_n = c^+ \phi_n^+ + c^- \phi_n^-,
\]
\nid
where $c^+$ and $c^-$ are complex parameters. We use $\phi_n$ 
to define a transformation matrix $\Ga_n$ by
\[
\Ga_n=\left(\begin{array}{cc} z+(1/z)a_n & b_n \cr c_n &-1/z+z d_n \cr
            \end{array} \right),
\]
\nid
where,
\begin{eqnarray*}
a_n &=& {z^{(d)} \over (\bar{z}^{(d)})^2\Dl_n}\bigg [|\phi_{n2}|^2+|z^{(d)}|^2|\phi_{n1}|^2
    \bigg ],\\
d_n &=& -{1 \over z^{(d)}\Dl_n}\bigg [|\phi_{n2}|^2+|z^{(d)}|^2|\phi_{n1}|^2
    \bigg ],\\
b_n &=& {|z^{(d)}|^4-1 \over (\bar{z}^{(d)})^2\Dl_n}\phi_{n1}\bar{\phi}_{n2}, \\
c_n &=& {|z^{(d)}|^4-1 \over z^{(d)}\bar{z}^{(d)}\Dl_n}\bar{\phi}_{n1}\phi_{n2}, \\
\Dl_n &=& -{1 \over \bar{z}^{(d)}}\bigg [|\phi_{n1}|^2+|z^{(d)}|^2|\phi_{n2}|^2
    \bigg ].
\end{eqnarray*}
\nid
From these formulae, we see that
\[
\bar{a}_n=-d_n,\ \ \bar{b}_n=c_n.
\]
\nid
Then we define $Q_n$ and $\Psi_n$ by
\begin{equation}
Q_n\equiv {i\over h}b_{n+1}-a_{n+1}q_n
\label{BD1}
\end{equation}
\nid
and
\begin{equation}
\Psi_n(t;z)\equiv \Ga_n(z;z^{(d)};\phi_n)\psi_n(t;z)
\label{BD2}
\end{equation}
\nid
where $\psi_n$ solves the discrete Lax pair (\ref{dLax1},\ref{dLax2}) 
at $(q_n(t),z)$. Formulas (\ref{BD1}) and (\ref{BD2}) are the 
B\"acklund-Darboux transformations for the potential and eigenfunctions, 
respectively. We have the following theorem \cite{Li92}.
\begin{theorem}
Let $q_n(t)$ denote a solution of the DNLS (\ref{DNLS}), for which 
(\ref{dLax1}) has a double point $z^{(d)}$ of geometric multiplicity 2, 
which is not on the unit circle. We denote two linearly independent 
solutions of the discrete Lax pair (\ref{dLax1},\ref{dLax2}) 
at $(q_n, z^{(d)})$ by $(\phi_n^+,\phi_n^-)$. We define $Q_n(t)$ and 
$\Psi_n(t;z)$ by (\ref{BD1}) and (\ref{BD2}). Then
\begin{enumerate}
\item $Q_n(t)$ is also a solution of the DNLS (\ref{DNLS}). (The eveness
of $Q_n$ can be obtained by choosing the complex B\"acklund parameter 
$c^+/c^-$ to lie on a certain curve, as shown in the example below.)
\item $\Psi_n(t;z)$ solves the discrete Lax pair (\ref{dLax1},\ref{dLax2}) 
at $(Q_n(t),z)$.
\item $\Dl(z;Q_n)=\Dl(z;q_n)$, for all $z\in C$.
\item $Q_n(t)$ is homoclinic to $q_n(t)$ in the sense that $Q_n(t) \ra 
e^{i\th_{\pm}}\ q_n(t)$, exponentially as $\exp (-\sg |t|)$ as $t \ra 
\pm \infty$. Here $\th_{\pm}$ are the phase shifts, $\sg$ is a 
nonvanishing growth rate associated to the double point $z^{(d)}$, and 
explicit formulas can be developed for this growth rate and 
for the phase shifts $\th_{\pm}$.
\end{enumerate}
\label{Backlund}
\end{theorem}

\nid
{\bf Example:} We start with the uniform solution of (\ref{DNLS})
\begin{equation}
q_n=q_c \ ,\ \forall n; \ \ \ \ q_c=a\exp \bigg \{-i[2(a^2-\om^2)t - \ga]
\bigg \}\ . 
\label{ucsl}
\end{equation}
We choose the amplitude $a$ in the range
\begin{eqnarray}
& & N\tan{\pi \over N}< a <N\tan{2\pi \over N}\ ,\ \ \ \mbox{when}\ N>3 \ ,
\nonumber \\ \label{constr} \\
& & 3\tan{\pi \over 3}< a < \infty\ ,\ \ \ \mbox{when}\ N=3 \ ;\nonumber
\end{eqnarray}
so that there is only one set of quadruplets of double points 
which are not on the unit circle, and denote one of them by $z=z_1^{(d)}=
z_1^{(c)}$ which corresponds to $\be = \pi / N$. The homoclinic orbit 
$Q_n$ is given by
\begin{equation}
Q_n = q_c (\hat{E}_{n+1})^{-1} \bigg [ \hat{A}_{n+1} - 2 \cos \be 
\sqrt{\rho \cos^2 \be -1}\hat{B}_{n+1} \bigg ]\ ,
\label{hetorb}
\end{equation}
where
\[
\hat{E}_n = ha\cos \be +\sqrt{\rho \cos^2 \be - 1} \ \mbox{sech}\ [
2\mu t +2p] \cos [(2n-1)\be +\vth ]\ ,
\]
\[
\hat{A}_{n+1} = ha\cos \be +\sqrt{\rho \cos^2 \be - 1} \ \mbox{sech}\ [
2\mu t +2p] \cos [(2n+3)\be +\vth ]\ ,
\]
\[
\hat{B}_{n+1} = \cos \varphi + i \sin \varphi \tanh [ 2\mu t +2p]
+ \ \mbox{sech}\ [ 2\mu t +2p] \cos [ 2(n+1)\be +\vth ]\ ,
\]
\[
\be = \pi / N\ , \ \ \rho = 1+h^2 a^2\ , \ \ \mu = 2h^{-2} \sqrt{\rho} 
\sin \be \sqrt{\rho \cos^2 \be -1}\ ,
\]
\[
h=1/N,\ \ c_+/c_- = i e^{2p} e^{i\vth}\ , \ \ \vth \in [0,2\pi]\ , \ \ 
p \in (-\infty, \infty)\ ,
\]
\[
z_1^{(d)}=\sqrt{\rho}\cos \be +\sqrt{\rho \cos^2 \be -1}\ , \ \ 
\th(t)=(a^2-\om^2)t - \ga/2\ ,
\]
\[
\sqrt{\rho \cos^2 \be -1} + i \sqrt{\rho} \sin \be = ha e^{i \varphi}\ ,
\]
where $\varphi=\sin^{-1} [\sqrt{\rho}(ha)^{-1} \sin \be ],
\ \ \varphi \in (0, \pi/2)$.

Next we study the ``evenness'' condition: $Q_{-n} = Q_n$. It turns out 
that the choices $\vth = - \be\ , \ - \be +\pi$ in the formula of 
$Q_n$ lead to the evenness of $Q_n$ in $n$. In terms of figure eight 
structure of $Q_n$, $\vth = - \be$ corresponds to one ear of the 
figure eight, and $\vth = - \be +\pi$ corresponds to the other 
ear. The even formula for $Q_n$ is given by,
\begin{equation}
Q_n = q_c \bigg [ \Ga / \La_n -1 \bigg ]\ , \label{ehetorb}
\end{equation}
where
\[
\Ga = 1-\cos 2 \varphi - i \sin 2 \varphi \tanh [ 2 \mu t + 2p]\ ,
\]
\[
\La_n = 1 \pm \cos \varphi [\cos \be ]^{-1}\ \mbox{sech}[2 \mu t + 2p] 
\cos [2n\be]\ ,
\]
where (`+' corresponds to $\vth = -\be$).

The heteroclinic orbit (\ref{ehetorb}) represents the figure eight 
structure. If we denote by $S$ the circle, we have the topological 
identification:
\[
\mbox{(figure 8)}\ \otimes S = \bigcup_{p \in (-\infty,\infty),\ 
\ga \in [0,2\pi]} Q_n(p, \ga, a, \om, \pm, N)\ .
\]

\section{Davey-Stewartson II (DSII) Equations}

Consider the Davey-Stewartson II equations (DSII),
\begin{equation}
  \left \{ \begin{array}{l}
i \partial_t q = [ \partial^2_x - \partial^2_y]q+ [ 2(
|q|^2 - \omega^2) + u_y ] q \, , \cr \cr 
[\partial^2_x + \partial^2_y] u = -4 \partial_y |q|^2 \, , \cr    
    \end{array} \right.
\label{fDS2}
\end{equation}
where $q$ and $u$ are respectively complex-valued and
real-valued functions of three variables $(t,x,y)$, and $\om$ is a
positive constant. We pose periodic boundary conditions,
\begin{eqnarray*}
& &  q(t,x + L_1,y) = q(t,x,y) = q(t,x,y+ L_2) \, , \\
& &  u(t,x + L_1,y) = u(t,x,y) = u(t,x,y+ L_2) \, ,
\end{eqnarray*}
and the even constraint,
\begin{eqnarray*}
& & q(t,-x,y) = q (t,x,y) = q(t,x,-y) \, ,\\
& & u(t,-x,y) = u(t,x,y) = u(t,x,-y) \, .
\end{eqnarray*}
Its Lax pair is defined as:
\begin{eqnarray}
         L \psi &=& \lambda \psi\,, \label{LP1} \\
       \partial_t \psi &=& A \psi\,,\label{LP2}
\end{eqnarray}
where $\psi = \left( \psi_1, \psi_2\right)$, and 
\[
  L = \left(
\begin{array}{lr}
D^{-} & q\\ \\
\bq & D^{+}
\end{array}
\right)\,,
\]
\[
A = i \left[
2 \left(
\begin{array}{cc}
- \partial^2_x & q \partial_x\\
\bq \partial_x & \partial^2_x
\end{array}
\right) \, + \,
\left(
\begin{array}{cc}
r_1 & (D^+ q)\\
-(D^{-} \bq ) & r_2
\end{array}
\right)
\right]\, ,
\]
\begin{equation}
  D^+ = \alpha \partial_y + \partial_x\,, \qquad D^{-} = \alpha
  \partial_y - \partial_x\, , \qquad \al^2 = -1\ .
\label{DD}
\end{equation}
$r_1$ and $r_2$ have the expressions,
\begin{equation} 
  r_1 = \frac{1}{2} [-w+iv] \ , \ \ \  r_2= \frac{1}{2} [w+iv] \, , 
\label{res1}
\end{equation}
where $u$ and $v$ are real-valued functions satisfying
\begin{eqnarray}
& &[\partial^2_x + \partial^2_y] w
= 2 [\partial^2_x - \partial^2_y] |q|^2 \, , \label{res2} \\
& &[\partial^2_x + \partial^2_y] v
= i4 \alpha \partial_x \partial_y |q|^2 \, , \label{res3}
\end{eqnarray}
and $w=2(|q|^2 - \om^2 )+u_y$.
Notice that DSII (\ref{fDS2}) is
invariant under the transformation $\sg$:
\begin{equation}
  \sigma \circ (q,\bq, r_1, r_2; \al) = (q,\bq, -r_2, -r_1; -\al)\,.
\label{IT}
\end{equation}
Applying the transformation $\sigma$ (\ref{IT}) to the Lax pair
(\ref{LP1}, \ref{LP2}), we have a congruent Lax pair for which
the compatibility condition gives the same DSII.
The congruent Lax pair is given as:
\begin{eqnarray}
 \hat{L} \hat{\psi} &=& \lambda \hat{\psi}
                               \,,\label{CLP1} \\
 \partial_t \hat{\psi} &=& \hat{A}
                                   \hat{\psi}
                                 \,,\label{CLP2}
\end{eqnarray}
where $\hat{\psi} = (\hat{\psi}_1, \hat{\psi}_2)$, and
\[
  \hat{L} =
  \left(
    \begin{array}{cc}
- D^+ & q\\ \\
\bq & -D^-
    \end{array}
  \right)\,,
\]
\[
\hat{A} = i \left[
  2 \left(
    \begin{array}{cc}
- \partial^2_x & q \partial_x\\
\bq \partial_x & \partial^2_x
    \end{array}
  \right) + 
  \left(
    \begin{array}{cc}
-r_2 & -(D^- q)\\
(D^+\bq ) & -r_1
    \end{array}
  \right)
\right]\,.
\]
The compatibility condition of the Lax pair (\ref{LP1}, \ref{LP2}),
\begin{displaymath}
  \partial_t L = [A, L]\ ,
\end{displaymath}
where $[A, L] = AL - LA$, and the compatibility condition of the
congruent Lax pair (\ref{CLP1}, \ref{CLP2}),
  \begin{displaymath}
    \partial_t \hat{L} = [\hat{A}, \hat{L}]
  \end{displaymath}
give the same DSII (\ref{fDS2}). Let $(q,u)$ be a solution to the DSII
(\ref{fDS2}), and let $\lambda_0$ be any value of $\lambda$. Let
$\psi = (\psi_1, \psi_2)$ be a solution to the Lax
pair (\ref{LP1}, \ref{LP2}) at $(q, \bar q, r_1, r_2;
\lambda_0)$. Define the matrix operator:
\begin{displaymath}
  \Gamma = 
\left[
  \begin{array}{cc}
             \wedge + a & b\\
             c & \wedge + d
  \end{array}
\right]\,,
\end{displaymath}
where $\wedge = \alpha \partial_y - \lambda$, and $a$, $b$, $c$,
$d$ are functions defined as:
\begin{eqnarray*}
  a &=& \frac{1}{\Delta} \left[ \psi_2 \wedge_2 \bar{\psi}_2 +
                  \bar{\psi}_1 \wedge_1 \psi_1 \right]\,,\\[2ex]
  b &=& \frac{1}{\Delta} \left[ \bar{\psi}_2 \wedge_1 \psi_1 -
                   \psi_1 \wedge_2 \bar{\psi}_2 \right]\,,\\[2ex]
  c &=& \frac{1}{\Delta} \left[ \bar{\psi}_1 \wedge_1 \psi_2
                     - \psi_2 \wedge_2 \bar{\psi}_1 \right]\,,\\[2ex]
  d &=&  \frac{1}{\Delta} \left[ \bar{\psi}_2 \wedge_1 \psi_2 +
                      \psi_1 \wedge_2 \bar{\psi}_1 \right]\,,
\end{eqnarray*}
in which $\wedge_1 = \alpha \partial_y - \lambda_0$, $\wedge_2 =
\alpha \partial_y + \bar{\lambda}_0$, and
\begin{displaymath}
  \Delta = - \left[ | \psi_1 |^2 + |\psi_2|^2 \right]\,.
\end{displaymath}
Define a transformation as follows:
\begin{displaymath}
  \left\{
    \begin{array}{ccc}
(q,r_1,r_2) &\rightarrow& (Q,R_1,R_2)\,, \\
\phi &\rightarrow& \Phi\,;
    \end{array}
\right.
\end{displaymath}
\begin{eqnarray}
                    Q   &=& q - 2b\,,\nonumber \\[2ex]
                    R_1 &=& r_1 + 2(D^+a)\,, \label{DSBT}\\[2ex]
                    R_2 &=& r_2 - 2 (D^- d)\,,\nonumber\\[2ex]
                   \Phi &=& \Gamma \phi\,;\nonumber
\end{eqnarray}
where $\phi$ is any solution to the Lax pair (\ref{LP1},
\ref{LP2}) at $(q, \bar{q}, r_1, r_2; \lambda)$, $D^+$
and $D^-$ are defined in (\ref{DD}), we have the following theorem 
\cite{Li00a}.
\begin{theorem}
The transformation (\ref{DSBT}) is a B\"acklund-Darboux
transformation. That is, the function $Q$ defined
through the transformation (\ref{DSBT}) is also a solution to the 
DSII (\ref{fDS2}). The function $\Phi$ defined through the transformation
(\ref{DSBT}) solves the Lax pair (\ref{LP1}, \ref{LP2}) at $(Q,
\bar{Q}, R_1, R_2; \lambda)$.
\label{DSTH}
\end{theorem}

\subsection{An Example \label{dsex}} 

Instead of using $L_1$ and $L_2$ to describe the periods of the 
periodic boundary condition, one can introduce $\k_1$ and $\k_2$ 
as $L_1 = \frac{2\pi }{\k_1}$ and $L_2 = \frac{2\pi }{\k_2}$. Consider 
the spatially independent solution,
\begin{equation}
q_c = \eta \exp \{ -2i [ \eta^2 - \om^2 ] t + i \ga \} \ .
\label{us}
\end{equation}
The dispersion relation for the linearized DSII at $q_c$ is 
\[
\Om = \pm \frac{|\xi_1^2 - \xi_2^2|}{\sqrt{\xi_1^2 +\xi_2^2}}
\sqrt{4 \eta^2 - (\xi_1^2 +\xi_2^2)}\ , \ \ \mbox{for} \ 
\dl q \sim q_c \exp \{ i (\xi_1 x +\xi_2 y) +\Om t \} \ ,
\]
where $\xi_1 = k_1 \k_1$, $\xi_2 = k_2 \k_2$, and $k_1$ and $k_2$ 
are integers. We restrict $\k_1$ and $\k_2$ as follows to have only 
two unstable modes ($\pm \k_1, 0$) and ($0, \pm \k_2$),
\[
\k_2 < \k_1 < 2 \k_2\ , \ \ 
\k_1^2 < 4 \eta^2 < \min \{ \k_1^2 + \k_2^2, 4 \k_2^2 \} \ ,
\]
or 
\[
\k_1 < \k_2 < 2 \k_1\ , \ \ 
\k_2^2 < 4 \eta^2 < \min \{ \k_1^2 + \k_2^2, 4 \k_1^2 \} \ .
\]
The Bloch eigenfunction of the Lax pair (\ref{LP1}) and (\ref{LP2})
is given as,
\begin{equation}
  \psi = c(t) \left[
    \begin{array}[]{c}
-q_c \\ \chi
    \end{array} \right]
\exp \left\{ i(\xi_1 x + \xi_2y) \right\} \, ,
\label{slLax}
\end{equation}
where
\begin{eqnarray*}
& & c(t) = c_0 \exp \left\{ \left[ 2\xi_1(i \alpha \xi_2 - \lambda )
            + ir_2 \right] t \right\} \, , \\
& & r_2 - r_1 = 2 ( \left| q_c \right|^2 - \omega^2 ) \, , \\
& & \chi = (i \alpha \xi_2- \lambda )-i\xi_1 \, , \\
& & (i \alpha \xi_2 - \lambda)^2 + \xi^2_1 = \eta^2 \, .
\end{eqnarray*}
For the iteration of the B\"acklund-Darboux transformations, one 
needs two sets of eigenfunctions. First, we choose
$\xi_1 = \pm \frac{1}{2} \k_1$, $\xi_2=0$, $\lambda_0 = \sqrt{\eta^2 -
\frac{1}{4}\k_1^2}$ (for a fixed branch),
\begin{eqnarray}
  \psi^{\pm} = c^{\pm} \left[
    \begin{array}{c}
      -q_c \\ \\ \chi^{\pm}
    \end{array} \right]
  \exp \left\{ \pm i \frac{1}{2}\k_1x \right\} \, , 
\label{efunc1}
\end{eqnarray}
where
\begin{eqnarray*}
& & c^{\pm} = c^{\pm}_0 \exp \left\{ \left[ \mp \k_1 \lambda_0 + ir_2
    \right] t \right\} \, , \\
& & \chi^{\pm} = - \lambda_0 \mp i \frac{1}{2} \k_1 =
  \eta e^{\mp i (\frac{\pi}{2} +\vth_1)} \, .
\end{eqnarray*}
We apply the B\"acklund-Darboux transformations with $\psi = 
\psi^+ + \psi^-$, which generates
the unstable foliation associated with the $(\k_1,0)$ and $(-\k_1,0)$
linearly unstable modes. Then, we choose $\xi_2 = \pm
\frac{1}{2}\k_2$, $\lambda =0$, $\xi^0_1 = \sqrt{\eta^2-\frac{1}{4}\k_2^2}$
(for a fixed branch),
\begin{equation}
  \phi_{\pm} = c_{\pm} \left[
    \begin{array}{c}
      -q_c \\ \\ \chi_{\pm}
    \end{array} \right] 
  \exp \left\{ i (\xi^0_1 x \pm \frac{1}{2}\k_2 y) \right\} \, , 
\label{efunc2}
\end{equation}
where
\begin{eqnarray*}
& & c_{\pm} = c^0_{\pm} \exp \left\{ \left[ \pm i \alpha \k_2\xi^0_1
      + ir_2 \right] t \right\} \, , \\
& & \chi_{\pm} = \pm i \alpha \frac{1}{2}\k_2 - i\xi^0_1 
=\pm \eta e^{\mp i \vth_2}\, .
\end{eqnarray*}
We start from these eigenfunctions $\phi_{\pm}$ to generate
$\Gamma \phi_{\pm}$ through \BD, and then iterate the \BD ~with
$\Gamma \phi_+ + \Gamma \phi_-$ to generate the unstable
foliation associated with all the linearly unstable modes $(\pm
\k_1,0)$ and $(0, \pm \k_2)$. It turns out that the following 
representations are convenient,
\begin{eqnarray}
\psi^\pm &=& \sqrt{c_0^+c_0^-}e^{ir_2 t}\left ( \begin{array}{c} 
v_1^\pm \cr v_2^\pm \cr  \end{array} \right ) \ ,
\label{rwf1} \\
\phi_\pm &=& \sqrt{c^0_+c^0_-}e^{i \xi_1^0 x + ir_2 t}\left ( \begin{array}{c} 
w_1^\pm \cr w_2^\pm \cr \end{array} \right ) \ ,
\label{rwf2} 
\end{eqnarray}
where
\[
v_1^\pm = -q_c e^{\mp \frac{\tau}{2} \pm i \tx} \ , \ \ 
v_2^\pm = \eta e^{\mp \frac{\tau}{2} \pm i \tz} \ ,
\]
\[
w_1^\pm = -q_c e^{\pm \frac{\htau}{2} \pm i \hy} \ , \ \ 
w_2^\pm = \pm \eta e^{\pm \frac{\htau}{2} \pm i \hz}\ ,
\]
and
\[
c_0^+/c_0^- = e^{\rho + i \vth }\ , \ \ 
\tau = 2\k_1 \la_0 t - \rho \ , \ \ 
\tx = \frac{1}{2} \k_1 x + \frac{\vth}{2}\ , 
\tz = \tx - \frac{\pi}{2} - \vth_1 \ ,
\]
\[
c^0_+/c^0_- = e^{\hrho + i \hvth }\ , \ \ 
\htau = 2i\al \k_2 \xi_1^0 t + \hrho \ , \ \ 
\hy = \frac{1}{2} \k_2 y + \frac{\hvth}{2}\ , 
\hz = \hy - \vth_2 \ .
\]
The following representations are also very useful,
\begin{eqnarray}
\psi &=& \psi^+ + \psi^- = 2 \sqrt{c_0^+c_0^-}e^{ir_2 t}
\left ( \begin{array}{c} 
v_1 \cr v_2 \cr  \end{array} \right ) \ ,
\label{rwf3} \\
\phi &=& \phi^+ + \phi^- = 2 \sqrt{c^0_+c^0_-}e^{i \xi_1^0 x + 
ir_2 t}\left ( \begin{array}{c} 
w_1 \cr w_2 \cr \end{array} \right ) \ ,
\label{rwf4} 
\end{eqnarray}
where
\[
v_1 = -q_c [ \cosh \frac{\tau}{2} \cos \tx - i \sinh \frac{\tau}{2} \sin \tx ]
\ , \ \ v_2 = \eta [ \cosh \frac{\tau}{2} \cos \tz - i \sinh \frac{\tau}{2} 
\sin \tz ] \ , 
\]
\[
w_1 = -q_c [ \cosh \frac{\htau}{2} \cos \hy + i \sinh \frac{\htau}{2} \sin 
\hy ]
\ , \ \ w_2 = \eta [ \sinh \frac{\htau}{2} \cos \hz + i \cosh \frac{\htau}{2} 
\sin \hz ] \ . 
\]
Applying the \BD ~(\ref{DSBT}) with $\psi$ given in (\ref{rwf3}), we
have the representations,
\begin{eqnarray}
a &=& -\lambda_0 \ \mbox{sech}\ \tau \sin (\tx + \tz) \sin (\tx -\tz)
\label{aexp} \\
& & \times \bigg [ 1 + \ \mbox{sech}\ \tau \cos (\tx + \tz) \cos (\tx -\tz) 
\bigg ]^{-1}\, , \nonumber \\
b &=& -q_c \tb = - \frac{\lambda_0 q_c}{\eta} \bigg [ \cos (\tx - \tz) -
i \tanh \tau \sin (\tx - \tz) \label{bexp}\\
& & +\ \mbox{sech}\ \tau \cos (\tx + \tz)\bigg ]
\bigg [ 1+ \ \mbox{sech}\ \tau  \cos (\tx + \tz) \cos (\tx - \tz)
\bigg ]^{-1} \, , \nonumber \\
& & c= \overline{b} \, , \ \ \ \ d= - \overline{a} =-a \, . \label{cdexp}
\end{eqnarray}
The evenness of $b$ in $x$ is enforced by the requirement that 
$\vth - \vth_1 = \pm \frac{\pi}{2}$, and
\begin{eqnarray}
a^{\pm} &=& \mp \lambda_0 \ \mbox{sech} \ \tau \cos \vth_1 \sin (\k_1 x)
\label{eaexp} \\
& & \times \bigg [ 1 \mp \ \mbox{sech}\ \tau \sin \vth_1 \cos (\k_1 x)
\bigg ]^{-1}\, , \nonumber \\
b^{\pm} &=& -q_c \tb^{\pm} = - \frac{\lambda_0 q_c}{\eta} \bigg [ 
-\sin \vth_1 -
i \tanh \tau \cos \vth_1 \label{ebexp}\\
& & \pm \ \mbox{sech}\ \tau \cos (\k_1 x) \bigg ] 
\bigg [ 1 \mp \ \mbox{sech}\ \tau  \sin \vth_1 \cos (\k_1 x)
\bigg ]^{-1} \, , \nonumber \\
& & c= \overline{b} \, , \ \ \ \ d= - \overline{a} =-a \, . \label{ecdexp}
\end{eqnarray}
Notice also that $a^{\pm}$ is an odd function in $x$.  Under 
the above \BD, the
eigenfunctions $\phi_{\pm}$ (\ref{efunc2}) are transformed into
\begin{equation}
  \varphi^{\pm} = \Gamma \phi_{\pm} \, , 
\label{tphi}
\end{equation}
where
\begin{eqnarray*}
  \Gamma = \left[
    \begin{array}{cc}
\Lambda + a & b \\ \\ 
\overline{b} & \Lambda -a
    \end{array} \right] \, ,
\end{eqnarray*}
and $\Lambda = \alpha \partial_y - \lambda$ with $\lambda$
evaluated at $0$. Let $\varphi = \varphi^+ + \varphi^-$
(the arbitrary constants $c^0_{\pm}$ are already included in
$\varphi^{\pm}$), $\varphi$ has the representation,
\begin{equation}
\varphi =  2 \sqrt{c_+^0 c_-^0} e^{i \xi_1^0 x + i r_2 t}\left[
    \begin{array}{c}
      -q_c W_1 \\ \\ \eta W_2
    \end{array} \right] \ , 
\label{ief}
\end{equation}
where
\[
W_1 = (\al \pa_y w_1) + a w_1 +\eta \tb w_2\ , \ \ 
W_2 = (\al \pa_y w_2) - a w_2 +\eta \overline{\tb} w_1\ .
\]
We generate the coefficients in the \BD ~
(\ref{DSBT}) with $\varphi$ (the iteration of the \BD),
\begin{eqnarray}
a^{(I)} &=& - \bigg [ W_2 (\al \pa_y \overline{W_2}) + 
\overline{W_1} (\al \pa_y W_1) \bigg ]\bigg [ |W_1|^2 
+|W_2|^2 \bigg ]^{-1}\ ,  \label{ria}\\
b^{(I)} &=& \frac{q_c}{\eta}\bigg [ \overline{W_2} (\al \pa_y W_1) - 
W_1 (\al \pa_y \overline{W_2}) \bigg ]\bigg [ |W_1|^2 
+|W_2|^2 \bigg ]^{-1}\ , \label{rib} \\
& & c^{(I)} = \overline{b^{(I)}} \, , \ \ \ \ \, d^{(I)} =-\overline{a^{(I)}} 
\, , \label{ricd}
\end{eqnarray}
where
\begin{eqnarray*}
& & W_2 (\al \pa_y \overline{W_2}) + \overline{W_1} (\al \pa_y W_1) \\
&=& \frac{1}{2} \al \k_2 \bigg \{ \cosh \htau \bigg [ -\al \k_2 a 
+ i a \eta (\tb + \overline{\tb}) \cos \vth_2 \bigg ]  \\
&+& \bigg [ \frac{1}{4}\k_2^2 - a^2 -\eta^2 |\tb|^2 \bigg ] 
\cos (\hy +\hz ) \sin \vth_2 + \sinh \htau \bigg [ a \eta  
(\tb - \overline{\tb}) \sin \vth_2 \bigg ]  \bigg \} \ , \\ \\
& & |W_1|^2 +|W_2|^2 \\
&=& \cosh \htau \bigg [ a^2 + \frac{1}{4}\k_2^2 + \eta^2 |\tb|^2  
+i\al \k_2 \eta \frac{1}{2} (\tb + \overline{\tb})\cos \vth_2 \bigg ] \\
&+& \bigg [ \frac{1}{4}\k_2^2 - a^2 -\eta^2 |\tb|^2 \bigg ]  
\sin (\hy +\hz ) \sin \vth_2 + \sinh \htau \bigg [ \al \k_2 \eta  
\frac{1}{2} (\tb - \overline{\tb}) \sin \vth_2 \bigg ]\ , \\ \\
& & \overline{W_2} (\al \pa_y W_1) - W_1 (\al \pa_y \overline{W_2}) \\ 
&=& \frac{1}{2} \al \k_2 \bigg \{ \cosh \htau \bigg [ -\al \k_2 \eta \tb  
+ i (-a^2 + \frac{1}{4}\k_2^2 + \eta^2 \tb^2 ) \cos \vth_2 \bigg ]  \\
&+& \sinh \htau (a^2 - \frac{1}{4}\k_2^2 + \eta^2 \tb^2 )\sin \vth_2 
\bigg \} \ . 
\end{eqnarray*}
The new solution to the focusing Davey-Stewartson~II equation
 (\ref{fDS2}) is given by
\begin{equation}
   Q= q_c -2b -2b^{(I)} \, .
\label{newsl} 
\end{equation}
The evenness of $b^{(I)}$ in $y$ is enforced by the requirement
that $\hvth - \vth_2 = \pm \frac{\pi}{2}$.  In fact, we have
\begin{lemma}
Under the requirements that $\vth - \vth_1 = \pm \frac{\pi}{2}$, and
$\hvth - \vth_2 = \pm \frac{\pi}{2}$,
\begin{equation}
  b(-x) = b(x) \, , \ \ \  b^{(I)}(-x,y)=b^{(I)}(x,y)=b^{(I)}(x,-y) \, , 
\end{equation}
and $Q = q_c -2b-2b^{(I)}$ is even in both $x$ and $y$.
\label{evenla}
\end{lemma}

Proof: It is a direct verification by noticing that under the
requirements, $a$ is an odd function in $x$. Q.E.D.

The asymptotic behavior of $Q$ can be computed directly.  In
fact, we have the asymptotic phase shift lemma.

\begin{lemma}[Asymptotic Phase Shift Lemma]
For $\la_0 > 0$, $\xi_1^0 > 0$, and $i\al =1$, as $t \ra \pm \infty$, 
\begin{equation}
    Q = q_c -2b-2b^{(I)} \ra q_c e^{i\pi } e^{\mp i 2 (\vth_1 -\vth_2 )}\ .
\label{ayp}  
\end{equation}
In comparison, the asymptotic phase shift of the
first application of the \BD ~ is given by
\[
q_c - 2b \ra  q_c e^{\mp i 2 \vth_1}\ .
\]
\end{lemma}

\section{Other Soliton Equations}

In general, one can classify soliton equations into two categories. 
Category I consists of those equations possessing instabilities, 
under periodic boundary condition. In their phase space, figure-eight 
structures (i.e. separatrices) exist. Category II consists of 
those equations possessing no instability, under periodic boundary 
condition. In their phase space, no figure-eight structure 
(i.e. separatrix) exists. Typical Category I soliton equations are, 
for example, focusing nonlinear Schr\"odinger equation, sine-Gordon 
equation \cite{Li03k}, modified KdV equation. Typical Category II 
soliton equations are, 
for example, KdV equation, defocusing nonlinear Schr\"odinger equation, 
sinh-Gordon equation, Toda lattice. In principle, figure-eight structures 
for Category I soliton equations can be constructed through B\"acklund-Darboux 
transformations, as illustrated in previous sections. It should be remarked 
that B\"acklund-Darboux transformations still exist for Category II soliton 
equations, but do not produce any figure-eight structure. A good reference 
on B\"acklund-Darboux transformations is \cite{MS91}.

\chapter{Melnikov Vectors}

\section{1D Cubic Nonlinear Schr\"odinger Equation \label{MVNLS}}

We select the NLS (\ref{NLS}) as our first example to show how 
to establish Melnikov vectors. We continue from Section \ref{1DCNSE}.
\begin{definition} 
Define the sequence of functionals $F_j$ as follows,
\begin{equation}
F_j(\vq) = \Delta(\lambda^c_j (\vec{q}),\vec{q}),   
\label{3.1}
\end{equation}
where $\la^c_j$'s are the critical points, $\vec{q}=(q, -\bq )$.
\end{definition}
We have the lemma \cite{LM94}:
\begin{lemma}
If $\lambda_j^c(\vec{q}) $ is a simple critical point of $\Delta 
(\lambda)$  [i.e., $\Delta''(\lambda^c_j) \neq 0$],   $F_j $ is 
analytic in a neighborhood of $\vec{q}$, 
with first derivative given by 
\begin{equation}
{\delta F_{j}\over {\delta q}} = \,{\delta\Delta\over{\delta 
q}}\bigg|_{\lambda = \lambda^c_{j}} \, +
{\partial\Delta\over {\partial\lambda}} 
\bigg|_{\lambda = \lambda^c_{j}} \, {\delta 
\lambda^c_{j}\over {\delta q}} 
= {\delta\Delta\over {\delta q}}
\bigg|_{\lambda = \lambda^c_{j}}\ ,    \label{gradf}
\end{equation}
where 
\begin{equation}
\frac{\delta}{\delta {\vec{q}}} \  \Delta(\lambda;\vq) = i
\frac{\sqrt{\Delta^2 - 4}}{W [ \psi ^{+}, \psi ^{-}]}
\left[ \begin{array}{c}
\psi^{+}_{2} (x;\lambda) \psi ^{-} _{2} (x; \lambda) \\
\\
\psi^{+}_{1} (x; \lambda) \psi ^{-} _{1} (x; \lambda)
\end{array}\right] \ ,
\label{3.2}
\end{equation}
and the Bloch eigenfunctions $\psi^\pm $ have the property that
\begin{eqnarray}
\psi^{\pm} (x+2\pi; \lambda) = \rho^{\pm 1} \psi^{\pm} (x; \lambda)\ ,
\label{3.3}
\end{eqnarray}
for some $\rho$, also the Wronskian is given by
\[
W [ \psi ^{+}, \psi ^{-}] = \psi ^{+}_1 \psi ^{-}_2 - 
\psi ^{+}_2 \psi ^{-}_1 \ .
\]
In addition, $\Delta'$ is given by
\begin{eqnarray}
\frac{d\Delta}{d \lambda}= -i \frac{\sqrt{\Delta ^{2} - 4}}{W 
[\psi
^{+},\psi ^{-}]} \int ^{2\pi}_{0} [\psi ^{+} _{1} \psi ^{-} _{2} + \psi _{2} 
^{+}
\psi ^{-} _{1} ] \  dx \ . 
\label{3.4}
\end{eqnarray}
\label{lemma 6}
\end{lemma}
Proof: To prove this lemma, one calculates using variation of 
parameters:
$$
\delta M (x; \lambda ) =  M (x) \, \int _{0} ^{x} M ^{-1} (x') \delta
\hat{Q}(x') M (x') dx',
$$
\noindent where
\begin{eqnarray*}
\delta  \hat{Q}   \equiv   i\left( \begin{array}{c}0 \quad 
\delta q\\
\\ \delta \bq   \quad   0
\end{array}\right) \ .
\end{eqnarray*} 
 Thus; one obtains the formula
$$
\delta \Delta (\lambda;\vq) = \ \mbox{trace}\ \bigg [M (2\pi) \int^{2\pi}_{0}
M^{-1} (x')  \ \delta \hat{Q}(x')  \ M (x') dx'\bigg ],
$$
\noindent which gives 

\begin{eqnarray}
\frac{\delta \Delta (\lambda)}{\delta q(x)} &=&  i \ \mbox{trace}\ 
\bigg[ M^{-1} (x)\Bigl(
\begin{array}{cc}
	0 & 1 \\ 0 & 0 
\end{array}
\Bigr)M (x) M (2\pi)\bigg ]\ , \nonumber \\
\label{derg} \\ 
\frac{\delta \Delta (\lambda)}{\delta \bq(x)} &=& i \ \mbox{trace}\ 
\bigg [ M^{-1} (x)\Bigl(
\begin{array}{cc}
	0 & 0 \\ 1 & 0 
\end{array}
\Bigr)M (x) M (2\pi)\bigg ] \ . \nonumber 
\end{eqnarray}
Next, we use the Bloch eigenfunctions $\{ \psi^{\pm} \} $ 
to form the matrix 
\[
N(x;\lambda) = \left( \begin{array}{cc}
	\psi^+_1 & \psi^-_1 \\
	\psi^+_2 & \psi^-_2 
\end{array} \right) \ .
\]
Clearly,
\[
N(x;\lambda) = M(x;\lambda) N(0;\lambda)\ ;
\]
or equivalently,
\begin{equation}
M(x;\lambda) = N(x;\lambda) [N(0;\lambda)]^{-1}\ .
\label{3.6}
\end{equation}
Since $\psi^\pm$ are Bloch eigenfunctions, one also has
$$
N (x + 2\pi; \lambda) = N (x; \lambda)   \left( \begin{array}{cc}
	\rho & 0 \\
	0 & \rho^{-1}
\end{array} \right)\ ,
$$
\noindent which implies
$$
N (2\pi; \lambda ) = M (2\pi; \lambda)  N (0; \lambda)  =  N (0; \lambda)
\left( \begin{array}{cc}
	\rho & 0 \\
	0 & \rho^{-1}
\end{array} \right)\ ,
$$
that is,
\begin{equation}
M (2\pi; \lambda )  =  N (0; \lambda) 
\left( \begin{array}{cc}
	\rho & 0 \\
	0 & \rho^{-1}
\end{array} \right) [N (0; \lambda)]^{-1}\ .
\label{3.7}
\end{equation}
For any $2 \times 2$ matrix $\sigma$, equations (\ref{3.6}) and (\ref{3.7}) 
imply
\[
\mbox{trace}\ \{[M(x)]^{-1} \ \sigma \ M(x) \ M(2\pi) \} = \mbox{trace}\  
\{[N(x)]^{-1} \  \sigma \  N(x) \ diag(\rho, \rho^{-1}) \},
\]
which, through an explicit evaluation of (\ref{derg}), proves formula 
(\ref{3.2}).  
Formula (\ref{3.4}) is established similarly.  These formulas, together 
with the fact that $\lambda^c(\vec{q})$ is differentiable because it 
is a simple zero of $\Delta'$, provide the representation of 
$\frac{\dl F_j}{\dl \vq}$. Q.E.D.
\begin{remark}
Formula (\ref{gradf}) for the $\frac{\dl F_j}{\dl \vq}$ is actually valid even 
if $\Dl''(\la^c_j(\vq);\vq)=0$. Consider a function $\vq_*$ at which
\[
\Dl''(\la^c_j(\vq_*);\vq_*)=0,
\]
\nid
and thus at which $\la^c_j(\vq)$ fails to be analytic. For $\vq$ near 
$\vq_*$, one has
\[
\Dl'(\la^c_j(\vq);\vq)=0;
\]
\[
{\dl \over \dl \vq}\Dl'(\la^c_j(\vq);\vq)=\Dl''{\dl \over \dl \vq}\la^c_j
+{\dl \over \dl \vq}\Dl'=0;
\]
\nid
that is,
\[
{\dl \over \dl \vq}\la^c_j=-{1 \over \Dl''}{\dl \over \dl \vq}\Dl' \ .
\]
\nid
Thus,
\[
{\dl \over \dl \vq} F_j = {\dl \over \dl \vq}\Dl+\Dl'{\dl \over \dl \vq}
\la^c_j = {\dl \over \dl \vq}\Dl-{\Dl' \over \Dl''}
{\dl \over \dl \vq}\Dl' \mid_{\la=\la^c_j(\vq)}.
\]
Since ${\Dl' \over \Dl''}\ra 0$, 
as $\vq \ra \vq_*$, one still has formula (\ref{gradf}) at $\vq = \vq_*$:
\[
{\dl \over \dl \vq} F_j={\dl \over \dl \vq}\Dl\mid_{\la=\la^c_j(\vq)}.
\]
\end{remark}
The NLS (\ref{NLS}) is a Hamiltonian system:
\begin{equation}
iq_t = \frac {\dl H} {\dl \bq } \ , 
\label{hNLS}
\end{equation}
where
\[
H = \int_{0}^{2\pi } \{ - |q_x|^2 + |q|^4 \} \ dx \ .
\]
\begin{corollary}
For any fixed $\la \in \CC$, $\Dl (\la, \vq)$ is a constant of motion 
of the NLS (\ref{NLS}). In fact,
\[
\{ \Dl (\la, \vq), H(\vq) \} = 0 \ , \ \ \
\{ \Dl (\la, \vq), \Dl (\la', \vq) \}  = 0 \ , \ \ \forall \la, 
\la' \in \CC \ ,
\]
where for any two functionals $E$ and $F$, their Poisson bracket is 
defined as 
\[
\{ E, F \} = \int_{0}^{2\pi } \left [ \frac{\dl E}{\dl q}
\frac{\dl F}{\dl \bq} - \frac{\dl E}{\dl \bq}
\frac{\dl F}{\dl q} \right ] dx \ .
\]
\end{corollary}
Proof: The corollary follows from a direction calculation 
from the spatial part (\ref{Lax1}) of the Lax pair and the 
representation (\ref{3.2}). Q.E.D.

For each fixed $\vq$, $\Dl$ is an entire function of 
$\la$; therefore, can be determined by its values at 
a countable number of values of $\la$. The invariance of $\Dl$ 
characterizes the isospectral nature of the NLS equation.

\begin{corollary}
The functionals $F_j$ are constants of motion of the NLS (\ref{NLS}). 
Their gradients provide Melnikov vectors:
\begin{equation}
\mbox{grad} \  F_j(\vq) = i
\frac{\sqrt{\Delta^2 - 4}}{W [ \psi ^{+}, \psi ^{-}]}
\left[ \begin{array}{c}
\psi^{+}_{2} (x;\lambda^c_j) \psi ^{-} _{2} (x; \lambda^c_j) \\
\\
\psi^{+}_{1} (x; \lambda^c_j) \psi ^{-} _{1} (x; \lambda^c_j)
\end{array}\right] \ .
\label{3.8}
\end{equation}
\end{corollary}
The distribution of the critical points $\la^c_j$ are described by the 
following counting lemma \cite{LM94},
\begin{lemma}[Counting Lemma for Critical Points]
For $q \in H^1$, set $N = N(\| q\|_1) \in \ZZ^+$ by 
$$
N(\|q\|_1) =  2 \bigg[ \|q\|^2_0 \cosh
\bigg(\|q\|_0\bigg)\ +\ 3 \|q\|_1 \sinh 
\bigg(\|q\|_0\bigg )\bigg],
$$
\noindent where $[x] =$ first integer greater than $x$.  
Consider
$$
\Delta'(\lambda; \vq)  = \frac d{d\lambda}\ \Delta (\lambda;\vq).              
$$  Then
\begin{enumerate}
\item $\Delta'(\lambda;\vq)$ has exactly $2N+1$ zeros (counted
according to multiplicity) in the interior of the 
disc $D = \{ \lambda\in \CC: \ |\lambda | 
< (2N+1)\frac \pi{2}\};$
\item $\forall k\, \in \ZZ,|k| > N, 
\Delta'(\lambda,\vq)$
has exactly one zero in each disc\newline
$\{\lambda \in \CC \colon\ |\lambda - k \pi | <
\frac\pi{4}\}$.
\item $\Delta' (\lambda; \vq)$ has no other zeros.
\item For $|\lambda | > (2N + 1)\frac\pi{2},$
the zeros of $\Delta',\{\lambda^c_j,|j| > N\}$, are all
real, simple, and satisfy the asymptotics
\end{enumerate}
$$
\lambda^c_j = j\pi + o(1) \ \ \mbox{ as }|j| \to
\infty .
$$
\label{countle1}
\end{lemma}

\subsection{Melnikov Integrals}

When studying perturbed integrable systems, the figure-eight structures 
often lead to chaotic dynamics through homoclinic bifurcations. An extremely 
powerful tool for detecting homoclinic orbits is the so-called Melnikov 
integral method \cite{Mel63}, which uses ``Melnikov integrals'' to 
provide estimates of the distance between the center-unstable manifold 
and the center-stable manifold of a normally hyperbolic invariant manifold. 
The Melnikov 
integrals are often integrals in time of the inner products of certain 
Melnikov vectors with the perturbations in the perturbed integrable 
systems. This implies that the Melnikov vectors play a key role in the 
Melnikov integral method. First, we consider the case of one 
unstable mode associated with a complex double point $\nu$, for which 
the homoclinic orbit is given by B\"acklund-Darboux formula (\ref{4.5}),
\begin{eqnarray*}
Q(x,t)  \equiv   q(x,t) \ + \ 
2 (\nu-\bar{\nu}) \  
\frac{\phi_1{\bar{\phi}}_2}{\phi_1 {\bar{\phi}}_1+ 
\phi_2{\bar{\phi}}_2}\ ,
\end{eqnarray*}
where $q$ lies in a normally hyperbolic invariant manifold and 
$\phi$ denotes a general solution to the Lax pair (\ref{Lax1}, \ref{Lax2}) 
at $(q, \nu)$, $\phi = c_+ \phi^+ + c_- \phi^-$, and $\phi^\pm$ are 
Bloch eigenfunctions. Next, we consider the perturbed NLS,
\[
iq_t =q_{xx} +2|q|^2q +i\e f(q,\bq)\ ,
\]
where $\e$ is the perturbation parameter.
The {\em Melnikov integral} can be defined using the constant of motion 
$F_j $, where $\lambda^c_j = \nu$ \cite{LM94}:
\begin{equation}
M_j \equiv \int_{-\infty}^{+\infty} \int_{0}^{2\pi}
\{ \frac {\dl F_j}{\dl q} f + \frac {\dl F_j}{\dl \bq} \bar{f} \}_{q=Q}
\ dx dt \ ,
\label{6.1}
\end{equation}
where the integrand is evaluated along the unperturbed homoclinic 
orbit $q = Q$, and the Melnikov vector $\frac{\dl F_j}{\dl \vq}$ 
has been given in the last section, which can be expressed rather 
explicitly using the B\"acklund-Darboux transformation \cite{LM94} .  
We begin with the expression (\ref{3.8}),
\begin{eqnarray}
\frac{\delta F_j}{\delta \vec{q}} = i \frac{\sqrt{\Delta^2 - 
4}}{W[\Phi^{+},\Phi^{-}]} \left( \begin{array}{c}
	\Phi_2^{+} \Phi_2^{-}  \\
	\Phi_1^{+}  \Phi_1^{-}
\end{array}
\right) 
\label{6.2}
\end{eqnarray}
where $\Phi^{\pm}$ are Bloch eigenfunctions at $(Q, \nu)$, which 
can be obtained from  B\"acklund-Darboux formula (\ref{4.5}):
\begin{eqnarray*}
\Phi^{\pm}(x,t; \nu) \  \equiv \   G(\nu; \nu ; 
\phi) \ \phi^{\pm}(x,t; \nu)\ ,
\end{eqnarray*}
with the transformation matrix $G$ given by
\begin{eqnarray*} 
 G  = G (\lambda ; \nu ; \phi)= N
\left( \begin{array}{cl}
 \lambda-\nu  & \quad	0\\
0 &  \lambda - \bar{\nu}
 \end{array}\right)
N^{-1},
\end{eqnarray*}
\begin{eqnarray*}
N \equiv
\bigg [ \begin{array}{cl}
 \phi_1  & -\bar{\phi}_2\\
\phi_2 &\ \   \bar{\phi}_1
 \end{array} \bigg ]\ .
\end{eqnarray*}
These B\"acklund-Darboux formulas are rather easy to manipulate to obtain 
explicit information.  For example, the transformation matrix 
$G(\lambda, \nu)$ has a simple limit as $\lambda \rightarrow \nu$:
\begin{eqnarray}
\lim_{\lambda \rightarrow \nu} G(\lambda, \nu) = \frac{\nu - 
\bar{\nu}}{|\phi|^2} \left( \begin{array}{cc}
	\phi_2 \bar{\phi}_2 & -\phi_1 \bar{\phi}_2  \\
	-\phi_2 \bar{\phi}_1 & \phi_1 \bar{\phi}_1
\end{array}
\right)
\label{6.3}
\end{eqnarray}
where $|\phi|^2$ is defined by
\[
|\vec{\phi}|^2 \equiv \phi_1 \bar{\phi}_1 +\phi_2 \bar{\phi}_2.
\]
With formula (\ref{6.3}) one quickly calculates 
\[
\Phi^\pm = \pm c_{\mp}  \ \ W[\phi^{+},\phi^{-}]  \ \      
\frac{\nu - {\bar{\nu}}}{| \phi |^2} 
\left( \begin{array}{c}
	{\bar {\phi}}_2   \\
	-{\bar {\phi}}_1 
\end{array} 
\right) 
\]
from which one sees that $\Phi^{+}$ and $\Phi^{-}$ 
are linearly dependent at $(Q,\nu)$,
\[
\Phi^{+} = -\frac{c_-}{c_+}  \  \Phi^{-}.
\]
\begin{remark}  
For $Q$ on the figure-eight, the two Bloch eigenfunctions 
$\Phi^\pm$ are linearly dependent. Thus, the geometric 
multiplicity of $\nu$ is only one, even though its 
algebraic multiplicity is two or higher. 
\end{remark}
Using L'Hospital's rule, one gets
\begin{eqnarray}
\frac{\sqrt{\Delta^2 - 4}}{W[\Phi^{+},\Phi^{-}]} = \frac{\sqrt{\Delta(\nu) 
\Delta''(\nu)}}{(\nu - \bar{\nu})  \ \ W[\phi^{+},\phi^{-}]}\ .
\label{6.4}
\end{eqnarray}  
With formulas (\ref{6.2}, \ref{6.3}, \ref{6.4}), one obtains the explicit 
representation of the $\mbox{grad}\ F_j$ \cite{LM94}:
\begin{eqnarray}
\frac{\delta F_j}{\delta \vec{q}} \ = \ C_\nu  \ \frac{c_+ c_-  
W[\psi^{(+)},\psi^{(-)}] }{|\phi|^4} 
\left( \begin{array}{c}
	{\bar{\phi}}^2_2   \\ \\ 
	-{\bar{\phi}}^2_1
\end{array}
\right)\ , 
\label{6.5}
\end{eqnarray}
where the constant $C_\nu$ is given by
\[
C_\nu \equiv  i (\nu - \bar{\nu}) \ \  \sqrt{\Delta(\nu) \Delta''(\nu)}\ .
\] 
With these ingredients, one obtains the following beautiful 
representation of 
the Melnikov function associated to the general complex double point 
$\nu$ \cite{LM94}:
\begin{equation}
M_j =  C_\nu \  c_+ c_- \int_{-\infty}^{+\infty} 
\int_0^{2\pi}  \ W[\phi^{+}, \phi^{-}]  \  
\left[\frac{(\bar{\phi}_2^2) f(Q,\bar{Q}) + (\bar{\phi}_1^2) 
\overline{f(Q,\bar{Q})}}{|\phi|^4} \right]  \ \ dx dt.
\label{6.6}
\end{equation}
In the case of several complex double points, each associated with 
an instability, one can iterate the B\"acklund-Darboux tranformations and use 
those functionals $F_j$ which are associated with each 
complex double point to obtain representations {\em Melnikov Vectors}.
In general, the relation between $\frac{\delta F_j}{\delta \vec{q}}$ and 
double points can be summarized in the following lemma \cite{LM94},
\begin{lemma}
Except for the trivial case $q = 0$,
\begin{eqnarray*}
(a). & & \ \ \frac{\delta F_j}{\delta q} = 0 \Leftrightarrow 
\frac{\delta F_j}{\delta \bar{q}} = 0 
\Leftrightarrow M(2\pi,\lambda^c_j;  \ \vec{q}_*) = \pm I. \\
(b). & & \ \  \frac{\delta F_j}{\delta \vq}\mid_{\vq_*}  = 0 \Rightarrow 
\Delta'(\lambda_j^c(\vec{q}_*); \ \vec{q}_*) = 0, 
\Rightarrow |F_j(\vec{q}_*)| = 2, \\
& &\ \ \Rightarrow  \lambda_j^c(\vec{q}_*)  \ \ \mbox{is a multiple point},
\end{eqnarray*}
where $I$ is the $2\times 2$ identity matrix.
\label{lemma 7}
\end{lemma}
The B\"acklund-Darboux transformation theorem indicates that 
the figure-eight structure is attached to a complex double point.
The above lemma shows that at the origin of the figure-eight, 
the gradient of $F_j$ vanishes. Together they indicate that the 
the gradient of $F_j$ along the figure-eight is a perfect Melnikov 
vector.

{\bf Example:} When $\frac{1}{2} < c < 1$ in (\ref{4.7}), and choosing 
$\vth = \pi $ in (\ref{4.13}), one can get the Melnikov vector field along 
the homoclinic orbit (\ref{4.13}),
\begin{eqnarray}
& & \frac{\dl F_1}{\dl q} = 2 \pi\  \sin^2 p
\hbox{ sech} \tau \;\frac{[(- \sin p + i \cos p
\tanh \tau) \cos x+\hbox{ sech} \tau ]} {[1 - \sin p \hbox{ sech } 
\tau \cos x ]^2}
\, c\; e^{i \theta}, \label{mvnls1} \\
& & \frac{\dl F_1}{\dl \bq} = 
\overline{\frac{\dl F_1}{\dl q}} \ .  \nonumber
\end{eqnarray}

\section{Discrete Cubic Nonlinear Schr\"odinger Equation}

The discrete cubic nonlinear Schr\"odinger equation (\ref{DNLS})
can be written in the Hamiltonian form \cite{AL76} \cite{Li92}:
\begin{equation}
i \dot{q}_n = \rho_n \pa H/ \pa \bq_n \ , \label{HDNLS}
\end{equation}
\\
where $\rho_n = 1 +h^2|q_n|^2$, and
\[
H={1 \over h^2}\sum^{N-1}_{n=0}
\bigg \{\bq_n(q_{n+1}+q_{n-1})-{2 \over h^2}(1+\om^2 h^2)\ln \rho_n
\bigg \} \ .
\]
$\sum^{N-1}_{n=0}\bigg \{\bq_n(q_{n+1}+q_{n-1})\bigg \}$ 
itself is also a constant of motion. This invariant, together with $H$,   
implies that $\sum^{N-1}_{n=0}\ln \rho_n$ is a constant of motion too. 
Therefore, 
\begin{equation}
D^2\equiv \prod^{N-1}_{n=0}\rho_n
\label{constD}
\end{equation}
is a constant of motion. We continue from Section \ref{DCNSE}. Using $D$, 
one can define a normalized Floquet 
discriminant $\tilde{\Dl}$ as
\[
\tilde{\Dl} = \Dl / D\ .
\]
\begin{definition}
The sequence of invariants $\tF_j$ is defined as:
\begin{equation}
\tF_j(\vq) = \tilde{\Dl}(z^{(c)}_j(\vq);\vq)\ ,
\label{coF}
\end{equation}
where $\vq = (q, -\bq)$, $q=(q_0,q_1,\cdot \cdot \cdot,q_{N-1})$.
\end{definition}
These invariants $\tF_j$'s are perfect candidate for building Melnikov 
functions. The Melnikov vectors are given by the gradients of these 
invariants.
\begin{lemma}
Let $z^{(c)}_j(\vq)$ be a simple critical point; then
\begin{equation}
{\dl \tF_j \over \dl \vq_n}(\vq) = {\dl \tilde{\Dl} \over \dl \vq_n}
(z^{(c)}_j(\vq);\vq)\ .
\label{derf}
\end{equation}
\begin{equation}
{\dl \tilde{\Dl} \over \dl \vq_n}(z;\vq) = {i h (\z -\z^{-1}) \over 
2 W_{n+1}} \left (\begin{array}{c} \psi^{(+,2)}_{n+1}\psi^{(-,2)}_{n}+ 
\psi^{(+,2)}_{n}\psi^{(-,2)}_{n+1} \cr \cr \psi^{(+,1)}_{n+1}\psi^{(-,1)}_{n}+ 
\psi^{(+,1)}_{n}\psi^{(-,1)}_{n+1} \cr \end{array} \right )\ , 
\label{fidr}
\end{equation}
where $\psi_n^\pm = (\psi_n^{(\pm,1)}, \psi_n^{(\pm,2)})^T$ are two 
Bloch functions of the Lax pair (\ref{Lax1},\ref{Lax2}), such that 
\[
\psi^\pm_n = D^{n/N} \z^{\pm n/N} \tilde{\psi}^\pm_n\ ,
\]
where $\tilde{\psi}^\pm_n$ are periodic in $n$ with period $N$,
$W_n = \ \mbox{det}\ ( \psi^+_n,\psi^-_n )$.
\end{lemma}
For $z^{(c)}_j = z^{(d)}$, the Melnikov vector field located on 
the heteroclinic orbit (\ref{BD1}) is given by
\begin{equation}
{\dl \tF_j \over \dl \vQ_n} = K {W_n \over E_n A_{n+1}} 
\left ( \begin{array}{c} [z^{(d)}]^{-2} \ \overline{\phi^{(1)}_n} 
\ \overline{\phi^{(1)}_{n+1}} \cr \cr [\overline{z^{(d)}}]^{-2} 
\ \overline{\phi^{(2)}_n} \ \overline{\phi^{(2)}_{n+1}} \cr \end{array} 
\right )\ , \label{fibd}
\end{equation}
where $\vQ_n =(Q_n, -\bar{Q_n})$,
\[
\phi_n = (\phi^{(1)}_n, \phi^{(2)}_n)^T= c_+ \psi_n^+ + c_- \psi_n^- \ ,
\]
\[
W_n = \left | \begin{array}{lr} \psi_n^+ & \psi_n^-  \cr \end{array} 
\right |\ ,
\]
\[
E_n = |\phi^{(1)}_n|^2 + |z^{(d)}|^2|\phi^{(2)}_n|^2\ ,
\]
\[
A_n = |\phi^{(2)}_n|^2 + |z^{(d)}|^2|\phi^{(1)}_n|^2\ ,
\]
\[
K=-{ihc_+c_- \over 2}|z^{(d)}|^4 (|z^{(d)}|^4 - 1)[\overline{z^{(d)}}]^{-1} 
\sqrt{ \tDl(z^{(d)};\vq) \tDl''(z^{(d)};\vq)}\ ,
\]
where
\[
\tDl''(z^{(d)};\vq) = {\pa^2 \tDl(z^{(d)};\vq) \over \pa z^2}\ .
\]

{\bf Example:} The Melnikov vector evaluated on the heteroclinic orbit 
(\ref{hetorb}) is given by
\begin{equation}
{\dl \tF_1 \over \dl \vQ_n} = \hat{K} \bigg [ \hat{E}_n \hat{A}_{n+1}
\bigg ]^{-1} \ \mbox{sech}\ [2\mu t +2p]\left ( \begin{array}{c} 
\hat{X}^{(1)}_n \cr - \hat{X}^{(2)}_n \cr \end{array} \right ) \ ,
\label{melv}
\end{equation}
where
\begin{eqnarray*}
\hat{X}^{(1)}_n &=& \bigg [ \cos \be \ \mbox{sech}\ [ 2\mu t +2p] 
+\cos [(2n+1)\be +\vth +\varphi] \\ 
& & - i \tanh [ 2\mu t +2p] 
\sin [(2n+1)\be +\vth +\varphi] \bigg ] e^{i2\th(t)}\ ,
\end{eqnarray*}
\begin{eqnarray*}
\hat{X}^{(2)}_n &=& \bigg [ \cos \be \ \mbox{sech}\ [ 2\mu t +2p] 
+\cos [(2n+1)\be +\vth -\varphi] \\
& & - i \tanh [ 2\mu t +2p] 
\sin [(2n+1)\be +\vth -\varphi] \bigg ] e^{-i2\th(t)}\ ,
\end{eqnarray*}
\[
\hat{K} = -2Nh^2a(1-z^4) [8 \rho^{3/2}z^2]^{-1}\sqrt{\rho \cos^2 \be -1}\ .
\]
Under the even constraint, the Melnikov vector evaluated on the 
heteroclinic orbit (\ref{ehetorb}) is given by
\begin{equation}
{ \dl \tF_1 \over \dl \vQ_n}\bigg |_{\mbox{even}} = \hat{K}^{(e)} 
\ \mbox{sech}[2\mu t + 2p] [\Pi_n]^{-1}\left ( \begin{array}{c} 
\hat{X}^{(1,e)}_n \cr - \hat{X}^{(2,e)}_n \cr \end{array} \right ) \ ,
\label{emelv}
\end{equation}
where
\[
\hat{K}^{(e)}= -2N (1-z^4) [8a\rho^{3/2} z^2]^{-1} 
\sqrt{\rho \cos^2\be - 1}\ ,
\]
\begin{eqnarray*}
\Pi_n &=& \bigg [ \cos \be \pm \cos \varphi \ \mbox{sech}[2\mu t +2p] 
\cos[2(n-1)\be]\bigg ] \times \\
& &\bigg [ \cos \be \pm \cos \varphi \ \mbox{sech}[2\mu t +2p] 
\cos[2(n+1)\be]\bigg ]\ ,
\end{eqnarray*}
\begin{eqnarray*}
\hat{X}^{(1,e)}_n &=& \bigg [ \cos \be \ \mbox{sech}\ [ 2\mu t +2p] 
\pm (\cos \varphi \\
& & -i \sin \varphi \tanh [ 2\mu t +2p]) \cos [2n\be]\bigg ] e^{i2\th(t)}\ ,
\end{eqnarray*} 
\begin{eqnarray*}
\hat{X}^{(2,e)}_n &=& \bigg [ \cos \be \ \mbox{sech}\ [ 2\mu t +2p] 
\pm (\cos \varphi \\
& & +i \sin \varphi \tanh [ 2\mu t +2p]) \cos [2n\be]\bigg ] e^{-i2\th(t)}\ .
\end{eqnarray*}

\section{Davey-Stewartson II Equations}

The DSII (\ref{fDS2}) can be written in the Hamiltonian form,
\begin{equation}
  \left\{
  \begin{array}{ccc}
    i q_t &=& \delta H / \delta \overline{q}  \ ,\cr
   i \overline{q}_t &=& - \delta H / \delta q \ , \cr 
  \end{array} \right. 
\label{fhDS2}
\end{equation}
where
\begin{displaymath}
  H= \int^{2 \pi}_0 \int^{2 \pi}_0 
  [\left| q_y \right|^2 - \left| q_x \right|^2 + 
       \frac{1}{2} (r_2-r_1) \left| q \right|^2] \, dx \, dy \, .
\end{displaymath}
We have the lemma \cite{Li00a}.
\begin{lemma}
The inner product of the vector
\begin{equation}
  \U= \left(
    \begin{array}{c}
      \psi_2  \hat{\psi}_2 \cr
      \psi_1  \hat{\psi}_1
    \end{array}\right)^- +S \left(
    \begin{array}{c}
      \psi_2  \hat{\psi}_2 \\
      \psi_1  \hat{\psi}_1
    \end{array}\right) \, ,
\nonumber
\end{equation}
where $\psi = (\psi_1 , \psi_2)$ is an eigenfunction solving the
Lax pair (\ref{LP1}, \ref{LP2}), and $\hat{\psi} = (\hat{\psi}_1 ,
\hat{\psi}_2)$ is an eigenfunction solving the corresponding 
congruent Lax pair (\ref{CLP1}, \ref{CLP2}),
and $S= \displaystyle{\left(
    \begin{array}{ccc}
      0 & 1 \\ 1 & 0
    \end{array}
\right)}$, with the vector field $J \na H$ given by the 
right hand side of (\ref{fhDS2}) vanishes,
\begin{displaymath}
  \langle \U\, , \,  J \na H \rangle =0 \, ,
\end{displaymath}
where $J= \displaystyle{\left(
    \begin{array}{ccc}
      0 & 1 \\ -1 & 0
    \end{array}
\right)}$.
\label{melem2}
\end{lemma}
If we only consider even functions, i.e., $q$ and $u=r_2 -r_1$
are even functions in both $x$ and $y$, then we can split $\U$
into its even and odd parts,
\begin{displaymath}
  \U=\U^{(e,x)}_{(e,y)} + \U^{(e,x)}_{(o,y)} + \U^{(o,x)}_{(e,y)} 
    +  \U^{(o,x)}_{(o,y)} \, , 
\end{displaymath}
where
\begin{eqnarray}
    \U^{(e,x)}_{(e,y)} &=& \frac{1}{4} \bigg [
      \U(x,y) + \U(-x,y) + \U(x,-y) + \U(-x,-y) \bigg ] \, , \nonumber\\
    \U^{(e,x)}_{(o,y)} &=& \frac{1}{4} \bigg [
      \U(x,y) + \U(-x,y) - \U(x,-y) - \U(-x,-y) \bigg ] \, , \nonumber\\
    \U^{(o,x)}_{(e,y)} &=& \frac{1}{4} \bigg [
      \U(x,y) - \U(-x,y) + \U(x,-y) - \U(-x,-y) \bigg ] \, , \nonumber\\
    \U^{(o,x)}_{(o,y)} &=& \frac{1}{4} \bigg [
      \U(x,y) - \U(-x,y) - \U(x,-y) + \U(-x,-y) \bigg ] \, . \nonumber
\end{eqnarray}
Then we have the lemma \cite{Li00a}.
\begin{lemma}
  When $q$ and $u=r_2-r_1$ are even functions in both $x$ and
  $y$, we have
\begin{displaymath}
    \langle \U^{(e,x)}_{(e,y)} \, , \, J \na H \rangle =0 \, .
\end{displaymath}
\label{melem3}
\end{lemma}

\subsection{Melnikov Integrals}

Consider the perturbed DSII equation,
\begin{equation}
\left\{
    \begin{array}{l}
      i \partial_t q  =  [\partial^2_x - \partial^2_y] q + 
      [2 ( \left| q \right|^2 - \om^2) + u_y] q + 
      \e i f \, , \\[1ex]
[\partial^2_x + \partial^2_y] u
       = -4 \partial_y \left| q \right|^2 \, ,
    \end{array} \right.
\label{PDS2}
\end{equation}
where $q$ and $u$ are respectively complex-valued and
real-valued functions of three variables $(t,x,y)$, and $G
= (f, \overline{f})$ are the perturbation terms which can
depend on $q$ and $\overline{q}$ and their derivatives and $t$, $x$
and $y$. The Melnikov integral is given by \cite{Li00a},
\begin{eqnarray}
  M &=&  \int^{\infty}_{- \infty} \langle \U ,
                 G \rangle \, dt \nonumber \\[1ex]
&=& 2 \int^{\infty}_{- \infty} \int^{2 \pi}_0 \int^{2 \pi}_0 
    R_e \left\{(\psi_2 \hat{\psi}_2)  f + (\psi_1 \hat{\psi}_1) 
      \overline{f} \right\} \, dx \, dy \, dt \, ,  \label{mlf2} 
\end{eqnarray}
where the integrand is evaluated on an unperturbed homoclinic
orbit in certain center-unstable ($=$ center-stable) manifold,
and such orbit can be obtained through the B\"acklund-Darboux
transformations given in Theorem \ref{DSTH}. A concrete example 
is given in section \ref{dsex}. When we only
consider even functions, i.e., $q$ and $u$ are even functions in
both $x$ and $y$, the corresponding Melnikov function is given
by \cite{Li00a},
\begin{eqnarray}
M^{(e)} &=&  \int^{\infty}_{- \infty} 
  \langle \U^{(e,x)}_{(e,y)} , \vec{G} \rangle \, dt \nonumber\\[1ex]
&=&  \int^{\infty}_{- \infty} 
          \langle \U , \vec{G} \rangle \, dt \, , \label{mlf3}
\end{eqnarray}
which is the same as expression (\ref{mlf2}).

\subsection{An Example \label{mids}}

We continue from the example in section \ref{dsex}. 
We generate the following eigenfunctions
corresponding to the potential $Q$ given in (\ref{newsl}) through 
the iterated \BD,
\begin{eqnarray}
\Psi^{\pm} &=& \Gamma^{(I)} \Gamma \psi^{\pm} \, , \ \ \ \ \mbox{ at } 
\ \lambda = \lambda_0 = \sqrt{\eta^2- \frac{1}{4}\k_1^2} \, , 
\label{nwef1} \\[1ex]
\Phi_{\pm} &=& \Gamma^{(I)} \Gamma \phi_{\pm} \, ,  \ \ \ \
\mbox{ at } \lambda =0 \, , \label{nwef2}
\end{eqnarray}
where
\[
  \Gamma = \left[
    \begin{array}{cc}
\Lambda +a & b\\ \\ 
\overline{b} & \Lambda -a
    \end{array} \right] \, , \ \ \ \Gamma^{(I)} = \left[
    \begin{array}{cc}
\Lambda + a^{(I)} & b^{(I)} \\ \\ 
\overline{b^{(I)}} & \Lambda - \overline{a^{(I)}}
    \end{array} \right] \, ,
\]
where $\Lambda = \alpha \partial_y - \lambda$ for general $\lambda$.
\begin{lemma}[see \cite{Li00a}]
The eigenfunctions $\Psi^{\pm}$ and $\Phi_{\pm}$ defined in
(\ref{nwef1}) and (\ref{nwef2}) have the representations,
\begin{eqnarray}
    \Psi^{\pm} &=& \frac{\pm 2 \lambda_0 W(\psi^+,
        \psi^-)}{\Delta}\left[ 
      \begin{array}{c}
        (- \lambda_0 + a^{(I)}) \overline{\psi}_2
        -b^{(I)} \overline{\psi}_1 \\ \\ 
\overline{b^{(I)}} \overline{\psi}_2 +
        (\lambda_0 + \overline{a^{(I)}}) \overline{\psi}_1
      \end{array}\right] \, ,\label{rnef1} \\
\Phi_{\pm} &=& \frac{\mp i \alpha \k_2}{\Delta^{(I)}}
\left[ 
  \begin{array}{c}
    \Xi_1 \\[1ex] \Xi_2
  \end{array} \right] \, , \label{rnef2}
\end{eqnarray}
where
\begin{eqnarray*}
& &  W(\psi^+,\psi^-) = \left|
    \begin{array}{cc}
\psi^+_1 & \psi^-_1 \\ \\ 
\psi^+_2 & \psi^-_2
    \end{array} \right| =
  -i \k_1 c^+_0 c^-_0 q_c \exp \left\{ i2r_2t \right\} \, , \\
& & \Delta = - \Bigl[ \left| \psi_1 \right|^2 +
     \left| \psi_2 \right|^2 \Bigr] \, , \\
& & \psi = \psi^+ + \psi^- \, , \\
& & \Delta^{(I)} = - \Bigl[ \left| \varphi_1 \right|^2
     + \left| \varphi_2 \right|^2 \Bigr]\, , \\
& & \varphi = \varphi^+ + \varphi^- \, , \\
& & \varphi^{\pm} = \Gamma \phi_{\pm} \hbox{ at }
     \lambda =0 \, , \\
& & \Xi_1 = \overline{\varphi}_1 (\varphi^+_1 \varphi^-_1)
+ \overline{\varphi^+_2} (\varphi^+_1 \varphi^-_2) +
\overline{\varphi^-_2} (\varphi^-_1 \varphi^+_2) \, , \\
& & \Xi_2 = \overline{\varphi}_2 (\varphi^+_2 \varphi^-_2)
+ \overline{\varphi^+_1} (\varphi^-_1 \varphi^+_2) +
\overline{\varphi^-_1} (\varphi^+_1 \varphi^-_2) \, . 
\end{eqnarray*}
If we take $r_2$ to be real (in the Melnikov vectors, $r_2$
appears in the form $r_2-r_1=2(\left|q_c \right|^2 - \omega^2))$, then
\begin{equation}
  \Psi^{\pm} \to 0 \, , \ \ \ \  \Phi_{\pm} \to 0 \, , \, 
  \hbox{ as } t \to \pm \infty \, .
\label{anef}
\end{equation}
\end{lemma}
Next we generate eigenfunctions solving the corresponding
congruent Lax pair (\ref{CLP1}, \ref{CLP2}) with the potential $Q$, through
the iterated \BD ~and the symmetry transformation (\ref{IT}) \cite{Li00a}.
\begin{lemma}
Under the replacements
\begin{displaymath}
    \alpha \longrightarrow - \alpha \, \ \ (\vth_2 \longrightarrow 
\pi - \vth_2 ), \ \ \  
    \hvth \longrightarrow \hvth +\pi -2 \vth_2 \, , \ \ \  
    \hat{\rho} \longrightarrow - \hat{\rho}\, , 
\end{displaymath}
the coefficients in the iterated \BD ~are transformed as follows,
\[
a^{(I)} \longrightarrow \overline{a^{(I)}} \, , \ \  
b^{(I)} \longrightarrow b^{(I)} \, , \,  
\]
\[
\bigg (c^{(I)} = \overline{b^{(I)}}\bigg ) \longrightarrow 
\bigg (c^{(I)} = \overline{b^{(I)}}\bigg ) \, , \, 
\bigg (d^{(I)} = - \overline{a^{(I)}}\bigg ) \longrightarrow 
\bigg (\overline{d^{(I)}} =-a^{(I)}\bigg ) \, .
\]
\label{rpl}
\end{lemma}
\begin{lemma}[see \cite{Li00a}]
Under the replacements
\begin{eqnarray}
& &   \alpha \mapsto - \alpha \, \ \ (\vth_2 \longrightarrow \pi - 
\vth_2 ), \ \  r_1 \mapsto -r_2 \, , \nonumber \\
 \label{rptr} \\
& &   r_2 \mapsto -r_1 \, , \ \ 
\hvth \longrightarrow \hvth +\pi -2 \vth_2 \, , \ \ \  
    \hat{\rho} \longrightarrow - \hat{\rho}\, , \nonumber
\end{eqnarray}
the potentials are transformed as follows,
\begin{eqnarray*}
& & Q  \longrightarrow Q \, , \\
& & (R=\overline{Q})  \longrightarrow (R= \overline{Q}) \, , \\
& & R_1 \longrightarrow -R_2 \, , \\
& & R_2 \longrightarrow -R_1 \, .
\end{eqnarray*}
\label{rppl}
\end{lemma}
The eigenfunctions $\Psi^{\pm }$ and $\Phi_{\pm}$ given in
(\ref{rnef1}) and (\ref{rnef2}) depend on the variables in the
replacement (\ref{rptr}):
\begin{eqnarray*}
  \Psi^{\pm} &=& \Psi^{\pm} (\alpha , r_1 , r_2 , \hvth , \hrho ) \, , \\
  \Phi_{\pm} &=& \Phi_{\pm} (\alpha , r_1 , r_2 , \hvth , \hrho ) \, .
\end{eqnarray*}
Under replacement (\ref{rptr}), $\Psi^{\pm}$ and $\Phi_{\pm}$ are 
transformed into
\begin{eqnarray}
  \widehat{\Psi}^{\pm} &=& \Psi^{\pm}
     (- \alpha , -r_2 , -r_1 , \hvth + \pi -2 \vth_2 , - \hrho )
     \, , \label{cref1}\\
  \widehat{\Phi}_{\pm} &=& \Phi_{\pm}
     (- \alpha , -r_2 , -r_1 , \hvth + \pi -2 \vth_2 , - \hrho )
     \, . \label{cref2}
\end{eqnarray}
\begin{corollary}[see \cite{Li00a}]
$\widehat{\Psi}^{\pm}$ and $ \widehat{\Phi}_{\pm}$ solve the
congruent Lax pair (\ref{CLP1}, \ref{CLP2}) at $(Q, \overline{Q}, 
R_1, R_2; \lambda_0)$ and $(Q, \overline{Q}, R_1, R_2; 0)$, respectively.
\label{ccor}
\end{corollary}
Notice that as a function of $\eta$, $\xi^0_1$ has two (plus and
minus) branches.  In order to construct Melnikov vectors, we need 
to study the effect of the replacement $\xi^0_1 \longrightarrow
-\xi^0_1$.
\begin{lemma}[see \cite{Li00a}]
Under the replacements
\begin{equation}
    \xi^0_1 \longrightarrow - \xi^0_1 \,  \ \ (\vth_2 \longrightarrow 
-\vth_2 ), \ \  
   \hvth \longrightarrow \hvth + \pi -2 \vth_2 , \ \ 
\hrho \longrightarrow - \hrho  ,
\label{krptr}  
\end{equation}
the coefficients in the iterated \BD ~are invariant,
\[
  a^{(I)} \mapsto a^{(I)} \, , \ \ \ \    b^{(I)} \mapsto b^{(I)} \, , 
\]
\[
( c^{(I)}= \overline{b^{(I)}}) \mapsto (c^{(I)} = \overline{b^{(I)}}) 
\, , \ \ \ \  
( d^{(I)}= - \overline{a^{(I)}}) \mapsto (d^{(I)} = -\overline{a^{(I)}}) 
\, ; 
\]
thus the potentials are also invariant,
\[
Q \longrightarrow Q \, , \ \ \ \  (R=\overline{Q}) \longrightarrow
(R=\overline{Q}) \, ,  
\]
\[
R_1 \longrightarrow R_1 \, , \ \ \ \  R_2 \longrightarrow R_2 \, .
\]
\label{lektr}
\end{lemma}
The eigenfunction $\Phi_{\pm}$ given in (\ref{rnef2}) depends on
the variables in the replacement (\ref{krptr}):
\begin{displaymath}
  \Phi_{\pm} = \Phi_{\pm} (\xi^0_1 , \hvth , \hrho ) \, .
\end{displaymath}
Under the replacement (\ref{krptr}), $\Phi_{\pm}$ is transformed
into
\begin{equation}
  \widetilde{\Phi}_{\pm} = \Phi_{\pm} 
  (-\xi^0_1 , \hvth + \pi -2 \vth_2, -\hrho )\ .
\label{kfr}
\end{equation}
\begin{corollary}
$\widetilde{\Phi}_{\pm}$ solves the Lax pair (\ref{LP1},\ref{LP2}) at
$(Q , \overline{Q} , R_1 , R_2 \, ; \, 0)$.
\label{kcor}
\end{corollary}

In the construction of the Melnikov vectors, we need to replace
$\Phi_{\pm}$ by $ \widetilde{\Phi}_{\pm}$ to guarantee the
periodicity in $x$ of period $L_1 =\frac{2\pi}{\k_1}$.

The Melnikov vectors for the Davey-Stewartson~II equations are
given by,
\begin{eqnarray}
  \U^{\pm} &=& \left( \begin{array}{c}
\Psi^{\pm}_2 \widehat{\Psi}^{\pm}_2 \\[1ex]
\Psi^{\pm}_1 \widehat{\Psi}^{\pm}_1    
    \end{array}\right)^- +S \left(
\begin{array}{c}
     \Psi^{\pm}_2 \widehat{\Psi}^{\pm}_2 \\[1ex]
\Psi^{\pm}_1 \widehat{\Psi}^{\pm}_1    
    \end{array} \right) \, , \label{mv1}\\[2ex]
  \U_{\pm} &=& \left(
\begin{array}{c}
      \widetilde{\Phi}_{\pm}^{(2)} \widehat{\Phi}_{\pm}^{(2)} \\[1ex]
\widetilde{\Phi}_{\pm}^{(1)} \widehat{\Phi}_{\pm}^{(1)}    
\end{array}\right)^- +S \left(
\begin{array}{c}
     \widetilde{\Phi}_{\pm}^{(2)} \widehat{\Phi}_{\pm}^{(2)} \\[1ex]
\widetilde{\Phi}_{\pm}^{(1)} \widehat{\Phi}_{\pm}^{(1)}    
    \end{array} \right) \, , \label{mv2}
\end{eqnarray}
where $S = \left ( \begin{array}{lr} 0 & 1 \\ 1 & 0 \end{array} \right )$. 
The corresponding Melnikov functions (\ref{mlf2}) are given by,
\begin{eqnarray}
M^{\pm} &=& \int^{\infty}_{- \infty} \langle \U^{\pm} \, , \, 
  \vec{G} \rangle \, dt \nonumber \\
&=& 2  \int^{\infty}_{- \infty} \int^{2\pi}_0  \int^{2\pi}_0 
  R_e \bigg \{[ \Psi^{\pm}_2  \widehat{\Psi}^{\pm}_2 ] 
    f(Q, \overline{Q}) \nonumber \\
& & + [ \Psi^{\pm}_1
    \widehat{\Psi}^{\pm}_1]
    \overline{f}(Q, \overline{Q}) \bigg \} \, dx \, dy \, dt \, 
  , \label{mf1}\\ 
M_{\pm} &=& \int^{\infty}_{- \infty} \langle \U_{\pm} \, , \, 
  \vec{G} \rangle \, dt \nonumber \\
&=&  2  \int^{\infty}_{- \infty} \int^{2\pi}_0  \int^{2\pi}_0 
R_e \bigg \{ [ \widetilde{\Phi}^{(2)}_{\pm}
  \widehat{\Phi}^{(2)}_{\pm}] f(Q, \overline{Q}) \nonumber \\
& & + [ \widetilde{\Phi}^{(1)}_{\pm}
  \widehat{\Phi}^{(1)}_{\pm}] \overline{f}(Q, \overline{Q})
  \bigg \}\, dx \, dy \, dt \, , \label{mf2}
\end{eqnarray}
where $Q$ is given in (\ref{newsl}), $\Psi^{\pm}$ is given in
(\ref{rnef1}), $\widetilde{\Phi}_{\pm}$ is given in (\ref{rnef2}) 
and (\ref{kfr}), $\widehat{\Psi}^{\pm}$ is given in (\ref{rnef1}) 
and (\ref{cref1}), and $\widehat{\Phi}_{\pm}$ is given in
(\ref{rnef2}) and (\ref{cref2}).  As given in (\ref{mlf3}), the
above formulas also apply when we consider even function $Q$ in
both $x$ and $y$.

\chapter{Invariant Manifolds}

Invariant manifolds have attracted intensive studies which led 
to two main approaches: Hadamard's method \cite{Had01} \cite{Fen71} 
and Perron's method \cite{Per30} \cite{CLL91}. For example, for a partial 
differential equation of the form
\[
\pa_t u = L u + N(u)\ ,
\]
where $L$ is a linear operator and $N(u)$ is the nonlinear term, if the 
following two ingredients
\begin{enumerate}
\item the gaps separating the unstable, center, and stable spectra 
of $L$ are large enough,
\item the nonlinear term $N(u)$ is Lipschitzian in $u$ with small 
Lipschitz constant,
\end{enumerate}
are available, then establishing the existence of unstable, center, and 
stable manifolds is rather straightforward. Building invariant manifolds 
when any of the above conditions fails, is a very challenging and interesting 
problem \cite{Li01b}.

There has been a vast literature on invariant manifolds. A good starting 
point of reading can be from the references \cite{Kel67} \cite{Fen71} 
\cite{CLL91}. Depending upon the emphasis on the specific problem,
one may establish invariant manifolds for a specific flow, or investigate 
the persistence of existing invariant manifolds under perturbations to 
the flow. 

In specific applications, most of the problems deal with manifolds with 
boundaries. In this context, the relevant concepts are overflowing invariant, 
inflowing invariant, and locally invariant submanifolds, defined in the 
Chapter on General Setup and Concepts.

\section{Nonlinear Schr\"odinger Equation Under Regular Perturbations
\label{rpsec}}

Persistence of invariant manifolds depends upon the nature of the 
perturbation. Under the so-called regular perturbations, i.e., the 
perturbed evolution operator is $C^1$ close to the unperturbed one,
for any fixed time; invariant manifolds persist ``nicely''. Under 
other singular perturbations, this may not be the case.

Consider the regularly perturbed nonlinear Schr\"odinger (NLS) equation 
\cite{LMSW96} \cite{LW97b},
\begin{equation}
iq_t = q_{xx} +2 [ |q|^2 - \om^2] q +i \e [\hat{\pa}^2_xq - \al q +\be ] \ ,
\label{rpnls}
\end{equation}
where $q = q(t,x)$ is a complex-valued function of the two real 
variables $t$ and $x$, $t$ represents time, and $x$ represents
space. $q(t,x)$ is subject to periodic boundary condition of period 
$2 \pi$, and even constraint, i.e., 
\[
q(t,x + 2 \pi) = q(t,x)\ , \ \ q(t,-x) = q(t,x)\ .
\]
$\om$ is a positive constant, $\al >0$ and $\be >0$ are constants, 
$\hat{\pa}^2_x$ is a bounded Fourier multiplier,
\[
\hat{\pa}^2_x q = -\sum_{k=1}^{N}k^2 \xi_k \tq_k \cos kx\ ,
\]
$\xi_k = 1$ when $k \leq N$, $\xi_k = 8k^{-2}$ when $k>N$, for some 
fixed large $N$, and $\e > 0$ is the perturbation parameter.

One can prove the following theorems \cite{LMSW96} \cite{LW97b}.
\begin{theorem}[Persistence Theorem]
For any integers $k$ and $n$ ($1 \leq k,n <\infty$), there exist a 
positive constant $\e_0$ and a neighborhood $\U$ of the circle 
$S_\om = \{ q \ | \ \pa_x q = 0, \ |q| = \om, \ 1/2 < \om <1 \}$ in 
the Sobolev space $H^k$, such 
that inside $\U$, for any $\e \in (-\e_0, \e_0)$, there exist $C^n$ 
locally invariant submanifolds $W^{cu}_\e$ and $W^{cs}_\e$ of 
codimension 1, and $W^c_\e$ ($=W^{cu}_\e \cap W^{cs}_\e$)
of codimension 2 under the evolution operator $F^t_\e$ given by 
(\ref{rpnls}). When $\e =0$, $W^{cu}_0$, $W^{cs}_0$, and $W^c_0$ are 
tangent to the center-unstable, center-stable, and center subspaces 
of the circle of fixed points $S_\om$, respectively.
$W^{cu}_\e$, $W^{cs}_\e$, and $W^c_\e$ 
are $C^n$ smooth in $\e$ for $\e \in (-\e_0,\e_0)$.
\label{Persthm}
\end{theorem}
$W^{cu}_\e$, $W^{cs}_\e$, and $W^c_\e$ are called persistent 
center-unstable, center-stable, and center submanifolds near $S_\om$ 
under the evolution operator $F^t_\e$ given by (\ref{rpnls}).
\begin{theorem}[Fiber Theorem]
Inside the persistent center-unstable submanifold $W^{cu}_\e$ near $S_\om$, 
there exists a family of $1$-dimensional $C^n$ smooth submanifolds (curves) 
$\{ \F^{(u,\e)}(q): q \in W^c_\e \}$, called unstable fibers:
\begin{itemize}
\item $W^{cu}_\e$ can be represented as a union of these fibers,
\[
W^{cu}_\e =\bigcup _{q \in W^c_\e} \F^{(u,\e)}(q).
\]
\item $\F^{(u,\e)}(q)$ depends $C^{n-1}$ smoothly on both $\e$ and $q$
for $\e \in (-\e_0,\e_0)$ and $q \in W^c_\e$, in the sense that $\W$ defined
by
\[
\W= \bigg \{ (q_1,q,\e)\ \bigg | \ q_1 \in \F^{(u,\e)}(q), 
\ q \in W^c_\e,\ \e \in (-\e_0,\e_0) \bigg \}
\]
is a $C^{n-1}$ smooth submanifold of $H^k \times H^k \times (-\e_0,\e_0)$.
\item Each fiber $\F^{(u,\e)}(q)$ intersects $W^c_\e$ transversally at 
$q$, two fibers $\F^{(u,\e)}(q_1)$ and $\F^{(u,\e)}(q_2)$ are either
disjoint or identical.
\item The family of unstable fibers $\{ \F^{(u,\e)}(q): q \in W^c_\e \}$ is
negatively invariant, in the sense that the family of fibers commutes 
with the evolution operator $F^t_\e$ in the following way:
\[
F^t_\e(\F^{(u,\e)}(q)) \subset \F^{(u,\e)}(F^t_\e(q))
\]
for all $q \in W^c_\e$ and all $t \leq 0$ such that 
$\bigcup_{\tau \in [t,0]}F^\tau_\e(q) \subset W^c_\e$.
\item There are positive constants $\k$ 
and $C$ which are independent of $\e$ such that if $q \in W^c_\e$ and 
$q_1 \in \F^{(u,\e)}(q)$, then
\[
\bigg \| F^t_\e(q_1) - F^t_\e(q) \bigg \| \leq C e^{\k t} 
\bigg \| q_1 -q \bigg \|,
\]
for all $t \leq 0$ such that $\bigcup_{\tau \in [t,0]}F^\tau_\e(q) 
\subset W^c_\e$, where $\| \ \|$ denotes $H^k$ norm of periodic 
functions of period $2\pi$.
\item For any $q, p \in W^c_\e$, $q \neq p$, any $q_1 \in 
\F^{(u,\e)}(q)$ and any $p_1 \in \F^{(u,\e)}(p)$; if
\[
F^t_\e(q), F^t_\e(p) \in W^c_\e,\ \ \forall t \in (-\infty, 0],
\]
and
\[
\| F^t_\e(p_1) -F^t_\e(q)\| \ra 0,\ \ \mbox{as}\ t\ra -\infty;
\]
then
\[
\bigg \{ {\| F^t_\e(q_1) -F^t_\e(q)\| \over \| F^t_\e(p_1) -F^t_\e(q)\|}
\bigg \} \bigg / e^{ {1\over 2} \k t} \ra 0, \ \ \mbox{as}\ t\ra -\infty.
\]
\end{itemize}
Similarly for $W^{cs}_\e$.
\label{fiberthm}
\end{theorem}
When $\e =0$, certain low-dimensional invariant submanifolds of 
the invariant manifolds, have explicit representations through 
Darboux transformations. Specifically, the periodic orbit (\ref{4.7})
where $1/2 < c < 1$, has two-dimensional stable and unstable manifolds 
given by (\ref{4.13}). Unstable and stable fibers with bases along 
the periodic orbit also have expressions given by (\ref{4.13}).

\section{Nonlinear Schr\"odinger Equation Under Singular Perturbations
\label{spsec}}

Consider the singularly perturbed nonlinear Schr\"odinger equation 
\cite{Li01b},
\begin{equation}
iq_t = q_{xx} +2 [|q|^2 - \om^2] q +i \e [q_{xx} - \al q +\be ] \ ,
\label{spnls}
\end{equation}
where $q = q(t,x)$ is a complex-valued function of the two real 
variables $t$ and $x$, $t$ represents time, and $x$ represents
space. $q(t,x)$ is subject to periodic boundary condition of period 
$2 \pi$, and even constraint, i.e., 
\[
q(t,x + 2 \pi) = q(t,x)\ , \ \ q(t,-x) = q(t,x)\ .
\]
$\om \in (1/2, 1)$ is a positive constant, $\al >0$ and $\be >0$ 
are constants, and $\e > 0$ is the perturbation parameter. 

Here the perturbation
term $\e \pa_x^2$ generates the semigroup $e^{\e t \pa_x^2}$, the 
regularity of the invariant mainfolds with respect to $\e$ at $\e =0$ 
will be closely related to the regularity of the semigroup $e^{\e t \pa_x^2}$
with respect to $\e$ at $\e =0$. Also the singular perturbation
term $\e \pa_x^2 q$ breaks the spectral gap separating the center 
spectrum and the stable spectrum. Therefore, standard invariant manifold
results do not apply. Invariant manifolds do not persist ``nicely''. 
Nevertheless, certain persistence results do hold.

One can prove the following unstable fiber theorem and center-stable 
manifold theorem \cite{Li01b}.
\begin{theorem}[Unstable Fiber Theorem]
Let ${\mathcal A}$ be the annulus: 
${\mathcal A} = \{ q \ | \ \pa_x q = 0, \ 1/2 < |q| < 1 \}$, for any 
$p\in {\mathcal A}$, there is an unstable 
fiber ${\mathcal F}^+_p$ which is a curve. ${\mathcal F}^+_p$ has
the following properties:
\begin{enumerate}
\item ${\mathcal F}^+_p$ is a $C^1$ smooth in 
$H^k$-norm, $k\geq 1$.
\item ${\mathcal F}^+_p$ is also $C^1$ smooth in $\epsilon$, $\alpha$, 
$\beta$, $\omega$, and $p$ in $H^k$-norm, $k \geq 1$,
$\epsilon \in [0,\epsilon_0)$ for some $\epsilon_0>0$.
\item ${\mathcal F}^+_p$ has the exponential decay property: $\forall
p_1\in {\mathcal F}^+_p$,
\[
\frac{\| F^tp_1-F^tp\|_k}{\| p_1-p\|_k}\leq 
Ce^{\mu t},\quad \forall t\leq 0,
\]
where $F^t$ is the evolution operator, $\mu > 0$.
\item $\{ {\mathcal F}^+_p\}_{p\in {\mathcal A}}$ forms an invariant 
family of unstable fibers,
\[
F^t{\mathcal F}^+_p\subset {\mathcal F}^+_{F^tp}\ ,\quad 
\forall t\in [-T,0],
\]
and $\forall T>0$ ($T$ can be $+\infty$), such that $F^tp\in 
{\mathcal A}$, $\forall t\in [-T,0]$.
\end{enumerate}
\label{UFT}
\end{theorem}
\begin{theorem}[Center-Stable Manifold Theorem]
There exists a $C^1$ smooth codimension 1 
locally invariant center-stable manifold $W^{cs}_k$ in a neighborhood of 
the annulus ${\mathcal A}$ (Theorem \ref{UFT}) in $H^k$ for any $k\geq 1$. 
\begin{enumerate}
\item At points in the subset $W^{cs}_{k+4}$ of 
$W^{cs}_k$, $W^{cs}_k$ is $C^1$ smooth in $\epsilon$ in $H^k$-norm 
for $\epsilon \in [0,\epsilon_0)$ and some $\epsilon_0 >0$. 
\item $W^{cs}_k$ is $C^1$ smooth in ($\alpha ,\beta ,\omega$).
\end{enumerate}
\label{CSM}
\end{theorem}
\begin{remark}\label{csnr} $C^1$ regularity in $\epsilon$ is 
crucial in locating a homoclinic orbit. As can be seen later, one
has detailed information on certain unperturbed (i.e. $\epsilon=0$) 
homoclinic orbit, which will be used in tracking candidates for a
perturbed homoclinic orbit. In particular, Melnikov measurement will 
be needed. Melnikov measurement measures zeros of
$\mathcal{O}(\epsilon)$ signed distances, thus, the perturbed orbit 
needs to be $\mathcal{O}(\epsilon)$ close to the unperturbed orbit in
order to perform Melnikov measurement.\end{remark}

\section{Proof of the Unstable Fiber Theorem \ref{UFT}}

Here we give the proof of the unstable fiber theorem \ref{UFT}, 
proofs of other fiber theorems in this chapter are easier.

\subsection{The Setup of Equations}

First, write $q$ as
\begin{equation}
q(t,x)=[\rho(t)+f(t,x)]e^{i\theta (t)},
\label{pcd} 
\end{equation}
where $f$ has zero spatial mean. We use the notation $\langle \cdot
\rangle$ to denote spatial mean,
\begin{equation}\label{mean} \langle q\rangle 
=\frac{1}{2\pi}\int^{2\pi}_0qdx.\end{equation}
Since the $L^2$-norm is an action variable when $\epsilon =0$, it is 
more convenient to replace $\rho$ by:
\begin{equation}\label{L2n} I=\langle |q|^2\rangle =\rho^2+\langle 
|f|^2\rangle.\end{equation}
The final pick is
\begin{equation}\label{Jpc}J=I-\omega^2.\end{equation}
In terms of the new variables $(J,\theta, f)$, Equation \eqref{spnls} 
can be rewritten as
\begin{align}\label{nc1} \dot{J}&=\epsilon \left[-2\alpha 
(J+\omega^2)+2\beta \sqrt{J+\omega^2}\cos \theta \right]+\epsilon
\mathcal{R}^J_2,\\
\label{nc2} \dot{\theta}&=-2J-\epsilon \beta \frac{\sin 
\theta}{\sqrt{J+\omega^2}}+\mathcal{R}^\theta_2,\\
\label{nc3}f_t&=L_\epsilon f+V_\epsilon 
f-i\mathcal{N}_2-i\mathcal{N}_3,\end{align}
where
\begin{align}\label{wnc1} L_\epsilon f&=-if_{xx}+\epsilon (-\alpha 
f+f_{xx})-i2\omega ^2(f+\bar{f}),\\
\label{wnc2}V_\epsilon f&=-i2J(f+\bar{f})+i\epsilon \beta f\frac{\sin 
\theta}{\sqrt{J+\omega^2}},\\
\label{wnc3}\mathcal{R}^J_2&=-2\langle |f_x|^2\rangle +2\beta \cos 
\theta \left[ \sqrt{J+\omega^2-\langle
|f|^2\rangle}-\sqrt{J+\omega^2}\right],\\
\begin{split}\label{wnc4}\mathcal{R}^\theta_2&=-\langle 
(f+\bar{f})^2\rangle -\frac{1}{\rho}\langle |f|^2(f+\bar{f})\rangle\\
&\quad  -\epsilon
\beta\sin
\theta
\left[
\frac{1}{\sqrt{J+\omega^2-\langle 
|f|^2\rangle}}-\frac{1}{\sqrt{J+\omega^2}}\right],\end{split}\\
\label{wnc5} \mathcal{N}_2&=2\rho [2(|f|^2-\langle |f|^2\rangle 
)+(f^2-\langle f^2\rangle )],\\
\begin{split}\label{wnc6} \mathcal{N}_3&=-\langle 
f^2+\bar{f}^2+6|f|^2\rangle f+2(|f|^2f-\langle |f|^2f\rangle )\\
&\quad -\frac{1}{\rho} \langle |f|^2(f+\bar{f})\rangle f-2\langle 
|f|^2\rangle \bar{f}\\
&\quad -\epsilon \beta \sin \theta \left[ \frac{1}{\sqrt{J+\omega^2-\langle
|f|^2\rangle}}-\frac{1}{\sqrt{J+\omega^2}}\right]f.\end{split}\end{align}

\begin{remark} The singular perturbation term ``$\epsilon \pa_x^2q$" 
can be seen at two locations, $L_\epsilon$ and $\mathcal{R}^J_2$
(\ref{wnc1},\ref{wnc3}). The singular perturbation term $\langle 
|f_x|^2\rangle $ in $\mathcal{R}^J_2$ does not create any difficulty.
Since $H^1$ is a Banach algebra~\cite{Ada75}, this term is still of 
quadratic order, $\langle |f_x|^2\rangle \sim \mathcal{O}(\|
f\|^2_1)$.\end{remark}
\begin{lemma} The nonlinear terms have the orders:
\begin{equation*}\begin{split}&|\mathcal{R}^J_2|\sim \mathcal{O}(\| 
f\|^2_s),\quad |\mathcal{R}^\theta_2|\sim \mathcal{O}(\|
f\|^2_s),\\
& \|\mathcal{N}_2\|_s\sim
\mathcal{O} (\| f\|^2_s),\quad \| \mathcal{N}_3\|_s\sim 
\mathcal{O}(\| f\|^3_s),\quad (s\geq 
1).\end{split}\end{equation*}\end{lemma}

Proof: The proof is an easy direct verification. Q.E.D.

\subsection{The Spectrum of $L_\epsilon$}

The spectrum of $L_\epsilon$ consists of only point spectrum. The 
eigenvalues of $L_\epsilon$ are:
\begin{equation}\label{leev} \mu^\pm_k=-\epsilon (\alpha+k^2)\pm 
k\sqrt{4\omega^2-k^2},\quad (k=1,2,\ldots );\end{equation}
where $\omega \in \left( \frac{1}{2},1\right)$, only $\mu^\pm_1$ are 
real, and $\mu^\pm_k$ are complex for $k>1$.

\setlength{\unitlength}{0.8in}
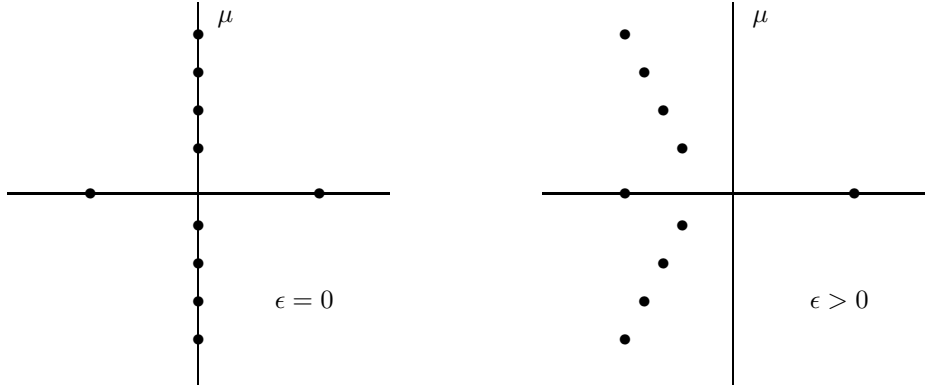
\begin{figure}
\begin{picture}(6.5,2.5)
\put(0,1.25){\line(2,0){2.5}}
\put(3.5,1.25){\line(2,0){2.5}}
\put(1.25,0){\line(0,1){2.5}}
\put(4.75,0){\line(0,1){2.5}}
\put(1.75,.5){$\epsilon =0$}
\put(5.25,.5){$\epsilon >0$}
\put(1.375,2.375){$\mu$}
\put(4.875,2.375){$\mu$}
\put(.5,1.21){$\bullet$}
\put(2,1.21){$\bullet$}
\put(4,1.21){$\bullet$}
\put(5.5,1.21){$\bullet$}
\put(1.21,.25){$\bullet$}
\put(1.21,.5){$\bullet$}
\put(1.21,.75){$\bullet$}
\put(1.21,1){$\bullet$}
\put(1.21,1.5){$\bullet$}
\put(1.21,1.75){$\bullet$}
\put(1.21,2){$\bullet$}
\put(1.21,2.25){$\bullet$}
\put(4,.25){$\bullet$}
\put(4.125,.5){$\bullet$}
\put(4.25,.75){$\bullet$}
\put(4.375,1){$\bullet$}
\put(4,2.25){$\bullet$}
\put(4.125,2){$\bullet$}
\put(4.25,1.75){$\bullet$}
\put(4.375,1.5){$\bullet$}
\end{picture}
\caption{The point spectra of the linear operator $L_\e$.}
\label{sgcb}
\end{figure}

The main difficulty introduced by the singular 
perturbation $\epsilon \partial^2_x f$ is the breaking of the 
spectral
gap condition. Figure~\ref{sgcb} shows the distributions of the 
eigenvalues when $\epsilon =0$ and $\epsilon \neq 0$. It clearly shows
the breaking of the stable spectral gap condition. As a result, 
center and center-unstable manifolds do not necessarily
persist. On the other hand, the unstable spectral gap condition is 
not broken. This gives the hope for the persistence of center-stable
manifold. Another case of persistence can be described as follows:
Notice that the plane $\Pi$,
\begin{equation}\label{Pi} \Pi=\{ q\mid \ \partial_xq=0\},\end{equation}
is invariant under the flow \eqref{spnls}. When $\epsilon =0$, there is an
unstable fibration with base points in a neighborhood of the circle 
$S_\omega$ of fixed points,
\begin{equation}
S_\om=\{ q\in \Pi \mid \ |q|=\om \},
\label{rcl} 
\end{equation}
in $\Pi$, as an invariant sub-fibration of the
unstable Fenichel fibration with base points in the center manifold. 
When $\epsilon >0$, the center manifold may not persist, but $\Pi$
persists, moreover, the unstable spectral gap condition is not 
broken, therefore, the unstable sub-fibration with base points in 
$\Pi$ may persist. Since the semiflow generated by \eqref{spnls} 
is not a $C^1$ perturbation of
that generated by the unperturbed NLS due to the singular 
perturbation $\epsilon \partial^2_x$, standard results on 
persistence can not be applied.

The eigenfunctions corresponding to the real eigenvalues are:
\begin{equation}
\label{leef} 
e^\pm_1=e^{\pm i\vth_1}\cos x,\quad e^{\pm i\vth_1}=\frac{1\mp
i\sqrt{4\omega^2-1}}{2\omega}.
\end{equation}
Notice that they are independent of $\epsilon$.
The eigenspaces corresponding to the complex conjugate pairs of 
eigenvalues are given by:
\begin{equation*}E_1=\ \mbox{span}_{\CC}\{ \cos  x \}.\end{equation*}
and have real dimension $2$.

\subsection{The Re-Setup of Equations}

For the goal of this subsection, we need to single out the 
eigen-directions \eqref{leef}. Let
\begin{equation*}f=\sum_{\pm }\xi_1^\pm e^\pm_1+h,\end{equation*}
where $\xi^\pm_1$ are real variables, and
\begin{equation*}\langle h\rangle=\langle h\cos x \rangle =0.\end{equation*}
In terms of the coordinates $(\xi_1^\pm , J,\theta ,h)$, 
\eqref{nc1}-\eqref{nc3} can be rewritten as:
\begin{align}\label{eig1} 
\dot{\xi}^+_1&=\mu^+_1\xi^+_1+V^+_1\xi^+_1+\mathcal{N}^+_1, \\
\label{eig2} \dot{J}&=\epsilon \left[ -2\alpha (J+\omega^2)+2\beta 
\sqrt{J+\omega^2}\cos \theta \right] +\epsilon \mathcal{R}^J_2,\\
\label{eig3} \dot{\theta} & = -2J-\epsilon \beta \frac{\sin 
\theta}{\sqrt{J+\omega^2}}+\mathcal{R}^\theta_2,\\
\label{eig4} h_t&=L_\epsilon h+V_\epsilon h+\tilde{\mathcal{N}},\\
\label{eig5} 
\dot{\xi}^-_1&=\mu^-_1\xi^-_1+V^-_1\xi^-_1+\mathcal{N}^-_1,\end{align}
where $\mu^\pm_1$ are given in \eqref{leev}, $\mathcal{N}^\pm_1$ and 
$\tilde{\mathcal{N}}$ are projections of $-i\mathcal{N}_2-i\mathcal{N}_3$ 
to the corresponding directions, and
\begin{align*}V^+_1\xi^+_1&=2c_1J(\xi^+_1+\xi^-_1)+\epsilon \beta 
\frac{\sin \theta}{\sqrt{J+\omega^2}}(c^+_1\xi^+_1-c^-_1\xi^-_1),\\
V^-_1\xi^-_1&=-2c_1J(\xi^+_1+\xi^-_1)+\epsilon \beta \frac{\sin 
\theta}{\sqrt{J+\omega^2}}(c_1^-\xi^+_1-c^+_1\xi^-_1),\\
c_1&=\frac{1}{\sqrt{4\omega^2-1}},\quad 
c^+_1=\frac{2\omega^2-1}{\sqrt{4\omega^2-1}},\quad
c^-_1=\frac{2\omega^2}{\sqrt{4\omega^2-1}}.\end{align*}

\subsection{A Modification}

\begin{definition} For any $\delta >0$, we define the annular 
neighborhood of the circle $S_\omega$ \eqref{rcl} as
\begin{equation*}\mathcal{A}(\delta)=\{ (J,\theta )\mid \ |J|<\delta 
\}.\end{equation*}
\end{definition}

To apply the Lyapunov-Perron's method, it is standard and necessary to 
modify the $J$ equation so that $\mathcal{A}(4\delta)$ is
overflowing invariant. Let $\eta \in C^\infty (R,R)$ be a ``bump" function:
\begin{equation*}\eta=\begin{cases} 0, & \text{in } (-2,2)\cup 
(-\infty,-6)\cup (6,\infty),\\
1, & \text{in } (3,5),\\
-1, & \text{in } (-5,-3),\end{cases}\end{equation*}
and $|\eta'|\leq 2$, $|\eta''|\leq C$. We 
modify the $J$ equation \eqref{eig2} as follows:
\begin{equation}\label{meig2}\dot{J} =\epsilon b\eta ( 
J/\delta)+\epsilon \left[ -2\alpha (J+\omega^2)+2\beta
\sqrt{J+\omega^2}\cos
\theta \right] +\epsilon \mathcal{R}^J_2,\end{equation}
where $b>2 (2\alpha \omega^2+2\beta \omega)$. Then 
$\mathcal{A}(4\delta)$ is overflowing invariant. There are two main 
points in
adopting the bump function:
\begin{enumerate}\item One needs $\mathcal{A}(4\delta)$ to be 
overflowing invariant so that a Lyapunov-Perron type integral
equation can be set up along orbits in $\mathcal{A}(4\delta)$ for 
$t\in (-\infty ,0)$.
\item One needs the vector field inside $\mathcal{A}(2\delta)$ to be 
unchanged so that results for the modified system can be claimed
for the original system in $\mathcal{A}(\delta)$.\end{enumerate}

\begin{remark} Due to the singular perturbation, the real part of 
$\mu^\pm_k$ approaches $-\infty$ as $k\to \infty$. Thus the $h$
equation \eqref{eig4} can not be modified to give overflowing flow. 
This rules out the construction of unstable fibers with base points
having general $h$ coordinates.\end{remark}

\subsection{Existence of Unstable Fibers}

For any $(J_0,\theta_0)\in \mathcal{A}(4\delta)$, let
\begin{equation}\label{borb} J=J_*(t),\quad \theta=\theta_*(t), \quad 
t\in (-\infty ,0],\end{equation}
be the backward orbit of the modified system \eqref{meig2} and 
\eqref{eig3} with the initial point $(J_0,\theta_0)$. If
\begin{equation*}(\xi^+_1(t),J_*(t)+\tilde{J}(t),\theta_*(t)+\tilde{\theta}(t),h(t),\xi^-_1(t))\end{equation*}
is a solution of the modified full system, then one has
\begin{eqnarray}\dot{\xi}^+_1 &=& \mu^+_1\xi^+_1+F^+_1, 
\label{meq1}\\
u_t &=& Au+F,\label{meq2} 
\end{eqnarray}
where
\begin{align*} u&=\begin{pmatrix} \tilde{J} \\ \tilde{\theta}\\ h\\ 
\xi^-_1 \end{pmatrix}, \quad A=\begin{pmatrix} 0 & 0& 0 & 0
\\
-2 & 0 & 0 & 0 \\ 0 & 0 & L_\epsilon & 0 \\
0 & 0 & 0 & \mu^-_1 \end{pmatrix},\quad F=\begin{pmatrix} F_J\\ 
F_\theta\\ F_h\\
F^-_1\end{pmatrix},\\
&F^+_1  = V^+_1\xi^+_1+\mathcal{N}^+_1,\\
&F_J=\epsilon b\left[ \eta (J/\delta)-\eta 
(J_*(t)/\delta)\right]+\epsilon \left[
-2\alpha \tilde{J} +2\beta \sqrt{J+\omega^2}\cos \theta \right.\\
&\quad -\left. 2\beta \sqrt{J_*(t)+\omega^2}\cos \theta_*(t)\right] +\epsilon
\mathcal{R}^J_2,\\
&F_\theta = -\epsilon \beta \frac{\sin 
\theta}{\sqrt{J+\omega^2}}+\epsilon \beta \frac{\sin
\theta_*(t)}{\sqrt{J_*(t)+\omega^2}}+\mathcal{R}^\theta_2,\\
&F_h=V_\epsilon h+\tilde{\mathcal{N}},\\
&F_1^-=V^-_1\xi^-_1+\mathcal{N}^-_1,\\
&J=J_*(t)+\tilde{J},\quad \theta=\theta_*(t)+\tilde{\theta}.
\end{align*}
\nid
System \eqref{meq1}-\eqref{meq2} can be written in the equivalent 
integral equation form:
\begin{align}\label{eit1} \xi^+_1(t)&= 
\xi^+_1(t_0)e^{\mu^+_1(t-t_0)}+\int^t_{t_0}e^{\mu^+_1(t-\tau)}F^+_1(\tau 
)d\tau, \\
\label{eit2} u(t)&= 
e^{A(t-t_0)}u(t_0)+\int^t_{t_0}e^{A(t-\tau)}F(\tau )d\tau.\end{align}
By virtue of the gap between $\mu^+_1$ and the real parts of the 
eigenvalues of $A$, one can introduce the following space: For $\sigma
\in \left( \frac{\mu^+_1}{100},\frac{\mu^+_1}{3}\right)$, and $n\geq 1$, let
\begin{equation*}\begin{split} G_{\sigma ,n}&=\bigg \{ 
g(t)=(\xi^+_1(t),u(t))\bigg | \  t \in (-\infty ,0], \ g(t)\text{ is
continuous} \\ &\quad \text{in } t\text{ in } H^n\text{ norm }, 
\ \| g\|_{\sigma ,n}=\sup_{t\leq 0}e^{-\sigma t} [
|\xi^+_1(t)|+\| u(t)\|_n ] <\infty\bigg \}\ . 
\end{split}\end{equation*}
$G_{\sigma ,n}$ is a Banach space under the norm $\| \cdot \|_{\sigma 
,n}$. Let $\mathcal{B}_{\sigma ,n}(r)$ denote the ball in
$G_{\sigma ,n}$ centered at the origin with radius $r$. Since $A$ 
only has point spectrum, the spectral mapping theorem is valid. It is
obvious that for $t\geq 0$,
\begin{equation*}\| e^{At}u\|_n\leq C(1+t)\| u\|_n,\end{equation*}
for some constant $C$. Thus, if $g(t)\in \mathcal{B}_{\sigma 
,n}(r)$, $r<\infty$ is a solution of \eqref{eit1}-\eqref{eit2}, by
letting $t_0\to -\infty$ in \eqref{eit2} and setting $t_0=0$ in 
\eqref{eit1}, one has
\begin{align}\label{per1} 
\xi^+_1(t)&=\xi^+_1(0)e^{\mu^+_1t}+\int^t_0e^{\mu^+_1(t-\tau)}F^+_1(\tau 
)d\tau,\\
\label{per2} u(t)&=\int^t_{-\infty}e^{A(t-\tau)}F(\tau )d\tau.\end{align}
For $g(t)\in \mathcal{B}_{\sigma ,n}(r)$, let $\Gamma(g)$ be the 
map defined by the right hand side of \eqref{per1}-\eqref{per2}.
Then a solution of \eqref{per1}-\eqref{per2} is a fixed point of 
$\Gamma$. For any $n\geq 1$ and $\epsilon <\delta^2$, and $\delta$ and
$r$ are small enough, $F^+_1$ and $F$ are Lipschitz in $g$ with small 
Lipschitz constants. Standard arguments of the Lyapunov-Perron's
method readily imply the existence of a fixed point $g_*$ of $\Gamma$ 
in $\mathcal{B}_{\sigma ,n}(r)$.

\subsection{Regularity of the Unstable Fibers in $\e$}

The difficulties lie in the investigation on the regularity of $g_*$ with 
respect to $(\epsilon ,\alpha ,\beta ,\omega ,J_0,\theta_0,\xi^+_1(0))$. 
The most difficult one is the regularity with respect to 
$\epsilon$ due to the singular perturbation.
Formally differentiating $g_*$ in \eqref{per1}-\eqref{per2} with 
respect to $\epsilon$, one gets
\begin{align}\label{dper1} \xi^+_{1,\epsilon}(t)&= 
\int^t_0e^{\mu^+_1(t-\tau)}\left[ \partial_uF^+_1\cdot u_\epsilon
+\partial_{\xi_1^+}F^+_1\cdot 
\xi^+_{1,\epsilon}\right](\tau )d\tau
+\mathcal{R}^+_1(t),\\
u_\epsilon (t)&=\int^t_{-\infty}e^{A(t-\tau)}\left[ \partial_uF\cdot 
u_\epsilon +\partial_{\xi_1^+}F\cdot
\xi^+_{1,\epsilon}\right](\tau )d\tau 
+\mathcal{R}(t),\label{dper2}\end{align}
where
\begin{align}\begin{split}\label{wdp1}
\mathcal{R}^+_1(t)&=\xi^+_1(0)\mu^+_{1,\epsilon}te^{\mu^+_1t}+\int^t_0\mu^+_{1,\epsilon}(t-\tau)e^{\mu^+_1(t-\tau)}F^+_1(\tau 
)d\tau\\
&\quad + \int^t_0e^{\mu^+_1(t-\tau)}[\partial_\epsilon 
F^+_1+\partial_{u_*}F^+_1\cdot u_{*,\epsilon}](\tau )d\tau,\end{split}\\
\begin{split}\label{wdp2} \mathcal{R}(t)&= 
\int^t_{-\infty}(t-\tau)A_\epsilon e^{A(t-\tau)}F(\tau )d\tau\\
&\quad +\int^+_{-\infty}e^{A(t-\tau)}[\partial_\epsilon 
F+\partial_{u_*}F\cdot u_{*,\epsilon}](\tau)d\tau,\end{split}\\
\label{wdp3} \mu^+_{1,\epsilon}&=-(\alpha +1),\\
A_\epsilon &=\begin{pmatrix} 0 & 0 & 0 & 0 \\
0 & 0 & 0 & 0 \\
0 & 0 & -\alpha+\partial^2_x & 0 \\
0 & 0 & 0 & -(\alpha+1) \end{pmatrix},\\
u_*&=(J_*,\theta_*,0,0)^T,\label{wdp4}\end{align}
where $T=$\ transpose, and $(J_*,\theta_*)$ are given in \eqref{borb}. 
The troublesome terms are the ones containing $A_\epsilon $ or
$u_{*,\epsilon}$ in \eqref{wdp1}-\eqref{wdp2}.
\begin{equation} 
\| A_\epsilon F\|_n\leq C\| F\|_{n+2}\leq \tilde{c} 
( \|u\|_{n+2}+|\xi^+_1|),
\label{upre}
\end{equation}
where $\tilde{c}$ is small when $(\ \cdot \ )$ on the right hand side 
is small.
\begin{align*}\begin{split} \partial_{J_*}F_J\cdot J_{*,\epsilon}&= 
\frac{\epsilon}{\delta}b[ \eta'(
J/\delta)-\eta'( J_*/\delta)] 
J_{*,\epsilon}\\
&\quad + \epsilon [ \beta \frac{\cos 
\theta}{\sqrt{J+\omega^2}}-\beta\frac{\cos 
\theta_*}{\sqrt{J_*+\omega^2}}]
J_{*,\epsilon}\\
&\quad +\text{easier terms}.\end{split}\\
\begin{split} |\partial_{J_*}F_J\cdot J_{*,\epsilon}|&\leq 
\frac{\epsilon}{\delta^2}b\sup_{0\leq \hat{\gamma}\leq 1}\left| \eta''(
[\hat{\gamma} J_*+(1-\hat{\gamma})J]/\delta)\right | \ |\tilde{J}| 
\ |J_{*,\epsilon}|\\
&\quad + \epsilon \beta 
C(|\tilde{J}|+|\tilde{\theta}|)|J_{*,\epsilon}|+\text{ easier terms}\\
& \leq C_1(|\tilde{J}|+|\tilde{\theta}|)|J_{*,\epsilon}|+\text{ 
easier terms,}\end{split}\\
\begin{split} \sup_{t\leq 0} e^{-\sigma t} |\partial_{J_*}F_J\cdot 
J_{*,\epsilon}|&\leq C_1\sup_{t\leq 0}[ e^{-(\sigma
+\tilde{\nu} )t}(|\tilde{J}|+|\tilde{\theta}|)]\sup_{t\leq 0}[e^{\tilde{\nu} 
t}|J_{*,\epsilon}|]\\
&\quad + \text{easier terms},\end{split}\end{align*}
where $\sup_{t\leq 0}e^{\tilde{\nu} t}|J_{*,\epsilon}|$ can be bounded when 
$\epsilon$ is sufficiently small for any fixed $\tilde{\nu} >0$, through a
routine estimate on Equations~\eqref{meig2} and \eqref{eig3} for 
$(J_*(t),\theta_*(t))$. Other terms involving $u_{*,\epsilon}$ can be
estimated similarly. Thus, the $\| \ \|_{\sigma ,n}$ norm of terms 
involving $u_{*,\epsilon}$ has to be bounded by $\|\ \|_{\sigma +\tilde{\nu}
,n}$ norms. This leads to the standard rate condition for the 
regularity of invariant manifolds. That is, the regularity is 
controlled by
the spectral gap. The $\| \ \|_{\sigma ,n}$ norm of the term involving 
$A_\epsilon$ has to be bounded by $\| \ \|_{\sigma ,n+2}$ norms. This
is a new phenomenon caused by the singular perturbation. This problem 
is resolved by virtue of a special property of the fixed point $g_*$
of $\Gamma$.
Notice that if $\sigma_2\geq \sigma_1$, $n_2\geq n_1$, then 
$G_{\sigma_2,n_2}\subset G_{\sigma_1,n_1}$. Thus by the uniqueness of 
the
fixed point, if $g_*$ is the fixed point of $\Gamma$ in 
$G_{\sigma_2,n_2}$, $g_*$ is also the fixed point of $\Gamma$ in
$G_{\sigma_1,n_1}$. Since $g_*$ exists in $G_{\sigma ,n}$ for an 
fixed $n\geq 1$ and $\sigma \in (
\frac{\mu^+_1}{100},\frac{\mu^+_1}{3}-10\tilde{\nu})$ where 
$\tilde{\nu}$ is small enough,
\begin{align*} \| \mathcal{R}^+_1\| _{\sigma,n}&\leq C_1+C_2\| 
g_*\|_{\sigma+\tilde{\nu},n},\\
\| \mathcal{R}\|_{\sigma ,n}&\leq C_3\| g_*\|_{\sigma ,n+2}+C_4\| 
g_*\|_{\sigma+\tilde{\nu},n}+C_5,\end{align*}
where $C_j\ (1\leq j\leq 5)$ depend upon $\| g_*(0)\|_n$ and $\| 
g_*(0)\|_{n+2}$. Let
\begin{equation*}M=2(\| \mathcal{R}^+_1\|_{\sigma 
,n}+\|\mathcal{R}\|_{\sigma, n}),\end{equation*}
and $\Gamma'$ denote the linear map defined by the right hand sides 
of \eqref{dper1} and \eqref{dper2}. Since the terms
$\partial_uF^+_1$, $\partial_{\xi^+_1}F^+_1$, $\partial_uF$, and 
$\partial_{\xi^+_1}F$ all have small $\|\ \|_n$ norms, $\Gamma '$
is a contraction map on $\mathcal{B}(M)\subset L(\mathcal{R}, G_{\sigma 
,n})$, where $\mathcal{B}(M)$ is the ball of radius $M$. Thus
$\Gamma '$ has a unique fixed point $g_{*,\epsilon}$. Next one needs 
to show that $g_{*,\epsilon}$ is indeed the partial derivative of
$g_*$ with respect to $\epsilon$. That is, one needs to show
\begin{equation}\label{ddef} \lim_{\Delta \epsilon \to 0}\frac{\| 
g_*(\epsilon +\Delta \epsilon)-g_*(\epsilon)-g_{*,\epsilon} \Delta
\epsilon \|_{\sigma ,n}}{\Delta \epsilon}=0.\end{equation}
\nid
This has to be accomplished directly from Equations 
\eqref{per1}-\eqref{per2}, \eqref{dper1}-\eqref{dper2} satisfied by 
$g_*$ and
$g_{*,\epsilon}$. The most troublesome estimate is still the one 
involving $A_\epsilon$. First, notice the fact that $e^{\epsilon
\partial^2_x}$ is holomorphic in $\epsilon$ when $\epsilon>0$, and 
not differentiable at $\epsilon=0$. Then, notice that $g_*\in
G_{\sigma ,n}$ for any $n\geq 1$, thus, $e^{\epsilon 
\partial_x^2}g_*$ is differentiable, up to certain order $m$, in 
$\epsilon$ at
$\epsilon=0$ from the right, i.e.
\begin{equation*}(d^+/d\epsilon)^me^{\epsilon 
\partial^2_x}g_*|_{\epsilon =0}\end{equation*}
exists in $H^n$. Let
\begin{equation*}\begin{split} z(t,\Delta \epsilon)&= e^{(\epsilon 
+\Delta \epsilon)t\partial^2_x}g_*-e^{\epsilon
t\partial^2_x}g_*-(\Delta \epsilon) t\partial ^2_x e^{\epsilon 
t\partial _x^2}g_*\\
&= e^{\epsilon t\partial^2_x}w(\Delta \epsilon),\end{split}\end{equation*}
where $t\geq 0$, $\Delta \epsilon >0$, and
\begin{equation*}w(\Delta \epsilon) =e^{(\Delta \epsilon ) t\partial 
_x^2}g_*-g_*-(\Delta \epsilon ) t\partial^2_x g_*.\end{equation*}
Since $w(0)=0$, by the Mean Value Theorem, one has
\begin{equation*}\| w(\Delta \epsilon )\|_n=\| w(\Delta \epsilon 
)-w(0)\|_n\leq \sup_{0\leq \lambda \leq 1}\| \frac{dw}{d\Delta
\epsilon}(\lambda
\Delta \epsilon)\|_n|\Delta \epsilon|,\end{equation*}
where at $\lambda=0$, $\frac{d}{d\Delta \epsilon}=\frac{d^+}{d\Delta 
\epsilon}$, and
\begin{equation*}\frac{dw}{d\Delta \epsilon}=t[e^{(\Delta \epsilon )
t\partial^2_x}\partial^2_x g_*-\partial^2_x g_*].
\end{equation*}
Since $\frac{dw}{d\Delta \epsilon}(0)=0$, by the Mean Value Theorem 
again, one has
\[
\| \frac {dw}{d\Dl \e }(\la \Dl \e )\|_n = 
\| \frac {dw}{d\Dl \e }(\la \Dl \e ) - 
\frac {dw}{d\Dl \e }(0)\|_n \leq \sup_{0 \leq \la_1 \leq 1}
\| \frac {d^2w}{d\Dl \e^2 }(\la_1 \la \Dl \e )\|_n |\la | |\Dl \e |\ ,
\]
where
\[
\frac {d^2w}{d\Dl \e^2 } = t^2 [e^{(\Dl \e )t \pa_x^2} \pa_x^4 g_* ]\ .
\]
Therefore, one has the estimate
\begin{equation}
\| z(t,\Dl \e )\|_n \leq |\Dl \e |^2 t^2 \| g_* \|_{n+4} \ .
\label{defes}
\end{equation}
This estimate is sufficient for handling the estimate involving 
$A_\e$. The estimate involving $u_{*,\e}$ can be handled in a similar 
manner. For instance, let 
\[
\tilde{z}(t,\Dl \e ) = F(u_*(t,\e +\Dl \e ))- F(u_*(t,\e ))
-\Dl \e \pa_{u_*}F \cdot u_{*,\e}\ ,
\]
then
\[
\| \tilde{z}(t,\Dl \e )\|_{\sg , n} \leq |\Dl \e |^2 
\sup_{0 \leq \la \leq 1} \| [u_{*,\e} \cdot \pa^2_{u_*}F \cdot u_{*,\e}
+\pa_{u_*}F \cdot u_{*,\e \e}](\la \Dl \e )\|_{\sg , n} \ .
\]
From the expression of $F$ (\ref{meq2}), one has
\begin{equation*}\begin{split} & \| u_{*,\epsilon}\cdot 
\partial^2_{u_*}F\cdot u_{*,\epsilon}+\partial_{u_*}F\cdot
u_{*,\epsilon\epsilon}\|_{\sigma,n}\\
&\quad \leq C_1\| g_*\|_{\sigma +2\tilde{\nu} ,n} [( \sup_{t\leq 
0}e^{\tilde{\nu} t}|u_{*,\epsilon}|)^2+\sup_{t\leq 0}e^{2\tilde{\nu}
t}|u_{*,\epsilon \epsilon}|],\end{split}\end{equation*}
and the term $[ \  ]$ on the right hand side can be easily shown to be 
bounded. In conclusion, let
\begin{equation*}h=g_*(\epsilon +\Delta 
\epsilon)-g_*(\epsilon)-g_{*,\epsilon}\Delta 
\epsilon,\end{equation*}
one has the estimate
\begin{equation*}\| h\|_{\sigma ,n}\leq \tilde{\k} \| 
h\|_{\sigma,n}+|\Delta \epsilon|^2\tilde{C}(\| g_*\|_{\sigma ,n+4};\|
g_*\|_{\sigma +2\tilde{\nu} ,n}),\end{equation*}
where $\tilde{\k}$ is small, thus
\begin{equation*}\| h\|_{\sigma ,n}\leq 2 |\Delta \epsilon 
|^2\tilde{C}(\| g_*\|_{\sigma,n+4};\| g_*\|_{\sigma +2\tilde{\nu}
,n}).\end{equation*} This implies that
\begin{equation*}\lim_{\Delta \epsilon \to 0}\frac{\| h\|_{\sigma 
,n}}{|\Delta \epsilon|}=0,\end{equation*}
which is \eqref{ddef}.

Let $g_*(t)=(\xi^+_1(t),u(t))$. First, let me comment on $\left. 
\frac{\partial u}{\partial
\xi^+_1(0)}\right|_{\xi^+_1(0)=0,\epsilon =0}=0$. From \eqref{per2}, one has
\begin{equation*}\| \frac{\partial u}{\partial 
\xi^+_1(0)}\| _{\sigma, n}\leq \k_1\| \frac{\partial 
g_*}{\partial \xi^+_1(0)}\|_{\sigma ,n},\end{equation*}
where by letting $\xi^+_1(0)\to 0$ and $\epsilon \to 0^+$, $\k_1\to 0$. Thus
\begin{equation*}\left. \frac{\partial u}{\partial 
\xi^+_1(0)}\right|_{\xi^+_1(0)=0,\epsilon =0}=0.\end{equation*}
I shall also comment on ``exponential decay" property. Since $\| 
g_*\|_{\frac{\mu^+_1}{3},n}\leq r$,
\begin{equation*}\| g_*(t)\|_n\leq re^{\frac{\mu^+_1}{3}t},\quad 
\forall t\leq 0.\end{equation*}
\begin{definition} Let $g_*(t)=(\xi^+_1(t), u(t))$, where
\begin{equation*}u(0)=\int^0_{-\infty}e^{A(t-\tau)}F(\tau)d\tau\end{equation*}
depends upon $\xi^+_1(0)$. Thus
\begin{equation*}u^0_*:\xi^+_1(0)\mapsto u(0),\end{equation*}
defines a curve, which we call an unstable fiber denoted by 
$\mathcal{F}^+_p$, where $p=(J_0,\theta_0)$ is the base point,
$\xi^+_1(0)\in [-r,r]\times [-r,r]$.
\end{definition}
Let $S^t$ denote the evolution operator of \eqref{meq1}-\eqref{meq2}, then
\begin{equation*}S^t\mathcal{F}_p^t\subset \mathcal{F}^t_{S^tp},\quad 
\forall t\leq 0.\end{equation*}
That is, $\{ \mathcal{F}^+_p\}_{p\in \mathcal{A}(4\delta)}$ is an 
invariant family of unstable fibers. The proof of the Unstable Fiber
Theorem is finished. Q.E.D.

\begin{remark} If one replaces the base orbit $(J_*(t),\theta_*(t))$ 
by a general orbit for which only $\|\ \|_n$ norm is bounded, then
the estimate \eqref{upre} will not be possible. The $\|\ \|_{\sigma 
,n+2}$ norm of the fixed point $g_*$ will not be bounded either. In
such case, $g_*$ may not be smooth in $\epsilon$ due to the singular 
perturbation.\end{remark}

\begin{remark} Smoothness of $g_*$ in $\epsilon$ at $\epsilon =0$ is 
a key point in locating homoclinic orbits as discussed in later sections.
From integrable theory, information is known at $\epsilon =0$. This 
key point will 
link ``$\epsilon =0$" information to ``$\epsilon \neq 0$" studies.
Only continuity in $\epsilon $ at $\epsilon =0$ is not enough for the 
study. The beauty of the entire theory is reflected by the fact
that although $e^{\epsilon \partial^2_x}$ is not holomorphic at 
$\epsilon =0$, $e^{\epsilon \partial^2_x}g_*$ can be smooth at
$\epsilon =0$ from the right, up to certain order depending upon the 
regularity of $g_*$. This is the beauty of the singular
perturbation.
\end{remark}

\section{Proof of the Center-Stable Manifold Theorem \ref{CSM}}

Here we give the proof of the center-stable manifold theorem \ref{CSM},
proofs of other invariant manifold theorems in this chapter are easier.

\subsection{Existence of the Center-Stable Manifold}

We start with Equations \eqref{eig1}-\eqref{eig5}, let
\begin{equation}\label{2cr} v=\begin{pmatrix} J\\ \theta \\ h \\ 
\xi^-_1\end{pmatrix},\quad \tilde{v}=\begin{pmatrix}
J\\ h\\ \xi^-_1\end{pmatrix},\end{equation}
and let $E_n(r)$ be the tubular neighborhood of $S_\om$ \eqref{rcl}:
\begin{equation}\label{defEn} E_n(r)=\{ (J,\theta 
,h,\xi^-_1)\in H^n\mid \ \| \tilde{v}\|_n\leq r\}.\end{equation}
$E_n(r)$ is of codimension $1$ in the entire phase space 
coordinatized by $(\xi^+_1,J,\theta
,h,\xi^-_1)$.

Let $\chi\in C^\infty(R,R)$ be a ``cut-off" function:
\begin{equation*}\chi=\begin{cases} 0, & \text{in } (-\infty ,-4)\cup 
(4,\infty),\\
1, & \text{in } (-2,2).\end{cases}\end{equation*}
We apply the cut-off \begin{equation*}\chi_\delta =\chi ( 
\|\tilde{v}\|_n/\delta) \chi (
\xi^+_1/\delta)\end{equation*} to 
Equations~\eqref{eig1}-\eqref{eig5}, so that the equations in a 
tubular neighborhood
of the circle
$S_\omega$ \eqref{rcl} are unchanged, and linear outside a 
bigger tubular neighborhood. The modified equations take the form:
\begin{align}\label{ceq1} 
\dot{\xi}_1^+&=\mu^+_1\xi^+_1+\tilde{F}^+_1, \\
\label{ceq2}v_t&=Av+\tilde{F},\end{align}
where $A$ is given in \eqref{meq2},
\begin{align*} \tilde{F}^+_1&=\chi_\delta[V^+_1\xi^+_1+\mathcal{N}^+_1],\\
\tilde{F}&=(\tilde{F}_J,\tilde{F}_\theta 
,\tilde{F}_h,\tilde{F}^-_1)^T,\quad 
T=\text{transpose},\\
\tilde{F}_J&=\chi_\delta \ \e \left[-2\alpha (J+\omega^2)+2\beta 
\sqrt{J+\omega^2}\cos \theta+\mathcal{R}^J_2\right],\\
\tilde{F}_\theta&=\chi _\delta \left[ -\epsilon \beta \frac{\sin 
\theta}{\sqrt{J+\omega^2}}+\mathcal{R}^\theta_2\right],\\
\tilde{F}_h&=\chi_\delta [V_\epsilon h+\tilde{\mathcal{N}}], \\
\tilde{F}_1^-&=\chi_\delta [V^-_1\xi^-_1+ \mathcal{N}^-_1],\end{align*}
Equations~\eqref{ceq1}-\eqref{ceq2} can be written in the equivalent 
integral equation form:
\begin{align}\label{csit1} 
\xi^+_1(t)&=\xi^+_1(t_0)e^{\mu^+_1(t-t_0)}+\int^t_{t_0}e^{\mu^+_1(t-\tau 
)}\tilde{F}^+_1(\tau 
)d\tau,\\
\label{csit2} 
v(t)&=e^{A(t-t_0)}v(t_0)+\int^t_{t_0}e^{A(t-\tau)}\tilde{F}(\tau 
)d\tau.\end{align}
We introduce the following space: For $\sigma \in \left( 
\frac{\mu^+_1}{100},\frac{\mu^+_1}{3}\right)$, 
and $n\geq 1$, let
\begin{equation*}\begin{split} \tilde{G}_{\sigma ,n}&= \bigg \{ 
g(t)=(\xi^+_1(t),v(t)) \bigg | \  t\in [0,\infty),g(t)\text{ is continuous 
in } t\\
&\quad \text{in } H^n \ \mbox{norm} ,\| g\|_{\sigma ,n}=\sup_{t\geq 
0}e^{-\sigma t}[ |\xi^+_1(t)|+\| v(t)\|_n ]
<\infty \bigg \}\ .\end{split}\end{equation*}
$\tilde{G}_{\sigma,n}$ is a Banach space under the norm $\| \cdot 
\|_{\sigma ,n}$. Let $\tilde{\mathcal{A}}_{\sigma ,n}(r)$ denote the
closed tubular neighborhood of $S_\omega$ \eqref{rcl}:
\begin{equation*} \tilde{\mathcal{A}}_{\sigma ,n}(r)= \bigg \{ g(t) 
=(\xi^+_1(t),v(t))\in \tilde{G}_{\sigma,n}\bigg | \ \sup_{t\geq
0}e^{-\sigma t}[ |\xi^+_1(t)|+\| 
\tilde{v}(t)\|_n] \leq r\bigg \}\ ,\end{equation*}
where $\tilde{v}$ is defined in \eqref{2cr}. If $g(t)\in 
\tilde{\mathcal{A}}_{\sigma ,n}(r)$, $r<\infty$, is a solution of
\eqref{csit1}-\eqref{csit2}, by letting $t_0\to +\infty$ in 
\eqref{csit1} and setting $t_0=0$ in \eqref{csit2}, one has
\begin{align}\label{cspe1} 
\xi^+_1(t)&=\int^t_{+\infty}e^{\mu^+_1(t-\tau)}\tilde{F}^+_1(\tau 
)d\tau, \\
\label{cspe2}v(t)&=e^{At}v(0)+\int^t_0e^{A(t-\tau)}\tilde{F}(\tau 
)d\tau .\end{align}
For any $g(t)\in \tilde{\mathcal{A}}_{\sigma ,n}(r)$, let 
$\tilde{\Gamma}(g)$ be the map defined by the right hand side of
\eqref{cspe1}-\eqref{cspe2}. In contrast to the map $\Gamma$ defined 
in \eqref{per1}-\eqref{per2}, $\tilde{\Gamma}$ contains constant
terms of order $\mathcal{O}(\epsilon)$, e.g. $\tilde{F}_J$ and 
$\tilde{F}_\theta$ both contain such terms. Also,
$\tilde{\mathcal{A}}_{\sigma, n}(r)$ is a tubular neighborhood of the 
circle $S_\omega$ \eqref{rcl} instead of the ball
$\mathcal{B}_{\sigma ,n}(r)$ for $\Gamma$. Fortunately, these facts 
will not create any difficulty in showing $\tilde{\Gamma}$ is a
contraction on $\tilde{\mathcal{A}}_{\sigma ,n}(r)$. For any $n\geq 
1$ and $\epsilon <\delta^2$, and $\delta $ and $r$ are small enough,
$\tilde{F}_1^+$ and $\tilde{F}$ are Lipschitz in $g$ with small 
Lipschitz constants. $\tilde{\Gamma}$ has a unique fixed point
$\tilde{g}_*$ in $\tilde{\mathcal{A}}_{\sigma ,n}(r)$, following from 
standard arguments. 

\subsection{Regularity of the Center-Stable Manifold in $\e$}

For the regularity of $\tilde{g}_*$ with
respect to $(\epsilon, \alpha ,\beta ,\omega ,v(0))$, the most 
difficult one is of course with respect to $\epsilon$ due to the 
singular
perturbation. Formally differentiating $\tilde{g}_*$ in 
\eqref{cspe1}-\eqref{cspe2} with respect to $\epsilon$, one gets
\begin{align}\label{dcsp1} 
\xi^+_{1,\epsilon}(t)&=\int^t_{+\infty}e^{\mu^+_1(t-\tau)}\left[
\partial_{\xi^+_1}\tilde{F}^+_1\cdot 
\xi^+_{1,\epsilon}+\partial_v\tilde{F}^+_1\cdot v_\epsilon 
\right](\tau
)d\tau +\tilde{R}^+_1(t),\\
\label{dcsp2}v(t)&=\int^t_0e^{A(t-\tau)}\left[ 
\partial_{\xi^+_1}\tilde{F}\cdot
\xi^+_{1,\epsilon}+\partial_v\tilde{F}\cdot v_\epsilon 
\right](\tau )d\tau +\tilde{R}(t),\end{align}
where
\begin{align}\label{wdcs1} 
\tilde{R}_1^+(t)&=\int^t_{+\infty}\mu^+_{1,\epsilon}(t-\tau)
e^{\mu^+_1(t-\tau)}\tilde{F}_1^+(\tau )d\tau
+\int^t_{+\infty}e^{\mu^+_1(t-\tau)}\partial _\epsilon 
\tilde{F}^+_1(\tau )d\tau, \\
\label{wdcs2} \tilde{R}(t)&= tA_\epsilon 
e^{At}v(0)+\int^t_0(t-\tau)A_\epsilon e^{A(t-\tau)}\tilde{F}(\tau 
)d\tau + \int^t_0e^{A(t-\tau )}\partial_\epsilon \tilde{F}(\tau 
)d\tau,\end{align}
and $\mu^+_{1,\epsilon}$ and $A_\epsilon$ are given in 
\eqref{wdp3}-\eqref{wdp4}. The troublesome terms are the ones 
containing
$A_\epsilon$ in \eqref{wdcs2}. These terms can be handled in the same 
way as in the Proof of the Unstable Fiber Theorem. The crucial
fact utilized is that if $v(0)\in H^{n_1}$, then $\tilde{g}_*$ is the 
unique fixed point of $\tilde{\Gamma}$ in both
$\tilde{G}_{\sigma,n_1}$ and $\tilde{G}_{\sigma ,n_2}$ for any $n_2\leq n_1$.

\begin{remark} In the Proof of the Unstable Fiber Theorem, the 
arbitrary initial datum in \eqref{per1}-\eqref{per2} is $\xi^+_1(0)$
which is a scalar. Here the arbitrary initial datum in 
\eqref{cspe1}-\eqref{cspe2} is $v(0)$ which is a function of $x$. 
If $v(0)\in H^{n_2}$ but not $H^{n_1}$ for some $n_1>n_2$, 
then $\tilde{g}_*\notin \tilde{G}_{\sigma ,n_1}$, in contrast to the 
case of
\eqref{per1}-\eqref{per2} where $g_*\in G_{\sigma ,n}$ for any fixed 
$n\geq 1$. The center-stable manifold $W^{cs}_n$ stated in the
Center-Stable Manifold Theorem will be defined through $v(0)$. This 
already illustrates why $W^{cs}_n$ has the regularity in $\epsilon$
as stated in the theorem.\end{remark}

We have
\begin{align*} &\| \tilde{R}^+_1\|_{\sigma ,n}\leq \tilde{C_1},\\
& \| \tilde{R}\|_{\sigma ,n}\leq \tilde{C}_2\| \tilde{g}_*\|_{\sigma 
,n+2}+\tilde{C}_3,\end{align*}
for $\tilde{g}_*\in \tilde{\mathcal{A}}_{\sigma , n+2}(r)$, where 
$\tilde{C}_j\ (j=1,2,3)$ are constants depending in particular upon the
cut-off in $\tilde{F}^+_1$ and $\tilde{F}$. Let $\tilde{\Gamma}'$ 
denote the linear map defined by the right hand sides of
\eqref{dcsp1}-\eqref{dcsp2}. If $v(0)\in H^{n+2}$ and $\tilde{g}_*\in 
\tilde{\mathcal{A}}_{\sigma,n+2}(r)$, standard argument shows that
$\tilde{\Gamma}'$ is a contraction map  on a closed ball in 
$L(R,\tilde{G}_{\sigma,n})$. Thus $\tilde{\Gamma}'$ has a unique 
fixed point
$\tilde{g}_{*,\epsilon}$. Furthermore, if $v(0)\in H^{n+4}$ and 
$\tilde{g}_*\in \tilde{\mathcal{A}}_{\sigma ,n+4}(r)$, one has that
$\tilde{g}_{*,\epsilon}$ is indeed the derivative of $\tilde{g}_*$ in 
$\epsilon$, following the same argument as in the Proof of the
Unstable Fiber Theorem. Here one may be able to replace the 
requirement $v(0)\in H^{n+4}$ and $\tilde{g}_*\in
\tilde{\mathcal{A}}_{\sigma ,n+4}(r)$ by just $v(0)\in H^{n+2}$ and 
$\tilde{g}_*\in \tilde{\mathcal{A}}_{\sigma ,n+2}(r)$. But we are
not interested in sharper results, and the current result is 
sufficient for our purpose.

\begin{definition} For any $v(0)\in E_n(r)$ where $r$ is sufficiently 
small and $E_n(r)$ is defined in \eqref{defEn}, let
$\tilde{g}_*(t)=(\xi^+_1(t),v(t))$ be the fixed point of 
$\tilde{\Gamma}$ in $\tilde{G}_{\sigma ,n}$, where one has
\begin{equation*}\xi^+_1(0)=\int^0_{+\infty}e^{\mu^+_1(t-\tau)}
\tilde{F}^+_1(\tau )d\tau ,\end{equation*}
which depend upon $v(0)$. Thus
\begin{equation*} \xi^+_*:v(0)\mapsto \xi^+_1(0),\end{equation*}
defines a codimension $1$ surface, which we call center-stable 
manifold denoted by $W^{cs}_n$.\end{definition}

The regularity of the fixed point $\tilde{g}_*$ immediately implies 
the regularity of $W^{cs}_n$. We have sketched the proof of the most
difficult regularity, i.e. with respect to $\epsilon$. Uniform 
boundedness of $\partial_\epsilon \xi^+_*$ in $v(0)\in E_{n+4}(r)$ and
$\epsilon \in [0,\epsilon _0)$, is obvious. Other parts of the 
detailed proof is completely standard. We have that $W^{cs}_n$ is a 
$C^1$
locally invariant submanifold which is $C^1$ in $(\alpha ,\beta, 
\omega)$. $W^{cs}_n$ is $C^1$ in $\epsilon$ at point in the subset
$W^{cs}_{n+4}$. Q.E.D.

\begin{remark}\label{reop} Let $S^t$ denote the evolution 
operator of the perturbed nonlinear Schr\"odinger equation
\eqref{spnls}. The proofs of the Unstable Fiber Theorem and the 
Center-Stable Manifold Theorem also imply the following: $S^t$ is a 
$C^1$
map on $H^n$ for any fixed $t>0$, $n\geq 1$. $S^t$ is also $C^1$ in 
$(\alpha ,\beta ,\omega)$. $S^t$ is $C^1$ in $\epsilon$ as a map
from $H^{n+4}$ to $H^n$ for any fixed $n\geq 1$, $\epsilon \in 
[0,\epsilon_0)$, $\epsilon_0>0$.\end{remark}

\section{Perturbed Davey-Stewartson II Equations \label{invds}}

Invariant manifold results in Sections \ref{spsec} and \ref{rpsec} also hold
for perturbed Davey-Stewartson II equations \cite{Li02b},
\begin{eqnarray*}
iq_t &=& \Upsilon q+ \bigg [2(|q|^2-\omega^2)+ u_y \bigg ]q
+i\epsilon f \ , \\
& & \ \ \Delta u = -4\partial_y |q|^2 \ ,
\end{eqnarray*}
where $q$ is a complex-valued function of the three variables ($t,x,y$), 
$u$ is a real-valued function of the three variables ($t,x,y$), 
$\Upsilon =\partial_{xx}-\partial_{yy}$, $\Delta=\partial_{xx}
+\partial_{yy}$, $\omega >0$ is a constant, and $f$ is the perturbation.
We also consider periodic boundary conditions.

Under singular perturbation 
\[
f = \Dl q - \al q + \be \ ,
\]
where $\alpha >0$, $\beta >0$ are constants, Theorems \ref{UFT} and 
\ref{CSM} hold for the perturbed Davey-Stewartson II equations \cite{Li02b}. 

When the singular perturbation $\Dl$ is mollified into a bounded 
Fourier multiplier
\[
\hat{\Dl} q = -\sum_{k \in Z^2} \be_k |k|^2 \tilde{q}_k \cos k_1 x 
\cos k_2 y \ ,
\]
in the case of periods ($2\pi , 2\pi$),
\[
\be_k = 1, \ \ |k| \leq N , \quad \be_k = |k|^{-2},\ \ |k| > N,
\]
for some large $N$, $|k|^2 = k_1^2 + k_2^2$, Theorems \ref{Persthm} 
and \ref{fiberthm} hold for the perturbed Davey-Stewartson II equations
\cite{Li02b}.

\section{General Overview \label{GOV}}

For discrete systems, i.e., the flow is given by a map, it is more 
convenient to use Hadamard's method 
to prove invariant manifold and fiber theorems \cite{HPS77}.
Even for continuous systems, Hadamard's method was often utilized
\cite{Fen71}. On the other hand, Perron's method provides shorter 
proofs. It involves manipulation of integral equations. This method 
should be a favorite of analysts. Hadamard's method deals with graph 
transform. The proof is often much longer, with clear geometric intuitions.
It should be a favorite of geometers. For finite-dimensional 
continuous systems, N. Fenichel proved persistence of invariant 
manifolds under $C^1$ perturbations of flow in a very general setting 
\cite{Fen71}. He then went on to prove the fiber theorems in \cite{Fen74} 
\cite{Fen77} also in this general setting. Finally, he applied this 
general machinery to a general system of ordinary differential equations
\cite{Fen79}. As a result, Theorems \ref{Persthm} and \ref{fiberthm} 
hold for the following perturbed discrete cubic nonlinear Schr\"odinger 
equations \cite{LM97}, 
\begin{eqnarray}
i\dot{q_n}&=&{1 \over h^2}\bigg[q_{n+1}-2q_n+q_{n-1}\bigg]+|q_n|^2(q_{n+1}+
           q_{n-1})-2\om^2 q_n \nonumber \\
        & &+i\e \bigg[-\al q_n +{1 \over h^2}(q_{n+1}-2q_n+q_{n-1})
           + \be \bigg], \label{PDNLS}
\end{eqnarray}
\nid
where $i=\sqrt{-1}$, $q_n$'s are complex variables,
\[
q_{n+N}=q_n, \ \ (\mbox{periodic}\  \mbox{condition}); \quad \mbox{and}\ 
q_{-n}=q_n, \ \ (\mbox{even}\  \mbox{condition});
\]
$h={1\over N}$, and 
\begin{eqnarray*}
& & N\tan{\pi \over N}< \om <N\tan{2\pi \over N},\ \ \mbox{for}\ N>3,\\
& & 3\tan{\pi \over 3}< \om < \infty, \ \ \mbox{for}\ N=3. \\
& & \e\in[0,\e_1),\ \alpha\ (>0), \ \be\ (>0) \ 
\mbox{are}\ \mbox{constants.}
\end{eqnarray*}
This is a $2(M+1)$ dimensional system, where
\[
M=N/2,\ \ (N\ \mbox{even}); \quad \mbox{and}\ 
M=(N-1)/2, \ \ (N\ \mbox{odd}).
\]
This system is a finite-difference discretization of the perturbed NLS 
(\ref{spnls}).

For a general system of ordinary differential equations, 
Kelley \cite{Kel67} used the Perron's method to give a very short proof 
of the classical unstable, stable, and center manifold theorem. This 
paper is a good starting point of reading upon Perron's method. In the book
\cite{HPS77}, Hadamard's method is mainly employed. This book is an
excellent starting point for a comprehensive reading on invariant 
manifolds. 

There have been more and more invariant manifold results for infinite 
dimensional systems \cite{LW97b}. For the employment of Perron's 
method, we refer the readers to \cite{CLL91}. For the employment of 
Hadamard's method, we refer the readers to \cite{BLZ98} \cite{BLZ99} 
\cite{BLZ00} which are terribly long papers.

\chapter{Homoclinic Orbits}

In terms of proving the existence of a homoclinic orbit, 
the most common tool is the so-called Melnikov integral method \cite{Mel63} 
\cite{Arn64}. This method was subsequently developed by Holmes and Marsden 
\cite{GH83}, and most recently by Wiggins \cite{Wig88}. For partial 
differential equations, this method was mainly developed by Li et al.
\cite{LMSW96} \cite{Li01b} \cite{LM97} \cite{Li02f}.

There are two derivations for the Melnikov integrals. One is the 
so-called geometric argument \cite{GH83} \cite{Wig88} \cite{LM97} 
\cite{LMSW96} \cite{Li01b}.
The other is the so-called Liapunov-Schmitt argument \cite{CHMP80} 
\cite{CH82}. The Liapunov-Schmitt argument is a fixed-point type argument 
which directly leads to the existence of a homoclinic orbit. The condition 
for the existence of a fixed point is the Melnikov integral. 
The geometric argument is a signed distance argument which applies 
to more general situations than the Liapunov-Schmitt argument. It turns out 
that the geometric argument is a much more powerful machinary than the 
Liapunov-Schmitt argument. In particular, the geometric argument can handle 
geometric singular perturbation problems. I shall also mention an 
interesting derivation in \cite{Arn64}.

In establishing the existence of homoclinic orbits in high dimensions, one 
often needs other tools besides the Melnikov analysis. For example, when 
studying orbits 
homoclinic to fixed points created through resonances in 
($n \geq 4$)-dimensional near-integrable systems, one often needs tools 
like Fenichel fibers, as presented in previous chapter, to set up geometric 
measurements for locating such 
homoclinci orbits. Such homoclinic orbits often have a geometric singular 
perturbation nature. In such cases, the Liapunov-Schmitt argument can not 
be applied. For such works on finite dimensional systems, see for 
example \cite{Kov92a} \cite{Kov92b} \cite{LM97}.
For such works on infinite dimensional systems, see for example 
\cite{LMSW96} \cite{Li01b}.

\section{Silnikov Homoclinic Orbits in NLS Under Regular 
Perturbations \label{horrnls}}

\begin{figure}
\includegraphics{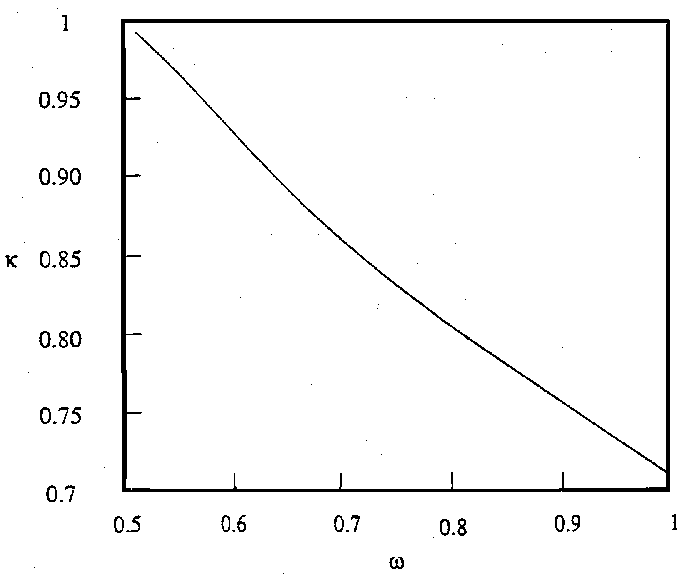}
\caption{The graph of $\k(\om)$.}
\label{kappa}
\end{figure}

We continue from section \ref{rpsec} and consider the regularly perturbed 
nonlinear Schr\"odinger (NLS) equation (\ref{rpnls}). The following theorem 
was proved in \cite{LMSW96}.
\begin{theorem}
There exists a $\e_0 > 0$, such that for any $\e \in (0, \e_0)$, there 
exists a codimension 1 surface in the external parameter space 
$(\alpha,\beta, \om) \in \RR^+\times \RR^+\times \RR^+$ where $\om \in 
(\frac{1}{2}, 1)$, and $\al \om < \be$. For any 
$(\alpha ,\beta, \omega)$ on the codimension 1
surface, the regularly perturbed nonlinear Schr\"odinger equation 
(\ref{rpnls}) possesses a symmetric pair of Silnikov homoclinic orbits 
asymptotic 
to a saddle $Q_\epsilon$. The codimension 1 surface has the approximate 
representation given by $\al = 1/\k(\om)$, where $\k(\om)$ is plotted 
in Figure \ref{kappa}.
\label{rhorbit}
\end{theorem}
The proof of this theorem is easier than that given in later sections.
To prove the theorem, one starts from the invariant plane 
\[
\Pi=\{ q\mid \ \partial_x q=0 \}.
\]
On $\Pi$, there is a saddle $Q_\e =\sqrt{I} e^{i\th}$ to which 
the symmetric pair of Silnikov homoclinic orbits will be asymptotic to, where
\begin{equation}
I=\omega^2-\epsilon \frac{1}{2\omega}\sqrt{\beta^2-\alpha^2\omega^2}+\cdots ,
\quad \cos \theta  =\frac{\alpha \sqrt{I}}{\beta}, \quad \theta \in 
(0,\frac{\pi}{2}).
\label{Qec}
\end{equation}
Its eigenvalues are
\begin{equation}
\la_n^\pm = -\e [\al +\xi_n n^2]\pm 2 \sqrt{(\frac{n^2}{2} +
\om^2-I)(3I -\om^2 -\frac{n^2}{2} )}\ , 
\label{Qev}
\end{equation}
where $n=0,1,2, \cdots $, $\om \in (\frac{1}{2}, 1)$, $\xi_n = 1$ when 
$n \leq N$, $\xi_n = 8n^{-2}$ when $n>N$, for some fixed large $N$,
and $I$ is given in 
(\ref{Qec}). The crucial points to notice are: (1). only $\la_0^+$ and 
$\la_1^+$ have positive real parts, $\mbox{Re}\{ \la_0^+\} < \mbox{Re}\{ 
\la_1^+\} $; (2). all the other eigenvalues have negative real parts among 
which the absolute value of $\mbox{Re}\{ \la_2^+\}=\mbox{Re}\{ \la_2^-\}$ 
is the smallest; (3). $|\mbox{Re}\{ \la_2^+\}| < \mbox{Re}\{ \la_0^+\}$. 
Actually, items (2) and (3) are the main characteristics of Silnikov 
homoclinic orbits. 

The unstable manifold $W^u(Q_\e)$ of $Q_\e$ has a fiber representation 
given by Theorem \ref{fiberthm}. The Melnikov measurement measures the 
signed distance between $W^u(Q_\e)$ and the center-stable manifold 
$W^{cs}_\e$ proved in Thoerem \ref{Persthm}. By virtue of the Fiber Theorem 
\ref{fiberthm}, one can show that, to the leading order in
$\e$, the signed distance is given by the Melnikov integral
\begin{eqnarray*}
M &=& \int^{+\infty}_{-\infty}\int^{2\pi}_0 
 [\partial_qF_1(q_0(t))(\hat{\partial}^2_xq_0(t)-\alpha
q_0(t)+\beta ) \\
& & \quad \quad + \partial_{\bar{q}}F_1(q_0(t))
(\hat{\partial}^2_x\overline{q_0(t)}-\alpha 
\overline{q_0(t)}+\beta )]dxdt,
\end{eqnarray*}
where $q_0(t)$ is given in section \ref{1DCNSE}, equation (\ref{4.13});
and $\partial_qF_1$ and $\partial_{\bar{q}}F_1$ are 
given in section \ref{MVNLS}, equation (\ref{6.5}). The zero of the 
signed distance implies the existence of an orbit in $W^u(Q_\e)\cap 
W^{cs}_\e$. The stable manifold $W^s(Q_\e)$ of $Q_\e$ is a codimension 
1 submanifold in $W^{cs}_\e$. To locate a homoclinic orbit, one needs 
to set up a second measurement measuring the signed distance between 
the orbit in $W^u(Q_\e)\cap W^{cs}_\e$ and $W^s(Q_\e)$ inside $W^{cs}_\e$.
To set up this signed distance, first one can rather easily track 
the (perturbed) orbit by an unperturbed orbit to an $\O (\e)$ neighborhood 
of $\Pi$, then one needs to prove the size of $W^s(Q_\e)$ to be $\O (\e^\nu)$
($\nu <1$) with normal form transform. To the leading order in
$\e$, the zero of the second signed distance is given by
\[
\be \cos \ga = \frac {\al \om (\Dl \ga )} {2 \sin \frac {\Dl \ga }{2}} \ ,
\]
where $\Dl \ga = -4 \vth_0$ and $\vth_0$ is given in (\ref{4.13}).
To the leading order in $\e$, the common zero of the two second signed 
distances satisfies $\al = 1/\k(\om)$, where $\k(\om)$ is plotted 
in Figure \ref{kappa}. Then the claim of the theorem is proved by virtue 
of the implicit function theorem. For rigorous details, see \cite{LMSW96}.
In the singular perturbation case as discussed in next section, 
the rigorous details are given in later sections.

\section{Silnikov Homoclinic Orbits in NLS Under Singular 
Perturbations \label{horsnls}}

We continue from section \ref{spsec} and consider the singularly perturbed 
nonlinear Schr\"odinger (NLS) equation (\ref{spnls}). The following theorem 
was proved in \cite{Li01b}.
\begin{theorem}
There exists a $\e_0 > 0$, such that for any $\e \in (0, \e_0)$, there 
exists a codimension 1 surface in the external parameter space 
$(\alpha,\beta, \om) \in \RR^+\times  \RR^+\times 
\RR^+$ where $\om \in (\frac{1}{2}, 1)/S$, $S$ is a finite subset, and 
$\al \om < \be$. For any $(\alpha ,\beta, \omega)$ on the codimension 1
surface, the singularly perturbed nonlinear Schr\"odinger equation 
(\ref{spnls}) possesses a symmetric pair of Silnikov homoclinic orbits 
asymptotic 
to a saddle $Q_\epsilon$. The codimension 1 surface has the approximate 
representation given by $\al = 1/\k(\om)$, where $\k(\om)$ is plotted 
in Figure \ref{kappa}.
\label{shorbit}
\end{theorem}
In this singular perturbation case, the persistence and fiber theorems 
are given in section \ref{spsec}, Theorems \ref{CSM} and \ref{UFT}. The 
normal form transform for proving the size estimate of the stable manifold 
$W^s(Q_\e)$ is still achievable. The proof of the theorem is also completed 
through two measurements: the Melnikov measurement and the second measurement.

\section{The Melnikov Measurement}

\subsection{Dynamics on an Invariant Plane}

\begin{figure}
\includegraphics{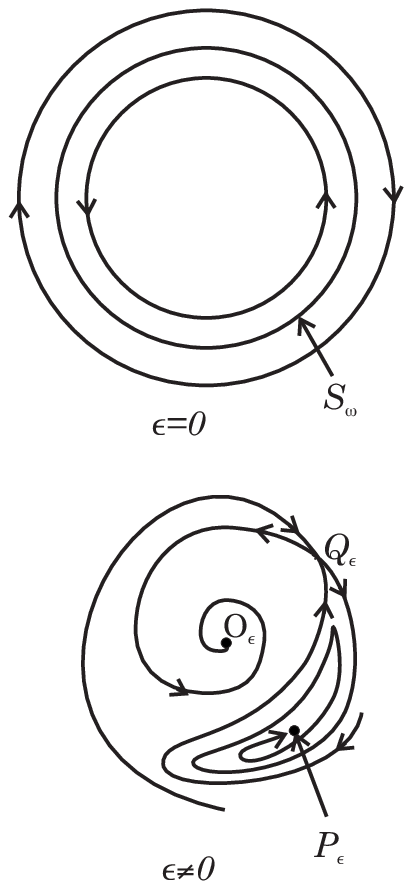}
\caption{Dynamics on the invariant plane $\Pi$.}
\label{figPi}
\end{figure}

\begin{figure}
\includegraphics{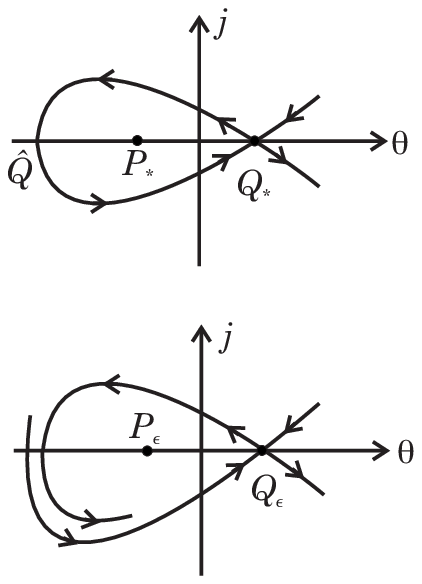}
\caption{The fish-like dynamics in the neighborhood of the resonant circle
$S_\om$.}
\label{fish}
\end{figure}

The 2D subspace $\Pi$,
\begin{equation}\label{sPi} \Pi=\{ q\mid \ \partial_xq=0\},\end{equation}
is an invariant plane under the flow (\ref{spnls}). The 
governing equation in $\Pi$ is
\begin{equation} i\dot{q}=2[|q|^2-\om^2]q+i\epsilon [-\alpha 
q+\beta],\label{Pie1}\end{equation}
where $\cdot =\frac{d}{dt}\ $.
Dynamics of this equation is shown in Figure \ref{figPi}. 
Interesting dynamics is created through resonance in the neighborhood of the
circle $S_\om$:
\begin{equation}\label{srcl} S_\om=\{ q\in \Pi \mid \ |q|=\om \}.\end{equation}
When $\epsilon =0$, $S_\om$ consists of fixed points. To explore the 
dynamics in this neighborhood better, one can make a series of
changes of coordinates. Let $q=\sqrt{I}e^{i\theta}$,
then \eqref{Pie1} can be rewritten as
\begin{align} \dot{I}&= \epsilon ( -2\alpha I+2\beta \sqrt{I}\cos 
\theta )\ ,\label{Ithe1}\\
\dot{\theta} &=-2(I-\om^2)-\epsilon \beta \frac{\sin 
\theta}{\sqrt{I}}\ .\label{Ithe2}\end{align}
There are three fixed points:
\begin{enumerate}\item The focus $O_\epsilon$ in the neighborhood of 
the origin,
\begin{equation}\begin{cases} I=\epsilon ^2\frac{\beta^2}{4\omega^4}+\cdots ,\\
\cos \theta=\frac{\alpha \sqrt{I}}{\beta}, & \theta \in \left( 
0,\frac{\pi}{2}\right).\end{cases}\label{Oec}\end{equation}
Its eigenvalues are
\begin{equation}\label{Oee} \mu_{1,2}=\pm 
i\sqrt{4(\omega^2-I)^2-4\epsilon \sqrt{I}\beta \sin \theta}-\epsilon 
\alpha,\end{equation}
where $I$ and $\theta$ are given in \eqref{Oec}.
\item The focus $P_\epsilon$ in the neighborhood of $S_\omega$ \eqref{srcl},
\begin{equation}\label{Pec}\begin{cases} I=\omega^2+\epsilon 
\frac{1}{2\omega}\sqrt{\beta^2-\alpha^2\omega^2}+\cdots ,\\
\cos \theta=\frac{\alpha \sqrt{I}}{\beta},& \theta \in \left( 
-\frac{\pi}{2}, 0\right).\end{cases}\end{equation}
Its eigenvalues are
\begin{equation}\label{Pee}\mu_{1,2}=\pm 
i\sqrt{\epsilon}\sqrt{-4\sqrt{I}\beta \sin \theta+\epsilon \left( 
\frac{\beta \sin
\theta}{\sqrt{I}}\right)^2}-\epsilon \alpha,\end{equation}
where $I$ and $\theta$ are given in \eqref{Pec}.
\item The saddle $Q_\epsilon$ in the neighborhood of $S_\omega$ \eqref{srcl},
\begin{equation}\label{Qecs}\begin{cases}I=\omega^2-\epsilon 
\frac{1}{2\omega}\sqrt{\beta^2-\alpha^2\omega^2}+\cdots ,\\
\cos \theta  =\frac{\alpha \sqrt{I}}{\beta},&\theta \in \left( 
0,\frac{\pi}{2}\right).\end{cases}\end{equation}
Its eigenvalues are
\begin{equation}\label{Qee} \mu_{1,2}=\pm 
\sqrt{\epsilon}\sqrt{4\sqrt{I}\beta\sin \theta -\epsilon \left( 
\frac{\beta \sin
\theta}{\sqrt{I}}\right)^2}-\epsilon \alpha,\end{equation}
where $I$ and $\theta $ are given in \eqref{Qecs}.
\end{enumerate}
\nid
Now focus our attention to order $\sqrt{\epsilon}$ neighborhood of 
$S_\omega$ \eqref{srcl} and let
\begin{equation*}J=I-\omega^2,\quad J=\sqrt{\epsilon}j,\quad 
\tau=\sqrt{\epsilon}t,\end{equation*}
we have
\begin{align}\label{je1} j'&= 2\left[ -\alpha 
(\omega^2+\sqrt{\epsilon}j) +\beta 
\sqrt{\omega^2+\sqrt{\epsilon}j}\cos \theta \right],\\
\label{je2}\theta '&= -2j-\sqrt{\epsilon}\beta \frac{\sin 
\theta}{\sqrt{\omega^2+\sqrt{\epsilon}j}},\end{align}
where $'=\frac{d}{d\tau}\ $. To leading order, we get
\begin{align}\label{je3} j'&=2[-\alpha \omega^2+\beta \omega \cos \theta]\ ,\\
\label{je4}\theta' & =-2j\ .\end{align}
There are two fixed points which are the counterparts of $P_\epsilon$ 
and $Q_\epsilon$ \eqref{Pec} and \eqref{Qecs}:
\begin{enumerate}\item The center $P_*$,
\begin{equation}\label{Pc} j=0,\quad \cos \theta=\frac{\alpha 
\omega}{\beta},\quad \theta \in
\left(-\frac{\pi}{2},0\right).\end{equation}
Its eigenvalues are
\begin{equation}\label{Pe} \mu_{1,2}=\pm 
i2\sqrt{\omega}(\beta^2-\alpha^2\omega^2)^{\frac{1}{4}}.\end{equation}
\item The saddle $Q_*$,
\begin{equation}\label{Qc} j=0,\quad \cos \theta=\frac{\alpha 
\omega}{\beta},\quad \theta \in \left(
0,\frac{\pi}{2}\right).\end{equation} Its eigenvalues are
\begin{equation}\label{Qe} \mu_{1,2}=\pm 
2\sqrt{\omega}(\beta^2-\alpha ^2\omega^2)^{\frac{1}{4}}.\end{equation}
\end{enumerate}
\nid
In fact, \eqref{je3} and \eqref{je4} form a Hamiltonian system with the 
Hamiltonian
\begin{equation}\label{fham} \mathcal{H}=j^2+2\omega (-\alpha \omega 
\theta+\beta \sin \theta).\end{equation}
Connecting to $Q_*$ is a fish-like singular level set of 
$\mathcal{H}$, which intersects the axis $j=0$ at $Q_*$ and
$\hat{Q}=(0,\hat{\theta})$,
\begin{equation}\label{head} \alpha \omega 
(\hat{\theta}-\theta_*)=\beta (\sin \hat{\theta}-\sin \theta_*),\quad 
\hat{\theta}\in (-\frac{3\pi}{2}, 0),\end{equation}
where $\theta_*$ is given in \eqref{Qc}. See Figure \ref{fish} for an 
illustration of the dynamics of \eqref{je1}-\eqref{je4}. 
\nid
For later use, we define
a piece of each of the stable and unstable manifolds of $Q_*$,
\begin{equation*}j=\phi_*^u(\theta),\quad j=\phi^s_*(\theta),\quad 
\theta\in [\hat{\theta}+\hat{\delta}, \theta_*+2\pi],\end{equation*}
for some small $\hat{\delta}>0$, and
\begin{equation}\begin{split}\label{cur} 
\phi^u_*(\theta)&=-\frac{\theta-\theta_*}{|\theta 
-\theta_*|}\sqrt{2\omega [\alpha \omega
(\theta-\theta_*)-\beta (\sin \theta-\sin \theta_*)]},\\
 \phi^s_*(\theta )&=-\phi^u_*(\theta).\end{split}\end{equation}
$\phi^u_*(\theta)$ and $\phi^s_*(\theta)$ perturb smoothly in $\theta 
$ and $\sqrt{\epsilon}$ into $\phi^u_{\sqrt{\epsilon}}$ and 
$\phi^s_{\sqrt{\epsilon}}$ for
\eqref{je1} and \eqref{je2}.

The homoclinic orbit to be located will take off from $Q_\epsilon$ 
along its unstable curve, flies away from and returns to $\Pi$, lands
near the stable curve of $Q_\epsilon$ and approaches $Q_\epsilon$ spirally.

\subsection{A Signed Distance}

\begin{figure}
\includegraphics{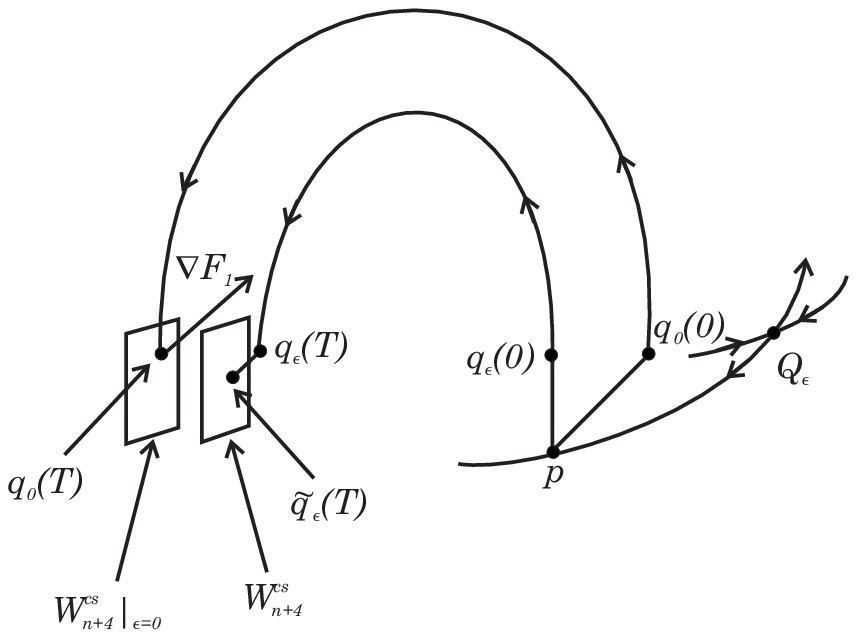}
\caption{The Melnikov measurement.}
\label{mms}
\end{figure}

Let $p$ be any point on $\phi^u_{\sqrt{\e}}$ \eqref{cur} which is the 
unstable curve of $Q_\epsilon $ in $\Pi$ (\ref{sPi}). Let $q_\epsilon (0)$ 
and $q_0(0)$ be any two points on the unstable fibers 
$\mathcal{F}_p^+\mid_\epsilon$ and 
$\mathcal{F}_p^+\mid_{\epsilon=0}$, with the same $\xi^+_1$
coordinate. See Figure \ref{mms} for an illustration.
By the Unstable Fiber Theorem \ref{UFT}, $\mathcal{F}^+_p$ is 
$C^1$ in $\epsilon $ for $\epsilon \in [0,\epsilon_0)$, $\epsilon_0
>0$, thus
\begin{equation*}\| q_\epsilon (0)-q_0(0)\|_{n+8}\leq C\epsilon.\end{equation*}
The key point here is that $\mathcal{F}^+_p\subset H^s$ for any fixed 
$s\geq 1$. By Remark~\ref{reop}, the evolution operator of the
perturbed NLS equation \eqref{spnls} $S^t$ is $C^1$ in $\epsilon$ as a 
map from $H^{n+4}$ to $H^n$ for any fixed $n\geq 1$, $\epsilon \in
[0,\epsilon_0)$, $\epsilon _0>0$. Also $S^t$ is a $C^1$ map on $H^n$ 
for any fixed $t>0$, $n\geq 1$. Thus
\begin{equation*}\| q_\epsilon (T)-q_0(T)\| _{n+4}=\| S^T(q_\epsilon 
(0))-S^T(q_0(0))\|_{n+4}\leq C_1\epsilon,\end{equation*}
where $T>0$ is large enough so that
\begin{equation*}q_0(T)\in W^{cs}_{n+4}\mid_{\epsilon=0}.\end{equation*}
Our goal is to determine when $q_\epsilon (T)\in W^{cs}_n$ through 
Melnikov measurement. Let $q_\epsilon (T)$ and $q_0(T)$ have the coordinate
expressions
\begin{equation}\label{sme} q_\epsilon 
(T)=(\xi^{+,\epsilon}_1,v_\epsilon),\quad 
q_0(T)=(\xi^{+,0}_1,v_0).\end{equation}
Let $\tilde{q}_\epsilon (T)$ be the unique point on 
$W^{cs}_{n+4}$, 
which has the same $v$-coordinate as $q_\epsilon (T)$,
\begin{equation*}\tilde{q}_\epsilon 
(T)=(\tilde{\xi}^{+,\epsilon}_1,v_\epsilon)\in 
W^{cs}_{n+4}.\end{equation*}
By the Center-Stable Manifold Theorem, at points in the subset 
$W^{cs}_{n+4}$, $W^{cs}_n$ is $C^1$ smooth in $\epsilon$ for $\epsilon
\in [0,\epsilon_0)$, $\epsilon_0 >0$, thus
\begin{equation}\label{sme1} \| q_\epsilon (T)-\tilde{q}_\epsilon 
(T)\|_n\leq C_2\epsilon .\end{equation}
Also our goal now is to determine when the signed distance
\begin{equation*}\xi^{+,\epsilon}_1-\tilde{\xi}^{+,\epsilon}_1,\end{equation*}
is zero through Melnikov measurement. Equivalently, one can define 
the signed distance
\begin{equation*}\begin{split} d_1&=\langle \nabla 
F_1(q_0(T)),q_\epsilon (T)-\tilde{q}_\epsilon (T)\rangle\\
&= \partial_qF_1(q_0(T)) (q_\epsilon (T)-\tilde{q}_\epsilon (T))
+ \partial_{\bar{q}}F_1(q_0(T)) (q_\epsilon 
(T)-\tilde{q}_\epsilon (T))^-, \end{split}\end{equation*}
where $\nabla F_1$ is given in (\ref{mvnls1}),
and $q_0(t)$ is the homoclinic orbit given in (\ref{4.13}).
In fact, $q_\epsilon (t)$, $\tilde{q}_\epsilon (t)$, 
$q_0(t)\in H^n$, for any fixed $n\geq 1$. The rest of the derivation
for Melnikov integrals is completely standard. For details, see 
\cite{LMSW96} \cite{LM97}.
\begin{equation}\label{Meln1} d_1=\epsilon M_1+o(\epsilon ),
\end{equation}
where
\begin{equation*}\begin{split} 
M_1&=\int^{+\infty}_{-\infty}\int^{2\pi}_0 
[\partial_qF_1(q_0(t))(\partial^2_xq_0(t)-\alpha
q_0(t)+\beta)\\
&\quad + 
\partial_{\bar{q}}F_1(q_0(t))(\partial^2_x\overline{q_0(t)}-\alpha 
\overline{q_0(t)}+\beta )]dxdt,\end{split}\end{equation*}
where again $q_0(t)$ is given in (\ref{4.13}), and 
$\partial_qF_1$, and $\partial_{\bar{q}}F_1$ are 
given in (\ref{mvnls1}). 

\section{The Second Measurement}

\subsection{The Size of the Stable Manifold of $Q_\e$}

Assume that the Melnikov measurement is successful, that is, the 
orbit $q_\epsilon(t)$ is in the intersection of the unstable manifold 
of $Q_\e$ and the center-stable manifold $W^u(Q_\epsilon )\cap W^{cs}_n$.
The goal of the second measurement is to determine when 
$q_\epsilon (t)$ is also in the codimension 2 stable manifold of $Q_\e$, 
$W^s_n(Q_\epsilon)$, in $H^n$. The existence of $W^s_n(Q_\epsilon)$
follows from standard stable manifold theorem.
$W^s_n(Q_\epsilon)$ can be visualized as a codimension 1 wall in 
$W^{cs}_n$ with base curve $\phi^u_{\sqrt{\e}}$ \eqref{cur} in $\Pi$.
\begin{theorem} \cite{Li01b} The size of $W^s_n(Q_\epsilon)$ off 
$\Pi$ is of order $\mathcal{O}(\sqrt{\epsilon})$ for $\omega \in
\left(\frac{1}{2},1 \right)/S$, where $S$ is a finite 
subset.\label{heiwall}
\end{theorem} 
Starting from the system 
\eqref{nc1}-\eqref{nc3}, one can only get the size of 
$W^s_n(Q_\epsilon)$ off $\Pi$
to be $\mathcal{O}(\epsilon)$ from the standard stable manifold theorem. 
This is not enough for the second measurement. An estimate of order 
$\mathcal{O}(\epsilon^\k)$, 
$\k<1$ can be achieved if the quadratic term $\mathcal{N}_2$
\eqref{wnc5} in \eqref{nc3} can be removed through a normal form 
transformation. Such a normal form transformation has been achieved 
\cite{Li01b}, see also the later section on Normal Form Transforms.

\subsection{An Estimate}

From the explicit expression of $q_0(t)$ (\ref{4.13}), we know that 
$q_0(t)$ approaches $\Pi$ at the 
rate $\mathcal{O}(e^{- \sqrt{4\omega^2-1}t})$. Thus
\begin{equation}\label{sarg1} \text{distance} \left\{ 
q_0(T+\frac{1}{\mu}|\ln \epsilon |), \Pi\right\}<C\epsilon.\end{equation}

\begin{lemma}\label{lesm} For all $t\in \left[T, T+\frac{1}{\mu}|\ln 
\epsilon |\right]$,
\begin{equation}\| q_\epsilon (t)-q_0(t)\|_n\leq \tilde{C}_1\epsilon |\ln 
\epsilon |^2,\end{equation}
where $\tilde{C}_1 =\tilde{C}_1(T)$.
\end{lemma}

Proof: We start with the system \eqref{cspe1}-\eqref{cspe2}. Let
\begin{align*} q_\epsilon (t)&= (\xi^{+,\epsilon}_1(t),J^\epsilon 
(t),\theta^\epsilon (t),h^\epsilon (t),\xi^{-,\epsilon}_1(t)),\\
q_0(t)&=(\xi^{+,0}_1(t),J^0(t),\theta^0(t),h^0(t),\xi^{-,0}_1(t)).\end{align*}
Let $T_1(>T)$ be a time such that
\begin{equation} \| q_\epsilon (t)-q_0(t)\| _n\leq \tilde{C}_2 \epsilon 
|\ln \epsilon |^2,\end{equation}
for all $t\in [T,T_1]$, where $\tilde{C}_2 =\tilde{C}_2(T)$ is 
independent of $\epsilon$. 
 From \eqref{sme1}, such a $T_1$ exists. The proof will be completed
through a continuation argument. For $t\in [T,T_1]$,
\begin{equation}\begin{split}\label{smcd}&
(|\xi^{+,0}_1(t)|+|\xi^{-,0}_1(t)|)+\|h^0(t)\|_n\leq 
C_3re^{-\frac{1}{2}\mu
(t-T)},\quad \ |J^0(t)|\leq C_4\sqrt{\epsilon},\\
& |J^\epsilon (t)|\leq |J^0(t)|+|J^\epsilon(t)-J^0(t)| \leq 
|J^0(t)|+\tilde{C}_2\epsilon |\ln
\epsilon |^2\leq C_5\sqrt{\epsilon},\\
&(|\xi^{+,\epsilon}_1(t)|+|\xi^{-,\epsilon}_1(t)|)+\|h^\epsilon 
(t)\|_n\leq C_3re^{-\frac{1}{2}\mu (t-T)} +\tilde{C}_2\epsilon
|\ln \epsilon |^2,\end{split}\end{equation}
where $r$ is small. Since actually $q_\epsilon (t)$, $q_0(t)\in H^n$ 
for any fixed $n\geq 1$, by Theorem~\ref{CSM},
\begin{equation} |\xi^{+,\epsilon}_1(t)-\xi^{+,0}_1(t)|\leq C_6\| 
v_\epsilon (t)-v_0(t)\|_n+C_7\epsilon ,\end{equation}
whenever $v_\epsilon (t)$, $v_0(t)\in E_{n+4}(r)$, where $v_\epsilon 
(T)=v_\epsilon $ and $v_0(T)=v_0$ are defined in \eqref{sme}. Thus
we only need to estimate $\| v_\epsilon (t)-v_0(t)\|_n$. From 
\eqref{cspe2}, we have for $t\in [T, T_1]$ that
\begin{equation}\label{smeq1} 
v(t)=e^{A(t-T)}v(T)+\int^t_Te^{A(t-\tau)}\tilde{F}(\tau 
)d\tau.\end{equation}
Let $\Delta v(t)=v_\epsilon (t)-v_0(t)$. Then
\begin{equation}\begin{split} \Delta 
v(t)&=[e^{A(t-T)}-e^{A\mid_{\epsilon 
=0}(t-T)}]v_0(T)+e^{A(t-T)}\Delta v(T)\\
&\quad + \int^t_Te^{A(t-\tau )}[\tilde{F}(\tau )-\tilde{F}(\tau 
)|_{\epsilon =0}]d\tau \\
&\quad + \int^t_T[e^{A(t-\tau )}-e^{A\mid_{\epsilon =0}(t-\tau 
)}]\ \tilde{F}(\tau )|_{\epsilon =0}d\tau.\end{split}\end{equation}
By the condition \eqref{smcd}, we have for $t\in [T, T_1]$ that
\begin{equation}\| 
\tilde{F}(t)-\tilde{F}(t)|_{\epsilon=0}\|_n\leq 
[C_8\sqrt{\epsilon}+C_9 re^{-\frac{1}{2}\mu (t-T)} ]\epsilon |\ln
\epsilon |^2.\end{equation}
Then
\begin{equation} \| \Delta v(t)\|_n\leq C_{10}\epsilon 
(t-T)+C_{11}r\epsilon |\ln \epsilon 
|^2+C_{12}\sqrt{\epsilon}(t-T)^2\epsilon
|\ln \epsilon |^2.\end{equation}
Thus by the continuation argument, for $t\in [T, T+\frac{1}{\mu}|\ln 
\epsilon |]$, there is a constant $\hat{C}_1=\hat{C}_1(T)$,
\begin{equation} \| \Delta v(t)\|_n\leq \hat{C}_1\epsilon |\ln 
\epsilon |^2.\end{equation}
Q.E.D.

By Lemma~\ref{lesm} and estimate \eqref{sarg1},
\begin{equation}\label{sarg2} \text{distance} \left\{ q_\epsilon 
( T+\frac{1}{\mu}|\ln \epsilon |), \Pi\right\}<\tilde{C}\epsilon 
|\ln \epsilon |^2.
\end{equation}
By Theorem \ref{heiwall}, the height of the wall $W^s_n(Q_\epsilon)$ 
off $\Pi$ is larger than the distance between $q_\epsilon 
( T+\frac{1}{\mu}|\ln \epsilon |)$ and $\Pi$. Thus, if $q_\epsilon 
( T+\frac{1}{\mu}|\ln \epsilon |)$ can move from one side of the wall 
$W^s_n(Q_\epsilon)$ to the other side, then by continuity $q_\epsilon 
( T+\frac{1}{\mu}|\ln \epsilon |)$ has to be on the wall 
$W^s_n(Q_\epsilon)$ at some values of the parameters. 

\subsection{Another Signed Distance}

\begin{figure}
\includegraphics{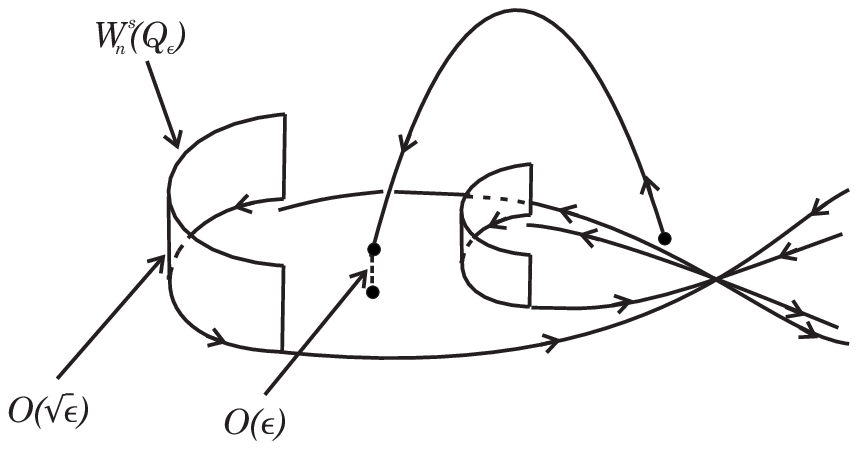}
\caption{The second measurement.}
\label{supsm}
\end{figure}

Recall the fish-like singular level set given by $\mathcal{H}$ 
\eqref{fham}, the width of the fish is of order
$\mathcal{O}(\sqrt{\epsilon })$, and the length of the fish is of 
order $\mathcal{O}(1)$. Notice also that $q_0(t)$ has a phase shift
\begin{equation} \theta^0_1=\theta^0( T+\frac{1}{\mu}|\ln 
\epsilon |)-\theta^0(0).\end{equation}
For fixed $\beta$, changing $\alpha$ can induce $\mathcal{O}(1)$ 
change in the length of the fish, $\mathcal{O}(\sqrt{\epsilon})$ 
change
in $\theta^0_1$, and $\mathcal{O}(1)$ change in $\theta^0(0)$. See 
Figure~\ref{supsm} for an illustration.
The leading order signed distance from $q_\epsilon ( 
T+\frac{1}{\mu}|\ln \epsilon |)$ to $W^s_n(Q_\epsilon )$ can be 
defined
as
\begin{equation}\begin{split} \tilde{d}
&=\mathcal{H}(j_0,\theta^0(0))-\mathcal{H}(j_0,\theta^0(0)+\theta^0_1)\\
&=2\omega \left [\alpha \omega \theta^0_1+\beta [\sin \theta^0(0)-\sin 
(\theta^0(0)+\theta^0_1)]\right ],\end{split}\label{dtd}\end{equation}
where $\mathcal{H}$ is given in \eqref{fham}. The common zero of 
$M_1$ \eqref{Meln1} and $\tilde{d}$ and the implicit function 
theorem
imply the existence of a homoclinic orbit asymptotic to $Q_\epsilon$. 
The common roots to $M_1$ and $\tilde{d}$ are given by $\al = 1/\k(\om)$, 
where $\k(\om)$ is plotted in Figure \ref{kappa}.

\section{Silnikov Homoclinic Orbits in Vector NLS Under Perturbations}

In recent years, novel results have been obtained on the solutions of 
the vector nonlinear Schr\"odinger equations \cite{AOT99} \cite{AOT00} 
\cite{YT01}. Abundant ordinary integrable results have been carried through 
\cite{WF00} \cite{FSW00}, including linear stability calculations 
\cite{FMMW00}. Specifically, the vector nonlinear 
Schr\"odinger equations can be written as
\begin{eqnarray*}
& & ip_t + p_{xx} + \frac{1}{2} (|p|^2 + \chi |q|^2) p = 0 , \\
& & iq_t + q_{xx} + \frac{1}{2} (\chi |p|^2 + |q|^2) q = 0 , 
\end{eqnarray*}
where $p$ and $q$ are complex valued functions of the two real variables 
$t$ and $x$, and $\chi$ is a positive constant. These equations describe
the evolution of two orthogonal pulse envelopes in birefringent optical 
fibers \cite{Men87} \cite{Men89}, with industrial applications in fiber 
communication systems \cite{HK95} and all-optical switching devices 
\cite{Isl92}. For linearly birefringent fibers \cite{Men87}, $\chi =2/3$.
For elliptically birefringent fibers, $\chi$ can take other positive 
values \cite{Men89}. When $\chi = 1$, these equations are first 
shown to be integrable by S. Manakov \cite{Man74}, and thus called Manakov 
equations. When $\chi$ is not 1 or 0, these equations are 
non-integrable. Propelled by the industrial applications, extensive 
mathematical studies on the vector nonlinear Schr\"odinger equations 
have been conducted. Like the scalar nonlinear Schr\"odinger equation, the 
vector nonlinear Schr\"odinger equations also possess figure eight 
structures in their phase space. Consider the singularly perturbed vector 
nonlinear Schr\"odinger equations,
\begin{eqnarray}
& & ip_t + p_{xx} + \frac{1}{2} [(|p|^2 + |q|^2)-\om^2] p = 
i \e [ p_{xx} -\al p - \be ]\ , \label{pnls1}\\
& & iq_t + q_{xx} + \frac{1}{2} [(|p|^2 + |q|^2)-\om^2] q = 
i \e [ q_{xx} -\al q - \be ]\  , \label{pnls2}
\end{eqnarray}
where $p(t,x)$ and $q(t,x)$ are subject to periodic boundary condition 
of period $2\pi$, and are even in $x$, i.e. 
\[
p(t,x + 2 \pi) = p(t,x)\ , \ \ p(t,-x) = p(t,x)\ , 
\]
\[
q(t,x + 2 \pi) = q(t,x)\ , \ \ q(t,-x) = q(t,x)\ , 
\]
$\om \in (1,2)$, $\al > 0$ and $\be$ are real constants, and 
$\e > 0$ is the perturbation parameter. We have
\begin{theorem}[\cite{Li02d}]
There exists a $\e_0 > 0$, such that 
for any $\e \in (0, \e_0)$, there exists 
a codimension 1 surface in the space of $(\alpha,\beta, \om) \in 
\RR^+\times \RR^+\times \RR^+$ where 
$\om \in (1, 2)/S$, $S$ is a finite subset, and 
$\al \om < \sqrt{2} \be$. For any $(\alpha ,\beta, \omega)$ on the 
codimension 1 surface, the singularly perturbed vector nonlinear 
Schr\"odinger equations (\ref{pnls1})-(\ref{pnls2}) possesses a 
homoclinic orbit asymptotic to a saddle
$Q_\epsilon$. This orbit is also the homoclinic orbit 
for the singularly perturbed scalar nonlinear Schr\"odinger equation
studied in last section, and is the only one asymptotic 
to the saddle $Q_\epsilon$ for the singularly perturbed 
vector nonlinear Schr\"odinger equations (\ref{pnls1})-(\ref{pnls2}).
The codimension 1 surface has the 
approximate representation given by $\al = 1/\k(\om)$, where $\k(\om)$ 
is plotted in Figure \ref{kappa}.
\end{theorem} 

\section{Silnikov Homoclinic Orbits in Discrete 
NLS Under Perturbations \label{hordnls}}

We continue from section \ref{GOV} and consider the perturbed discrete
nonlinear Schr\"odinger equation (\ref{PDNLS}). The following theorem 
was proved in \cite{LM97}.

\begin{figure}
\includegraphics{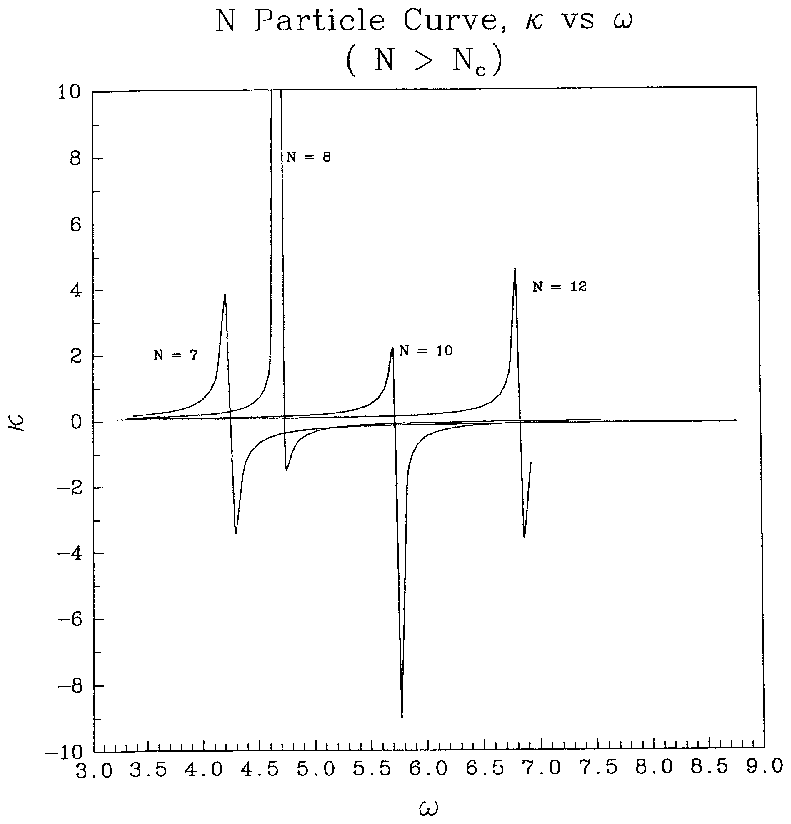}
\includegraphics{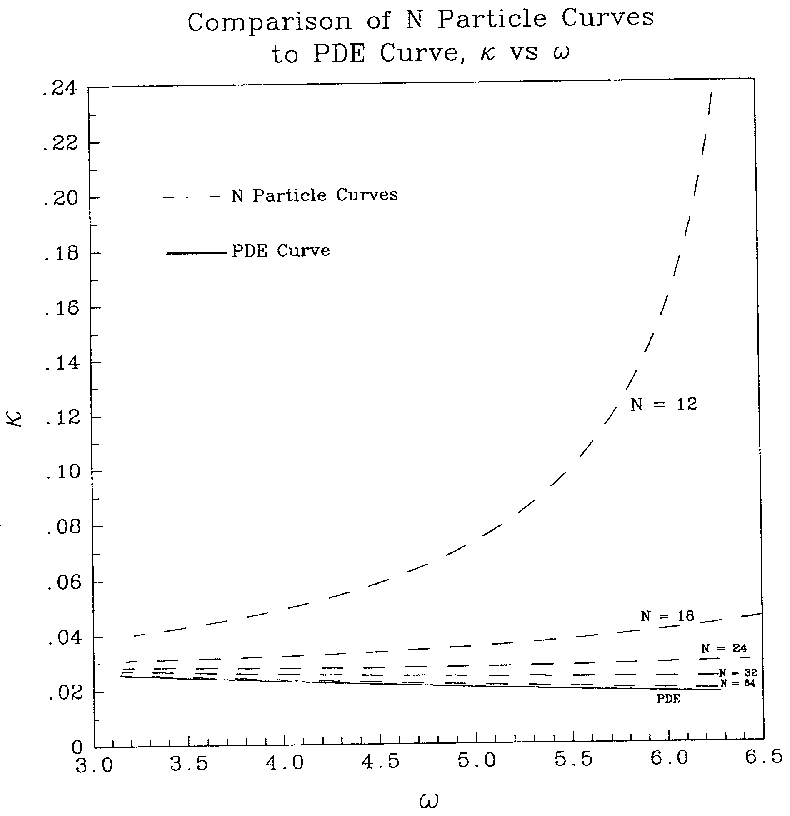}
\caption{The graph of $\k(\om;N)$.}
\label{nkappa}
\end{figure}

Denote by $\Sg_N\ (N\geq 7)$ the external parameter space,
\begin{eqnarray*}
\Sg_N&=&\bigg\{ (\om,\al,\be)\ \bigg | \ \om \in (N\tan{\pi \over N}, 
               N\tan{2\pi \over N}),\\ 
     & &\al\in (0,\al_0), \be\in (0,\be_0); \\
     & &\mbox{where}\ \al_0\ \mbox{and}\ \be_0\ \mbox{are}\ \mbox{any}
        \ \mbox{fixed}\ \mbox{positive}\ \mbox{numbers}. \bigg\}
\end{eqnarray*}
\begin{theorem}
For any $N$ ($7\leq N<\infty$), there exists a positive number $\e_0$, 
such that for any $\e \in (0,\e_0)$, there exists a 
codimension $1$ surface $E_\e$ in $\Sg_N$; for any
external parameters ($\om,\al,\be$) on $E_\e$, there exists a 
homoclinic orbit asymptotic to a saddle $Q_\e$.
The codimension $1$ surface $E_\e$ has the approximate expression
$\al=1/\k$, where $\k=\k(\om;N)$
is shown in Fig.\ref{nkappa}.
\label{dhorbit}
\end{theorem}
In the cases ($3\leq N \leq 6$), $\k$ is always negative. For 
$N \geq 7$, $\k$ 
can be positive as shown in Fig.\ref{nkappa}. When $N$ is even and 
$\geq 7$, there is in fact a symmetric pair of homoclinic orbits asymptotic to 
a fixed point $Q_\e$ at the same values of the external parameters; 
since for even $N$, we have the symmetry:
If $q_n=f(n,t)$ solves (\ref{PDNLS}),
then $q_n=f(n+N/2,t)$ also solves (\ref{PDNLS}). When $N$ is odd
and $\geq 7$, the study can not guarantee that two homoclinic orbits 
exist at the same value of the external parameters.

\section{Comments on DSII Under Perturbations}

We continue from section \ref{invds}. Under both regular and singular 
perturbations, the rigorous Melnikov measurement can be established 
\cite{Li02b}. It turns out that only local well-posedness is necessary 
for rigorously setting up the Melnikov measurement, thanks to the fact 
that the unperturbed homoclinic orbits given in section \ref{dsex} are 
classical solutions. Thus the Melnikov integrals given in section \ref{mids} 
indeed rigorously measure signed distances. For details, see \cite{Li02b}.

The obstacle toward proving the existence of homoclinic 
orbits comes from a technical difficulty in solving a linear system 
to get the normal form for proving the size estimate of the stable 
manifold of the saddle. For details, see \cite{Li02b}.

\section{Normal Form Transforms}

Consider the singularly perturbed Davey-Stewartson II equation in section 
\ref{invds},
\begin{equation}
\left \{ \begin{array}{l} iq_t=\Upsilon q+ [2(|q|^2-\omega^2)+ 
u_y ]q +i\epsilon [\Delta q-\alpha q+\beta ]\ , \cr
\Delta u = -4\partial_y |q|^2 \ . \cr \end{array} \right.
\label{PDS}
\end{equation}
Let
\[
q(t,x,y)=[\rho(t) +f(t,x,y)]e^{i\theta(t)},\quad \lag f \rag = 0,
\]
where $\lag \ , \  \rag$ denotes spatial mean. Let 
\[
I = \lag |q|^2 \rag = \rho^2 + \lag |f|^2 \rag , \quad  J=I-\om^2.
\]
In terms of the new variables ($J, \th, f$), Equation (\ref{PDS}) can 
be rewritten as 
\begin{eqnarray}
\dot{J} &=& \epsilon \bigg [ -2\alpha(J+\omega^2)+2\beta\sqrt{J+\omega^2}
\cos \theta \bigg ] +\epsilon \R_2^J, \label{cc1} \\
\dot{\th} &=& -2J - \epsilon \beta \frac {\sin \theta}{\sqrt{J+\omega^2}}
+\R_2^\th, \label{cc2} \\
f_t &=& L_\epsilon f+V_\epsilon f-i \N_2 -i \N_3, \label{cc3}
\end{eqnarray} 
where 
\begin{eqnarray*}
L_\epsilon f &=& -i\Upsilon f+\epsilon (\Delta-\alpha )f-2i\omega^2
\Delta^{-1}\Upsilon (f+\bar f), \\
\N_2 &=& 2\om \bigg [ \Delta^{-1}\Upsilon |f|^2
+f\Delta^{-1} \Upsilon (f+\bar f) 
- \lag f\Delta^{-1}\Upsilon (f+\bar f)\rag \bigg ],
\end{eqnarray*}
and $\R_2^J$, $\R_2^\th$, $V_\epsilon f$, and $\N_3$ are higher order 
terms. It is the quadratic term $\N_2$ that blocks the size estimate 
of the stable manifold of the saddle \cite{Li02b}.Thus, our goal is to find a normal form transform $g = f + K(f,f)$ 
where $K$ is a bilinear form, that transforms the equation
\[
f_t=L_\epsilon f-i\N_2,
\]
into an equation with a cubic nonlinearity
\[
g_t=L_\epsilon g+{\mathcal O}(\| g\|^3_s),\quad 
(s\geq 2),
\]
where $L_\epsilon$ is given in (\ref{cc3}). In terms of Fourier transforms,
\[
f=\sum_{k\neq 0}\hat{f}(k)e^{ik\cdot \xi },\quad 
\bar{f}=\sum_{k\neq 0}\overline{\hat{f}(-k)}e^{ik\cdot \xi }\ ,
\]
where $k=(k_1,k_2) \in \ZZ^2$, $\xi = (\k_1x, \k_2y)$. The terms in 
$\N_2$ can be written as
\[
\Dl^{-1}\Upsilon |f|^2 = \frac {1}{2} \sum_{k+\ell \neq 0} a(k+\ell) 
\bigg [ \hat{f}(k)\overline{\hat{f}(-\ell )}+\hat{f}(\ell )
\overline{\hat{f}(-k)} \bigg ] e^{i(k+ \ell ) \cdot \xi }\ ,
\]
\[
f\Delta^{-1} \Upsilon f - \lag f\Delta^{-1}\Upsilon f \rag  
= \frac {1}{2} \sum_{k+\ell \neq 0} [a(k)+a( \ell )] \hat{f}(k)\hat{f}(\ell)
e^{i(k+ \ell ) \cdot \xi }\ ,
\]
\[
f\Delta^{-1} \Upsilon \bar{f} - \lag f\Delta^{-1}\Upsilon \bar{f} \rag  
= \frac {1}{2} \sum_{k+\ell \neq 0} \bigg [ a(\ell)\hat{f}(k)
\overline{\hat{f}(-\ell )}+a(k)\hat{f}(\ell)
\overline{\hat{f}(-k)}\bigg ] e^{i(k+ \ell ) \cdot \xi }\ ,
\]
where
\[
a(k) = \frac {k_1^2 \k_1^2 - k_2^2 \k_2^2}{k_1^2 \k_1^2 + k_2^2 \k_2^2}\ .
\]
We will search for a normal form transform of the 
general form,
\[
g=f+K(f,f),
\]
where
\begin{eqnarray*}
K(f,f) &=& \sum_{k+\ell\neq 0}\left[
\hat{K}_1(k,\ell)\hat{f}(k) 
\hat{f}(\ell)+\hat{K}_2(k,\ell)\hat{f}(k)\overline{\hat{f}(-\ell)}\right.\\
& &\quad \left.+\hat{K}_2(\ell,k)\overline{\hat{f}(-k)}\hat{f}(\ell)+
\hat{K}_3(k,\ell)\overline{\hat{f}(-k)}\overline{\hat{f}(-\ell)}\right]
e^{i(k+\ell)x},
\end{eqnarray*}
where $\hat{K}_j(k,\ell)$, $(j=1,2,3)$ are the unknown coefficients to 
be determined, and $\hat{K}_j(k,\ell)=\hat{K}_j(\ell,k)$, $(j=1,3)$.
To eliminate the quadratic terms, we first need to set
\[
iL_\epsilon K(f,f)-iK(L_\epsilon 
f,f)-iK(f,L_\epsilon f)=\N_2,
\]
which takes the explicit form:
\begin{eqnarray}
& &(\sigma_1+i\sigma)\hat{K}_1(k,\ell)+B(\ell)\hat{K}_2(k,\ell)+B(k)
\hat{K}_2(\ell,k)\nonumber \\
& & \quad +B(k+\ell)\overline{\hat{K}_3(k,\ell)}=\frac{1}{2\om} 
[B(k)+B(\ell)],\label{nore1}\\
& &-B(\ell)\hat{K}_1(k,\ell)+(\sigma_2+i\sigma)\hat{K}_2(k,\ell)+ B(k+\ell)
\overline{\hat{K}_2(\ell,k)}\nonumber \\
& & \quad +B(k)\hat{K}_3(k,\ell)=\frac{1}{2\om} 
[B(k+\ell)+B(\ell)],\label{nore2}\\
& &-B(k)\hat{K}_1(k,\ell)+B(k+\ell)
\overline{\hat{K}_2(k,\ell)}+(\sigma_3+i\sigma)\hat{K}_2(\ell,k)\nonumber \\
& & \quad +B(\ell)\hat{K}_3(k,\ell)= \frac{1}{2\om} 
[B(k+\ell)+B(k)],\label{nore3}\\
& &B(k+\ell) \overline{\hat{K}_1(k,\ell)}-B(k)\hat{K}_2(k,\ell) 
-B(\ell)\hat{K}_2(\ell,k)\nonumber \\
& & \quad +(\sigma _4+i\sigma )\hat{K}_3(k,\ell)=0,\label{nore4}
\end{eqnarray}
where $B(k)=2\om^2 a(k)$, and 
\begin{eqnarray*} 
& & \sg = \e \bigg [ \al -2(k_1\ell_1\k_1^2+k_2\ell_2\k_2^2)\bigg ]\ , \\
& & \sg_1= 2(k_2\ell_2\k_2^2-k_1\ell_1\k_1^2) +B(k+\ell)-B(k)-B(\ell)\ , \\
& & \sg_2= 2[(k_2+\ell_2)\ell_2\k_2^2-(k_1+\ell_1)\ell_1\k_1^2] 
+B(k+\ell)-B(k)+B(\ell)\ , \\
& & \sg_3= 2[(k_2+\ell_2)k_2\k_2^2-(k_1+\ell_1)k_1\k_1^2] 
+B(k+\ell)+B(k)-B(\ell)\ , \\
& & \sg_4= 2[(k_2^2+k_2\ell_2+\ell_2^2)\k_2^2-(k_1^2+k_1\ell_1+\ell_1^2)
\k_1^2] +B(k+\ell)+B(k)+B(\ell)\ . 
\end{eqnarray*}
Since these coefficients are even in $(k,\ell)$, we will search for 
even solutions, i.e.
\[
\hat{K}_j(-k,-\ell)=\hat{K}_j(k,\ell),\quad j=1,2,3.
\]

The technical difficulty in the normal form transform comes from not 
being able to answer the following two questions in solving the linear 
system (\ref{nore1})-(\ref{nore4}):
\begin{enumerate}
\item Is it true that for all $k, \ell \in \ZZ^2/\{ 0\}$, there 
is a solution ?
\item What is the asymptotic behavior of the solution as $k$ and/or 
$\ell \ra \infty$ ? In particular, is the asymptotic behavior like 
$k^{-m}$ and/or $\ell^{-m}$ ($m\geq 0$) ?
\end{enumerate}

\nid
Setting $\pa_y = 0$ in (\ref{PDS}), the singularly perturbed 
Davey-Stewartson II equation reduces to the singularly perturbed 
nonlinear Schr\"odinger (NLS) equation (\ref{spnls}) in section \ref{spsec}.
Then the above two questions can be answered \cite{Li01b}. For the 
regularly perturbed nonlinear Schr\"odinger (NLS) equation (\ref{rpnls})
in section \ref{rpsec}, the answer to the above two questions is even 
simplier \cite{LMSW96} \cite{Li01b}.

\section{Transversal Homoclinic Orbits in a Periodically Perturbed 
SG \label{PPSGE}}

Transversal homoclinic orbits in continuous systems often appear 
in two types of systems: (1). periodic systems where the Poincar\'e 
period map has a transversal homoclinic orbit, (2). autonomous 
systems where the homoclinic orbit is asymptotic to a hyperbolic 
limit cycle. 

Consider the periodically perturbed sine-Gordon (SG) equation,
\begin{equation}
u_{tt}=c^2 u_{xx}+\sin u+\epsilon [-a u_t+u^3 \chi(\| u\|)\cos t],
\label{PSG}
\end{equation} 
where
\[
\chi(\| u\|)=\left\{ \begin{array}{ll} 1, & \| 
u\| \leq M,\\ 0, & \| u\| \geq 2M,\end{array}\right.
\]
for $M<\| u\| <2M$, $\chi (\| u\|)$ is a smooth bump function, 
under odd periodic boundary condition,
\[
u(x+2\pi ,t)=u(x,t),\quad 
u(x,t)=-u(x,t),
\]
$\frac{1}{4}<c^2<1$, $a>0$, $\epsilon$ is a small 
perturbation parameter.
\begin{theorem}[\cite{LMSW96}, \cite{SZ00}] There exists an 
interval $I\subset \RR^{+}$ such that for any $a\in I$, there exists a
transversal homoclinic orbit
$u=\xi (x,t)$ asymptotic to $0$ in $H^{1}$.
\end{theorem}

\section{Transversal Homoclinic Orbits in a Derivative NLS \label{ADNS}}

Consider the derivative nonlinear Schr\"odinger equation,
\begin{equation}
i q_t = q_{xx} + 2 |q|^2 q +i \e \bigg [ (\frac{9}{16}-|q|^2 )q +\mu 
|\hat{\pa}_x q|^2 \bar{q} \bigg ]\ , \label{derNLS}
\end{equation}
where $q$ is a complex-valued function of two real variables $t$ and $x$,
$\e > 0$ is the perturbation parameter, $\mu$ is a real constant, and
$\hat{\pa}_x $ is a bounded Fourier multiplier,
\[
\hat{\pa}_x q = -\sum_{k=1}^K k \tq_k \sin kx\ , \quad 
\mbox{for} \ q = \sum_{k=0}^\infty \tq_k \cos kx\ ,
\]
and some fixed $K$. Periodic boundary condition 
and even constraint are imposed,
\[
q(t,x+2\pi ) = q(t,x)\ , \ \ q(t,-x)=q(t,x) \ . 
\] 
\begin{theorem}[\cite{Li02a}]
There exists a $\e_0 > 0$, such that 
for any $\e \in (0, \e_0)$, and $|\mu | > 5.8$,
there exist two transversal homoclinic orbits asymptotic to 
the limit cycle $q_c = \frac{3}{4} \exp \{ -i [ \frac{9}{8} t + \ga ]\}$.
\label{thmdns}
\end{theorem}

\chapter{Existence of Chaos}

The importance of homoclinic orbits with respect to chaotic dynamics was 
first realized by Poincar{\'e} \cite{Poi99}. In 1961, Smale constructed 
the well-known horseshoe in the neighborhood of a transversal homoclinic 
orbit \cite{Sma61} \cite{Sma65} \cite{Sma67}. In particular, 
Smale's theorem implies Birkhoff's theorem on the existence of a 
sequence of structurely stable periodic orbits in the neighborhood 
of a transversal homoclinic orbit \cite{Bir12}. In 1984 and 1988 \cite{Pal84}
\cite{Pal88}, Palmer gave a beautiful proof of Smale's theorem using a 
shadowing lemma. Later, this proof was generalized to infinite dimensions 
by Steinlein and Walther \cite{SW89} \cite{SW90} and Henry \cite{Hen94}. 
In 1967, Silnikov proved Smale's theorem for autonomous systems in finite 
dimensions using a fixed point argument \cite{Sil67b}. In 1996, Palmer proved 
Smale's theorem for autonomous systems in finite dimensions using shadowing 
lemma \cite{CKP95} \cite{Pal96}. In 2002, Li proved Smale's theorem for 
autonomous systems in infinite dimensions using shadowing lemma \cite{Li02a}.

For nontransversal homoclinic orbits, the most well-known type which leads 
to the existence of Smale horseshoes is the so-called Silnikov homoclinic 
orbit \cite{Sil65} \cite{Sil67a} \cite{Sil70} \cite{Den89} \cite{Den93}. 
Existence of Silnikov homoclinic orbits and new constructions of Smale 
horseshoes for concrete nonlinear wave systems have been established 
in finite dimensions \cite{LM97} \cite{LW97} and in infinite dimensions 
\cite{LMSW96} \cite{Li99a} \cite{Li01b}. 

\section{Horseshoes and Chaos}

\subsection{Equivariant Smooth Linearization}

Linearization has been a popular topics in dynamical systems. It asks the 
question whether or not one can transform a nonlinear system into its 
linearized system at a fixed point, in a small neighborhood of the fixed point.
For a sample of references, see \cite{Har63} \cite{Sel85}. Here we 
specifically consider the singularly perturbed nonlinear Schr\"odinger (NLS) 
equation (\ref{spnls}) in sections \ref{spsec} and \ref{horsnls}. 
The symmetric pair of Silnikov homoclinic orbits is asymptotic to the 
saddle $Q_\e =\sqrt{I} e^{i\th}$, where
\begin{equation}
I=\omega^2-\epsilon \frac{1}{2\omega}\sqrt{\beta^2-\alpha^2\omega^2}+\cdots ,
\quad \cos \theta  =\frac{\alpha \sqrt{I}}{\beta}, \quad \theta \in 
(0,\frac{\pi}{2}).
\label{sQec}
\end{equation}
Its eigenvalues are
\begin{equation}
\la_n^\pm = -\e [\al +n^2]\pm 2 \sqrt{(\frac{n^2}{2} +\om^2-I)(3I -\om^2 -
\frac{n^2}{2} )}\ , 
\label{sQev}
\end{equation}
where $n=0,1,2, \cdots $, $\om \in (\frac{1}{2}, 1)$, and $I$ is given in 
(\ref{sQec}). 
In conducting linearization, the crucial factor is the so-callled nonresonance 
conditions. 
\begin{lemma}[\cite{Li99a} \cite{Li02c}]
For any fixed $\e \in (0,\e_0)$, let $E_\e$ be the codimension 1 surface in 
the external parameter space, on which the symmetric pair of Silnikov 
homoclinic orbits are supported (cf: Theorem \ref{shorbit}). For almost 
every $(\al,\be,\om)\in E_\e$, the eigenvalues $\la_n^\pm$ (\ref{Qev}) 
satisfy the nonresonance condition of Siegel type: There exists a natural 
number $s$ such that for any integer $n \geq 2$,
\[
\bigg | \La_n - \sum_{j=1}^r \La_{l_j} \bigg | \geq  1/r^s\ ,
\]
for all $r =2,3, \cdots, n$ and all $l_1, l_2, \cdots, l_r \in \ZZ$, 
where $\La_n = \la_n^+$ for $n \geq 0$, and $\La_n = \la_{-n-1}^-$ for $n <0$.
\end{lemma}
Thus, in a neighborhood of $Q_\e$, the singularly perturbed NLS (\ref{spnls}) 
is analytically equivalent to its linearization at $Q_\e$ \cite{Nik86}.
In terms of eigenvector basis, (\ref{spnls}) can be rewritten as 
\begin{eqnarray}
\dot{x} &=& -a x - b y +\N_x(\vec{Q}), \nonumber \\
\dot{y} &=& b x - a y +\N_y(\vec{Q}), \nonumber \\
\dot{z}_1 &=& \ga_1 z_1 +\N_{z_1}(\vec{Q}), \label{nfeq} \\
\dot{z}_2 &=& \ga_2 z_2 +\N_{z_2}(\vec{Q}), \nonumber \\
\dot{Q} &=& LQ +\N_{Q}(\vec{Q}); \nonumber
\end{eqnarray}
where $a= -\mbox{Re}\{\la^+_2\}$, $b=\mbox{Im}\{\la^+_2\}$, 
$\ga_1 = \la^+_0$, $\ga_2 = \la^+_1$; $\N$'s vanish identically 
in a neighborhood $\Om$ of $\vec{Q} =0$, $\vec{Q} = (x,y,z_1,z_2,Q)$, $Q$ 
is associated with the rest of eigenvalues, $L$ is given as
\[
LQ = -i Q_{\z\z} -2 i [(2|Q_\e|^2-\om^2)Q+Q^2_\e\bar{Q}]+
\e [-\al Q+Q_{\z\z}]\ ,
\]
and $Q_\e$ is given in (\ref{sQec}).

\subsection{Conley-Moser Conditions}

We continue from last section. Denote by $h_k$ ($k=1,2$) the symmetric 
pair of Silnikov homoclinic orbits. 
The symmetry $\sg$ of half spatial period shifting has the new 
representation in terms of the new coordinates
\begin{equation}
\sg \circ (x,y,z_1,z_2,Q) = (x,y,z_1,-z_2,\sg \circ Q).
\label{symm}
\end{equation}
\begin{definition}
The Poincar\'e section $\Sg_0$ is defined by the constraints:
\begin{eqnarray*}
& & y=0,\ \eta \exp \{ -2 \pi a/b \} < x < \eta; \\
& & 0 < z_1 <\eta,\ -\eta < z_2 <\eta, \ \|Q\| < \eta;
\end{eqnarray*}
where $\eta$ is a small parameter.
\end{definition}
The horseshoes are going to be constructed on this Poincar\'e section.
\begin{definition}
The auxiliary Poincar\'e section $\Sg_1$ is defined by the constraints:
\begin{eqnarray*}
& & z_1 =\eta ,\quad  -\eta < z_2 <\eta ,  \\
& & \sqrt{x^2+y^2} < \eta , \quad \|Q\| < \eta . 
\end{eqnarray*}
\end{definition}
Maps from a Poincar\'e section to a Poincar\'e section are induced by the 
flow. Let $\vQ^0$ and $\vQ^1$ be the coordinates on $\Sg_0$ and $\Sg_1$ 
respectively, then the map $P_0^1$ from $\Sg_0$ to $\Sg_1$ has the 
simple expression:
\begin{eqnarray*}
x^1 &=& \bigg (\frac{z_1^0}{\eta}\bigg )^{\frac{a}{\ga_1}}x^0 \cos \bigg 
[ \frac{b}{\ga_1}\ln \frac{\eta}{z_1^0} \bigg ]\ , \\
y^1 &=& \bigg (\frac{z_1^0}{\eta}\bigg )^{\frac{a}{\ga_1}}x^0 \sin \bigg 
[ \frac{b}{\ga_1}\ln \frac{\eta}{z_1^0} \bigg ]\ , \\
z_2^1 &=& \bigg (\frac{\eta}{z_1^0}\bigg )^{\frac{\ga_2}{\ga_1}}z_2^0\ , \\
Q^1 &=& e^{t_0 L} Q^0\ .
\end{eqnarray*}
Let $\vQ_*^0$ and $\vQ_*^1$ be the intersection points of the homoclinic 
orbit $h_1$ with $\overline{\Sg_0}$ and $\Sg_1$ respectively, and let  
$\vtQ^0 = \vQ^0 - \vQ^0_*$, and $\vtQ^1 = \vQ^1 - \vQ^1_*$. One can define
slabs in $\Sg_0$. 
\begin{definition}
For sufficiently large natural number $l$, we define slab $S_l$ in $\Sg_0$ 
as follows: 
\begin{eqnarray*}
S_l &\equiv& \bigg \{ \vQ \in \Sg_0 \ \bigg | \ \eta \exp \{ -\ga_1 
(t_{0,2(l+1)}-{\pi \over 2b})\} \leq \\
& & \tz^0_1(\vQ) \leq \eta \exp \{ -\ga_1 (t_{0,2l}-{\pi \over 2b})\}, \\
& & |\tx^0(\vQ)| \leq \eta \exp \{ -{1\over 2}a \ t_{0,2l} \}, \\
& & |\tz^1_2(P^1_0(\vQ))| \leq \eta \exp \{ -{1\over 2}a \ t_{0,2l} \}, \\
& & \|\tQ^1(P^1_0(\vQ))\| \leq \eta \exp \{ -{1\over 2}a \ t_{0,2l} \} 
\bigg \},
\end{eqnarray*}
where 
\[
t_{0,l} = {1 \over b} [l\pi -\varphi_1] +o(1)\ ,
\]
as $l \ra +\infty$, are the time that label the fixed points of the 
Poincar\'e map $P$ from $\Sg_0$ to $\Sg_0$ \cite{Li99a} \cite{Li02c},
and the notations $\tx^0(\vQ)$, $\tz^1_2(P^1_0(\vQ))$, etc. denote the 
$\tx^0$ coordinate of the point $\vQ$, the $\tz^1_2$ coordinate of the 
point $P^1_0(\vQ)$, etc..
\label{dfslab}
\end{definition}
$S_l$ is defined so that it includes two fixed points $p^+_l$ and $p^-_l$ 
of $P$.
Let $S_{l,\sg}= \sg \circ S_l$ where the symmetry $\sg$ is defined in 
(\ref{symm}). We need to define a larger slab $\hS_l$ such that 
$S_l \cup S_{l,\sg} \subset \hS_l$.
\begin{definition}
The larger slab $\hS_l$ is defined as
\begin{eqnarray*}
\hS_l &=& \bigg \{ \vQ \in \Sg_0 \ \bigg | \ \eta \exp \{ -\ga_1 
(t_{0,2(l+1)} -{\pi \over 2b})\} \leq \\
& & z^0_1(\vQ) \leq \eta \exp \{ -\ga_1 (t_{0,2l} -{\pi \over 2b})\},\\
& & |x^0(\vQ)-x^0_*| \leq \eta \exp \{ -{1\over 2} a\ t_{0,2l}\}, \\
& & |z^1_2(P^1_0(\vQ))| \leq |z^1_{2,*}| + \eta \exp \{ -{1\over 2} 
a\ t_{0,2l}\},\\
& & \|Q^1(P^1_0(\vQ))\| \leq \eta \exp \{ -{1\over 2}a\ t_{0,2l}\} \bigg \},
\end{eqnarray*}
where $z^1_{2,*}$ is the $z^1_2$-coordinate of $\vQ^1_*$. 
\end{definition}
\begin{definition}
In the coordinate system $\{ \tx^0,\tz^0_1,\tz^0_2,\tQ^0 \}$, the stable 
boundary of $\hS_l$, denoted by $\pa_s \hS_l$, is defined to be the boundary 
of $\hS_l$ along ($\tx^0,\tQ^0$)-directions, and the unstable 
boundary of $\hS_l$, denoted by $\pa_u \hS_l$, is defined to be the boundary 
of $\hS_l$ along ($\tz^0_1,\tz^0_2$)-directions. A stable slice $V$ in 
$\hS_l$ is a subset of $\hS_l$, defined as the region swept out through 
homeomorphically moving and deforming 
$\pa_s \hS_l$ in such a way that the part
\[
\pa_s \hS_l \cap \pa_u \hS_l
\]
of $\pa_s \hS_l$ only moves and deforms inside $\pa_u \hS_l$. The 
new boundary obtained through such moving and deforming of 
$\pa_s \hS_l$ is called the stable boundary of $V$, which is denoted by 
$\pa_s V$. The rest of the boundary of $V$ is called its unstable 
boundary, which is denoted by $\pa_u V$. An unstable slice of $\hS_l$,
denoted by $H$, is defined similarly.
\end{definition}
As shown in \cite{Li99a} and \cite{Li02c}, under certain generic 
assumptions, when $l$ is sufficiently 
large, $P(S_l)$ and $P(S_{l,\sg})$ intersect $\hS_l$ into four disjoint 
stable slices 
$\{ V_1,V_2 \}$ and $\{ V_{-1},V_{-2}\}$ in $\hS_l$. $V_j$'s 
($j=1,2,-1,-2$) do not intersect $\pa_s\hS_l$; moreover,
\begin{equation}
\pa_s V_i \subset P(\pa_sS_l), (i=1,2);\ 
\pa_s V_i \subset P(\pa_sS_{l,\sg}), (i=-1,-2).
\label{bdc}
\end{equation}
Let
\begin{equation}
H_j = P^{-1}(V_j),\ \ (j=1,2,-1,-2), \label{defhs}
\end{equation}
where and for the rest of this article, $P^{-1}$ denotes preimage of $P$. 
Then $H_j$ ($j=1,2,-1,-2$) are unstable slices. More importantly, the 
Conley-Moser conditions are satisfied. Specifically,
Conley-Moser conditions are:
\framebox[1.8in][l]{Conley-Moser condition (i):}
\[
\left \{ \begin{array}{l} V_j = P(H_j), \\ \pa_sV_j = P(\pa_sH_j), \ 
\ (j=1,2,-1,-2) \\ \pa_uV_j = P(\pa_uH_j).
\end{array}\right.
\]
\framebox[1.8in][l]{Conley-Moser condition (ii):} There exists a constant 
$0< \nu <1$, such that for any stable slice $V \subset V_j\ \ (j=1,2,-1,-2)$,
the diameter decay relation
\[
d(\tilde{V}) \leq \nu d(V)
\]
holds, where $d(\cdot)$ denotes the diameter \cite{Li99a},
and $\tilde{V}=P(V\cap H_k), \ \ (k=1,2,-1,-2)$;
for any unstable slice $H \subset H_j\ \ (j=1,2,-1,-2)$, the diameter 
decay relation
\[
d(\tilde{H}) \leq \nu d(H)
\]
holds, where $\tilde{H}=P^{-1}(H\cap V_k), \ \ (k=1,2,-1,-2)$.

The Conley-Moser conditions are sufficient conditions for establishing
the topological conjugacy between the Poincare map $P$ restricted to
a Cantor set in $\Sg_0$, and the shift automorphism on symbols. It also 
display the horseshoe nature of the intersection of $S_l$ and $S_{l,\sg}$ with 
their images $P(S_l)$ and $P(S_{l,\sg})$.

\subsection{Shift Automorphism}

Let $\W$ be a set which consists of elements of the doubly infinite 
sequence form:
\[
a =(\cdot \cdot \cdot  a_{-2} a_{-1} a_0, a_1 a_2 \cdot \cdot \cdot ),
\]
where $a_k \in \{ 1, 2, -1, -2\}$; $k\in \ZZ$. We introduce a topology in $\W$
by taking as neighborhood basis of 
\[
a^* =( \cdot \cdot \cdot a^*_{-2} a^*_{-1} a^*_0, a^*_1 a^*_2 
\cdot \cdot \cdot ),
\]
the set 
\[
W_j =  \bigg \{ a\in \W \ \bigg | \ a_k=a^*_k\ (|k|<j) \bigg \}
\]
\nid
for $j=1,2,\cdot \cdot \cdot $. This makes $\W$ a topological space.
The shift automorphism $\chi$ is defined on $\W$ by
\begin{eqnarray*}
\chi &:& \W \mapsto \W, \\
  & & \forall a \in \W,\ \chi(a) = b,\ \mbox{where}\ b_k=a_{k+1}.
\end{eqnarray*}
The shift automorphism $\chi$ exhibits {\em{sensitive dependence on 
initial conditions}}, which is a hallmark of {\em{chaos}}.

Let
\[
a =(\cdot \cdot \cdot  a_{-2} a_{-1} a_0, a_1 a_2 \cdot \cdot \cdot ),
\]
\nid
be any element of $\W$. Define inductively for $k \geq 2$ the stable
slices 
\begin{eqnarray*}
& & V_{a_0 a_{-1}}=P(H_{a_{-1}}) \cap H_{a_0}, \\
& & V_{a_0 a_{-1} ... a_{-k}}=P(V_{a_{-1} ... a_{-k}}) \cap H_{a_0} .
\end{eqnarray*}
\nid
By Conley-Moser condition (ii),
\[
d(V_{a_0 a_{-1} ... a_{-k}}) \leq \nu_1 d(V_{a_0 a_{-1} ... a_{-(k-1)}})
\leq ... \leq \nu_1^{k-1} d(V_{a_0 a_{-1}}).
\]
\nid
Then,
\[
V(a)=\bigcap^{\infty}_{k=1}V_{a_0 a_{-1} ... a_{-k}}
\]
\nid
defines a 2 dimensional continuous surface in $\Sg_0$; moreover,
\begin{equation}
\pa V(a) \subset \pa_u \hS_l. \label{inter1}
\end{equation}
\nid
Similarly, define inductively for $k \geq 1$ the unstable slices
\begin{eqnarray*}
& & H_{a_0 a_1}=P^{-1}(H_{a_1} \cap V_{a_0}), \\
& & H_{a_0 a_1 ... a_k}=P^{-1}(H_{a_1 ... a_k} \cap V_{a_0}) .
\end{eqnarray*}
\nid
By Conley-Moser condition (ii),
\[
d(H_{a_0 a_1 ... a_k}) \leq \nu_2 d(H_{a_0 a_1 ... a_{k-1}})
\leq ... \leq \nu_2^k d(H_{a_0}).
\]
\nid
Then,
\[
H(a)=\bigcap^{\infty}_{k=0}H_{a_0 a_1 ... a_k}
\]
\nid
defines a codimension 2 continuous surface in $\Sg_0$; moreover,
\begin{equation}
\pa H(a) \subset \pa_s \hS_l. \label{inter2}
\end{equation}
\nid
By (\ref{inter1};\ref{inter2}) and dimension count,
\[
V(a) \cap H(a) \neq \emptyset
\]
\nid
consists of points. Let 
\[
p\in V(a) \cap H(a)
\]
\nid
be any point in the intersection set. Now we define the mapping
\begin{eqnarray*}
\phi &:& \W \mapsto \hS_l, \\
  & & \phi(a) = p. 
\end{eqnarray*}
\nid
By the above construction,
\[
P(p)=\phi (\chi(a)).
\]
\nid
That is,
\[
P\circ \phi =\phi \circ \chi .
\]
\nid
Let 
\[
\La \equiv \phi (\W ),
\]
\nid
then $\La$ is a compact Cantor subset of $\hS_l$, and invariant
under the Poincare map $P$. 
Moreover, with the topology inherited from $\hS_l$ for $\La$,
$\phi$ is a homeomorphism from $\W$ to $\La$. 

\section{Nonlinear Schr\"odinger Equation Under Singular Perturbations
\label{shsnls}}

Finally, one has the theorem on the existence of chaos in the singularly
perturbed nonlinear Schr\"odinger system (\ref{spnls}).
\begin{theorem}[Chaos Theorem, \cite{Li02c}]
Under certain generic assumptions, for the 
perturbed nonlinear Schr\"odinger system (\ref{spnls}), there exists a 
compact Cantor subset 
$\La$ of $\hS_l$, $\La$ consists of points, and is invariant under $P$.
$P$ restricted to $\La$, is topologically conjugate to the shift 
automorphism $\chi$ on four symbols $1, 2, -1, -2$. That is, there exists
a homeomorphism
\[
\phi \ : \ \W \mapsto  \La,
\]
\nid
such that the following diagram commutes:
\begin{equation} 
\begin{array}{ccc}
\W &\maprightu{\phi} & \Lambda\\
\mapdownl{\chi} & & \mapdownr{P}\\
\W & \maprightd{\phi} & \Lambda
\end{array} 
\nonumber
\end{equation}
\label{horseshthm}
\end{theorem}
Although the symmetric pair of Silnikov homoclinic orbits is not structurally 
stable, the Smale horseshoes are structurally stable. Thus, the Cantor 
sets and the conjugacy to the shift automorphism are structurally stable.

\section{Nonlinear Schr\"odinger Equation Under Regular Perturbations
\label{shrnls}}

We continue from section \ref{horrnls} and consider the regularly perturbed 
nonlinear Schr\"odinger (NLS) equation (\ref{rpnls}). Starting from the 
Homoclinic Orbit Theorem \ref{rhorbit}, the arguments in previous sections 
equally apply. In fact, under the regular perturbation, the argument can be 
simplified due to the fact that the evolution operator is a group.
Thus, the Chaos Theorem \ref{horseshthm} holds for the regularly perturbed 
nonlinear Schr\"odinger (NLS) equation (\ref{rpnls}).

\section{Discrete Nonlinear Schr\"odinger Equation Under Perturbations
\label{shdnls}}

We continue from section \ref{hordnls}. Starting from the 
Homoclinic Orbit Theorem \ref{dhorbit}, the arguments in previous sections 
equally apply and are easier. Thus, the Chaos Theorem \ref{horseshthm} 
holds for the perturbed discrete nonlinear Schr\"odinger equation 
(\ref{PDNLS}). In the case $N \geq 7$ and is odd, we only define one 
slab $S_l$. Consequently, the Poincar\'e map is topologically conjugate 
to the shift automorphism on two symbols. 

\section{Numerical Simulation of Chaos}

The finite-difference discretization of both the regularly and the singularly 
perturbed nonlinear Schr\"odinger equations (\ref{rpnls}) and (\ref{spnls})
leads to the same discrete perturbed nonlinear Schr\"odinger equation 
(\ref{PDNLS}).
 
\begin{figure}
\includegraphics{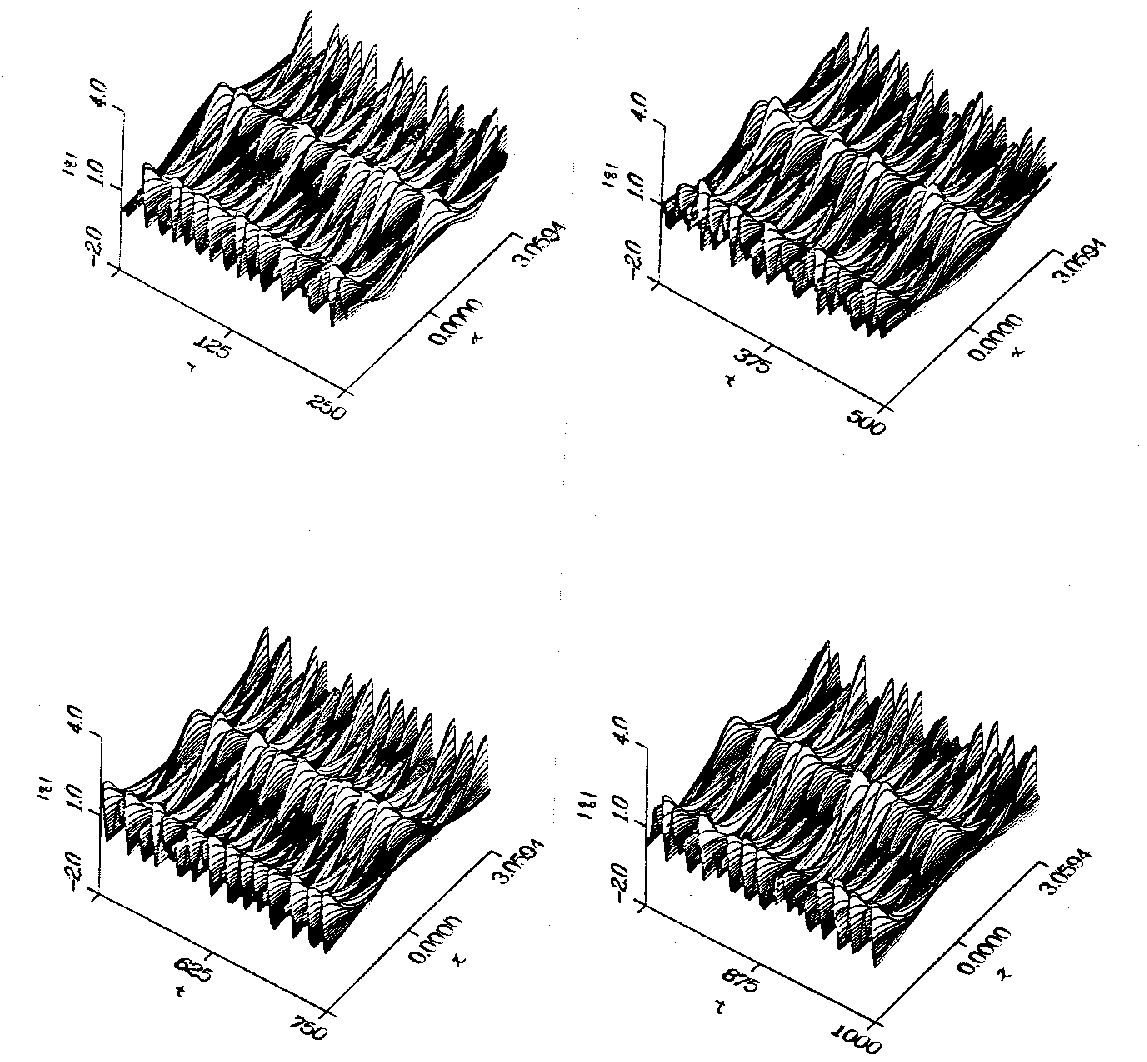}
\caption{A chaotic solution in the discrete perturbed NLS 
system (\ref{PDNLS}).}
\label{numo}
\end{figure}

\begin{figure}
\includegraphics{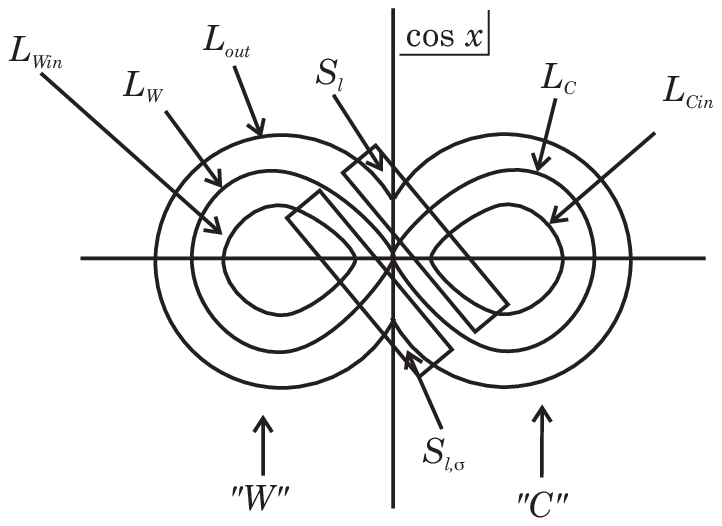}
\caption{Figure-eight structure and its correspondence with the chaotic
center-wing jumping.}
\label{pfi}
\end{figure}

\begin{figure}
\includegraphics{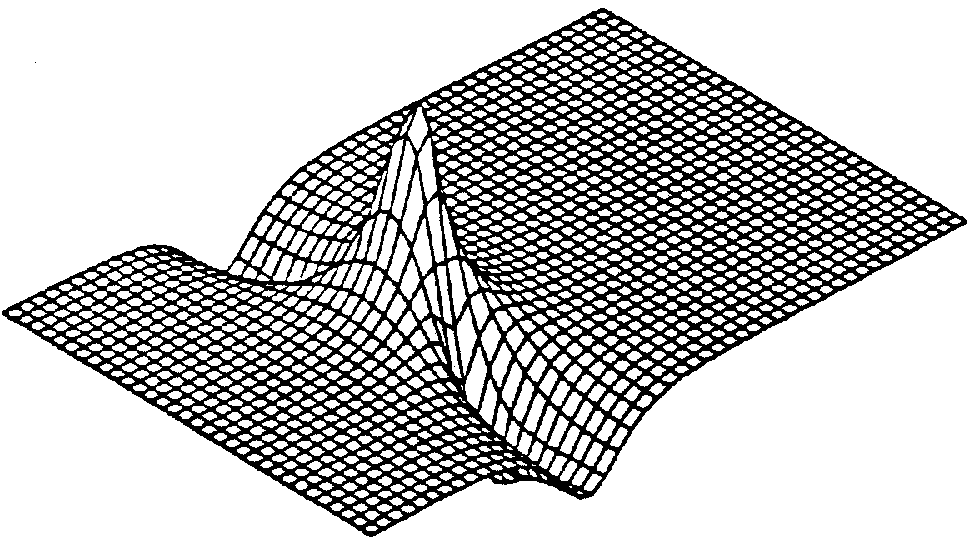}
\caption{Spatial-temporal profile realization of $L_C$ in 
Figure \ref{pfi} (coordinates are the same as in Figure \ref{numo}).}
\label{figlc}
\end{figure}

\begin{figure}
\includegraphics{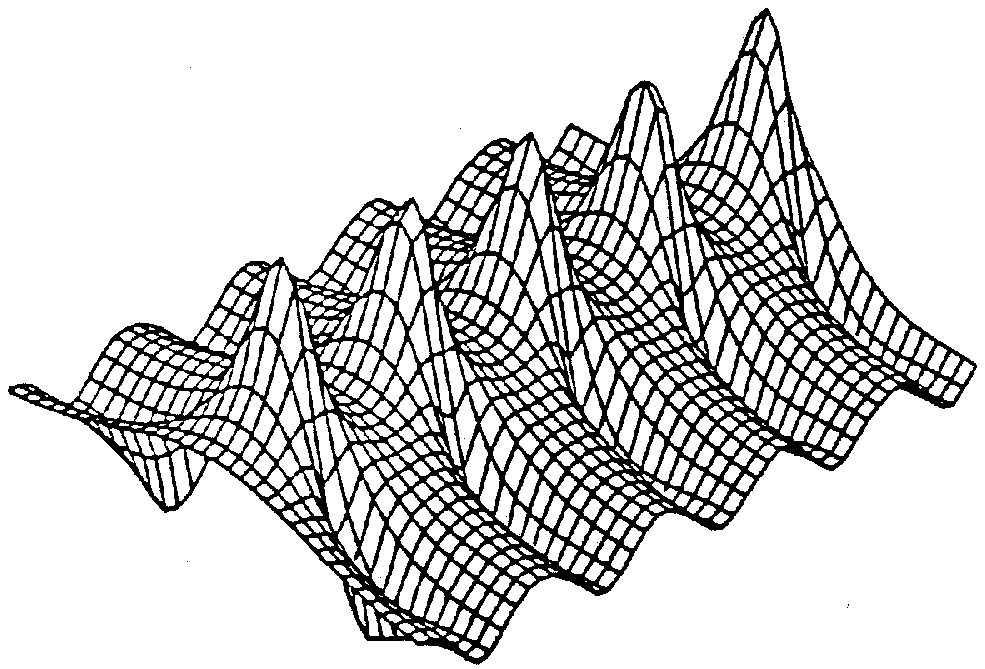}
\caption{Spatial-temporal profile realization of $L_{Cin}$ in 
Figure \ref{pfi} (coordinates are the same as in Figure \ref{numo}).}
\label{figlcin}
\end{figure}

\begin{figure}
\includegraphics{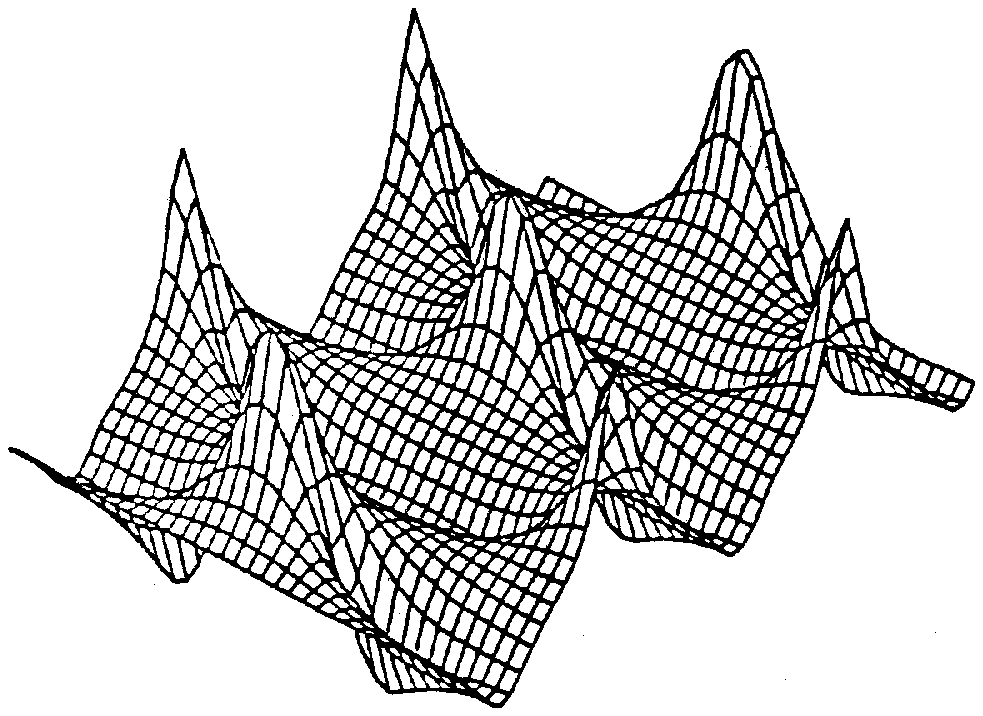}
\caption{Spatial-temporal profile realization of $L_{out}$ in 
Figure \ref{pfi} (coordinates are the same as in Figure \ref{numo}).}
\label{figlout}
\end{figure}

In the chaotic regime, typical numerical output is shown in 
Figure \ref{numo}. Notice that there are two typical profiles at a fixed 
time: (1). a breather type profile with its hump located at the center of 
the spatial period interval, (2). a breather type profile with its hump 
located at the boundary (wing) of the spatial period interval. These two 
types of profiles are half spatial period translate of each other.
If we label the profiles, with their humps at the center of the spatial 
period interval, by ``C''; those profiles, with their humps at the wing
of the spatial period interval, by ``W'', then 
\begin{equation}
``W\mbox{''}=\sg \circ ``C\mbox{''}, \label{CWJ}
\end{equation}
where $\sg$ is the symmetry group element representing half spatial period 
translate. The time series of the output in Figure \ref{numo} is a 
chaotic jumping between ``C'' and ``W'', which we call ``chaotic center-wing 
jumping''. By the relation (\ref{CWJ}) and the study 
in previous sections, we interpret the chaotic center-wing jumping as the 
numerical realization of the shift automorphism $\chi$ on four 
symbols $1, 2, -1, -2$. We can make this more precise in terms of the 
phase space geometry. The figure-eight structure (\ref{4.13}) projected 
onto the plane of the Fourier component $\cos x$ is illustrated in 
Figure \ref{pfi}, and labeled by $L_C$ and $L_W$. From 
the symmetry, we know that
\[
L_W = \sg \circ L_C.
\]
$L_C$ has the spatial-temporal profile realization
as in Figure \ref{figlc} with the hump located at the center. $L_W$ 
corresponds to the half spatial period translate 
of the spatial-temporal profile realization as in Figure \ref{figlc},
with the hump located at the boundary (wing).
An orbit inside $L_C$, $L_{Cin}$ has a spatial-temporal profile realization as
in Figure \ref{figlcin}. The half period translate of 
$L_{Cin}$, $L_{Win}$ is 
inside $L_W$. An orbit outside $L_C$ and $L_W$, $L_{out}$ has the
spatial-temporal profile realization as in Figure \ref{figlout}.
The two slabs (unstable slices of $\hS_l$) $S_l$ and $S_{l,\sg}$
projected onto the plane of ``$\cos x$'', are also illustrated in 
Figure \ref{pfi}. From these figures, we can see clearly that 
the chaotic center-wing jumping 
(Figure \ref{numo}) is the realization of the shift automorphism on four 
symbols $1, 2, -1, -2$ in $S_l \cup S_{l,\sg}$.

\section{Shadowing Lemma and Chaos in Finite-D Periodic Systems \label{Palmer}}

Since its invention \cite{Ano67}, shadowing lemma has been a useful tool
for solving many dynamical system problems. Here we only focus upon its 
use in proving the existence of chaos in a neighborhood of a transversal 
homoclinic orbit. According to the type of the system, the level of 
difficulty in proving the existence of chaos with a shadowing lemma
is different. 

A finite-dimensional periodic system can be written in the 
general form
\[
\dot{x} = F(x,t)\ ,
\]
where $x \in \RR^n$, and $F(x,t)$ is periodic in $t$. Let $f$ be the 
Poincar\'e period map.
\begin{definition}
A doubly infinite sequence $\{ y_j \}$ in $\RR^n$ is a $\dl$ pseudo-orbit 
of a $C^1$ diffeomorphism $f : \RR^n \mapsto \RR^n$ if for all integers $j$
\[
|y_{j+1}-f(y_j)| \leq \dl \ .
\]
An orbit $\{ f^j(x) \}$ is said to $\e$-shadow the $\dl$ pseudo-orbit 
$\{ y_j \}$ if for all integers $j$
\[
|f^j(x)-y_j| \leq \e \ .
\]
\end{definition}
\begin{definition}
A compact invariant set $S$ is hyperbolic if there are positive 
constants $K$, $\al$ and a projection matrix valued function $P(x)$, 
$x \in S$, of constant rank such that for all $x$ in $S$
\[
P(f(x))Df(x) = Df(x)P(x)\ ,
\]
\[
|Df^j(x)P(x)| \leq Ke^{-\al j}\ ,\quad (j \geq 0 )\ , 
\]
\[
|Df^j(x)(I-P(x))| \leq Ke^{\al j}\ ,\quad (j \leq 0 )\ . 
\] 
\end{definition}
\begin{theorem}[Shadowing Lemma \cite{Pal88}]
Let $S$ be a compact hyperbolic set for the $C^1$ diffeomorphism 
$f : \RR^n \mapsto \RR^n$. Then given $\e >0$ sufficiently small there 
exists $\dl >0$ such that every $\dl$ pseudo-orbit in $S$ has a 
unique $\e$-shadowing orbit.
\label{shal}
\end{theorem}
The proof of this theorem by Palmer \cite{Pal88} is overall a 
fixed point argument with the help of Green functions for linear maps.

Let $y_0$ be a transversal homoclinic point asymptotic to a saddle $x_0$ 
of a $C^1$ diffeomorphism $f : \RR^n \mapsto \RR^n$. Then the set
\[
S= \{ x_0 \} \cup \{ f^j(y_0): j \in Z\}
\]
is hyperbolic. Denote by $A_0$ and $A_1$ the two orbit segments of 
length $2m+1$
\[
A_0= \{ x_0,x_0, \cdots, x_0 \}\ , \quad 
A_1= \{ f^{-m}(y_0), f^{-m+1}(y_0), \cdots, f^{m-1}(y_0), f^{m}(y_0) \} \ .
\]
Let 
\[
a=(\cdots, a_{-1},a_0, a_1, \cdots ) \ , 
\]
where $a_j \in \{ 0,1\}$, be any doubly infinite binary sequence. Let 
$A$ be the doubly infinite sequence of points in $S$, associated with 
$a$
\[
A=\{ \cdots, A_{a_{-1}},A_{a_0}, A_{a_1}, \cdots \} \ .
\]
When $m$ is sufficiently large, $A$ is a $\dl$ pseudo-orbit in $S$.
By the shadowing lemma (Theorem \ref{shal}), there is a unique 
$\e$-shadowing orbit that shadows $A$. In this manner, Palmer \cite{Pal88}
gave a beautiful proof of Smale's horseshoe theorem.
\begin{definition}
Denote by $\Sg$ the set of doubly infinite binary sequences
\[
a=(\cdots, a_{-1},a_0, a_1, \cdots ) \ , 
\]
where $a_j \in \{ 0,1\}$. We give the set $\{ 0,1\}$ the discrete topology 
and $\Sg$ the product topology. The Bernoulli shift $\chi$ is defined by
\[
[\chi(a)]_j = a_{j+1}\ .
\]
\end{definition}
\begin{theorem}
Let $y_0$ be a transversal homoclinic point asymptotic to a saddle $x_0$ 
of a $C^1$ diffeomorphism $f : \RR^n \mapsto \RR^n$. Then there is a 
homeomorphism $\phi$ of $\Sg$ onto a compact subset of $\RR^n$ which is 
invariant under $f$ and such that when $m$ is sufficiently large
\[
f^{2m+1} \circ \phi = \phi \circ \chi\ ,
\]
that is, the action of $f^{2m+1}$ on $\phi(\Sg)$ is topologically conjugate 
to the action of $\chi$ on $\Sg$.
\end{theorem}
Here one can define $\phi(a)$ to be the point on the shadowing orbit that 
shadows the midpoint of the orbit segment $A_{a_0}$, which is either $x_0$ 
or $y_0$. The topological conjugacy can be easily verified. For details, 
see \cite{Pal88}.
Other references can be found in \cite{Pal84} \cite{Zen95}.

\section{Shadowing Lemma and Chaos in Infinite-D Periodic Systems}

An infinite-dimensional periodic system defined in a Banach space $X$ 
can be written in the general form
\[
\dot{x} = F(x,t)\ ,
\]
where $x \in X$, and $F(x,t)$ is periodic in $t$. Let $f$ be the 
Poincar\'e period map. When $f$ is a $C^1$ map which needs not to be 
invertible, shadowing lemma and symbolic dynamics around a transversal 
homoclinic orbit can both be established \cite{SW89} \cite{SW90} \cite{Hen94}.
Other references can be found in \cite{HL86} \cite{CLP89} \cite{Zen97} 
\cite{Bla86}. There exists also a work on horseshoe construction without
shadowing lemma for 
sinusoidally forced vibrations of buckled beam \cite{HM81}.

\section{Periodically Perturbed Sine-Gordon (SG) Equation}

We continue from section \ref{PPSGE}, and use the notations in  
section \ref{Palmer}. For the periodically perturbed sine-Gordon 
equation (\ref{PSG}), the Poincar\'e period map is a $C^1$ diffeomorphism 
in $H^1$. As a corollary of the result in last section, we have the theorem 
on the existence of chaos.
\begin{theorem} There is an integer $m$ and a homeomorphism $\phi$ 
of $\Sigma$ onto a compact Cantor subset $\Lambda $ of
$H^{1}$. $\Lambda$ is invariant under the Poincar\'e period-$2\pi$ map $P$ of 
the periodically perturbed sine-Gordon equation (\ref{PSG}). The 
action of $P^{2m+1}$ on $\Lambda$ is topologically conjugate to
the action of $\chi$ on
$\Sigma: P^{2m+1} \circ \phi =\phi \circ
\chi$. That is, the following diagram commutes:
\[
\begin{array}{ccc}
\Sg &\maprightu{\phi} & \Lambda\\
\mapdownl{\chi} & & \mapdownr{P^{2m+1}}\\
\Sg & \maprightd{\phi} & \Lambda
\end{array} 
\]
\end{theorem}

\section{Shadowing Lemma and Chaos in Finite-D Autonomous Systems}

A finite-dimensional autonomous system can be written in the 
general form
\[
\dot{x} = F(x)\ ,
\]
where $x \in \RR^n$. In this case, a transversal homoclinic orbit 
can be an orbit asymptotic to a normally hyperbolic limit cycle.
That is, it is an orbit in the intersection of the stable and 
unstable manifolds of a normally hyperbolic limit cycle. Instead of 
the Poincar\'e period map as for periodic system, one may want 
to introduce the so-called Poincar\'e return map which is a map induced by 
the flow on a codimension 1 section which is transversal to the limit 
cycle. Unfortunately, such a map is not even well-defined in the 
neighborhood of the homoclinic orbit. This poses a challenging 
difficulty in extending the arguments as in the case of a Poincar\'e 
period map. In 1996, Palmer \cite{Pal96} completed a proof of a 
shadowing lemma and existence of chaos using Newton's method. It will 
be difficult to extend this method to infinite dimensions, since it 
used heavily differentiations in time. 
Other references can be found in \cite{CKP95} \cite{CKP97} \cite{Sil67b} 
\cite{FS77}.

\section{Shadowing Lemma and Chaos in Infinite-D Autonomous Systems}

An infinite-dimensional autonomous system defined in a Banach space $X$ 
can be written in the general form
\[
\dot{x} = F(x)\ ,
\]
where $x \in X$. In 2002, the author \cite{Li02a} completed a proof of a 
shadowing lemma and existence of chaos using Fenichel's persistence of 
normally hyperbolic invariant manifold idea. The setup is as follows,
\begin{itemize}
\item {\bf Assumption (A1):} There exist a hyperbolic limit cycle
$S$ and a transversal homoclinic orbit $\xi$ 
asymptotic to $S$. As curves, $S$ and $\xi$ are $C^{3}$.
\item {\bf Assumption (A2):} The Fenichel fiber theorem is valid at $S$. 
That is, there exist a family of unstable Fenichel fibers
$\{ {\mathcal F}^{u}(q): \  q \in S\}$ and a family of stable Fenichel 
fibers $\{ {\mathcal F}^{s}(q): \  q\in S\}$. For each fixed $q\in S$,
${\mathcal F}^{u}(q)$ and ${\mathcal F}^{s}(q)$ are $C^{3}$ submanifolds.
${\mathcal F}^{u}(q)$ and ${\mathcal F}^{s}(q)$ are $C^{2}$ in $q,
\forall q\in S$. The unions $\bigcup_{q\in S}{\mathcal F}^{u}(q)$ and
$\bigcup_{q\in S}{\mathcal F}^{s}(q)$ are the unstable and stable 
manifolds of $S$. Both families are invariant, i.e.
\[
F^{t}({\mathcal F}^{u}(q))\subset
{\mathcal F}^{u}(F^{t}(q)),  \forall\ t \leq 0, q\in S,
\]
\[
F^{t}({\mathcal F}^{s}(q))\subset {\mathcal F}^{s}(F^{t}(q)), 
\forall\ t \geq 0,  q \in S, 
\]
where $F^{t}$ is the evolution operator. There are positive constants
$\k$ and $\widehat{C}$ such that $\forall q\in S$, $\forall 
q^{-}\in {\mathcal F}^{u}(q)$ and $\forall q^{+}\in
{\mathcal F}^{s}(q)$,
\[
\| F^{t}(q^{-})-F^{t}(q)\| \leq
\widehat{C}e^{\k t}\| q^{-}-q\|,  \forall \ t \leq 0\ ,
\]
\[
\| F^{t}(q^{+})-F^{t}(q)\| \leq \widehat{C}e^{-\k t}\| q^{+}-q\|, 
\forall \ t \geq 0\ .
\]
\item {\bf Assumption (A3):} $F^{t}(q)$ is $C^{0}$ in $t$, for
$t\in (-\infty ,\infty)$, $q\in X$. For any fixed $t\in 
(-\infty ,\infty )$, $F^{t}(q)$ is a $C^{2}$ diffeomorphism on
$X$.
\end{itemize}
\begin{remark}
Notice that we do not assume that as functions of time, $S$ and $\xi$ 
are $C^3$ , and we only assume that as curves, $S$ and $\xi$ 
are $C^3$.
\end{remark}

Under the above setup, a shadowing lemma and existence of chaos can be
proved \cite{Li02a}. Another crucial element in the argument is the
establishment of a $\la$-lemma (also called inclination lemma) which 
will be discussed in a later section.

\section{A Derivative Nonlinear Schr\"odinger Equation}

We continue from section \ref{ADNS}, and consider the derivative nonlinear 
Schr\"odinger equation (\ref{derNLS}). The transversal homoclinic orbit 
given in Theorem \ref{thmdns} is a classical solution. Thus, Assumption (A1) 
is valid. Assumption (A2) follows from the standard arguments 
in \cite{LW97} \cite{LMSW96} \cite{Li01b}. Since the perturbation 
in (\ref{derNLS}) is bounded, Assumption (A3) follows from standard 
arguments. Thus there exists chaos in the derivative nonlinear 
Schr\"odinger equation (\ref{derNLS}) \cite{Li02a}.

\section{$\la$-Lemma}

$\la$-Lemma (inclination lemma) has been utilized in proving many 
significant theorems in dynamical system \cite{PM82} \cite{Pal83} 
\cite{Wal87}. 
Here is another example, in \cite{Li02a}, it is shown that $\la$-Lemma 
is crucial for proving a shadowing lemma in an infinite dimensional 
autonomous system. Below we give a brief introduction to $\la$-Lemma 
\cite{PM82}.

Let $f$ be a $C^r$ ($r \geq 1$) diffeomorphism in $\RR^m$ with $0$ as 
a saddle. Let $E^s$ and $E^u$ be the stable and unstable subspaces. 
Through changing coordinates, one can obtain that for the local 
stable and unstable manifolds,
\[
W^s_{\mbox{loc}} \subset E^s\ , \quad W^u_{\mbox{loc}} \subset E^u\ .
\]
Let $B^s \subset W^s_{\mbox{loc}}$ and $B^u \subset W^u_{\mbox{loc}}$
be balls, and $V=B^s \times B^u$. Let $q \in W^s_{\mbox{loc}}/\{0\}$,
and $D^u$ be a disc of the same dimension as $B^u$, which is transversal to 
$W^s_{\mbox{loc}}$ at $q$.
\begin{lemma} [The $\la$-Lemma]
Let $D_n^u$ be the connected component of $f^n(D^u) \cap V$ to which 
$f^n(q)$ belongs. Given $\e > 0$, there exists $n_0$ such that if $n > n_0$, 
then $D_n^u$ is $\e$ $C^1$-close to $B^u$. 
\end{lemma}
The proof of the lemma is not complicated, yet nontrivial \cite{PM82}.
In different situations, the claims of the $\la$-lemma needed may be 
different \cite{PM82} \cite{Li02a}. Nevertheless, the above simple version
illustrated the spirit of $\la$-lemmas.

\section{Homoclinic Tubes and Chaos Cascades}

When studying high-dimensional systems, 
instead of homoclinic orbits one is more interested in the 
so-called homoclinic tubes \cite{Li03j} \cite{Li03k} \cite{Li99b} 
\cite{Li02f} \cite{Li02g}.
The concept of a homoclinic tube was introduced by Silnikov 
\cite{Sil68b} in a study on the structure of the neighborhood 
of a homoclinic tube asymptotic to an invariant torus $\sg$ 
under a diffeomorphism $F$ in a finite dimensional phase space. 
The asymptotic torus is of saddle type. The homoclinic tube 
consists of a doubly infinite sequence of tori $\{ \sg_j,\ j=0, 
\pm 1, \pm 2, \cdot \cdot \cdot \}$ in the transversal 
intersection of the stable and unstable manifolds of $\sg$, 
such that $\sg_{j+1} = F \circ \sg_j$ for any $j$. It is a 
generalization of the concept of a transversal homoclinic 
orbit when the points are replaced by tori. 

We are interested in homoclinic tubes for several reasons \cite{Li03j} 
\cite{Li03k} \cite{Li99b} \cite{Li02f} \cite{Li02g}: 1. 
Especially in high dimensions, dynamics inside each invariant 
tubes in the neighborhoods of homoclinic tubes are often 
chaotic too. We call such chaotic dynamics ``{\em{chaos in 
the small}}'', and the symbolic dynamics of the invariant 
tubes ``{\em{chaos in the large}}''. Such cascade structures 
are more important than the structures in a neighborhood of a 
homoclinic orbit, when high or infinite dimensional dynamical 
systems are studied. 2. Symbolic dynamics structures in the 
neighborhoods of homoclinic tubes are more observable than 
in the neighborhoods of homoclinic orbits in numerical and 
physical experiments. 3. When studying high or infinite 
dimensional Hamiltonian system (for example, the cubic nonlinear 
Schr\"odinger equation under Hamiltonian perturbations), each 
invariant tube contains both KAM tori and stochastic layers 
(chaos in the small). Thus, not only dynamics inside each 
stochastic layer is chaotic, all these stochastic layers also 
move chaotically under Poincar\'e maps.

Using the shadowing lemma technique developed in \cite{Li02a}, 
we obtained in \cite{Li03j} a theorem on the symbolic dynamics 
of submanifolds in a neighborhood of a homoclinic tube under a 
$C^3$-diffeomorphism defined in a Banach space. Such a proof removed 
an uncheckable assumption in \cite{Sil68b}. The result of \cite{Sil68b}
gives a symbolic labeling of all the invariant tubes around a 
homoclinic tube. Such a symbolic labeling does not imply the 
symbolic dynamics of a single map proved in \cite{Li03j}.

Then in \cite{Li03k}, as an example, the following sine-Gordon equation 
under chaotic perturbation is studied,
\begin{equation}
u_{tt} = c^2 u_{xx} + \sin u +\e [ -a u + f(t, \th_2, \th_3, \th_4) 
(\sin u - u)]\ , \label{ls1} 
\end{equation}
which is subject to periodic boundary condition and odd constraint
\begin{equation}
u(t, x+2\pi ) = u(t, x)\ , \quad u(t, -x) = - u(t, x)\ ,
\label{obc}
\end{equation}
where $u$ is a real-valued function of two real variables $t \geq 0$ 
and $x$, c is a parameter, $\frac{1}{2} < c < 1$, $a>0$ is a parameter,
$\e \geq 0$ is a small parameter, $f$ is periodic in $t, \th_2, \th_3,
\th_4$, and $(\th_2, \th_3, \th_4) \in \mathbb{T}^{3}$, by introducing 
the extra variable $\th_1 = \om_1 t +\th_1^0$, $f$ takes the form 
\[
f(t) = \sum_{n=1}^4 a_n \cos [ \th_n(t)]\ ,
\]
and $a_n$'s are parameters. Let $\th_n = \om_n t + \th_n^0 +\e^\mu \vth_n$, 
$n=2,3,4$, $\mu > 1$, and $\vth_n$'s are given by the ABC flow 
\cite{DFGHMS86} which is verified numerically to be chaotic,
\begin{eqnarray}
\dot{\vth}_2 &=& A \sin \vth_4 + C \cos \vth_3 \ , \nonumber \\
\dot{\vth}_3 &=& B \sin \vth_2 + A \cos \vth_4 \ , \nonumber \\
\dot{\vth}_4 &=& C \sin \vth_3 + B \cos \vth_2 \ , \nonumber 
\end{eqnarray}
where $A$, $B$, and $C$ are real parameters. Existence of a homoclinic tube 
is proved. As a corollary of the theorem proved in \cite{Li03j}, Symbolic 
dynamics of tori around the homoclinic tube is established, which is the 
``chaos in the large''. The chaotic dynamics of ($\th_2, \th_3, \th_4$) 
is the ``chaos in the small''.

Here we see the embedding of smaller scale chaos in larger scale chaos. 
By introducing more variables, such embedding can be continued with even 
smaller scale chaos. This leads to a chain of embeddings. We call this 
chain of embeddings of smaller scale chaos in larger scale chaos, a ``chaos 
cascade''. We hope that such ``chaos cascade'' will be proved important.

\chapter{Stabilities of Soliton Equations in $\RR^n$}

Unlike soliton equations under periodic boundary conditions, phase space 
structures, especially hyperbolic structures, of soliton equations under 
decay boundary conditions are not well understood. The application of 
B\"{a}cklund-Darboux transformations for generating hyperbolic foliations 
is not known in decay boundary condition case. Nevertheless, we believe 
that B\"{a}cklund-Darboux transformations may still have great potentials 
for understanding phase space structures. This can be a great area for 
interested readers to work in. The common use of B\"{a}cklund-Darboux 
transformations is to generate multi-soliton solutions from single soliton 
solutions. The relations between different soliton solutions in phase 
spaces (for example certain Sobolev spaces) are not clear yet. There are 
some studies on the linear and nonlinear stabilities of traveling-wave 
(soliton) solutions \cite{PW92} \cite{Liu94}. Results concerning soliton 
equations are only stability results. Instability results are obtained 
for non-soliton equations. On the other hand, non-soliton equations are 
not integrable, have no Lax pair structures, and their phase space 
structures are much more difficult to be understood. The challenging 
problem here is to identify which soliton equations possess traveling-wave 
solutions which are unstable. The reason why we emphasize instabilities is 
that they are the sources of chaos. Below we are going to review two types 
of studies on the phase space structures of soliton equations under decay 
boundary conditions.

\section{Traveling Wave Reduction}

If we only consider traveling-wave solutions to the soliton equations or 
generalized ( or perturbed ) soliton equations, the resulting equations 
are ordinary differential equations with parameters. These ODEs often 
offer physically meaningful and mathematically pleasant problems to be 
studied. One of the interesting features is that the traveling-wave 
solutions correspond to homoclinic orbits to such ODEs. We also naturally 
expect that such ODEs can have chaos. Consider the KdV equation
\begin{equation}
u_t +6uu_x +u_{xxx} = 0 \ ,
\label{kdv}
\end{equation}
and let $u(x,t)= U(\xi)$, $\xi = x -c t$ ($c$ is a real parameter), we have 
\[
U''' + 6U U' -c U' = 0 \ .
\]
After one integration, we get
\begin{equation}
U''+3U^2-cU+c_1 = 0 \ ,
\label{tveq}
\end{equation}
where $c_1$ is a real integration constant. This equation can be rewritten 
as a system,
\begin{equation}
\left \{ \begin{array}{l} U'=V\ , \cr V'=-3U^2+cU-c_1\ . 
\cr \end{array}\right .
\label{tvsys}
\end{equation}
For example, when $c_1=0$ and $c > 0$ we have the soliton,
\begin{equation}
U_s = {c \over 2} \ \mbox{sech}^2[\pm {\sqrt{c} \over 2} \xi ] \ ,
\label{solit}
\end{equation}
which is a homoclinic orbit asymptotic to $0$.

We may have an equation of the following form describing a real physical phenomenon,
\[
u_t + [f(u)]_x +u_{xxx}+\e g(u) = 0\ ,
\]
where $\e$ is a small parameter, and traveling waves are physically important. Then we have 
\[
U'''+[f(U)]'-cU'+\e g(U)=0\ ,
\]
which can be rewritten as a system,
\[
\left \{ \begin{array}{l} U'= V\ , \cr V'=W\ , \cr 
W'=-f'V+cV-\e g(U)\ .
\cr \end{array}\right .
\]
This system may even have chaos. Such low dimensional systems are very popular in 
current dynamical system studies and are mathematically very pleasant \cite{JKL91} \cite{JK91}. Nevertheless, 
they represent a very narrow class of solutions to the original PDE, and in no way 
they can describe the entire phase space structures of the original PDE.

\section{Stabilities of the Traveling-Wave Solutions}

The first step toward understanding the phase space structures of soliton equations 
under decay boundary conditions is to study the linear or nonlinear stabilities of 
traveling-wave solutions \cite{PW92} \cite{Liu94} \cite{GSS87} \cite{KS98}. Consider the KdV equation 
(\ref{kdv}), and study the linear stability of the traveling wave (\ref{solit}). 
First we change the variables from ($x,t$) into ($\xi,t$) where $\xi=x-ct$; then 
in terms of the new variables ($\xi,t$), the traveling wave (\ref{solit}) is a fixed 
point of the KdV equation (\ref{kdv}). Let $u(x,t)=U_s(\xi)+v(\xi,t)$ and linearize 
the KdV equation (\ref{kdv}) around $U_s(\xi)$, we have
\[
\pa_t v -c \pa_\xi v +6 [U_s \pa_\xi v +v \pa_\xi U_s]+\pa_\xi^3 v = 0\ .
\]
We seek solutions to this equation in the form $v= e^{\la t}Y(\xi)$ and we have the 
eigenvalue problem:
\begin{equation}
\la Y - c \pa_\xi Y + 6 [U_s \pa_\xi Y + Y \pa_\xi U_s ] +\pa_\xi^3 Y = 0 \ .
\label{kdvl}
\end{equation}
One method for studying such eigenvalue problem is carried in two steps \cite{PW92}: 
First, one studies the $|\xi| \ra \infty$ limit system 
\[
\la Y - c \pa_\xi Y + \pa_\xi^3 Y = 0 \ ,
\]
and finds that it has solutions $Y(\xi)=e^{\mu_j \xi}$ for $j=1,2,3$ where the $\mu_j$ satisfy 
\[
\mbox{Re} \{ \mu_1 \} < 0 < \mbox{Re} \{ \mu_l \} \ \ \ \ \mbox{for} \ 
l=2,3.
\]
Thus the equation (\ref{kdvl}) has a one-dimensional subspace of solutions which decay as $x \ra 
\infty$, and a two-dimensional subspace of solutions which decay as $x \ra -\infty$. $\la$ will 
be an eigenvalue when these subspaces meet nontrivially. In step 2, one measures the angle between 
these subspaces by a Wronskian-like analytic function $D(\la)$, named Evans's function. One 
interpretation of this function is that it is like a transmission coefficient, in the sense 
that for the solution of (\ref{kdvl}) satisfying
\[
Y(\xi) \sim e^{\mu_1 \xi}\ \ \ \ \mbox{as}\ \xi \ra \infty \ ,
\]
we have 
\[
Y(\xi) \sim D(\la) e^{\mu_1 \xi}\ \ \ \ \mbox{as}\ \xi \ra -\infty \ .
\]
In equation (\ref{kdvl}), for $\mbox{Re}\{\la \} > 0$, if $D(\la)$ vanishes, then $\la$ is 
an eigenvalue, and conversely. The conclusion for the equation (\ref{kdvl}) is that there 
is no eigenvalue with $\mbox{Re}\{\la \} > 0$. In fact, for the KdV equation (\ref{kdv}), 
$U_s$ is $H^1$-orbitally stable \cite{Ben72} \cite{Bon75} \cite{Wei86} \cite{BSS87}.

For generalized soliton equations, linear instability results have been established \cite{PW92}. 
Let $f(u)=u^{p+1}/(p+1)$, and 
\[
U_s(\xi)=a\ \mbox{sech}^{2/p}(b\xi) 
\]
for appropriate constants $a$ and $b$. For the generalized KdV equation \cite{PW92}
\[
\pa_t u +\pa_x f(u) +\pa_x^3 u = 0 \ ,
\]
if $p>4$, then $U_s$ is linearly unstable for all $c > 0$. If $1 < p< 4$, then $U_s$ is 
$H^1$-orbitally stable. For the generalized Benjamin-Bona-Mahoney equation \cite{PW92}
\[
\pa_t u +\pa_x u +\pa_x f(u) - \pa_t \pa_x^2 u = 0 \ ,
\]
if $p> 4$, there exists a positive number $c_0(p)$, such that $U_s$ with $1 < c< c_0(p)$ are 
linearly unstable. For the generalized regularized Boussinesq equation \cite{PW92}
\[
\pa_t^2 u -\pa_x^2 u -\pa_x^2 f(u)-\pa_t^2 \pa_x^2 u = 0 \ ,
\]
if $p > 4$, $U_s$ is linearly unstable if $1 < c^2< c^2_0(p)$, where
\[
c^2_0(p) = 3p /(4+2p)\ .
\]

\section{Breathers}

Breathers are periodic solutions of soliton equations under decay boundary conditions. The 
importance of breathers with respect to the phase space structures of soliton equations is 
not clear yet. B\"{a}cklund-Darboux transformations may play a role for this. For example, 
starting from a breather with one frequency, B\"{a}cklund-Darboux transformations can generate 
a new breather with two frequencies, i.e. the new breather is a quasiperiodic solution. The 
stability of breathers will be a very important subject to study.

First we study the reciprocal relation between homoclinic solutions generated through 
B\"{a}cklund-Darboux transformations for soliton equations under periodic boundary conditions 
and breathers for soliton equations under decay boundary conditions. Consider the sine-Gordon 
equation
\begin{equation}
u_{tt}-u_{xx}+\sin u = 0 \ ,
\label{sG}
\end{equation}
under the periodic boundary condition
\begin{equation}
u(x+L,t)=u(x,t)\ ,
\label{pbc}
\end{equation}
and the decay boundary condition
\begin{equation}
u(x,t) \ra 0\ \ \ \ \mbox{as}\ |x| \ra \infty \ .
\label{dbc}
\end{equation}
Starting from the trivial solution $u = \pi$, one can generate the homoclinic solution to the 
Cauchy problem (\ref{sG}) and (\ref{pbc}) through B\"{a}cklund-Darboux transformations 
\cite{EFM90},
\begin{equation}
u_H(x,t) = \pi + 4 \tan^{-1} \bigg [ {\tan \nu \cos [ (\cos \nu ) x]  \over \cosh [ 
(\sin \nu) t ] } \bigg ] \ , 
\label{sghorb}
\end{equation}
where $L = 2 \pi /\cos \nu$. Applying the sine-Gordon symmetry
\[
(x,t,u) \longrightarrow (t,x,u-\pi)\ ,
\]
one can immediately generate the breather solution to the Cauchy problem (\ref{sG}) and 
(\ref{dbc}) \cite{EFM90},
\begin{equation}
u_B(x,t) = 4 \tan^{-1} \bigg [ {\tan \nu \cos [ (\cos \nu ) t]  \over \cosh [ (\sin \nu) x ] } 
\bigg ] \ . 
\label{sgbr}
\end{equation}
The period of this breather is $L = 2 \pi /\cos \nu$. 
The linear stability can be investigated through studying the Floquet theory for the linear 
partial differential equation which is the linearized equation of (\ref{sG}) at $u_B$ 
(\ref{sgbr}):
\begin{equation}
u_{tt}-u_{xx}+(\cos u_B)u=0\ , \label{lsG}
\end{equation}
under the decay boundary condition
\begin{equation}
u(x,t) \ra 0 \quad \mbox{as}\ |x| \ra \infty\ .
\label{lsGbc}
\end{equation}
Such studies will be very important in terms of understanding the phase space structures 
in the neighborhood of the breather and developing infinite dimensional Floquet theory. 
In fact, McLaughlin and Scott \cite{MS78} had studied the equation (\ref{lsG}) through 
infinitessimal B\"acklund transformations, which lead to the so-called radiation of 
breather phenomenon. In physical variables only, the B\"acklund transformation for the 
sine-Gordon equation (\ref{sG}) is given as,
\begin{eqnarray*}
{\pa \over \pa \xi} \bigg [ {u_+ + u_- \over 2} \bigg ] &=& -{i \over \z} \sin 
\bigg [ {u_+ - u_- \over 2} \bigg ] \ , \\
{\pa \over \pa \eta} \bigg [ {u_+ - u_- \over 2} \bigg ] &=& i\z  \sin 
\bigg [ {u_+ + u_- \over 2} \bigg ] \ , 
\end{eqnarray*}
where $\xi = \frac {1}{2} (x+t)$, $\eta = \frac {1}{2} (x-t)$, and $\z$ is a complex 
B\"acklund parameter. Through the nonlinear superposition principle \cite{RS82} \cite{AI79},
\begin{equation}
\tan \bigg [ {u_+ - u_- \over 4} \bigg ] = \frac {\z_1+\z_2}{\z_1-\z_2} 
\tan \bigg [ {u_1 - u_2 \over 4} \bigg ]
\label{nsup}
\end{equation}
built upon the Bianchi diagram \cite{RS82} \cite{AI79}.
The breather $u_B$ (\ref{sgbr}) is produced by starting with $u_-=0$, and choosing 
$\z_1 = \cos \nu +i \sin \nu$ and $\z_2 = -\cos \nu +i \sin \nu$. Let $\tu_-$ be a 
solution to the linearized sine-Gordon equation (\ref{lsG}) with $u_B$ replaced by $u_-=0$,
\[
\tu_- = e^{i(kt +\sqrt{k^2-1}x)}\ .
\]
Then variation of the nonlinear superposition principle leads to the solution to the 
linearized sine-Gordon equation (\ref{lsG}):
\begin{eqnarray*}
& & \tu_+ = \tu_- +\frac {i\al}{\be}\cos^2 (u_B/4) \bigg [ \cos^2 [(u_1-u_2)/4]\bigg 
]^{-1}(\tu_1 -\tu_2) \\
&=& (1+A^2)^{-1}\bigg [ a_1 +a_2 \frac {\sin^2\be t}{\cosh^2 \al x}+\frac {a_3}{2}\frac 
{(\cosh^2 \al x)'}{\cosh^2 \al x}+\frac {a_4}{2}\frac {(\sin^2\be t)'}{\cosh^2 \al x}
\bigg ]\tu_-\ ,
\end{eqnarray*}
where
\begin{eqnarray*}
& & \al =\sin \nu\ , \ \ \ \be =\cos \nu\ , \ \ \ A= \frac {\al}{\be}\frac {\sin \be t}
{\cosh \al x}\ , \\
& & a_1 =\frac {1}{2} (k^2-\be^2)-\al^2\ , \ \ \  a_2 =\frac {1}{2} (k^2-\be^2)\frac{\al^2}
{\be^2}+\al^2\ ,  \\
& & a_3=i\sqrt{k^2-1}\ , \ \ \ a_4 = ik \frac{\al^2}{\be^2}\ ,
\end{eqnarray*}
which represents the radiation of the breather.

Next we briefly survey some interesting studies on breathers. The rigidity of sine-Gordon 
breathers was studied by Birnir, Mckean and Weinstein \cite{BMW94} \cite{Bir94}. Under the scaling 
\[
x \ra x'=(1+\e a)^{-1/2} x\ , \ \ t \ra t'=(1+\e a)^{-1/2}t\ , \ \ u \ra u'=(1+\e b) u 
\]
the sine-Gordon equation (\ref{sG}) is transformed into
\[
u_{tt}-u_{xx}+\sin u +\e (a \sin u +b u \cos u) +O(\e^2) = 0 \ .
\]
The breather (\ref{sgbr}) is transformed into a breather for the new equation. The 
remarkable fact is the rigidity of the breather (\ref{sgbr}) as stated in the following theorem.
\begin{theorem}[Birnir, Mckean and Weinstein \cite{BMW94}]
If $f(u)$ vanishes at $u=0$ and is holomorphic in the open strip $|\mbox{Re}\{u\}| < \pi$, 
then $u_{tt}-u_{xx}+\sin u +\e f(u) +O(\e^2) = 0$ has breathers $u=u_B +\e \hat{u}_1 +
O(\e^2)$, one to each $\sin \nu$ from an open subinterval of $(0, 1/\sqrt{2}]$ issuing 
smoothly from $u=u_B$ at $\e =0$, with $\hat{u}_1$ vanishing at $x = \pm \infty$, 
if and only if $f(u)$ is a linear combination of $\sin u$, $u\cos u$, and/or $1 +3 \cos u 
-4 \cos (u/2) +4 \cos u \ln \cos (u/4)$.
\end{theorem}
In fact, it is proved that sine-Gordon equation (\ref{sG}) is the only nonlinear wave 
equation possessing small analytic breathers \cite{Kic91}. These results lend color to 
the conjecture that, for the general nonlinear wave equation, breathing is an 
``arithmetical'' phenomenon in the sense that it takes place only for isolated 
nonlinearities, scaling and other trivialities aside.

\chapter{Lax Pairs of Euler Equations of Inviscid Fluids}

The governing equations for the incompressible viscous fluid flow are the 
Navier-Stokes equations. Turbulence occurs in the regime of high Reynolds 
number. By formally setting the Reynolds number equal to infinity, the 
Navier-Stokes equations reduce to the Euler equations of incompressible 
inviscid fluid flow. One may view the Navier-Stokes equations with large 
Reynolds number as a singular perturbation of the Euler equations.

Results of T. Kato show that 2D Navier-Stokes equations are globally 
well-posed in $C^0([0, \infty); H^s(R^2)), \ s>2$, and for any 
$0 < T < \infty$, the mild solutions of the 2D Navier-Stokes equations
approach those of the 2D Euler equations in $C^0([0, T]; H^s(R^2))$ 
\cite{Kat86}. 3D Navier-Stokes equations are locally well-posed in 
$C^0([0, \tau]; H^s(R^3)), \ s>5/2$, and the mild solutions of the 3D 
Navier-Stokes equations approach those of the 3D Euler equations in 
$C^0([0, \tau]; H^s(R^3))$, where $\tau$ depends on the norms of the initial 
data and the external force \cite{Kat72} \cite{Kat75}. Extensive studies on 
the inviscid limit have been carried by J. Wu et al. \cite{Wu96} 
\cite{CW96} \cite{Wu98} \cite{BW99}. There is no doubt that mathematical 
study on Navier-Stokes (Euler) equations is one of the most important 
mathematical problems. In fact, Clay Mathematics Institute has posted the 
global well-posedness of 3D Navier-Stokes equations as one of the one 
million dollars problems.

V. Arnold \cite{Arn66} realized that 2D Euler equations are a Hamiltonian 
system. Extensive studies on the symplectic structures of 2D Euler equations
have been carried by J. Marsden, T. Ratiu et al. \cite{Mar92}.

Recently, the author found Lax pair structures for Euler 
equations \cite{Li01a} \cite{LY01} \cite{Li02e} \cite{Li02f}.
Understanding the structures of solutions to Euler equations is of 
fundamental interest. Of particular interest is the question on the 
global well-posedness of 3D Navier-Stokes and Euler equations. Our number 
one hope is that the Lax pair structures can be useful in investigating 
the global well-posedness. Our secondary hope is that the Darboux 
transformation \cite{LY01} associated with the Lax pair can generate 
explicit representation of homoclinic structures \cite{Li00a}. 

The philosophical significance of the existence of Lax pairs for Euler 
equations is even more important. If one defines integrability of an equation 
by the existence of a Lax pair, then both 2D and 3D Euler equations 
are integrable. More importantly, both 2D and 3D Navier-Stokes equations 
at high Reynolds numbers are singularly perturbed integrable systems. 
Such a point of view changes our old ideology on Euler and Navier-Stokes 
equations.

\section{A Lax Pair for 2D Euler Equation}

The 2D Euler equation can be written in the vorticity form,
\begin{equation}
\pa_t \Om + \{ \Psi, \Om \} = 0 \ ,
\label{euler}
\end{equation}
where the bracket $\{\ ,\ \}$ is defined as
\[
\{ f, g\} = (\pa_x f) (\pa_y g) - (\pa_y f) (\pa_x g) \ ,
\]
$\Om$ is the vorticity, and $\Psi$ is the stream function given by,
\[
u=- \pa_y \Psi \ ,\ \ \ v=\pa_x \Psi \ ,
\]
and the relation between vorticity $\Om$ and stream 
function $\Psi$ is,
\[
\Om =\pa_x v - \pa_y u =\Dl \Psi \ .
\]
\begin{theorem}[Li, \cite{Li01a}]
The Lax pair of the 2D Euler equation (\ref{euler}) is given as
\begin{equation}
\left \{ \begin{array}{l} 
L \varphi = \la \varphi \ ,
\\
\pa_t \varphi + A \varphi = 0 \ ,
\end{array} \right.
\label{laxpair}
\end{equation}
where
\[
L \varphi = \{ \Om, \varphi \}\ , \ \ \ A \varphi = \{ \Psi, \varphi \}\ ,
\]
and $\la$ is a complex constant, and $\varphi$ is a complex-valued function.
\label{2dlp}
\end{theorem}

\section{A Darboux Transformation for 2D Euler Equation}

Consider the Lax pair (\ref{laxpair}) at $\la =0$, i.e.
\begin{eqnarray}
& & \{ \Om, p \} = 0 \ , \label{d1} \\
& & \pa_t p + \{ \Psi, p \} = 0 \ , \label{d2} 
\end{eqnarray}
where we replaced the notation $\varphi$ by $p$.
\begin{theorem}[\cite{LY01}]
Let $f = f(t,x,y)$ be any fixed solution to the system 
(\ref{d1}, \ref{d2}), we define the Gauge transform $G_f$:
\begin{equation}
\tilde{p} = G_f p = \frac {1}{\Om_x}[p_x -(\pa_x \ln f)p]\ ,
\label{gauge}
\end{equation}
and the transforms of the potentials $\Om$ and $\Psi$:
\begin{equation}
\tilde{\Psi} = \Psi + F\ , \ \ \ \tilde{\Om} = \Om + \Dl F \ ,
\label{ptl}
\end{equation}
where $F$ is subject to the constraints
\begin{equation}
\{ \Om, \Dl F \} = 0 \ , \ \ \ \{ \Dl F, F \} = 0\ .
\label{constraint}
\end{equation}
Then $\tilde{p}$ solves the system (\ref{d1}, \ref{d2}) at 
$(\tilde{\Om}, \tilde{\Psi})$. Thus (\ref{gauge}) and 
(\ref{ptl}) form the Darboux transformation for the 2D 
Euler equation (\ref{euler}) and its Lax pair (\ref{d1}, \ref{d2}).
\label{dt}
\end{theorem}
\begin{remark}
For KdV equation and many other soliton equations, the 
Gauge transform is of the form \cite{MS91},
\[
\tilde{p} =  p_x -(\pa_x \ln f)p \ .
\]
In general, Gauge transform does not involve potentials.
For 2D Euler equation, a potential factor $\frac {1}{\Om_x}$
is needed. From (\ref{d1}), one has
\[
\frac{p_x}{\Om_x} = \frac{p_y}{\Om_y} \ .
\]
The Gauge transform (\ref{gauge}) can be rewritten as
\[
\tilde{p} = \frac{p_x}{\Om_x} - \frac{f_x}{\Om_x} \frac{p}{f}
=\frac{p_y}{\Om_y} - \frac{f_y}{\Om_y} \frac{p}{f}\ .
\]
The Lax pair (\ref{d1}, \ref{d2}) has a symmetry, i.e. it is 
invariant under the transform $(t,x,y) \ra (-t,y,x)$. The form 
of the Gauge transform (\ref{gauge}) resulted from the inclusion 
of the potential factor $\frac {1}{\Om_x}$, is consistent with 
this symmetry.
\end{remark}
Our hope is to use the Darboux transformation to generate homoclinic 
structures for 2D Euler equation \cite{Li00a}.

\section{A Lax Pair for Rossby Wave Equation}

The Rossby wave equation is
\[
\pa_t \Om + \{ \Psi , \Om \} + \be \pa_x \Psi = 0 \ ,
\]
where $\Om = \Om (t,x,y)$ is the vorticity, 
$\{ \Psi , \Om \} = \Psi_x \Om_y - \Psi_y \Om_x $, 
and $\Psi = \Dl^{-1} \Om$ 
is the stream function. Its Lax pair can be obtained
by formally conducting the transformation, $\Om = \tilde{\Om} +\be y$,
to the 2D Euler equation \cite{Li01a},
\[
\{ \Om , \varphi \} - \be \pa_x \varphi = \la \varphi \ ,
\quad \pa_t \varphi + \{ \Psi , \varphi \} = 0 \ ,
\]
where $\varphi$ is a complex-valued function, and $\la$ is 
a complex parameter. 

\section{Lax Pairs for 3D Euler Equation}

The 3D Euler equation can be written in vorticity form,
\begin{equation}
\pa_t \Om + (u \cdot \na) \Om - (\Om \cdot \na) u = 0 \ ,
\label{3deuler}
\end{equation}
where $u = (u_1, u_2, u_3)$ is the velocity, $\Om = (\Om_1, \Om_2, \Om_3)$
is the vorticity, $\na = (\pa_x, \pa_y, \pa_z)$, 
$\Om = \na \times u$, and $\na \cdot u = 0$. $u$ can be 
represented by $\Om$ for example through Biot-Savart law.
\begin{theorem}
The Lax pair of the 3D Euler equation (\ref{3deuler}) is given as
\begin{equation}
\left \{ \begin{array}{l} 
L \phi = \la \phi \ ,
\\
\pa_t \phi + A \phi = 0 \ ,
\end{array} \right.
\label{alaxpair}
\end{equation}
where
\[
L \phi = (\Om \cdot \na )\phi \ , 
\ \ \ A \varphi = (u \cdot \na )\phi \ , 
\]
$\la$ is a complex constant, and $\phi$ is a complex scalar-valued function.
\end{theorem}
\begin{theorem}[\cite{Chi00}]
Another Lax pair of the 3D Euler equation (\ref{3deuler}) is given as
\begin{equation}
\left \{ \begin{array}{l} 
L \varphi = \la \varphi \ ,
\\
\pa_t \varphi + A \varphi = 0 \ ,
\end{array} \right.
\label{3dlaxpair}
\end{equation}
where
\[
L \varphi = (\Om \cdot \na )\varphi - (\varphi \cdot \na )\Om \ , 
\ \ \ A \varphi = (u \cdot \na )\varphi - (\varphi \cdot \na ) u \ , 
\]
$\la$ is a complex constant, and $\varphi = (\varphi_1, \varphi_2, 
\varphi_3)$ is a complex 3-vector valued function.
\end{theorem}
Our hope is that the infinitely many conservation laws generated by $\la 
\in C$ can provide a priori estimates for the global well-posedness of 
3D Navier-Stokes equations, or better understanding on the global 
well-posedness. For more informations on the topics, 
see \cite{LY01}.

\chapter{Linearized 2D Euler Equation at a Fixed Point}

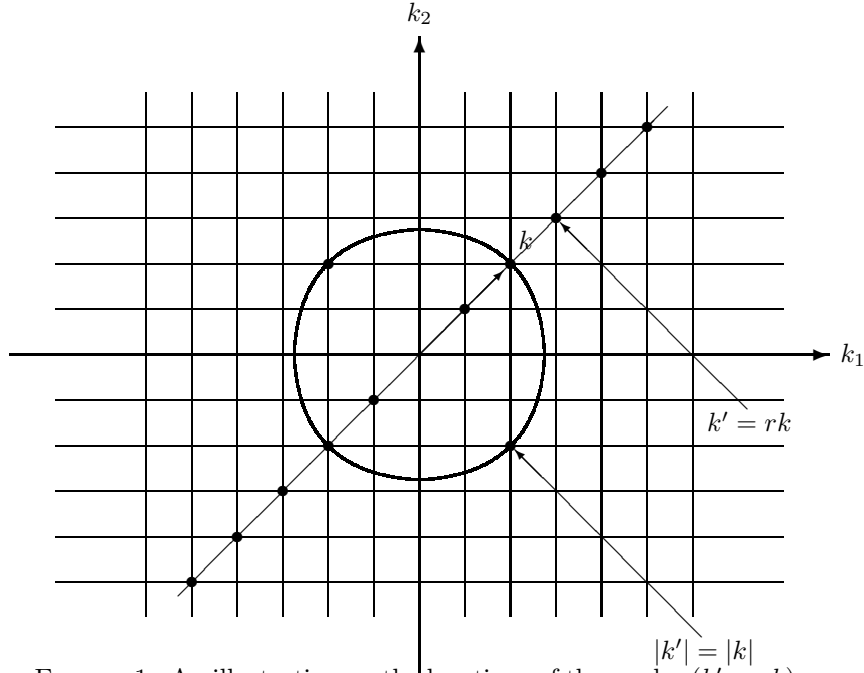
\begin{figure}[ht]
  \begin{center}
    \leavevmode
      \setlength{\unitlength}{2ex}
  \begin{picture}(36,27.8)(-18,-12)
    \thinlines
\multiput(-12,-11.5)(2,0){13}{\line(0,1){23}}
\multiput(-16,-10)(0,2){11}{\line(1,0){32}}
    \thicklines
\put(0,-14){\vector(0,1){28}}
\put(-18,0){\vector(1,0){36}}
\put(0,15){\makebox(0,0){$k_2$}}
\put(18.5,0){\makebox(0,0)[l]{$k_1$}}
\qbezier(-5.5,0)(-5.275,5.275)(0,5.5)
\qbezier(0,5.5)(5.275,5.275)(5.5,0)
\qbezier(5.5,0)(5.275,-5.275)(0,-5.5)
\qbezier(0,-5.5)(-5.275,-5.275)(-5.5,0)
    \thinlines
\put(4,4){\circle*{0.5}}
\put(0,0){\vector(1,1){3.7}}
\put(4,-4){\circle*{0.5}}
\put(-4,4){\circle*{0.5}}
\put(2,2){\circle*{0.5}}
\put(6,6){\circle*{0.5}}
\put(8,8){\circle*{0.5}}
\put(10,10){\circle*{0.5}}
\put(-2,-2){\circle*{0.5}}
\put(-4,-4){\circle*{0.5}}
\put(-6,-6){\circle*{0.5}}
\put(-8,-8){\circle*{0.5}}
\put(-10,-10){\circle*{0.5}}
\put(-10.6,-10.6){\line(1,1){21.5}}
\put(4.35,4.65){$k$}
\put(12.5,-12.5){\makebox(0,0)[t]{$|k'|=|k|$}}
\put(12.4,-12.4){\vector(-1,1){8.2}}
\put(14.5,-2.5){\makebox(0,0)[t]{$k'=rk$}}
\put(14.4,-2.4){\vector(-1,1){8.2}}
  \end{picture}
  \end{center}
\caption{An illustration on the locations of the modes ($k'=rk$) and 
($|k'|=|k|$) in the definitions of $E^1_k$ and $E^2_k$ 
(Proposition \ref{eman}).}
\label{eulman}
\end{figure}

To begin an infinite dimensional dynamical system study, we investigate 
the linearized 2D Euler equation at a fixed point \cite{Li00}. We consider the 
2D Euler equation (\ref{euler}) under periodic boundary condition in 
both $x$ and $y$ directions with period $2\pi$. Expanding $\Om$ into 
Fourier series,
\[
\Om =\sum_{k\in \Z} \om_k \ e^{ik\cdot X}\ ,
\]
where $\om_{-k}=\overline{\om_k}\ $, $k=(k_1,k_2)^T$, 
and $X=(x,y)^T$. The 2D Euler equation
can be rewritten as 
\begin{equation}
\dot{\om}_k = \sum_{k=p+q} A(p,q) \ \om_p \om_q \ ,
\label{Keuler}
\end{equation}
where $A(p,q)$ is given by,
\begin{eqnarray}
A(p,q)&=& {1\over 2}[|q|^{-2}-|p|^{-2}](p_1 q_2 -p_2 q_1) \nonumber \\
\label{Af} \\      
      &=& {1\over 2}[|q|^{-2}-|p|^{-2}]\left | \begin{array}{lr} 
p_1 & q_1 \\ p_2 & q_2 \\ \end{array} \right | \ , \nonumber
\end{eqnarray}
where $|q|^2 =q_1^2 +q_2^2$ for $q=(q_1,q_2)^T$, similarly for $p$.
The 2D Euler equation (\ref{Keuler}) has huge dimensional equilibrium 
manifolds. 
\begin{proposition}
For any $k \in Z^2/\{0\}$, the infinite dimensional space
\[
E^1_k \equiv \bigg \{ \{ \om_{k'}\} \ \bigg | \ \om_{k'}=0,
\ \mbox{if}\ k' \neq rk,\ \forall r \in R \bigg \} \ ,
\]
and the finite dimensional space 
\[
E^2_k \equiv \bigg \{ \{ \om_{k'}\} \ \bigg | \ \om_{k'}=0,
\ \mbox{if}\ |k'| \neq |k| \bigg \} \ ,
\]
entirely consist of fixed points of the system (\ref{Keuler}).
\label{eman}
\end{proposition}
Fig.\ref{eulman} shows an example on the locations of the modes 
($k'=rk$) and ($|k'|=|k|$) in the definitions of $E_k^1$ and $E_k^2$
(Proposition \ref{eman}).

\section{Hamiltonian Structure of 2D Euler Equation}

For any two functionals $F_1$ and $F_2$ of $\{ \om_k \}$, 
define their Lie-Poisson bracket:
\begin{equation}
\{ F_1,F_2 \} = \sum_{k+p+q=0} \left | \begin{array}{lr} 
q_1 & p_1 \\ q_2 & p_2 \\ \end{array} \right | \ \om_k \
{\pa F_1 \over \pa \overline{\om_p}} \ {\pa F_2 \over \pa \overline{\om_q}}\ .
\label{Liebr}
\end{equation}
Then the 2D Euler equation (\ref{Keuler}) is a Hamiltonian system \cite{Arn66},
\begin{equation}
\dot{\om}_k = \{ \om_k, H\}, \label{hEft}
\end{equation}
where the Hamiltonian $H$ is the kinetic energy,
\begin{equation}
H= {1\over 2} \sum_{k \in Z^2/\{0\}} |k|^{-2} |\om_k |^2. 
\label{Ih}
\end{equation}
Following are Casimirs (i.e. invariants that Poisson commute with 
any functional) of the Hamiltonian system (\ref{hEft}):
\begin{equation}
J_n = \sum_{k_1 + \cdot \cdot \cdot +k_n =0} \om_{k_1} 
\cdot \cdot \cdot \om_{k_n}. \label{Ic}
\end{equation}

\section{Linearized 2D Euler Equation at a Unimodal Fixed Point}

\begin{figure}[ht]
  \begin{center}
    \leavevmode
      \setlength{\unitlength}{2ex}
  \begin{picture}(36,27.8)(-18,-12)
    \thinlines
\multiput(-12,-11.5)(2,0){13}{\line(0,1){23}}
\multiput(-16,-10)(0,2){11}{\line(1,0){32}}
    \thicklines
\put(0,-14){\vector(0,1){28}}
\put(-18,0){\vector(1,0){36}}
\put(0,15){\makebox(0,0){$k_2$}}
\put(18.5,0){\makebox(0,0)[l]{$k_1$}}
\qbezier(-5.5,0)(-5.275,5.275)(0,5.5)
\qbezier(0,5.5)(5.275,5.275)(5.5,0)
\qbezier(5.5,0)(5.275,-5.275)(0,-5.5)
\qbezier(0,-5.5)(-5.275,-5.275)(-5.5,0)
    \thinlines
\put(4,4){\circle*{0.5}}
\put(0,0){\vector(1,1){3.7}}
\put(4.35,4.35){$p$}
\put(4,-4){\circle*{0.5}}
\put(8,0){\circle*{0.5}}
\put(-8,0){\circle*{0.5}}
\put(-8,-2){\circle*{0.5}}
\put(-12,-4){\circle*{0.5}}
\put(-12,-6){\circle*{0.5}}
\put(-4,2){\circle*{0.5}}
\put(-4,4){\circle*{0.5}}
\put(0,6){\circle*{0.5}}
\put(0,8){\circle*{0.5}}
\put(4,10){\circle*{0.5}}
\put(12,4){\circle*{0.5}}
\put(0,-8){\circle*{0.5}}
\put(-4,-12){\line(1,1){17.5}}
\put(-13.5,-7.5){\line(1,1){19.5}}
\put(-13.5,-5.5){\line(1,1){17.5}}
\put(-3.6,1.3){$\hat{k}$}
\put(-7,12.1){\makebox(0,0)[b]{$(-p_2, p_1)^T$}}
\put(-6.7,12){\vector(1,-3){2.55}}
\put(6.5,13.6){\makebox(0,0)[l]{$\Sg_{\hat{k}}$}}
\put(6.4,13.5){\vector(-2,-3){2.0}}
\put(7,-12.1){\makebox(0,0)[t]{$(p_2, -p_1)^T$}}
\put(6.7,-12.25){\vector(-1,3){2.62}}
\put(-4.4,-13.6){\makebox(0,0)[r]{$\bar{D}_{|p|}$}}
\put(-4.85,-12.55){\vector(1,3){2.45}}
\end{picture}
\end{center}
\caption{An illustration of the classes $\Sg_{\hk}$ and the disk 
$\bar{D}_{|p|}$.}
\label{class}
\end{figure}
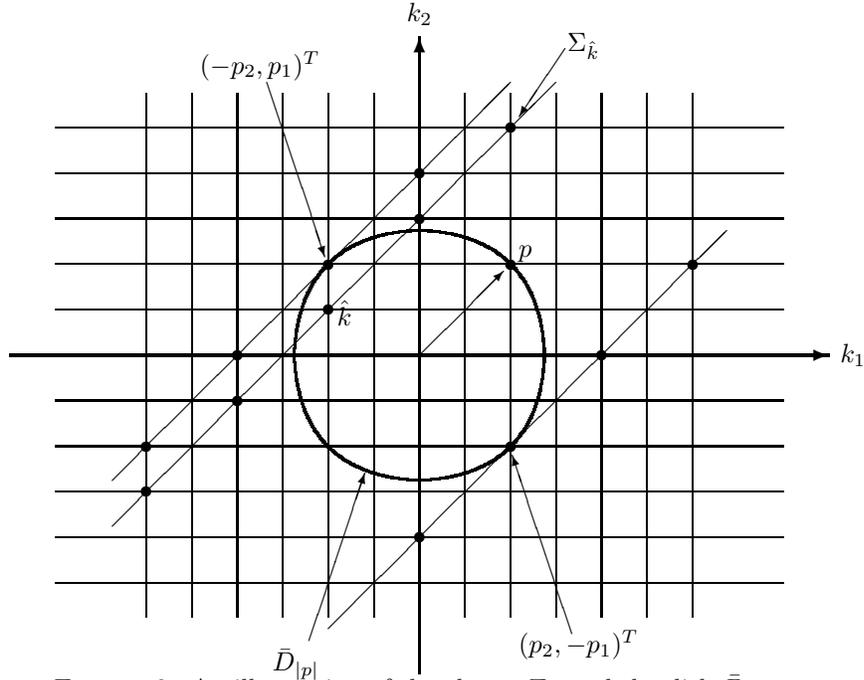

Denote $\{ \om_k \}_{k\in \Z}$ by $\om$. We consider the simple fixed point 
$\om^*$ \cite{Li00}:
\begin{equation}
\om^*_p = \Ga,\ \ \ \om^*_k = 0 ,\ \mbox{if} \ k \neq p \ \mbox{or}\ -p,
\label{fixpt}
\end{equation}
of the 2D Euler equation (\ref{Keuler}), where 
$\Ga$ is an arbitrary complex constant. 
The {\em{linearized two-dimensional Euler equation}} at $\om^*$ is given by,
\begin{equation}
\dot{\om}_k = A(p,k-p)\ \Ga \ \om_{k-p} + A(-p,k+p)\ \bar{\Ga}\ \om_{k+p}\ .
\label{LE}
\end{equation}
\begin{definition}[Classes]
For any $\hk \in \Z$, we define the class $\Sg_{\hk}$ to be the subset of 
$\Z$:
\[
\Sg_{\hk} = \bigg \{ \hk + n p \in \Z \ \bigg | \ n \in Z, \ 
\ p \ \mbox{is specified in (\ref{fixpt})} \bigg \}.
\]
\label{classify}
\end{definition}
\nid
See Fig.\ref{class} for an illustration of the classes. 
According to the classification 
defined in Definition \ref{classify}, the linearized two-dimensional Euler 
equation (\ref{LE}) decouples into infinitely many {\em{invariant subsystems}}:
\begin{eqnarray}
\dot{\omega}_{\hat{k} + np} &=& A(p, \hat{k} + (n-1) p) 
     \ \Gamma \ \omega_{\hat{k} + (n-1) p} \nonumber \\
& & + \ A(-p, \hat{k} + (n+1)p)\ 
     \bar{\Gamma} \ \omega_{\hat{k} +(n+1)p}\ . \label{CLE}
\end{eqnarray}

\subsection{Linear Hamiltonian Systems}

Each invariant subsystem (\ref{CLE}) can be rewritten as a linear 
Hamiltonian system as shown below.
\begin{definition}[The Quadratic Hamiltonian] 
The quadratic Hamiltonian $\HH_{\hat{k}}$ is defined as:
\begin{eqnarray}
\HH_{\hat{k}} &=& -2 \ \im \bigg \{ \sum_{n \in Z} \rho_n \ \Gamma 
\ A(p, \hat{k} + (n-1)p)\ \omega_{\hat{k} + (n-1)p} \ 
\bar{\omega}_{\hat{k} +np} \bigg \} \nonumber \\
\label{CHAM} \\             
&=& - \left| 
      \begin{array}{cc}
          p_1 & \hat{k}_1\\
          p_2 & \hat{k}_2
      \end{array}
      \right| \ \im
      \bigg \{ \sum_{n \in Z} \Gamma \ \rho_n \ \rho_{n-1}
        \ \omega_{\hat{k} + (n-1)p}
        \ \bar{\omega}_{\hat{k} + np} 
      \bigg \}, \nonumber
\end{eqnarray}
where $\rho_n = [ |\hat{k} + np |^{-2}-|p|^{-2}]$,
``\ $\im$\ '' denotes `` imaginary part ''. 
\end{definition}
Then the invariant subsystem (\ref{CLE}) can be rewritten as a linear 
Hamiltonian system \cite{Li00},
\begin{equation}
i \ \dot{\omega}_{\hat{k} +n p} = \rho^{-1}_n \ \frac{\pa \HH_{\hat{k}}}
         {\partial \bar{\omega}_{\hat{k} + np}}\ .
\label{CHAF}
\end{equation}
For a finite dimensional linear Hamiltonian system, it is well-known
that the eigenvalues are of 
four types: real pairs ($c, -c$), purely imaginary pairs ($id, -id$), 
quadruples ($\pm c \pm id$), and zero eigenvalues \cite{Poi99} \cite{Lia49} 
\cite{Arn80}. There is also a complete theorem on the normal forms of such
Hamiltonians \cite{Arn80}. For the above infinite dimensional system 
(\ref{CLE}), the classical proofs can not be applied. Nevertheless, the 
conclusion is still true with different proof \cite{Li00} \cite{Li02m}.
Let $\LL_{\hk}$ be the linear operator defined by the right hand side of 
(\ref{CLE}), and $H^s$ be the Sobolev space where $s \geq 0$ is an integer and 
$H^0=\ell_2$. 
\begin{theorem}[\cite{Li00} \cite{Li02m}]
The eigenvalues of the linear operator $\LL_{\hk}$ in $H^s$ are of 
four types: real pairs ($c, -c$), purely imaginary pairs ($id, -id$), 
quadruples ($\pm c \pm id$), and zero eigenvalues.
\end{theorem}

\subsection{Liapunov Stability}

\begin{definition}[An Important Functional]
For each invariant subsystem (\ref{CLE}), we define the functional 
$I_{\hat{k}}$,
\begin{eqnarray}
    I_{\hat{k}} &=& I_{(\hbox{\textup{\small restricted to }} 
                       \Sg_{\hat{k}})} \nonumber \\
\label{invIk}   \\            
&=& \sum_{n \in Z} \{ | \hat{k} + np |^{-2}
                    - | p |^{-2} \} \left|
                    \omega_{\hat{k} + np} \right|^2 \, . \nonumber
\end{eqnarray}
\end{definition}
\begin{lemma}
$I_{\hat{k}}$ is a constant of motion for the system (\ref{CHAF}).
\label{const}
\end{lemma}
\begin{definition}[The Disk]
The disk of radius $| p |$ in $\Z$, denoted by
$\bar{D}_{| p |}$, is defined as
\[ 
 \bar{D}_{| p |} = \bigg \{ k \in \Z \ \bigg| 
     \ | k | \leq | p | \bigg \} \, .
\]
\end{definition}
\nid
See Fig.\ref{class} for an illustration. 
\begin{theorem}[Unstable Disk Theorem, \cite{Li00}]
If $\ \Sigma_{\hat{k}} \cap \bar{D}_{\left| p \right|} = \emptyset,\ $
then the invariant subsystem (\ref{CLE}) is Liapunov stable
for all $t \in R$, in fact, 
\[
  \sum_{n \in Z} \left|
  \omega_{\hat{k}+np}(t) \right|^2 \leq \sigma \ \sum_{n \in Z} \left|
  \omega_{\hat{k}+np}(0) \right|^2 \, , \quad \quad \forall t \in R \, , 
\]
where 
\[
  \sigma = \left[ \max_{n \in Z} 
              \left\{ - \rho_n \right\}
           \right] \, 
           \left[ \min_{n \in Z} 
               \left\{ -\rho_n \right\}
           \right]^{-1} \, , \quad 0< \sigma < \infty \, .
\]
\label{UDT}
\end{theorem}

\subsection{Spectral Theorems}

Again denote by $\LL_{\hk}$ the linear operator defined by the right hand 
side of (\ref{CLE}).
\begin{theorem}[The Spectral Theorem, \cite{Li00} \cite{Li02m}] We have 
the following claims on the spectrum of the linear operator $\LL_{\hk}$:
\begin{enumerate}
\item If $\Sg_{\hat{k}} \cap \bar{D}_{|p|} = \emptyset$, then the entire
$H^s$ spectrum of the linear operator $\LL_{\hk}$ 
is its continuous spectrum. See Figure \ref{splb}, where
$b= - \frac{1}{2}|\Gamma | |p|^{-2} 
\left|
  \begin{array}{cc}
p_1 & \hat{k}_1 \\
p_2 & \hat{k}_2
  \end{array}
\right| \ .$
That is, both the residual and the point spectra of $\LL_{\hk}$ are empty.
\item If $\Sg_{\hat{k}} \cap \bar{D}_{|p|} \neq \emptyset$, then the entire
essential $H^s$ spectrum of the linear operator $\LL_{\hk}$ is its 
continuous spectrum. 
That is, the residual 
spectrum of $\LL_{\hk}$ is empty. The point 
spectrum of $\LL_{\hk}$ is symmetric with respect to both real and 
imaginary axes. 
See Figure \ref{spla2}.
\end{enumerate}
\label{SST}
\end{theorem}
\begin{figure}[ht]
  \begin{center}
    \leavevmode
      \setlength{\unitlength}{2ex}
  \begin{picture}(36,27.8)(-18,-12)
    \thicklines
\put(0,-14){\vector(0,1){28}}
\put(-18,0){\vector(1,0){36}}
\put(0,15){\makebox(0,0){$\Im \{ \la \}$}}
\put(18.5,0){\makebox(0,0)[l]{$\Re \{ \la \}$}}
\put(0.1,-7){\line(0,1){14}}
\put(.2,-.2){\makebox(0,0)[tl]{$0$}}
\put(-0.2,-7){\line(1,0){0.4}}
\put(-0.2,7){\line(1,0){0.4}}
\put(2.0,-6.4){\makebox(0,0)[t]{$-i2|b|$}}
\put(2.0,7.6){\makebox(0,0)[t]{$i2|b|$}}
\end{picture}
  \end{center}
\caption{The spectrum of $\LL_{\hk}$ in case (1).}
\label{splb}
\end{figure}
\begin{figure}[ht]
  \begin{center}
    \leavevmode
      \setlength{\unitlength}{2ex}
  \begin{picture}(36,27.8)(-18,-12)
    \thicklines
\put(0,-14){\vector(0,1){28}}
\put(-18,0){\vector(1,0){36}}
\put(0,15){\makebox(0,0){$\Im \{ \la \}$}}
\put(18.5,0){\makebox(0,0)[l]{$\Re \{ \la \}$}}
\put(0.1,-7){\line(0,1){14}}
\put(.2,-.2){\makebox(0,0)[tl]{$0$}}
\put(-0.2,-7){\line(1,0){0.4}}
\put(-0.2,7){\line(1,0){0.4}}
\put(2.0,-6.4){\makebox(0,0)[t]{$-i2|b|$}}
\put(2.0,7.6){\makebox(0,0)[t]{$i2|b|$}}
\put(2.4,3.5){\circle*{0.5}}
\put(-2.4,3.5){\circle*{0.5}}
\put(2.4,-3.5){\circle*{0.5}}
\put(-2.4,-3.5){\circle*{0.5}}
\put(5,4){\circle*{0.5}}
\put(-5,4){\circle*{0.5}}
\put(5,-4){\circle*{0.5}}
\put(-5,-4){\circle*{0.5}}
\put(8,6){\circle*{0.5}}
\put(-8,6){\circle*{0.5}}
\put(8,-6){\circle*{0.5}}
\put(-8,-6){\circle*{0.5}}
\end{picture}
  \end{center}
\caption{The spectrum of $\LL_{\hk}$ in case (2).}
\label{spla2}
\end{figure}
For a detailed proof of this theorem, see \cite{Li00} \cite{Li02m}.
Denote by $L$ the right hand side of (\ref{LE}), i.e. the whole 
linearized 2D Euler operator, the spectral mapping theorem 
holds.
\begin{theorem}[\cite{LLM01}]
$$\sigma(e^{tL})=e^{t\sigma(L)}, t\neq 0.$$
\end{theorem}

\subsection{A Continued Fraction Calculation of Eigenvalues} 

Since the introduction of continued fractions for calculating 
the eigenvalues of steady fluid flow, by Meshalkin and Sinai \cite{MS61},
this topics had been extensively explored \cite{Yud65} \cite{Liu92a} 
\cite{Liu92b} \cite{Liu93} \cite{Liu94a} \cite{Liu95} \cite{BFY99} 
\cite{Li00}. Rigorous justification on the continued fraction calculation 
was given in \cite{Li00} \cite{Liu95}.

Rewrite the equation (\ref{CLE}) as follows,
\begin{equation}
\rho^{-1}_n \dot{\tz}_n = a\ \bigg [ \tz_{n+1} - \tz_{n-1} \bigg ]\ ,
\label{cfr1}
\end{equation}
where $\tz_n = \rho_n e^{in (\th +\pi /2)} \om_{\hat{k}+np}$, $\th +\ga 
=\pi /2$, $\Ga = |\Ga| e^{i\ga}$, $a = {1 \over 2} |\Ga| \left | \begin{array}
{lr} p_1 & \hat{k}_1 \\ p_2 & \hat{k}_2 \\ \end{array} \right |$, 
$\rho_n = | \hat{k}+np|^{-2} - |p|^{-2}$. Let $\tz_n = e^{\la t} z_n$, where 
$\la \in C$; then $z_n$ satisfies 
\begin{equation}
a_n z_n +z_{n-1} - z_{n+1} = 0 \ , 
\label{cfr2}
\end{equation}
where $a_n = \la (a \rho_n)^{-1}$. Let $w_n = z_n / z_{n-1}$ \cite{MS61}; 
then $w_n$ satisfies
\begin{equation}
a_n + {1 \over w_n} = w_{n+1}\ .
\label{cfr3}
\end{equation}
Iteration of (\ref{cfr3}) leads to the continued fraction solution \cite{MS61},
\begin{equation}
w_n^{(1)}=a_{n-1} +{1 \over a_{n-2} + {1 \over a_{n-3}+_{\ \ddots}}}\ \ .
\label{cfr4}
\end{equation}
Rewrite (\ref{cfr3}) as follows,
\begin{equation}
w_n = {1 \over -a_n +w_{n+1}}\ . 
\label{cfr5}
\end{equation}
Iteration of (\ref{cfr5}) leads to the continued fraction solution 
\cite{MS61},
\begin{equation}
w_n^{(2)}=-{1 \over a_{n} + {1 \over a_{n+1}+{1 \over a_{n+2}
+_{\ \ddots}}}}\ \ .
\label{cfr6}
\end{equation}
The eigenvalues are given by the condition $w_1^{(1)}=w_1^{(2)}$, 
i.e. 
\begin{equation}
f= a_0 + \bigg ( {1 \over a_{-1} + {1 \over a_{-2} +{1 \over a_{-3}
+_{\ \ddots}}}} \bigg ) + \bigg ( {1 \over a_{1} + {1 \over a_{2} +
{1 \over a_{3}+_{\ \ddots}}}} \bigg ) = 0 \ ,
\label{cfr17}
\end{equation}
where $f = f(\tla,\hat{k},p)$, $\tla = \la /a$.

As an example, we take $p=(1,1)^T$. When $\Ga \neq 0$, the fixed point 
has $4$ eigenvalues which form a 
quadruple. These four eigenvalues appear in the invariant 
linear subsystem labeled by $\hk = (-3,-2)^T$. One of them is \cite{Li00}:
\begin{equation}
\tla=2 \lambda / | \Gamma | = 0.24822302478255 \ + \ i \ 0.35172076526520\ .
\label{evun}
\end{equation}
See Figure \ref{figev} for an illustration. The essential spectrum 
(= continuous spectrum) of $\LL_{\hk}$ with $\hk = (-3,-2)^T$ is the segment 
on the imaginary axis shown in Figure \ref{figev}, where 
$b = -\frac{1}{4} \Ga$. The essential spectrum 
(= continuous spectrum) of the linear 2D Euler operator at this fixed point 
is the entire imaginary axis. 
\begin{figure}[ht]
  \begin{center}
    \leavevmode
      \setlength{\unitlength}{2ex}
  \begin{picture}(36,27.8)(-18,-12)
    \thicklines
\put(0,-14){\vector(0,1){28}}
\put(-18,0){\vector(1,0){36}}
\put(0,15){\makebox(0,0){$\Im \{ \la \}$}}
\put(18.5,0){\makebox(0,0)[l]{$\Re \{ \la \}$}}
\put(2.4,3.5){\circle*{0.5}}
\put(-2.4,3.5){\circle*{0.5}}
\put(2.4,-3.5){\circle*{0.5}}
\put(-2.4,-3.5){\circle*{0.5}}  
\put(0.1,-10){\line(0,1){20}}
\put(.2,-.2){\makebox(0,0)[tl]{$0$}}
\put(-0.2,-10){\line(1,0){0.4}}
\put(-0.2,10){\line(1,0){0.4}}
\put(2.0,-9.4){\makebox(0,0)[t]{$-i2|b|$}}
\put(2.0,10.6){\makebox(0,0)[t]{$i2|b|$}}
\end{picture}
  \end{center}
\caption{The spectrum of $\LL_{\hk}$ with $\hk = (-3,-2)^T$, when $p=(1,1)^T$.}
\label{figev}
\end{figure}
Denote by $L$ the right hand side of (\ref{LE}), i.e. the whole 
linearized 2D Euler operator. Let $\z$ denote the number of points $q \in
\Z$ that belong to the open disk of radius $|p|$, 
and such that $q$ is not parallel to $p$. 
\begin{theorem}[\cite{LLM01}]
The number of nonimaginary eigenvalues of $L$ (counting the multiplicities) 
does not exceed $2\z$.
\end{theorem}
Another interesting discussion upon the discrete spectrum can be found in 
\cite{Fad71}.

A rather well-known open problem is proving the existence of unstable, stable,
and center manifolds. The main difficulty comes from the fact that the 
nonlinear term is non-Lipschitzian.

\chapter{Arnold's Liapunov Stability Theory}

\section{A Brief Summary}

Let $D$ be a region on the ($x,y$)-plane bounded 
by the curves $\Ga_i$ ($i=1,2$), an ideal fluid flow in $D$ is governed 
by the 2D Euler equation written in the stream-function form:
\begin{equation}
{\pa \over \pa t} \Dl \psi = [ \na \psi, \na \Dl \psi ]\ , \label{sfef}
\end{equation}
where 
\[
[ \na \psi, \na \Dl \psi ] = {\pa \psi \over \pa x}
{\pa \Dl \psi \over \pa y}
- {\pa \psi \over \pa y}{\pa \Dl \psi \over \pa x}\ ,
\]
with the boundary conditions,
\[
\psi|_{\Ga_i}=c_i(t)\ ,\ \ c_1 \equiv 0\ ,\ \ {d \over dt} \oint_{\Ga_i}
{\pa \psi \over \pa n} ds = 0\ .
\]
For every function $f(z)$, the functional 
\begin{equation}
F= \int\int_{D} f(\Dl \psi) \ dxdy \label{brs1}
\end{equation}
is a constant of motion (a Casimir) for (\ref{sfef}). The conditional 
extremum of the kinetic energy
\begin{equation}
E={1 \over 2}\int\int_{D}\na \psi \cdot \na \psi  \  dxdy \label{brs2}
\end{equation}
for fixed $F$ is given by the Lagrange's formula \cite{Arn65},
\begin{equation}
\dl H = \dl (E+\la F) =0\ ,\ \ \ \ \Rightarrow \ \ \psi_0 = 
\la f'(\Dl \psi_0)\ . 
\label{brs3}
\end{equation}
where $\la$ is the Lagrange multiplier. Thus, $\psi_0$ is the stream 
function of a stationary flow, which satisfies 
\begin{equation}
\psi_0 = \Phi(\Dl \psi_0)\ , \label{brs4}
\end{equation}
where $\Phi = \la f'$. The second variation is given by \cite{Arn65},
\begin{equation}
\dl^2 H = {1 \over 2} \int\int_{D} \bigg \{ \na \phi \cdot \na \phi + 
\Phi'(\Dl \psi_0) \ (\Dl \phi)^2 \bigg \} dxdy\ . \label{brs5}
\end{equation}
Let $\psi = \psi_0 + \varphi$ be a solution to the 2D Euler equation 
(\ref{sfef}), Arnold proved the estimates \cite{Arn69}: (a). when $c \leq 
\Phi'(\Dl \psi_0) \leq C$, $0 < c \leq C <\infty$,
\[
\int\int_{D} \bigg \{ \na \varphi(t) \cdot \na \varphi(t) + c 
[\Dl \varphi(t)]^2 \bigg \} \ dxdy
\leq \int\int_{D} \bigg \{ \na \varphi(0) \cdot \na \varphi(0) + C 
[\Dl \varphi(0)]^2 \bigg \} \ dxdy,
\]
for all $t \in (-\infty, +\infty)$, (b). when $c \leq 
-\Phi'(\Dl \psi_0) \leq C$, $0 < c < C <\infty$,
\[
\int\int_{D} \bigg \{ c [\Dl \varphi(t)]^2 - \na \varphi(t) \cdot 
\na \varphi(t)\bigg \} \ dxdy
\leq \int\int_{D} \bigg \{ C [\Dl \varphi(0)]^2 - \na \varphi(0) 
\cdot \na \varphi(0) \bigg \} \ dxdy,
\]
for all $t \in (-\infty, +\infty)$. Therefore, when the second variation 
(\ref{brs5}) is positive definite, or when 
\[
\int\int_{D} \bigg \{ \na \phi \cdot \na \phi + 
[\max \Phi'(\Dl \psi_0)] \ (\Dl \phi)^2 \bigg \} \ dxdy
\]
is negative definite, the stationary flow (\ref{brs4}) is nonlinearly stable 
(Liapunov stable). 

Arnold's Liapunov stability theory had been extensively explored, see e.g.
\cite{HMRW85} \cite{Mar92} \cite{WS00}. 

\section{Miscellaneous Remarks}

Establishing Liapunov instability 
along the line of the above, has not been successful. Some rather technical, 
with no clear physical meaning as above, argument showing nonlinear 
instability starting from linear instability, has been established 
by Guo et al. \cite{Guo96} \cite{ABG97} \cite{FSV97}.

Yudovich \cite{Yud00} had been promoting the importance of the 
so-called ``slow collapse'', that is, not finite time blowup, rather 
growing to infinity in time. The main thought is that if derivatives 
growing to infinity in time, the function itself should gain randomness. 
Yudovich \cite{Yud65} had been studying bifurcations of fluid flows.

There are also interests \cite{GMW01} in studying structures of divergent free 
2D vector fields on 2-tori.

\chapter{Miscellaneous Topics}

This chapter serves as a guide to other interesting topics.
Some topics are already well-developed. Others are poorly developed 
in terms of partial differential equations.

\section{KAM Theory}

KAM (Kolmogorov-Arnold-Moser) theory in finite dimensions has 
been a well-known topic \cite{Arn63}. It dealt with the persistence 
of Liouville tori in integrable Hamiltonian systems under Hamiltonian 
perturbations. A natural idea of constructing such tori in perturbed 
systems is conducting canonical transformations which lead to small 
divisor problem. To overcome such difficulties, Kolmogorov introduced 
the Newton's method to speed up the rate of convergence of the 
canonical transformation series. Under certain non-resonance condition 
and non-degeneracy condition of certain Hessian, Arnold completed 
the proof of a rather general theorem \cite{Arn63}. Arnold proved the 
case that the Hamiltonian is an analytic function. Moser was able to 
prove the theorem for the case that the Hamiltonian is 333-times 
differentiable \cite{Mos66a} \cite{Mos66b}, with the help of Nash 
implicit function theorem. Another related topic is the Arnold theorem 
on circle map \cite{Arn61}. It answers the question when a circle map 
is equivalent to a rotation. Yoccoz \cite{Yoc92} was able to prove
an if and only if condition for such equivalence, using Brjuno number.

KAM theory for partial differential equations has also been studied
\cite{Way84} \cite{Kuk93} \cite{Kuk98} \cite{Bou96} \cite{Bou98}.
For partial differential equations, KAM theory is studied on a case 
by case base. There is no general theorem. So far the common types of 
equations studied are nonlinear wave equations as perturbations of certain 
linear wave equations, and solition equations under Hamiltonian perturbations.
Often the persistent Liouville tori are limited to finite dimensional,
sometimes, even one dimensional, i.e. periodic solutions \cite{CW93}.

\section{Gibbs Measure}

Gibbs measure is one of the important concepts in thermodynamic and 
statistical mechanics. In an effort to understand the statistical 
mechanics of nonlinear wave equations, Gibbs measure was introduced 
\cite{MV94} \cite{Bou94} \cite{Bou96b}, which is built upon the Hamiltonians 
of such systems. The calculation of a Gibbs measure is 
similar to that in quantum field theory. In terms of classical analysis, 
the Gibbs measure is not well-defined. One of the central questions is 
which space such Gibbs measure is supported upon. Bourgain \cite{Bou94} 
\cite{Bou96b} was able to give a brilliant answer. For example, 
for periodic nonlinear Schr\"odinger equation, it is supported on $H^{-1/2}$
\cite{Bou94}. For the thermodynamic formalism of infinite dynamical 
systems, Gibbs measure should be an important concept in the future.

\section{Inertial Manifolds and Global Attractors}

Global attractor is a concept for dissipative systems, and inertial manifold
is a concept for strongly dissipative systems. A global attractor is a set 
in the phase space, that attracts all the big balls to it as time approaches 
infinity. An inertial manifold is an invariant manifold that attracts its 
neighborhood 
exponentially. A global attractor can be just a point, and often it is 
just a set. There is no manifold structure with it. Based upon the idea 
of reducing the complex infinite dimensional flows, like Navier-Stokes flow, 
to finite dimensional flows, the concept of inertial manifold is introduced. 
Often inertial manifolds are finite dimensional, have manifold structures,
and most importantly attract their neighborhoods exponentially. Therefore, one 
hopes that the complex infinite dimensional dynamics is slaved by the 
finite dimensional dynamics on the inertial manifolds. More ambitiously, 
one hopes that a finite system of ordinary differential equations can be 
derived to govern the dynamics on the inertial manifold. Under certain 
spectral gap conditions, inertial manifolds can be obtained for many 
evolution equations \cite{Con89} \cite{CFNT89}. Usually, global attractors
can be established rather easily. Unfortunately, inertial manifolds for 
either 2D or 3D Navier-Stokes equations have not been established.

\section{Zero-Dispersion Limit}

Take the KdV equation as an example
\[
u_t-6uu_x +\e^2 u_{xxx}=0\ ,
\]
Lax \cite{LL83} asked the question: what happens to the dynamics as 
$\e \ra 0$, i.e. what is the 
zero-dispersion limit ? One can view the KdV equation as a singular 
perturbation of the corresponding inviscid Burgers equation, 
do the solutions of the KdV equation converge strongly, or weakly, or 
not at all to those of the Burgers equation ? In a series of three 
papers \cite{LL83}, Lax and Levermore investigated these questions.
It turns out that in the zero-dispersion limit, fast oscillations 
are generated instead of shocks or multi-valuedness. Certain 
weak convergences can also established.

In comparison with the singular perturbation studies of soliton 
equations in previous sections, our interests are focused upon 
dynamical systems objects like invariant manifolds and homoclinic 
orbits. In the zero-parameter limit, regularity of invariant manifolds
changes \cite{Li01b}.

\section{Zero-Viscosity Limit}

Take the viscous Burgers equation as an example
\[
u_t+uu_x = \mu u_{xx}\ ,
\]
through the Cole-Hopf transformation \cite{Hop50} \cite{Col51}, 
this equation can be transformed into the heat equation which leads 
to an explicit expression of the solution to the Burgers equation.
Hopf \cite{Hop50}, Cole \cite{Col51}, and Whitham \cite{Whi74} asked the 
question on the zero-viscosity limit. It turns out that the limit 
of a solution to the viscous Burgers equation can form shocks instead 
of the multi-valuedness of the solution to the inviscid Burgers equation
\[
u_t+uu_x = 0\ .
\]
That is, strong convergence does not happen. Nevertheless, the location 
of the shock is determined by the multi-valued portion of the solution to 
the inviscid Burgers equation.  

One can also ask the zero-viscosity limit question for Navier-Stokes equations.
In fact, this question is the core of the studies on fully developed 
turbulence. In most of Kato's papers on fluids \cite{Kat72} \cite{Kat75} 
\cite{Kat86}, he studied the zero-viscosity limits of the solutions to 
Navier-Stokes equations in finite (or small) time interval. It turns 
out that strong convergence can be established for 2D in finite time interval
\cite{Kat86}, and 3D in small time interval \cite{Kat75}. Order $\sqrt{\nu}$ 
rate of convergence was obtained \cite{Kat72}. More recently, Constantin 
and Wu \cite{CW96} had investigated the zero-viscosity limit problem for vortex
patches. 

\section{Finite Time Blowup}

Finite time blowup and its general negative Hamiltonian criterion 
for nonlinear Schr\"odinger equations have been well-known \cite{SS99}. 
The criterion was obtained from a variance relation found by Zakharov
\cite{SS99}. Extension of such criterion to nonlinear Schr\"odinger equations
under periodic boundary condition, is also obtained \cite{Kav87}. 
Extension of such criterion to Davey-Stewartson type equations 
is also obtained \cite{GS90}.

An explicit finite time blowup solution to the integrable Davey-Stewartson 
II equation, was obtained by Ozawa \cite{Oza92} by inventing an extra 
conservation law due to a symmetry. This shows that integrability
and finite time blowup are compatible. The explicit solution $q(t)$ has 
the property that 
\[
\| q(t) \|_{L^2} = 2 \sqrt{\pi}\ ,\ \ \forall t\ , 
\ \ \ \ q(t) \not\in H^1(R^2)\ ,\ \ \forall t\ .
\]
\[
q(t) \in H^s(R^2)\ ,\ \ s \in (0,1)\ , \ \ t\in [0,T]\ ,
\]
for some $T > 0$. When $t \ra T$,
\[
\| q(t) \|_{H^s} \geq C |T-t|^{-s} \ra \infty\ , \ \ s \in (0,1)\  . 
\]
The question of global well-posedness of Davey-Stewartson 
II equation in $H^s(R^2)$ $(s >1)$ is open.

Unlike the success in nonlinear wave equations, the search for 
finite time blowup solutions for 3D Euler equations and other 
equations of fluids has not been successful. The well-known result is the 
Beale-Kato-Majda necessary condition \cite{BKM84}. There are results 
on non-existence of finite time blowup \cite{CF01} \cite{CF02}.

\section{Slow Collapse}

Since 1960's, Yudovich \cite{Yud00} had been promoting the idea of 
slow collapse. That is, although there is no finite time blowup, if 
the function's derivative grows to infinity in norm as time approaches 
infinity, then the function itself should gain randomness in space.

The recent works of Fefferman and Cordoba \cite{CF01} \cite{CF02} 
are in resonance with the slow collapse idea. Indeed, they found 
the temporal growth can be as fast as $e^{e^t}$ and beyond.

\section{Burgers Equation}

Burgers equation
\[
u_t+uu_x = \mu u_{xx} + f(t,x)\ ,
\]
was introduced by J. M. Burgers \cite{Bur39} as a simple model of 
turbulence. By 1950, E. Hopf \cite{Hop50} had conducted serious mathematical 
study on the Burgers equation
\[
u_t+uu_x = \mu u_{xx} \ ,
\]
and found interesting mathematical structures of this equation including
the so-called Cole-Hopf transformation \cite{Hop50} \cite{Col51} and 
Legendre transform. Moreover,
Hopf investigated the limits $\mu \ra 0$ and $t\ra \infty$. It turns out 
that the order of taking the two limits is important. Especially with the 
works of Lax \cite{Lax53} \cite{Lax54}, this led to a huge interests of 
studies on conservation laws for many years. Since 1992, Sinai \cite{Sin92} 
had led a study on Burgers equation with random data which can be random 
initial data or random forcing \cite{EKMS00}. 

Another old model introduced by E. Hopf \cite{Hop48} did not catch 
too much attention. As a turbulence model, the main drawback of Burgers 
equation is lack of 
incompressibility condition. In fact, there are studies focusing upon 
only incompressibility \cite{GMW01}. To remedy this drawback, other models are 
necessary.

\section{Other Model Equations}

A model to describe the hyperbolic structures in a neighborhood of 
a fixed point of 2D Euler equation under periodic boundary condition,
was introduced in \cite{Li02l}  \cite{Li02m} \cite{Li02e} \cite{Li02f}.
It is a good model in terms of capturing the linear instability. 
At special value of a parameter, the explicit expression for the 
hyperbolic structure can be calculated \cite{Li02l} \cite{Li02e}. 
One fixed point's unstable manifold is the stable manifold of another
fixed point, and vice versa. All are 2D ellipsoidal surfaces. Together,
they form a lip shape hyperbolic structure.

Another model is the so-called shell model \cite{KLWB95} \cite{SKL95} 
\cite{KLS97} which model the energy transfer in the spectral space to 
understand for example Kolmogorov spectra.

\section{Kolmogorov Spectra and An Old Theory of Hopf}

The famous Kolmogorov $-5/3$ law of homogeneous isotropic turbulence 
\cite{Kol41a} \cite{Kol41b} \cite{Kol41c} still fascinates a lot of 
researchers even nowadays. On the contrary, a statistical theory of 
Hopf on turbulence \cite{Hop52} \cite{HT53} \cite{Hop57} \cite{Hop62}
\cite{VF86} has almost been forgotten. By introducing initial probability 
in a function space, and realizing the conservation of probability under 
the Navier-Stokes 
flow, Hopf derived a functional equation for the characteristic functional 
of the probability, around 1940, published in \cite{Hop52}. Initially,
Hopf tried to find some near Gaussian solution \cite{HT53}, then he 
knew the work of Kolmogorov on the $-5/3$ law which is far from Gaussian
probability, and verified by experiments. So the topic was not pursued much
further.

\section{Onsager Conjecture}

In 1949, L. Onsager \cite{Ons49} conjectured that solutions of
the incompressible Euler equation with H\"older continuous velocity of 
order $\nu > 1/3$ conserves the energy, but not necessarily if 
$\nu \leq 1/3$. In terms of Besov spaces, if the weak solution of 
Euler equation has certain regularity, it can be proved that energy 
indeed conserves \cite{Eyi94} \cite{CET94}.

\section{Weak Turbulence}

Motivated by a study on 2D Euler equation \cite{Zak90}, Zakharov 
led a study on infinte dimensional Hamiltonian systems in the 
spectral space \cite{Zak98}. Under near Gaussian and small amplitude 
assumptions, 
Zakharov heuristically gave a closure relation for the averaged equation.
He also heuristically found some stationary solution to the averaged equation
which leads to power-type energy spectra all of which he called Kolmogorov 
spectra. One of the mathematical manipulations he often used is the 
canonical transformation, based upon which he classified the kinetic 
Hamiltonian systems into 3-wave and 4-wave resonant systems.

\section{Renormalization Idea}

Renormalization group approach has been very successful in proving 
universal property, especially Feigenbaum constants, of one dimensional
maps \cite{CEL80}. Renormalization-group-type idea has also been
applied to turbulence \cite{OY87}, although not very successful. 

\section{Random Forcing}

As mentioned above, there are studies on Burgers equation under random 
forcing \cite{EKMS00}, and studies on the structures of incompressible 
vector fields \cite{GMW01}. Take any steady solution of 2D Euler equation, 
it defines a Hamiltonian system with the stream function being the Hamiltonian.
There have been studies on such systems under random forcings \cite{FP94} 
\cite{FK02}.

\section{Strange Attractors and SBR Invariant Measure}

Roughly speaking, strange attractors are attractors in which dynamics has 
sensitive dependence upon initial data. If the Cantor sets proved in 
previous sections are also attractors, then they will be strange attractors.
In view of the obvious fact that often the attractor may contain many 
different objects rather than only the Cantor set, global 
strange attractors are the ideal cases. For one dimensional logistic-type 
maps, and two dimensional H\'enon map in the small parameter range such that 
the 2D map can viewed as a perturbation of a 1D logistic map, there have been 
proofs on the existence of strange attractors and SBR (Sinai-Bowen-Ruelle) 
invariant measures \cite{BC91} \cite{BY92} \cite{BY93}.

\section{Arnold Diffusions}

The term Arnold diffusion started from the paper \cite{Arn64}. This paper 
is right after Arnold 
completed the proof on the persistence of KAM tori in \cite{Arn63}. If 
the degree of freedom is 2, the persistent KAM tori are 2 dimensional, 
and the level set of the perturbed Hamiltonian is 3 dimensional, then 
the 2 dimensional KAM tori will isolate the level set, and diffusion is 
impossible. In \cite{Arn64}, Arnold gave an explicit example to show that 
diffusion is possible when the degree of freedom is more than 2. Another 
interesting point is that in \cite{Arn64} Arnold also derived an integral 
expression for a distance measurement, which is the so-called Melnikov 
integral \cite{Mel63}. Arnold's derivation was quite unique. There have 
been a lot of studies upon Arnold diffusions \cite{Loc99}, unfortunately, 
doable examples are not much beyond that of Arnold \cite{Arn64}.
Sometimes, the diffusion can be very slow as the famous Nekhoroshev's theorem
shows \cite{Nek77} \cite{Nek79}.
 
\section{Averaging Technique}

For systems with fast small oscillations, long time dynamics are 
governed by their averaged systems. There are quite good estimates 
for long time deviation of solutions of the averaged systems from those 
of the original systems \cite{Arn65b} \cite{Arn80}. Whether or not 
the averaging method can be useful in chaos in partial differential 
equations is still to be seen.

\backmatter

\bibliographystyle{amsalpha}

\begin{thebibliography}{A}

\bibitem{AC91}
M.~J. Ablowitz and P.~A. Clarkson.
\newblock {\em Solitons, {N}onlinear {E}volution {E}quations and {I}nverse
  {S}cattering}.
\newblock London Math. Soc. Lect. Note Ser. 149, Cambridge Univ. Press, 1991.

\bibitem{AL76}
M.~J. Ablowitz and J.~F. Ladik.
\newblock A {N}onlinear {D}ifference {S}cheme and {I}nverse {S}cattering.
\newblock {\em Stud. Appl. Math.}, 55:213, 1976.

\bibitem{AOT99}
M.~J. Ablowitz, Y.~Ohta, and A.~D. Trubatch.
\newblock On {D}iscretizations of the {V}ector {N}onlinear {S}chr{\"{o}}dinger
  {E}quation.
\newblock {\em Phys. Lett. A}, 253, no.5-6:287--304, 1999.

\bibitem{AOT00}
M.~J. Ablowitz, Y.~Ohta, and A.~D. Trubatch.
\newblock On {I}ntegrability and {C}haos in {D}iscrete {S}ystems.
\newblock {\em Chaos Solitons Fractals}, 11, 1-3:159--169, 2000.

\bibitem{AS81}
M.~J. Ablowitz and H.~Segur.
\newblock {\em Solitons and the {I}nverse {S}cattering {T}ransform}.
\newblock SIAM, Philadelphia, 1981.

\bibitem{Ada75}
R.~Adams.
\newblock {\em Sobolev {S}pace}.
\newblock Academic Press, New York, 1975.

\bibitem{ABG97}
L.~Almeida, F.~Bethuel, and Y.~Guo.
\newblock A {R}emark on {I}nstability of the {S}ymmetric {V}ortices with 
  {L}arge {C}oupling {C}onstant.
\newblock {\em Comm. Pure Appl. Math.}, L:1295--1300, 1997.

\bibitem{AI79}
R.~L. Anderson and N.~H. Ibragimov.
\newblock {\em Lie-{B}{\"{a}}cklund {T}ransformations in {A}pplications}.
\newblock SIAM, Philadelphia, 1979.

\bibitem{Ano67}
D.~V. Anosov.
\newblock Geodesic {F}lows on {C}ompact {R}iemannian {M}anifolds of {N}egative
  {C}urvature.
\newblock {\em Proc. Steklov Inst. Math.}, 90, 1967.

\bibitem{Arn61}
V.~I. Arnold.
\newblock Small {D}enominators. {I}. {M}apping the {C}ircle onto {I}tself.
\newblock {\em Izv. Akad. Nauk SSSR Ser. Mat.}, 25:21--86, 1961.

\bibitem{Arn63}
V.~I. Arnold.
\newblock Proof of a {T}heorem of {A}. {N}. {K}olmogorov on the {P}reservation 
of {C}onditionally {P}eriodic {M}otions under a {S}mall {P}erturbation of the
  {H}amiltonian.
\newblock {\em Uspehi Mat. Nauk}, 18, no.5(113):13--40, 1963.

\bibitem{Arn64}
V.~I. Arnold.
\newblock Instability of {D}ynamical {S}ystems with {M}any {D}egrees of
  {F}reedom.
\newblock {\em Sov. Math. Dokl.}, 5 No. 3:581--585, 1964.

\bibitem{Arn65b}
V.~I. Arnold.
\newblock Applicability {C}onditions and an {E}rror {B}ound for the 
{A}veraging {M}ethod for {S}ystems in the {P}rocess of {E}volution 
{T}hrough a {R}esonance.
\newblock {\em Dokl. Akad. Nauk SSSR}, 161:9--12, 1965.

\bibitem{Arn65}
V.~I. Arnold.
\newblock Conditions for {N}onlinear {S}tability of {S}tationary {P}lane
  {C}urvilinear {F}lows of an {I}deal {F}luid.
\newblock {\em Sov. Math. Dokl.}, 6:773--776, 1965.

\bibitem{Arn66}
V.~I. Arnold.
\newblock Sur la {G}eometrie {D}ifferentielle des {G}roupes de {L}ie de
  {D}imension {I}nfinie et ses {A}pplications a {L}'hydrodynamique des
  {F}luides {P}arfaits.
\newblock {\em Ann. Inst. Fourier, Grenoble}, 16,1:319--361, 1966.

\bibitem{Arn69}
V.~I. Arnold.
\newblock On an {A}priori {E}stimate in the {T}heory of {H}ydrodynamical
  {S}tability.
\newblock {\em Amer. Math. Soc. Transl., Series 2}, 79:267--269, 1969.

\bibitem{Arn80}
V.~I. Arnold.
\newblock {\em Mathematical {M}ethods of {C}lassical {M}echanics}.
\newblock Springer-Verlag, 1980.

\bibitem{BLZ98}
P.~Bates, K.~Lu, and C.~Zeng.
\newblock Existence and {P}ersistence of {I}nvariant {M}anifolds for
  {S}emiflows in {B}anach {S}pace.
\newblock {\em Mem. Amer. Math. Soc.}, 135, no.645, 1998.

\bibitem{BLZ99}
P.~Bates, K.~Lu, and C.~Zeng.
\newblock Persistence of {O}verflowing {M}anifolds for {S}emiflow.
\newblock {\em Comm. Pure Appl. Math.}, 52, no.8:983--1046, 1999.

\bibitem{BLZ00}
P.~Bates, K.~Lu, and C.~Zeng.
\newblock Invariant {F}oliations {N}ear {N}ormally {H}yperbolic {I}nvariant
  {M}anifolds for {S}emiflows.
\newblock {\em Mem. Amer. Math. Soc.}, 352, no.10:4641--4676, 2000.

\bibitem{BKM84}
J.~T. Beale, T.~Kato, and A.~Majda.
\newblock Remarks on the {B}reakdown of {S}mooth {S}olutions for the 
3-{D} {E}uler {E}quations.
\newblock {\em Comm. Math. Phys.}, 94, no.1:61--66, 1984.

\bibitem{BFY99}
L.~Belenkaya, S.~Friedlander, and V.~Yudovich.
\newblock The {U}nstable {S}pectrum of {O}scillating {S}hear {F}lows.
\newblock {\em SIAM J. Appl. Math.}, 59, no.5:1701--1715, 1999.

\bibitem{BC91}
M.~Benedicks and L.~Carleson.
\newblock The dynamics of the {H}enon map.
\newblock {\em Ann. of Math.}, 133, no.1:73--169, 1991.

\bibitem{BY92}
M.~Benedicks and L.-S. Young.
\newblock Absolutely {C}ontinuous {I}nvariant {M}easures and {R}andom 
{P}erturbations for {C}ertain {O}ne-{D}imensional {M}aps.
\newblock {\em Ergodic Theory Dynam. Systems}, 12, no.1:13--37, 1992.

\bibitem{BY93}
M.~Benedicks and L.-S. Young.
\newblock Sinai-{B}owen-{R}uelle {M}easures for {C}ertain {H}enon {M}aps.
\newblock {\em Invent. Math.}, 112, no.3:541--576, 1993.

\bibitem{Ben72}
T.~B. Benjamin.
\newblock The {S}tability of {S}olitary {W}aves.
\newblock {\em Proc. R. Soc. Lond.}, A328:153--183, 1972.

\bibitem{Bir12}
G.~D. Birkhoff.
\newblock Some {T}heorems on the {M}otion of {D}ynamical {S}ystems.
\newblock {\em Bull. Soc. Math. France}, 40:305--323, 1912.

\bibitem{Bir94}
B.~Birnir.
\newblock Qualitative {A}nalysis of {R}adiating {B}reathers.
\newblock {\em Comm. Pure Appl. Math.}, XLVII:103--119, 1994.

\bibitem{BMW94}
B.~Birnir, H.~Mckean, and A.~Weinstein.
\newblock The {R}igidity of {S}ine-{G}ordon {B}reathers.
\newblock {\em Comm. Pure Appl. Math.}, XLVII:1043--1051, 1994.

\bibitem{Bla86}
C.~M. Blazquez.
\newblock Transverse {H}omoclinic {O}rbits in {P}eriodically {P}erturbed
  {P}arabolic {E}quations.
\newblock {\em Nonlinear Analysis, Theory, Methods and Applications}, 10,
  no.11:1277--1291, 1986.

\bibitem{BW99}
J.~Bona and J.~Wu.
\newblock Zero {D}issipation {L}imit for {N}onlinear {W}aves.
\newblock {\em Preprint}, 1999.

\bibitem{Bou93}
J.~Bourgain.
\newblock Fourier {R}estriction {P}henomena for {C}ertain {L}attice {S}ubsets
  and {A}pplications to {N}onlinear {E}volution {E}quations.
\newblock {\em Geom. and Funct. Anal.}, 3, No 2:107--156, 209--262, 1993.

\bibitem{Bou94}
J.~Bourgain.
\newblock Periodic {N}onlinear {S}chroedinger {E}quation and {I}nvariant
  {M}easures.
\newblock {\em Comm. Math. Phys.}, 166:1--26, 1994.

\bibitem{Bou96}
J.~Bourgain.
\newblock Construction of {A}pproximative and {A}lmost-{P}eriodic {S}olutions
  of {P}erturbed {L}inear {S}chroedinger and {W}ave {E}quations.
\newblock {\em Geom. Func. Anal.}, 6(2):201--230, 1996.

\bibitem{Bou96b}
J.~Bourgain.
\newblock Invariant measures for the 2{D}-defocusing nonlinear {S}chroedinger
  equation.
\newblock {\em Comm. Math. Phys.}, 176, no.2:421--445, 1996.

\bibitem{Bou98}
J.~Bourgain.
\newblock Quasi-{P}eriodic {S}olutions of {H}amiltonian {P}erturbations of 2{D}
  {L}inear {S}chroedinger {E}quations.
\newblock {\em Ann. of Math.}, 148, no.2:363--439, 1998.

\bibitem{Bur39}
J.~M. Burgers.
\newblock Mathematical {E}xamples {I}llustrating {R}elations {O}ccuring 
in the {T}heory of {T}urbulent {F}luid {M}otion.
\newblock {\em Verh. Nederl. Akad. Wetensch. Afd. Natuurk. Sect. 1.}, 17,
  no.2:53 pp., 1939.

\bibitem{Caz89}
T.~Cazenave.
\newblock {\em An {I}ntroduction to {N}onlinear {S}chrodinger {E}quations}.
\newblock IMUFRJ Rio De Janeiro volume 22, 1989.

\bibitem{Chi00}
S.~Childress.
\newblock A {L}ax pair of 3{D} {E}uler equation.
\newblock {\em personal communication}, 2000.

\bibitem{CH82}
S.-N. Chow and J.~K. Hale.
\newblock {\em Methods of {B}ifurcation {T}heory}.
\newblock Springer-Verlag, New York-Berlin, 1982.

\bibitem{CHMP80}
S.-N. Chow, J.~K. Hale, and J.~Mallet-Paret.
\newblock An {E}xample of {B}ifurcation to {H}omoclinic {O}rbits.
\newblock {\em J. Differential Equations}, 37, no.3:351--373, 1980.

\bibitem{CLL91}
S.-N. Chow, X.-B. Lin, and K.~Lu.
\newblock Smooth {I}nvariant {F}oliations in {I}nfinite-{D}imensional {S}paces.
\newblock {\em J. Differential Equations}, 94, no.2:266--291, 1991.

\bibitem{CLP89}
S.~N. Chow, X.~B. Lin, and K.~J. Palmer.
\newblock A {S}hadowing {L}emma with {A}pplications to {S}emilinear {P}arabolic
  {E}quations.
\newblock {\em SIAM J. Math. Anal.}, 20, no.3:547--557, 1989.

\bibitem{Col51}
J.~Cole.
\newblock On a {Q}uasi-{L}inear {P}arabolic {E}quation {O}ccurring in 
{A}erodynamics.
\newblock {\em Quart. Appl. Math.}, 9:225--236, 1951.

\bibitem{CEL80}
P.~Collet, J.-P. Eckmann, and O.~E. Lanford.
\newblock Universal {P}roperties of {M}aps on an {I}nterval.
\newblock {\em Comm. Math. Phys.}, 76, no.3:211--254, 1980.

\bibitem{Con89}
P.~Constantin.
\newblock A {C}onstruction of {I}nertial {M}anifolds.
\newblock {\em Contemporary Mathematics}, 99, 1989.

\bibitem{CET94}
P.~Constantin, W.~E, and E.~Titi.
\newblock Onsager's {C}onjecture on the {E}nergy {C}onservation for 
{S}olutions of {E}uler's {E}quation.
\newblock {\em Comm. Math. Phys.}, 165, no.1:207--209, 1994.

\bibitem{CFNT89}
P.~Constantin, C.~Foias, B.~Nicolaenko, and R.~Temam.
\newblock {\em Integral {M}anifolds and {I}nertial {M}anifolds for
  {D}issipative {P}artial {D}ifferential {E}quations}.
\newblock Applied Mathematical Sciences, 70. Springer-Verlag, New York, 1989.

\bibitem{CW96}
P.~Constantin and J.~Wu.
\newblock The {I}nviscid {L}imit for {N}on-smooth {V}orticity.
\newblock {\em Indiana Univ. Math. J.}, 45, no.1:67--81, 1996.

\bibitem{CKP95}
B.~Coomes, H.~Kocak, and K.~Palmer.
\newblock A {S}hadowing {T}heorem for {O}rdinary {D}ifferential {E}quations.
\newblock {\em Z. Angew. Math. Phys.}, 46, no.1:85--106, 1995.

\bibitem{CKP97}
B.~Coomes, H.~Kocak, and K.~J. Palmer.
\newblock Long {P}eriodic {S}hadowing.
\newblock {\em Numerical Algorithms}, 14:55--78, 1997.

\bibitem{CF01}
D.~Cordoba and C.~Fefferman.
\newblock Behavior of {S}everal {T}wo-{D}imensional {F}luid {E}quations 
in {S}ingular {S}cenarios.
\newblock {\em Proc. Natl. Acad. Sci. USA}, 98, no.8:4311--4312, 2001.

\bibitem{CF02}
D.~Cordoba and C.~Fefferman.
\newblock Growth of {S}olutions for {Q}{G} and 2{D} {E}uler {E}quations.
\newblock {\em J. Amer. Math. Soc.}, 15, no.3:665--670, 2002.

\bibitem{CW93}
W.~Craig and C.~E. Wayne.
\newblock Newton's {M}ethod and {P}eriodic {S}olutions of {N}onlinear 
{W}ave {E}quations.
\newblock {\em Comm. Pure Appl. Math.}, 46, no.11:1409--1498, 1993.

\bibitem{Den89}
B.~Deng.
\newblock Exponential {E}xpansion with {S}ilnikov's {S}addle-{F}ocus.
\newblock {\em J. Differential Equations}, 82, no.1:156--173, 1989.

\bibitem{Den93}
B.~Deng.
\newblock On {S}ilnikov's {H}omoclinic-{S}addle-{F}ocus {T}heorem.
\newblock {\em J. Differential Equations}, 102, no.2:305--329, 1993.

\bibitem{DFGHMS86} 
T. Dombre et al. 
\newblock Chaotic {S}treamlines in the {A}{B}{C} {F}lows.
\newblock {\em J. Fluid Mech.}, 167, 353--391, 1986

\bibitem{EKMS00}
W.~E, K.~Khanin, A.~Mazel, and Ya. Sinai.
\newblock Invariant {M}easures for {B}urgers {E}quation with {S}tochastic 
{F}orcing.
\newblock {\em Ann. of Math.}, 151, no.3:877--960, 2000.

\bibitem{EFM90}
N.~M. Ercolani, M.~G. Forest, and D.~W. McLaughlin.
\newblock {G}eometry of the {M}odulational {I}nstability, {P}art {I}{I}{I}:
  {H}omoclinic {O}rbits for the {P}eriodic {S}ine- {G}ordon {E}quation.
\newblock {\em Physica D}, 43:349--84, 1990.

\bibitem{Eyi94}
G.~L. Eyink.
\newblock Energy dissipation without viscosity in ideal hydrodynamics. {I}.
  {F}ourier analysis and local energy transfer.
\newblock {\em Phys. D}, 78, no.3-4:222--240, 1994.

\bibitem{Fad71}
L.~D. Faddeev.
\newblock On the {S}tability {T}heory for {S}tationary {P}lane-{P}arallel
  {F}lows of {I}deal {F}luid.
\newblock {\em Kraevye Zadachi Mat. Fiziki (Zapiski Nauchnykh Seminarov LOMI,
  v.21), Moscow, Leningrad: Nauka}, 5:164--172, 1971.

\bibitem{FK02}
A.~Fannjiang and T.~Komorowski.
\newblock Diffusive and {N}ondiffusive {L}imits of {T}ransport in {N}onmixing 
Flows.
\newblock {\em SIAM J. Appl. Math.}, 62, no.3:909--923, 2002.

\bibitem{FP94}
A.~Fannjiang and G.~Papanicolaou.
\newblock Convection Enhanced Diffusion for Periodic Flows.
\newblock {\em SIAM J. Appl. Math.}, 54, no.2:333--408, 1994.

\bibitem{Fen71}
N.~Fenichel.
\newblock Persistence and {S}moothness of {I}nvariant {M}anifolds for {F}lows.
\newblock {\em Ind. University Math. J.}, 21:193--225, 1971.

\bibitem{Fen74}
N.~Fenichel.
\newblock Asymptotic {S}tability with {R}ate {C}onditions.
\newblock {\em Ind. Univ. Math. J.}, 23:1109--1137, 1974.

\bibitem{Fen77}
N.~Fenichel.
\newblock Asymptotic {S}tability with {R}ate {C}onditions {I}{I}.
\newblock {\em Ind. Univ. Math. J.}, 26:81--93, 1977.

\bibitem{Fen79}
N.~Fenichel.
\newblock Geometric {S}ingular {P}erturbation {T}heory for {O}rdinary
  {D}ifferential {E}quations.
\newblock {\em J. Diff. Eqns.}, 31:53--98, 1979.

\bibitem{FM76}
H.~Flaschka and D.~W. McLaughlin.
\newblock Canonically {C}onjugate {V}ariables for the {K}orteweg-de {V}ries
  {E}quation and the {T}oda {L}attice with {P}eriodic {B}oundary {C}onditions.
\newblock {\em Progress of Theoretical Physics}, 55, No. 2:438--465, 1976.

\bibitem{FMMW00}
M.~G. Forest, D.~W. McLaughlin, D.~J. Muraki, and O.~C. Wright.
\newblock Nonfocusing {I}nstabilities in {C}oupled, {I}ntegrable {N}onlinear
  {S}chr{\"{o}}dinger pdes.
\newblock {\em J. Nonlinear Sci.}, 10, no.3:291--331, 2000.

\bibitem{FSW00}
M.~G. Forest, S.~P. Sheu, and O.~C. Wright.
\newblock On the {C}onstruction of {O}rbits {H}omoclinic to {P}lane {W}aves in
  {I}ntegrable {C}oupled {N}onlinear {S}chr{\"{o}}dinger {S}ystem.
\newblock {\em Phys. Lett. A}, 266, no.1:24--33, 2000.

\bibitem{FS77}
J.~E. Franke and J.~F. Selgrade.
\newblock Hyperbolicity and {C}hain {R}ecurrence.
\newblock {\em J. Diff. Eq.}, 26:27--36, 1977.

\bibitem{FSV97}
S.~Friedlander, W.~Strauss, and M.~Vishik.
\newblock Nonlinear Instability in an Ideal Fluid.
\newblock {\em Ann. Inst. Henri Poincare}, 14, no.2:187--209, 1997.

\bibitem{GS90}
J.~M. Ghidaglia and J.~C. Saut.
\newblock On the {I}nitial {V}alue {P}roblem for the {D}avey-{S}tewartson
  {S}ystems.
\newblock {\em Nonlinearity}, 3:475--506, 1990.

\bibitem{GMW01}
M.~Ghil, T.~Ma, and S.~Wang.
\newblock Structural Bifurcation of 2-{D} Incompressible Flows.
\newblock {\em Indiana Univ. Math. Journal}, 50, no.1:159--180, 2001.

\bibitem{GSS87}
M.~Grillakis, J.~Shatah, and W.~Strauss.
\newblock Stability {T}heory of {S}olitary {W}aves in the {P}resence of
  {S}ymmetry, {I}.
\newblock {\em J. Funct. Anal.}, 74:160--197, 1987.

\bibitem{GH83}
J.~Guckenheimer and P.~J. Holmes.
\newblock {\em Nonlinear {O}scillations, {D}ynamical {S}ystems, and
  {B}ifurcations of {V}ector {F}ields}.
\newblock Springer-Verlag, New York, 1983.

\bibitem{Guo96}
Y.~Guo.
\newblock Instability of Symmetric Vortices with Large Charge and Coupling
  constant.
\newblock {\em Comm. Pure Appl. Math.}, XLIX:1051--1080, 1996.

\bibitem{Had01}
J.~Hadamard.
\newblock Sur {L}iteration et les {S}olution {A}symptotiques des {E}quations
  {D}ifferentielles.
\newblock {\em Bull. Soc. Math. France}, 29, 1901.

\bibitem{HL86}
J.~K. Hale and X.~B. Lin.
\newblock Symbolic {D}ynamics and {N}onlinear {S}emiflows.
\newblock {\em Ann. Mat. Pura Appl.}, 144:229--259, 1986.

\bibitem{Har63}
P.~Hartman.
\newblock On the {L}ocal {L}inearization of {D}ifferential {E}quations.
\newblock {\em Proc. Amer. Math. Soc.}, 14:568--573, 1963.

\bibitem{HK95}
A.~Hasegawa and Y.~Kodama.
\newblock {\em Solitons in {O}ptical {C}ommunications}.
\newblock Academic Press, San Diego, 1995.

\bibitem{Has72}
H.~Hasimoto.
\newblock A {S}oliton on a {V}ortex {F}ilament.
\newblock {\em J. Fluid Mech.}, 51 (3):477--485, 1972.

\bibitem{Hen94}
D.~Henry.
\newblock Exponential {D}ichotomies, the {S}hadowing {L}emma and {H}omoclinic
  {O}rbits in {B}anach {S}paces.
\newblock {\em Resenhas}, 1, no.4:381--401, 1994.

\bibitem{HPS77}
M.~W. Hirsch, C.~C. Pugh, and M.~Shub.
\newblock {\em Invariant {M}anifolds}.
\newblock Springer-Verlag, New York, 1977.

\bibitem{HMRW85}
D.~D. Holm, J.~E. Marsden, T.~Ratiu, and A.~Weinstein.
\newblock Nonlinear {S}tability of {F}luid and {P}lasma {E}quilibria.
\newblock {\em Physics Reports}, 123:1--116, 1985.

\bibitem{HM81}
P.~J. Holmes and J.~E. Marsden.
\newblock A {P}artial {D}ifferential {E}quation with {I}nfinitely {M}any
  {P}eriodic {O}rbits: {C}haotic {O}scillations of a {F}orced {B}eam.
\newblock {\em Arch. Rat. Mech. Anal.}, 76:135--166, 1981.

\bibitem{Hop48}
E.~Hopf.
\newblock A Mathematical Example Displaying Features of Turbulence.
\newblock {\em Comm. Pure Appl. Math.}, 1:303--322, 1948.

\bibitem{Hop50}
E.~Hopf.
\newblock The Partial Differential Equation $u_t+uu_x=\mu u_{xx}$.
\newblock {\em Comm. Pure Appl. Math.}, 3:201--230, 1950.

\bibitem{Hop52}
E.~Hopf.
\newblock Statistical Hydromechanics and Functional Calculus.
\newblock {\em J. Rational Mech. Anal.}, 1:87--123, 1952.

\bibitem{Hop57}
E.~Hopf.
\newblock On the Application of Functional Calculus to the Statistical Theory
  of Turbulence.
\newblock {\em Applied probability, Proceedings of Symposia in Applied
  Mathematics, McGraw-Hill Book Co., for the American Mathematical Society,
  Providence, R. I.}, 7:41--50, 1957.

\bibitem{Hop62}
E.~Hopf.
\newblock Remarks on the Functional-Analytic Approach to Turbulence.
\newblock {\em 1962 Proc. Sympos. Appl. Math., American Mathematical Society,
  Providence, R.I.}, 8:157--163, 1962.

\bibitem{HT53}
E.~Hopf and E.~W. Titt.
\newblock On Certain Special Solutions of the $\phi$-Equation of Statistical
  Hydrodynamics.
\newblock {\em J. Rational Mech. Anal.}, 2:587--591, 1953.

\bibitem{Isl92}
M.~N. Islam.
\newblock {\em Ultrafast {F}iber {S}witching {D}evices and {S}ystems}.
\newblock Cambridge University Press, New York, 1992.

\bibitem{JKL91}
C.~Jones, N.~Kopell, and R.~Langer.
\newblock Construction of the {F}itz{H}ugh-{N}agumo {P}ulse {U}sing
  {D}ifferential {F}orms.
\newblock {\em in Patterns and Dynamics in Reactive Media, H. Swinney, G. Aris,
  and D. Aronson eds., IMA Volumes in Mathematics and its Applications,
  Springer-Verlag, New York}, 37, 1991.

\bibitem{JK91}
C.K.R.T. Jones and N.~Kopell.
\newblock Tracking {I}nvariant {M}anifolds with {D}ifferential {F}orms in
  {S}ingularly {P}erturbed {S}ystems.
\newblock {\em J. Diff. Eq.}, 108:64--88, 1994.

\bibitem{KLS97}
L.~Kadanoff, D.~Lohse, and N.~Schoerghofer.
\newblock Scaling and Linear Response in the {G}{O}{Y} Turbulence Model.
\newblock {\em Phys. D}, 100, no.1-2:165--186, 1997.

\bibitem{KLWB95}
L.~Kadanoff, D.~Lohse, J.~Wang, and R.~Benzi.
\newblock Scaling and Dissipation in the {G}{O}{Y} Shell Model.
\newblock {\em Phys. Fluids}, 7, no.3:617--629, 1995.

\bibitem{KS98}
T.~Kapitula and B.~Sandstede.
\newblock Stability of {B}right {S}olitary-{W}aves {S}olutions to {P}erturbed
  {N}onlinear {S}chr{\"{o}}dinger {E}quations.
\newblock {\em Physica D}, 124:58--103, 1998.

\bibitem{Kat72}
T.~Kato.
\newblock Nonstationary {F}lows of {V}iscous and {I}deal {F}luids in ${R}^3$.
\newblock {\em Journal of Functional Analysis}, 9:296--305, 1972.

\bibitem{Kat75}
T.~Kato.
\newblock Quasi-{L}inear {E}quations of {E}volution, with {A}pplications to
  {P}artial {D}ifferential {E}quations.
\newblock {\em Lecture Notes in Mathematics}, 448:25--70, 1975.

\bibitem{Kat86}
T.~Kato.
\newblock Remarks on the {E}uler and {N}avier-{S}tokes {E}quations in ${R}^2$.
\newblock {\em Proc. Symp. Pure Math., Part 2}, 45:1--7, 1986.

\bibitem{Kav87}
O.~Kavian.
\newblock A Remark on the Blowing-Up of Solutions to the {C}auchy Problem for
  Nonlinear {S}chroedinger Equations.
\newblock {\em Trans. Amer. Math. Soc.}, 299, no.1:193--203, 1987.

\bibitem{Kel67}
A.~Kelley.
\newblock The {S}table, {C}enter-{S}table, {C}enter, {C}enter-{U}nstable,
  {U}nstable {M}anifolds.
\newblock {\em J. Diff. Eqns.}, 3:546--570, 1967.

\bibitem{Kic91}
S.~Kichsnassamy.
\newblock Breather {S}olutions of the {N}onlinear {W}ave {E}quation.
\newblock {\em Comm. Pure Appl. Math.}, 44:789--818, 1991.

\bibitem{Kol41c}
A.~N. Kolmogoroff.
\newblock Dissipation of Energy in the Locally Isotropic Turbulence.
\newblock {\em C. R. (Doklady) Acad. Sci. URSS (N.S.)}, 32:16--18, 1941.

\bibitem{Kol41a}
A.~N. Kolmogoroff.
\newblock The Local Structure of Turbulence in Incompressible Viscous Fluid for
  Very Large Reynold's Numbers.
\newblock {\em C. R. (Doklady) Acad. Sci. URSS (N.S.)}, 30:301--305, 1941.

\bibitem{Kol41b}
A.~N. Kolmogoroff.
\newblock On Degeneration of Isotropic Turbulence in an Incompressible Viscous
  Liquid.
\newblock {\em C. R. (Doklady) Acad. Sci. URSS (N.S.)}, 31:538--540, 1941.

\bibitem{Kov92a}
G.~Kovacic.
\newblock Dissipative {D}ynamics of {O}rbits {H}omoclinic to a {R}esonance
  {B}and.
\newblock {\em Phys Lett A}, 167:143--150, 1992.

\bibitem{Kov92b}
G.~Kovacic.
\newblock Hamiltonian {D}ynamics of {O}rbits {H}omoclinic to a {R}esonance
  {B}and.
\newblock {\em Phys Lett A}, 167:137--142, 1992.

\bibitem{Kuk93}
S.~B. Kuksin.
\newblock {\em Nearly {I}ntegrable {I}nfinite-{D}imensional {H}amiltonian
  {S}ystems}.
\newblock Lecture Notes in Mathematics, 1556. Springer-Verlag, Berlin, 1993.

\bibitem{Kuk98}
S.~B. Kuksin.
\newblock A {K}{A}{M}-Theorem for Equations of the {K}orteweg-de {V}ries Type.
\newblock {\em Rev. Math. Math. Phys.}, 10, no.3, 1998.

\bibitem{LLM01}
Y.~Latushkin, Y.~Li, and M.~Stanislavova.
\newblock The {S}pectrum of a {L}inearized 2{D} {E}uler {O}perator.
\newblock {\em Submitted to Studies in Appl. Math., available at:
http://www.math.missouri.edu/\~{}cli}, 2003.

\bibitem{Lax53}
P.~Lax.
\newblock Nonlinear Hyperbolic Equations.
\newblock {\em Comm. Pure Appl. Math.}, 6:231--258, 1953.

\bibitem{Lax54}
P.~Lax.
\newblock {\em The {I}nitial {V}alue {P}roblem for {N}onlinear {H}yperbolic
  {E}quations in {T}wo {I}ndependent {V}ariables, {C}ontributions to the
  {T}heory of {P}artial {D}ifferential {E}quations}.
\newblock Annals of Math. Studies, Vol. 33, pp.211-229, Princeton Univ. Press,
  Princeton, 1954.

\bibitem{LL83}
P.~Lax and C.~D. Levermore.
\newblock The {S}mall {D}ispersion {L}imit of the {K}orteweg-de {V}ries
  {E}quation, {P}art {I}, {I}{I}, {I}{I}{I}.
\newblock {\em Comm. Pure App. Math.}, 36:253--290, 571--593, 809--830, 1983.

\bibitem{Bon75}
J.~L.Bona.
\newblock On the {S}tability of {S}olitary {W}aves.
\newblock {\em Proc. R. Soc. Lond.}, A344:363--374, 1975.

\bibitem{BSS87}
J.~L.Bona, P.~E. Souganidis, and W.~A. Strauss.
\newblock Stability and {I}nstability of {S}olitary {W}aves of {K}d{V} {T}ype.
\newblock {\em Proc. R. Soc. Lond.}, A411:395--412, 1987.

\bibitem{Li92}
Y.~Li.
\newblock Backlund {T}ransformations and {H}omoclinic {S}tructures for the
  {N}{L}{S} {E}quation.
\newblock {\em Phys. Letters A}, 163:181--187, 1992.

\bibitem{Li99b}
Y.~Li.
\newblock Homoclinic Tubes in Nonlinear {S}chr{\"{o}}dinger Equation Under
  {H}amiltonian Perturbations.
\newblock {\em Progress of Theoretical Physics}, 101, no.3:559--577, 1999.

\bibitem{Li99a}
Y.~Li.
\newblock Smale {H}orseshoes and {S}ymbolic {D}ynamics in {P}erturbed
  {N}onlinear {S}chr{\"{o}}dinger {E}quations.
\newblock {\em J. Nonlinear Sciences}, 9:363--415, 1999.

\bibitem{Li00a}
Y.~Li.
\newblock B{\"{a}}cklund-{D}arboux {T}ransformations and {M}elnikov {A}nalysis
  for {D}avey-{S}tewartson {I}{I} {E}quations.
\newblock {\em J. Nonlinear Sci.}, 10:103--131, 2000.

\bibitem{Li00}
Y.~Li.
\newblock On 2{D} {E}uler {E}quations: {P}art {I}. {O}n the {E}nergy-{C}asimir
  {S}tabilities and {T}he {S}pectra for {L}inearized 2{D} {E}uler {E}quations.
\newblock {\em J. Math. Phys.}, 41, No.2:728--758, 2000.

\bibitem{Li01a}
Y.~Li.
\newblock A {L}ax {P}air for the {T}wo {D}imensional {E}uler {E}quation.
\newblock {\em J. Math. Phys.}, 42, No.8:3552--3553, 2001.

\bibitem{Li01b}
Y.~Li.
\newblock Persistent {H}omoclinic {O}rbits for {N}onlinear {S}ch{\"{o}}dinger
  {E}quation {U}nder {S}ingular {P}erturbation.
\newblock {\em In Press, International Journal of Mathematics and 
Mathematical Sciences}, 2003.

\bibitem{Li02a}
Y.~Li.
\newblock Chaos and {S}hadowing {L}emma for {A}utonomous {S}ystems of
  {I}nfinite {D}imensions.
\newblock {\em In Press, J. Dynamics and Differential Equations}, 2003.

\bibitem{Li03j}
Y.~Li.
\newblock Chaos and {S}hadowing {A}round a {H}omoclinic {T}ube.
\newblock {\em In Press, Abstract and Applied Analysis}, 2003.

\bibitem{Li03k}
Y.~Li.
\newblock Homoclinic {T}ubes and {C}haos in {P}erturbed {S}ine-{G}ordon 
{E}quation. 
\newblock {\em In Press, Chaos, Solitons and Fractals}, 2003.

\bibitem{Li02f}
Y.~Li.
\newblock Chaos in Partial Differential Equations.
\newblock {\em Contemporary Mathematics}, 301: 93--115, 2002.

\bibitem{Li02c}
Y.~Li.
\newblock Existence of {C}haos for a {S}ingularly {P}erturbed {N}{L}{S}
  {E}quation.
\newblock {\em In Press, International Journal of Mathematics and 
Mathematical Sciences}, 2003.

\bibitem{Li02g}
Y.~Li.
\newblock Homoclinic Tubes in Discrete Nonlinear {S}chroedinger Equation Under
  {H}amiltonian Perturbations.
\newblock {\em Nonlinear Dynamics}, 31, No.4: 393--434, 2003.

\bibitem{Li02e}
Y.~Li.
\newblock Integrable Structures for 2{D} {E}uler Equations of Incompressible
  Inviscid Fluids.
\newblock {\em Proc. Institute Of Mathematics of NAS of Ukraine}, 43, part
  1:332--338, 2002.

\bibitem{Li02b}
Y.~Li.
\newblock Melnikov {A}nalysis for {S}ingularly {P}erturbed {D}{S}{I}{I}
  {E}quation.
\newblock {\em Submitted, available at:
  http://xxx.lanl.gov/abs/math.AP/0206272, or
  http://www.math.missouri.edu/\~{}cli}, 2003.

\bibitem{Li02l}
Y.~Li.
\newblock On 2{D} {E}uler equations: {I}{I}. {L}ax Pairs and Homoclinic
  Structures.
\newblock {\em Communications on Applied Nonlinear Analysis}, 10, no.1: 
1--43, 2003.

\bibitem{Li02m}
Y.~Li.
\newblock Euler {E}quations of {I}nviscid {F}luids. 
\newblock {\em Submitted to Advances in Mathematical Sciences, 
Nova Science Publishers, 
available at: http://www.math.missouri.edu/\~{}cli}, 2003.

\bibitem{Li02d}
Y.~Li.
\newblock Singularly Perturbed Vector and Scalar Nonlinear {S}chroedinger
  Equations with Persistent Homoclinic Orbits.
\newblock {\em Studies in Applied Mathematics}, 109:19--38, 2002.

\bibitem{LMSW96}
Y.~Li, D.~McLaughlin, J.~Shatah, and S.~Wiggins.
\newblock Persistent {H}omoclinic {O}rbits for a {P}erturbed {N}onlinear
  {S}chr{\"{o}}dinger equation.
\newblock {\em Comm. Pure Appl. Math.}, XLIX:1175--1255, 1996.

\bibitem{LM94}
Y.~Li and D.~W. McLaughlin.
\newblock Morse and {M}elnikov {F}unctions for {N}{L}{S} {P}des.
\newblock {\em Comm. Math. Phys.}, 162:175--214, 1994.

\bibitem{LM97}
Y.~Li and D.~W. McLaughlin.
\newblock Homoclinic {O}rbits and {C}haos in {P}erturbed {D}iscrete {N}{L}{S}
  {S}ystem. {P}art {I} {H}omoclinic {O}rbits.
\newblock {\em Journal of Nonlinear Sciences}, 7:211--269, 1997.

\bibitem{LW97}
Y.~Li and S.~Wiggins.
\newblock Homoclinic {O}rbits and {C}haos in {P}erturbed {D}iscrete {N}{L}{S}
  {S}ystem. {P}art {II} {S}ymbolic {D}ynamics.
\newblock {\em Journal of Nonlinear Sciences}, 7:315--370, 1997.

\bibitem{LW97b}
Y.~Li and S.~Wiggins.
\newblock {\em Invariant {M}anifolds and {F}ibrations for {P}erturbed
  {N}onlinear {S}chr{\"{o}}dinger {E}quations}, volume 128.
\newblock Springer-Verlag, Applied Mathematical Sciences, 1997.

\bibitem{LY01}
Y.~Li and A.~Yurov.
\newblock Lax {P}airs and {D}arboux {T}ransformations for {E}uler {E}quations.
\newblock {\em Studies in Appl. Math.}, 111: 101--113, 2003.

\bibitem{Lia49}
A.~M. Liapunov.
\newblock {\em Probl{\`{e}}me {G}{\'{e}}n{\'{e}}ral de la {S}tabilit{\'{e}} du
  {M}ouvement}.
\newblock Annals of Math. Studies, Vol. 17, Princeton Univ. Press, Princeton,
  1949.

\bibitem{Liu92a}
V.~X. Liu.
\newblock An {E}xample of {I}nstability for the {N}avier-{S}tokes {E}quations
  on the 2-{D}imensional {T}orus.
\newblock {\em Comm. Partial Differential Equations}, 17, no.11-12:1995--2012,
  1992.

\bibitem{Liu92b}
V.~X. Liu.
\newblock Instability for the {N}avier-{S}tokes {E}quations on the
  2-{D}imensional {T}orus and a {L}ower {B}ound for the {H}ausdorff {D}imension
  of {T}heir {G}lobal {A}ttractors.
\newblock {\em Comm. Math. Phys.}, 147, no.2:217--230, 1992.

\bibitem{Liu93}
V.~X. Liu.
\newblock A {S}harp {L}ower {B}ound for the {H}ausdorff {D}imension of the
  {G}lobal {A}ttractors of the 2{D} {N}avier-{S}tokes {E}quations.
\newblock {\em Comm. Math. Phys.}, 158, no.2:327--339, 1993.

\bibitem{Liu94a}
V.~X. Liu.
\newblock Remarks on the {N}avier-{S}tokes {E}quations on the {T}wo- and
  {T}hree-{D}imensional {T}orus.
\newblock {\em Comm. Partial Differential Equations}, 19, no.5-6:873--900,
  1994.

\bibitem{Liu95}
V.~X. Liu.
\newblock On {U}nstable and {N}eutral {S}pectra of {I}ncompressible {I}nviscid
  and {V}iscid {F}luids on the 2{D} {T}orus.
\newblock {\em Quart. Appl. Math.}, 53, no.3:465--486, 1995.

\bibitem{Liu94}
Y.~Liu.
\newblock Instability of {S}olutions for {G}eneralized {B}oussinesq
  {E}quations.
\newblock {\em Ph. D. Thesis, Brown University}, 1994.

\bibitem{Loc99}
P.~Lochak.
\newblock Arnold Diffusion, a Compendium of Remarks and Questions.
\newblock {\em Hamiltonian Systems with Three or More Degrees of Freedom
  (S'Agaro, 1995), NATO Adv. Sci. Inst. Ser. C Math. Phys. Sci., Kluwer Acad.
  Publ., Dordrecht}, 533:168--183, 1999.

\bibitem{Man74}
S.~V. Manakov.
\newblock On the {T}heory of {T}wo-{D}imensional {S}tationary {S}elf-{F}ocusing
  of {E}lectromagnetic {W}aves.
\newblock {\em Soviet Physics JETP}, 38, no.2:248--253, 1974.

\bibitem{Mar92}
J.~E. Marsden.
\newblock {\em Lectures on {M}echanics, {L}ond. {M}ath. {S}oc. {L}ect. {N}ote,
  {S}er. 174}.
\newblock Cambridge Univ. Press, 1992.

\bibitem{MS91}
V.~B. Matveev and M.~A. Salle.
\newblock {\em Darboux {T}ransformations and {S}olitons}, volume~5.
\newblock Springer Series in Nonlinear Dynamics, 1991.

\bibitem{MT76}
H.~P. Mckean and E.~Trubowitz.
\newblock Hill's {O}perator and {H}yperelliptic {F}unction {T}heory in the
  {P}resence of {I}nfinitely {M}any {B}ranch {P}oints.
\newblock {\em Comm. Pure. App. Math.}, 29:143--226, 1976.

\bibitem{MM75}
H.~P. Mckean and P.~van Moerbeke.
\newblock The {S}pectrum of {H}ill's {E}quation.
\newblock {\em Inventiones Math.}, 30:217--274, 1975.

\bibitem{MV94}
H.~P. McKean and K.~L. Vaninsky.
\newblock Statistical Mechanics of Nonlinear Wave Equations.
\newblock {\em Appl. Math. Sci., 100, Springer, New York}, 100:239--264, 1994.

\bibitem{MS78}
D.~McLaughlin and A.~Scott.
\newblock Perturbation {A}nalysis of {F}luxon {D}ynamics.
\newblock {\em Phys. Rev. A}, 18:1652--1680, 1978.

\bibitem{MO95}
D.W. McLaughlin and E.~A. Overman.
\newblock Whiskered {T}ori for {I}ntegrable {P}des and {C}haotic {B}ehavior in
  {N}ear {I}ntegrable {P}des.
\newblock {\em Surveys in Appl Math 1}, 1995.

\bibitem{Mel63}
V.~K. Melnikov.
\newblock On the {S}tability of the {C}enter for {T}ime {P}eriodic
  {P}erturbations.
\newblock {\em Trans. Moscow Math. Soc.}, 12:1--57, 1963.

\bibitem{Men87}
C.~R. Menyuk.
\newblock Nonlinear {P}ulse {P}ropagation in {B}irefringent {O}ptical {F}ibers.
\newblock {\em IEEE J. Quantum Electron}, 23, no.2:174--176, 1987.

\bibitem{Men89}
C.~R. Menyuk.
\newblock Pulse {P}ropagation in an {E}lliptically {B}irefringent {K}err
  {M}edium.
\newblock {\em IEEE J. Quantum Electron}, 25, no.12:2674--2682, 1989.

\bibitem{MS61}
L.~D. Meshalkin and Ia.~G. Sinai.
\newblock Investigation of the {S}tability of a {S}tationary {S}olution of a
  {S}ystem of {E}quations for the {P}lane {M}ovement of an {I}ncompressible
  {V}iscous {L}iquid.
\newblock {\em J. Appl. Math. Mech. (PMM)}, 25:1140--1143, 1961.

\bibitem{Mos66a}
J.~Moser.
\newblock A Rapidly Convergent Iteration Method and Non-linear Partial
  Differential Equations. {I}.
\newblock {\em Ann. Scuola Norm. Sup. Pisa}, 20:265--315, 1966.

\bibitem{Mos66b}
J.~Moser.
\newblock A Rapidly Convergent Iteration Method and Non-linear Partial
  Differential Equations. {I}{I}.
\newblock {\em Ann. Scuola Norm. Sup. Pisa}, 20:499--535, 1966.

\bibitem{Nek77}
N.~N. Nekhoroshev.
\newblock An Exponential Estimate of the Time of Stability of Nearly Integrable
  {H}amiltonian Systems.
\newblock {\em Uspehi Mat. Nauk}, 32, no.6:5--66, 1977.

\bibitem{Nek79}
N.~N. Nekhoroshev.
\newblock An Exponential Estimate of the Time of Stability of Nearly Integrable
  {H}amiltonian Systems {I}{I}.
\newblock {\em Trudy Sem. Petrovsk}, 5:5--50, 1979.

\bibitem{Nik86}
N.~V. Nikolenko.
\newblock The {M}ethod of {P}oincare {N}ormal {F}orms in {P}roblems of
  {I}ntegrability of {E}quations of {E}volution {T}ype.
\newblock {\em Russian Math. Surveys}, 41:5:63--114, 1986.

\bibitem{Ons49}
L.~Onsager.
\newblock Statistical Hydrodynamics.
\newblock {\em Nuovo Cimento}, 6, Supplemento, no.2:279--287, 1949.

\bibitem{OY87}
S.~A. Orszag and V.~Yakhot.
\newblock Renormalization Group Analysis of Turbulence.
\newblock {\em Proceedings of the International Congress of Mathematicians,
  (Berkeley, Calif., 1986), Amer. Math. Soc., Providence, RI}, 1,2:1395--1399,
  1987.

\bibitem{Oza92}
T.~Ozawa.
\newblock Exact {B}low-{U}p {S}olutions to the {C}auchy {P}roblem for the
  {D}avey-{S}tewartson {S}ystems.
\newblock {\em Proc. R. Soc. Lond.}, 436:345--349, 1992.

\bibitem{Pal83}
J.~Palis.
\newblock A {N}ote on the {I}nclination {L}emma ($\lambda$-{L}emma) and
  {F}eigenbaum's {R}ate of {A}pproach.
\newblock {\em Lect. Notes in Math.}, 1007:630--635, 1983.

\bibitem{PM82}
J.~Palis and W.~de~Melo.
\newblock {\em Geometric {T}heory of {D}ynamical {S}ystems}.
\newblock Springer-Verlag, 1982.

\bibitem{Pal84}
K.~Palmer.
\newblock Exponential {D}ichotomies and {T}ransversal {H}omoclinic {P}oints.
\newblock {\em J. Differential Equations}, 55, no.2:225--256, 1984.

\bibitem{Pal88}
K.~Palmer.
\newblock Exponential {D}ichotomies, the {S}hadowing {L}emma and {T}ransversal
  {H}omoclinic {P}oints.
\newblock {\em Dynamics Reported}, 1:265--306, 1988.

\bibitem{Pal96}
K.~Palmer.
\newblock Shadowing and {S}ilnikov {C}haos.
\newblock {\em Nonlinear Anal.}, 27, no.9:1075--1093, 1996.

\bibitem{PW92}
R.~L. Pego and M.~I. Weinstein.
\newblock Eigenvalues, and {I}nstabilities of {S}olitary {W}aves.
\newblock {\em Phil. Trans. R. Soc. Lond.}, A340:47--94, 1992.

\bibitem{Per30}
O.~Perron.
\newblock Die {S}tabilitatsfrage bei {D}ifferentialgleichungssysteme.
\newblock {\em Math. Zeit.}, 32:703--728, 1930.

\bibitem{Poi99}
H.~Poincar{\'{e}}.
\newblock {\em Les {M}ethodes {N}ouvelles de la {M}ecanique {C}eleste, {V}ols.
  1-3. {E}nglish {T}ranslation: {N}ew {M}ethods of {C}elestial {M}echanics.
  {V}ols. 1-3, edited by {D}aniel {L}. {G}oroff, {A}merican {I}nstitute of
  {P}hysics, {N}ew {Y}ork, 1993}.
\newblock Gauthier-Villars, Paris, 1899.

\bibitem{PT87}
J.~Poschel and E.~Trubowitz.
\newblock {\em Inverse {S}pectral {T}heory}.
\newblock Academic Press, 1987.

\bibitem{RS82}
C.~Rogers and W.~F. Shadwick.
\newblock {\em B{\"{a}}cklund {T}ransformations and {T}heir {A}pplications}.
\newblock Academic Press, 1982.

\bibitem{SZ87}
D.~H. Sattinger and V.~D. Zurkowski.
\newblock Gauge {T}heory of {B}{\"{a}}cklund {T}ransformations.
\newblock {\em Physica D}, 26:225--250, 1987.

\bibitem{SKL95}
N.~Schoerghofer, L.~Kadanoff, and D.~Lohse.
\newblock How the Viscous Subrange Determines Inertial Range Properties in
  Turbulence Shell Models.
\newblock {\em Phys. D}, 88, no.1:40--54, 1995.

\bibitem{Sel85}
G.~R. Sell.
\newblock Smooth {L}inearization {N}ear a {F}ixed {P}oint.
\newblock {\em Amer. J. Math.}, 107, no.5:1035--1091, 1985.

\bibitem{SZ00}
J.~Shatah and C.~Zeng.
\newblock Homoclinic {O}rbits for the {P}erturbed {S}ine-{G}ordon {E}quation.
\newblock {\em Comm. Pure Appl. Math.}, 53, no.3:283--299, 2000.

\bibitem{Sil65}
L.~P. Silnikov.
\newblock A {C}ase of the {E}xistence of a {C}ountable {N}umber of {P}eriodic
  {M}otions.
\newblock {\em Soviet Math. Doklady}, 6:163--166, 1965.

\bibitem{Sil67a}
L.~P. Silnikov.
\newblock The {E}xistence of a {D}enumerable {S}et of {P}eriodic {M}otions in
  {F}our-dimensional {S}pace in an {E}xtended {N}eighborhood of a
  {S}addle-{F}ocus.
\newblock {\em Soviet Math. Doklady}, 8:54--58, 1967.

\bibitem{Sil67b}
L.~P. Silnikov.
\newblock On a {P}oincare-{B}irkoff {P}roblem.
\newblock {\em Math. USSR Sb.}, 3:353--371, 1967.

\bibitem{Sil68b}
L.~P. Silnikov.
\newblock Structure of the {N}eighborhood of a {H}omoclinic {T}ube of an
  {I}nvariant {T}orus.
\newblock {\em Soviet Math. Dokl.}, 9, 1968.

\bibitem{Sil70}
L.~P. Silnikov.
\newblock A {C}ontribution to the {P}roblem of the {S}tructure of an {E}xtended
  {N}eighborhood of a {R}ough {E}quilibrium {S}tate of {S}addle-focus {T}ype.
\newblock {\em Math. USSR Sb.}, 10:91--102, 1970.

\bibitem{Sin92}
Ya. Sinai.
\newblock Statistics of Shocks in Solutions of Inviscid {B}urgers Equation.
\newblock {\em Comm. Math. Phys.}, 148, no.3:601--621, 1992.

\bibitem{Sma61}
S.~Smale.
\newblock A {S}tructurally {S}table {D}ifferentiable {H}omeomorphism with an
  {I}nfinite {N}umber of {P}eriodic {P}oints.
\newblock {\em Qualitative Methods in the Theory of Non-linear Vibrations
  (Proc. Internat. Sympos. Non-linear Vibrations)}, II:365--366, 1961.

\bibitem{Sma65}
S.~Smale.
\newblock Diffeomorphisms with {M}any {P}eriodic {P}oints.
\newblock {\em Differential and Combinatorial Topology (A Symposium in Honor of
  Marston Morse), Princeton Univ. Press, Princeton, N.J.}, pages 63--80, 1965.

\bibitem{Sma67}
S.~Smale.
\newblock Differentiable {D}ynamical {S}ystems.
\newblock {\em Bull. Amer. Math. Soc.}, 73:747--817, 1967.

\bibitem{SW89}
H.~Steinlein and H.~Walther.
\newblock Hyperbolic {S}ets and {S}hadowing for {N}oninvertible {M}aps.
\newblock {\em Advanced Topics in the Theory of Dynamical Systems, Academic
  Press, Boston, MA}, pages 219--234, 1989.

\bibitem{SW90}
H.~Steinlein and H.~Walther.
\newblock Hyperbolic {S}ets, {T}ransversal {H}omoclinic {T}rajectories, and
  {S}ymbolic {D}ynamics for $c^1$-{M}aps in {B}anach {S}paces.
\newblock {\em J. Dynamics Differential Equations}, 2, no.3:325--365, 1990.

\bibitem{SS99}
C.~Sulem and P.-L. Sulem.
\newblock {\em The {N}onlinear {S}chroedinger {E}quation, self-focusing and
  wave collapse}.
\newblock Applied Mathematical Sciences, 139, Springer-Verlag, New York, 1999.

\bibitem{VF86}
M.~I. Vishik and A.~V. Fursikov.
\newblock {\em Mathematical {P}roblems of {S}tatistical {H}ydromechanics}.
\newblock Akademische Verlagsgesellschaft Geest and Portig K.-G., Leipzig,
  1986.

\bibitem{Wal87}
H.-O. Walther.
\newblock Inclination {L}emmas with {D}ominated {C}onvergence.
\newblock {\em ZAMP}, pages 327--337, 1987.

\bibitem{Way84}
C.~E. Wayne.
\newblock The KAM Theory of Systems with Short Range Interactions. I, II.
\newblock {\em Comm. Math. Phys.}, 96, no. 3:311--329, 331--344, 1984.

\bibitem{Wei86}
M.~I. Weinstein.
\newblock Lyapunov {S}tability of {G}round {S}tates of {N}onlinear {D}ispersive
  {E}volution {E}quations.
\newblock {\em Comm. Pure Appl. Math.}, 39:51--68, 1986.

\bibitem{Whi74}
G.~B. Whitham.
\newblock {\em Linear and {N}onlinear {W}aves}.
\newblock John Wiley and Sons, 1974.

\bibitem{Wig88}
S.~Wiggins.
\newblock {\em Global {B}ifurcations and {C}haos: {A}nalytical {M}ethods}.
\newblock Springer-Verlag, New York, 1988.

\bibitem{WS00}
D.~Wirosoetisno and T.~G. Shepherd.
\newblock On the Existence of Two-Dimensional {E}uler Flows Satisfying
  Energy-{C}asimir Stability Criteria.
\newblock {\em Phys.Fluids}, 12:727--730, 2000.

\bibitem{WF00}
O.~C. Wright and M.~G. Forest.
\newblock On the {B}{\"{a}}cklund-{G}auge {T}ransformation and {H}omoclinic
  {O}rbits of a {C}oupled {N}onlinear {S}chr{\"{o}}dinger {S}ystem.
\newblock {\em Phys. D}, 141, no.1-2:104--116, 2000.

\bibitem{Wu96}
J.~Wu.
\newblock The {I}nviscid {L}imits for {I}ndividual and {S}tatistical
  {S}olutions of the {N}avier-{S}tokes {E}quations.
\newblock {\em Ph.D. Thesis, Chicago University}, 1996.

\bibitem{Wu98}
J.~Wu.
\newblock The {I}nviscid {L}imit of the {C}omplex {G}inzburg-{L}andau
  {E}quation.
\newblock {\em J. Diff. Equations}, 142, No.2:413--433, 1998.

\bibitem{YT01}
J.~Yang and Y.~Tan.
\newblock Fractal {D}ependence of {V}ector-{S}oliton {C}ollisions in
  {B}irefringent {F}ibers.
\newblock {\em Phys. Lett. A}, 280:129--138, 2001.

\bibitem{Yoc92}
J.-C. Yoccoz.
\newblock An Introduction to Small Divisors Problems.
\newblock {\em From number theory to physics ({L}es {H}ouches, 1989), Springer,
  Berlin}, pages 659--679, 1992.

\bibitem{Yud65}
V.~I. Yudovich.
\newblock Example of the {G}eneration of a {S}econdary {S}tationary or
  {P}eriodic {F}low {W}hen {T}here is {L}oss of {S}tability of the {L}aminar
  {F}low of a {V}iscous {I}ncompressible {F}luid.
\newblock {\em J. Appl. Math. Mech. (PMM)}, 29:527--544, 1965.

\bibitem{Yud00}
V.~I. Yudovich.
\newblock On the Loss of Smoothness of the Solutions of {E}uler Equations and
  the Inherent Instability of Flows of an Ideal Fluid.
\newblock {\em Chaos}, 10, no.3:705--719, 2000.

\bibitem{Zak90}
V.~E. Zakharov.
\newblock On the Algebra of Integrals of Motion in Two-Dimensional
  Hydrodynamics in {C}lebsch Variables.
\newblock {\em Funct. Anal. Appl.}, 23, no.3:189--196, 1990.

\bibitem{Zak98}
V.~E. Zakharov.
\newblock {\em Nonlinear {W}aves and {W}eak {T}urbulence}.
\newblock American Mathematical Society Translations, Series 2, 182. Advances
  in the Mathematical Sciences, 36. American Mathematical Society, Providence,
  RI, 1998.

\bibitem{ZS72}
V.~E. Zakharov and A.~B. Shabat.
\newblock Exact {T}heory of {T}wo-dimensional {S}elf-focusing and
  {O}ne-dimensional {S}elf-modulation of {W}aves in {N}onlinear {M}edia.
\newblock {\em Sov. Phys. JETP}, 43(1):62--69, 1972.

\bibitem{Zen95}
W.~Zeng.
\newblock Exponential {D}ichotomies and {T}ransversal {H}omoclinic {O}rbits in
  {D}egenerate {C}ases.
\newblock {\em J. Dyn. Diff. Eq.}, 7, no.4:521--548, 1995.

\bibitem{Zen97}
W.~Zeng.
\newblock Transversality of {H}omoclinic {O}rbits and {E}xponential
  {D}ichotomies for {P}arabolic {E}quations.
\newblock {\em J. Math. Anal. Appl.}, 216:466--480, 1997.

\end{thebibliography}

%

\begin{theindex}

\item Asymptotic phase shift
\subitem for DSII, 25
\subitem for NLS, 14

\indexspace

\item B\"acklund-Darboux transformation
\subitem for discrete NLS, 18
\subitem for DSII, 21
\subitem for NLS, 13
\subitem for 2D Euler equation, 92
\item Breather solution, 3

\indexspace

\item Casimir, 96
\item Chaos theorem, 76
\item Conley-Moser conditions, 74
\item Continued fraction, 100
\item Counting lemma, 29

\indexspace

\item Equivariant smooth linearization, 71
\item Evolution operator, 2

\indexspace

\item Floquet discriminant
\subitem for discrete NLS, 17
\subitem for NLS, 12
\item Fiber theorem
\subitem for perturbed DSII, 53
\subitem for regularly perturbed NLS, 40
\subitem for singularly perturbed NLS, 41

\indexspace

\item Hadamard's method, 39
\item Heteroclinic orbit, 2
\item Homoclinic orbit, 2

\indexspace

\item Inflowing invariance, 3

\indexspace

\item $\la$-lemma, 83
\item Lax pair 
\subitem for discrete NLS, 16
\subitem for DSII, 20
\subitem for NLS, 11
\subitem for Rossby wave equation, 93
\subitem for 3D Euler equation, 93
\subitem for 2D Euelr equation, 92
\item Lie-Poisson bracket, 95
\item Local invariance, 3

\indexspace

\item Overflowing invariance, 3

\indexspace

\item Perron's method, 39
\item Persistence theorem
\subitem for perturbed DSII, 53
\subitem for regularly perturbed NLS, 40
\subitem for singularly perturbed NLS, 42
\item Pseudo-orbit, 79

\indexspace

\item Quadratic products of eigenfunctions
\subitem for DSII, 34
\subitem for NLS, 16

\indexspace

\item Shadowing lemma, 80
\item Shift automorphism, 75
\item Silnikov homoclinic orbit
\subitem for discrete NLS under perturbations, 65
\subitem for regularly perturbed NLS, 55
\subitem for singularly perturbed NLS, 57
\subitem for vector NLS under perturbations, 65
\item Spectral mapping theorem, 99

\indexspace

\item Transversal homoclinic orbit
\subitem for a derivative NLS, 69
\subitem for periodically perturbed SG, 69

\end{theindex}


\end{document}